\documentclass[11pt]{cernrep}

\usepackage{graphicx}

\usepackage{cite,./mcite}
\usepackage{psfrag}
\usepackage{epsfig}

\usepackage{amssymb,amscd}
\usepackage{amsmath}
\usepackage{amsfonts,amsbsy}
\usepackage[figuresright]{rotating}
\usepackage{array}
\usepackage{graphics}
\usepackage{longtable}
\usepackage{dcolumn}
\usepackage{xspace}
\usepackage{tabularx}
\usepackage{color}
\usepackage{calc}
\usepackage{floatflt}
\usepackage{wrapfig,rotating}
\usepackage{comment} 
\usepackage{vmargin,times,graphicx,amsmath,amssymb,color}

\usepackage{epsfig}
\usepackage{graphicx}
\usepackage{amssymb}
\usepackage{a4}
\usepackage{cite,./mcite}
\usepackage{graphicx}
\usepackage{epic}
\usepackage{floatflt}
\usepackage{amsmath}
\usepackage{a4wide}
\usepackage{xspace}

\usepackage{vmargin,times,graphicx,amsmath,amssymb,color}
\usepackage{latexsym} 

\setmarginsrb{22mm}{10mm}{20mm}{10mm}{12pt}{11mm}{0pt}{11mm}

\newcommand{\ggev}{\ensuremath{\mathrm{\,Ge\kern -0.1em V}}\xspace}
\newcommand{\mev}{\ensuremath{\mathrm{\,Me\kern -0.1em V}}\xspace}
\usepackage{url}  

\newcommand{\bs}{\begin{slide}}
\newcommand{\es}{\end{slide}}
\newcommand{\be}{\begin{equation}}
\newcommand{\ee}{\end{equation}}
\newcommand{\bea}{\begin{eqnarray}}
\newcommand{\eea}{\end{eqnarray}}

\newcommand{\bc}{\begin{center}}
\newcommand{\ec}{\end{center}}

\newcommand{\bi}{\begin{itemize}}
\newcommand{\ei}{\end{itemize}}
\def\epm#1#2{\hbox{${\lower1pt\hbox{$\scriptstyle +~#1$}}
\atop {\raise1pt\hbox{$\scriptstyle -~#2$}}$}}

\newcommand{\asmz}{\alpha_s(M_Z^2)}
\newcommand{\msbar}{\mbox{$\overline{\rm{MS}}$}\ }

\newcommand{\gsim}{\raisebox{-0.07cm}{$\:\:\stackrel{>}{{\scriptstyle
 \sim}}\:\: $} }
\newcommand{\lsim}{\raisebox{-0.07cm}{$\:\:\stackrel{<}{{\scriptstyle
 \sim}}\:\: $} }
\newcommand{\beq}{\begin{equation}}
\newcommand{\eeq}{\end{equation}}

\def\z#1{{\zeta_{#1}^{}}}

\def\pqq(#1){p_{\rm{qq}}(#1)}
\def\H(#1){{\rm{H}}_{#1}^{}}
\def\Hh(#1,#2){{\rm{H}}_{#1,#2}^{}}
\def\Hhh(#1,#2,#3){{\rm{H}}_{#1,#2,#3}^{}}
\def\Hhhh(#1,#2,#3,#4){{\rm{H}}_{#1,#2,#3,#4}^{}}

     \def\s{\sigma} 

\def\lsim{\mathrel{
   \rlap{\raise 0.511ex \hbox{$<$}}{\lower 0.511ex \hbox{$\sim$}}}}
\def\gsim{\mathrel{
   \rlap{\raise 0.511ex \hbox{$>$}}{\lower 0.511ex \hbox{$\sim$}}}}
\newcommand{\bq}{\begin{equation}}

\def\X{\boldsymbol{X}}
\def\x{\boldsymbol{x}}
\def\y{\boldsymbol{y}}
\def\z{\boldsymbol{z}}

\newcommand{\as}{\alpha_\mathrm{s}} 
\newcommand{\GeV}{\,\text{GeV}}
\def\lsim{\mathrel{\rlap{\lower4pt\hbox{\hskip1pt$\sim$}}
    \raise1pt\hbox{$<$}}}         

\newcommand{\MS}{$\overline{\rm MS}$}
\newcommand{\abar}{\ensuremath{\bar{\alpha}_S}}
\newcommand{\bb}{\ensuremath{\bar{\beta}_0}}

\begin{document}
\title{\sc 
Summary Report for the HERA - LHC Workshop proceedings\\
  WORKING GROUP I: Parton Distributions}

\author{
{\sc Conveners:}
\\
M.~Dittmar$~^{1}$,
S.~Forte$~^{2}$,
A.~Glazov$~^{3}$,
S.~Moch$~^{4}$
\\
{\sc Contributing authors:}
\\
G.~Altarelli$^{5,6}$, 
J.~Anderson$^{7}$, 
R.~D.~Ball$^{8}$, 
G.~Beuf~$^9$, 
M.~Boonekamp$^{10}$, 
H.~Burkhardt$^{11}$, 
F.~Caola$^2$, 
M.~Ciafaloni$^{12}$, 
D.~Colferai$^{12}$, 
A.~Cooper-Sarkar$^{13}$,
A.~de~Roeck$^{14}$,
L.~Del~Debbio$^8$, 
J.~Feltesse$^{15,16}$, 
F.~Gelis~$^5$, 
J.~Grebenyuk$^{3}$, 
A.~Guffanti$^{17}$, 
V. Halyo$^{l8}$, 
J.~I.~Latorre$^{19}$, 
V.~Lendermann$^{20}$,
G.~Li\,$^{21}$,
L.~Motyka~$^{22,23}$, 
T.~Petersen$^{14}$ 
A.~Piccione$^2$, 
V.~Radescu$^3$, 
M.~Rogal$^4$,
J.~Rojo$^{2,24}$, 
C.~Royon~$^{10}$, 
G.~P.~Salam~$^{24}$, 
D.~\v S\'alek~$^{25}$, 
A.~M.~Sta\'sto~$^{26,27,28}$,
R.~S.~Thorne$^{29}$, 
M.~Ubiali$^{8}$, 
J.~A.~M.~Vermaseren$^{30}$,
A.~Vogt$^{31}$,
G.~Watt$^{29}$,
C.~D.~White$^{30}$
}
\institute{
$^{1}$~Institute for Particle Physics, ETH-Z\"urich H\"onggerberg, CH 8093 Z\"urich, Switzerland
\\
$^{2}$~Dipartimento di Fisica, Universit\`a di Milano, INFN Sezione di
Milano, Via Celoria 16, \\ $^{\phantom{2}}$~I-20133 Milan, Italy 
\\
$^{3}$~DESY, Notkestrasse 85, D-22603 Hamburg, Germany 
\\
$^{4}$~DESY, Platanenallee 6, D-15738 Zeuthen, Germany 
\\$^5$~CERN, Department of Physics, Theory Division, CH-1211 Geneva 23, Switzerland 
\\$^{6}$~Dipartimento di Fisica ``E.Amaldi'', 
Universit\`a Roma Tre and INFN, Sezione di Roma Tre,
\\ $^{\phantom{6}}$~via della Vasca Navale 84, I-00146 Rome, Italy
\\$^7$~School of Physics, University College Dublin, Ireland
\\
$^{8}$~School of Physics, University of Edinburgh, Edinburgh EH9 3JZ, UK
\\$^{9}$~Institut de Physique Th\'eorique, CEA--Saclay F-91191
Gif-sur-Yvette, France 
\\$^{10}$~IRFU, Service de Physique des Particules, CEA--Saclay F-91191
Gif-sur-Yvette, France 
\\$^{11}$~CERN-AB, CH-1211 Geneva 23, Switzerland 
\\$^{12}$~Dipartimento di Fisica, Universit\`a di Firenze and INFN,
Sezione di Firenze, I-50019 \\ $^{\phantom{11}}$~Sesto Fiorentino, Italy
\\$^{13}$~Department of Physics, Nuclear and Astrophysics Lab., Keble Road, Oxford, OX1 3RH, UK
\\$^{14}$~CERN-PH, CH-1211 Geneva 23, Switzerland 
\\$^{15}$~IRFU, CEA--Saclay F-91191
Gif-sur-Yvette, France 
\\$^{16}$~University of
Hamburg, Luruper Chaussee 149, Hamburg, D-22761 Germany
\\$^{17}$~Physikalisches Institut, Albert-Ludwigs-Universit\"at
Freiburg, Hermann-Herder-Stra\ss e 3, \\ $^{\phantom{14}}$~D-79104 Freiburg i. B., Germany  
\\$^{18}$~Department of Physics, Princeton University, Princeton, NJ08544, USA
\\$^{19}$~Departament d'Estructura i Constituents de la Mat\`eria, Universitat de Barcelona, Diagonal 647, 
\\ $^{\phantom{17}}$~E~08028 Barcelona, Spain
\\$^{20}$~Kirchhoff-Institut f\"ur Physik, Universit\"at Heidelberg,
 Im Neuenheimer Feld 227,\\$^{\phantom{17}}$~D-69120 Heidelberg, Germany
\\$^{21}$~Laboratoire de l'Acc\'{e}l\'{e}rateur Lin\'{e}aire,
Universit\'{e} Paris-Sud, Orsay, France
\\$^{22}$~II Institute for  Theoretical Physics, University of
Hamburg, Luruper Chaussee 149, \\$^{\phantom{17}}$~Hamburg, D-22761 Germany
\\$^{23}$~Institute of Physics, Jagellonian University 
Reymonta 4, 30-059 Cracow, Poland
\\$^{24}$~LPTHE,  UPMC -- Paris 6,
   Paris-Diderot -- Paris 7,
   CNRS UMR 7589, F-75005 Paris, France 
\\$^{25}$~Institute of Particle and Nuclear Physics, Charles University, Prague, Czech Republic 
\\$^{26}$~Physics Department, Penn State University, 104 Davey
Laboratory, University Park, PA 16802, USA
\\$^{27}$~H. Niewodnicza\'nski Institute of Nuclear Physics, Polish
Academy of Science, 
\\ $^{\phantom{11}}$~ul.Radzikowskiego 152, 31-342 Cracow, Poland 
\\$^{28}$~Brookhaven National Laboratory, Upton, NY-11073, USA
\\$^{29}$~Department of Physics and Astronomy, University College,
London, WC1E~6BT, UK 
\\$^{30}$~NIKHEF Theory Group, Kruislaan 409, NL 1098 SJ Amsterdam, The Netherlands 
\\$^{31}$~Department of Mathematical Sciences, University of
Liverpool, Liverpool, L69 3BX, UK}
\maketitle

\begin{abstract}
We provide an assessment of the state of the art in various issues
related to experimental measurements, phenomenological methods and
theoretical results relevant for the determination of parton
distribution functions (PDFs) and their uncertainties, with the specific aim of
providing benchmarks of different existing approaches and results in
view of their application to physics at the LHC.\\
We discuss higher
order corrections, we review and compare different approaches to small
$x$ resummation, and we assess the possible relevance of parton saturation in
the determination of PDFS at HERA and its possible study in LHC
processes. We provide various benchmarks of PDF fits, with the
specific aim of studying issues of  error propagation, non-gaussian
uncertainties, choice of functional forms of PDFs, and combination of
data from different experiments and different processes. We study the
impact of combined HERA (ZEUS-H1) structure function data, their
impact on PDF uncertainties, and their implications for the
computation of standard candle processes, and we review the recent
$F_L$ determination at HERA. Finally, we
 compare and assess methods for  luminosity
measurements at the LHC and the impact of PDFs on them.

\end{abstract}
\pagebreak
\tableofcontents
\pagebreak

\section{INTRODUCTION}
With the start of data--taking at the LHC getting closer, the
importance of a
detailed understanding of the physics of parton distributions (PDFs) has
increased considerably, along with the awareness of the LHC community
for the importance of the issues related to it.
Clearly, the main reason why PDFs are important at the LHC is that 
at a hadron collider a
detailed understanding of PDFs is needed in order to obtain accurate
predictions for both signal and background processes. Indeed,  for many
physical processes at the LHC,   PDFs are 
the dominant source of uncertainty. On the other hand, an accurate
control of PDF uncertainties allows one to use selected processes as
``standard candles'', for instance in the determination of
luminosities. However, this also means that experimentation at the LHC
will provide a considerable amount of new experimental
information on PDFs, and it
will enable us to test the adequacy of their current theoretical
understanding.

The main aim of this document is to provide a state of the art
assessment of our understanding of PDFs at the dawn of the
LHC.
Since the  previous HERA-LHC
workshop~\cite{Dittmar:2005ed}, we have witnessed  several important
directions of progress in the physics of PDFs.
On the theoretical side there has been conclusive progress in
extending the treatment of perturbative QCD beyond the current
default, namely, the next--to--leading perturbative order. 
On the phenomenological side there has been a joint effort between
experimental and theoretical groups involved in the extraction of
PDFs, specifically from global fits, in agreeing on common procedures,
benchmarks and standards. On the experimental side, new
improved results from the HERA runs are being  finalized: these
include both the construction of a joint determination of structure
function which combines the result of the ZEUS and H1 experiments, and
the first direct measurements of the structure function $F_L$ which
have been made possible by running HERA at a reduced proton beam energy in
2007. Also, the LHC experiments (ATLAS, CMS and LHCb) are now
assessing the use of standard candle processes for luminosity
measurements.

All these issues are discussed in this document. In each case, our
main goal has been to provide as much as possible a joint treatment
by the various groups involved, as well as a comparison of different
approaches and benchmarking of results.
In particular, in Sect.~\ref{sec:theory}, after briefly  reviewing
(Sect.~\ref{sec:nnloprecision}) the
current status of higher--order calculations for DIS, we 
provide (Sect.~\ref{smallx-intro}) detailed comparisons of techniques and results of different
existing approaches to small $x$ resummation, and then we summarize
(Sect.~\ref{sec:satgs}) the current status of studies of parton
saturation at HERA, their possible impact on current PDF extraction and the
prospects of future studies at the LHC.
In Sect.~\ref{sec:pdfunc} we  discuss methods and results for the
benchmarking of PDF fits: with specific reference to two benchmark
fits based on a common agreed set of data, we  discuss issues
related to error propagation and non-gaussian errors, to the choice of
functional form and corresponding bias, to possible incompatibilities
between different data sets.
In Sect.~\ref{sec:pdfdet} we turn to recent progress in the
extraction of PDFs from HERA data, specifically
the
impact of 
combined ZEUS-H1 structure function data on PDF determination and the
ensuing calculation of $W$ and $Z$ cross-sections (Sect.~\ref{sec:combfit}) 
and the recent first determination of the structure function $F_L$ 
(Sect.~\ref{sec:fldet}). In Sect.~\ref{sec:lumi} we 
discuss and compare  luminosity measurements based on absolute
proton--proton luminosity measurements to those based on
the use of standard candle
processes, and the impact on all of them of PDF uncertainties.
Finally, in Sect.~\ref{sec:pdflhc} we  present the PDF4LHC
initiative, which will provide a framework for the continuation of PDF
studies for the LHC.

{\it Note:} Most of the contributions to this workshop are the result
of collaboration between various groups. The common set of authors
given for each  section or subsection has read and approved the entire
content of that section or subsection; however, when a subset of these 
authors is given for a specific part of the section or subsection, they
are  responsible for it.

\pagebreak
\section{THEORETICAL ISSUES}
\label{sec:theory}
\subsection{Precision calculations for inclusive DIS: an update\protect\footnote{Contributing authors: S.~Moch, M.~Rogal, J.~A.~M.~Vermaseren, A.~Vogt}}
\label{sec:nnloprecision}

\noindent
With high-precision data from HERA and in view of the outstanding importance of
hard scattering cross sections at the LHC, a quantitative understanding of 
deep-inelastic processes is indispensable, necessitating calculations beyond 
the standard next-to-leading order of perturbative QCD. 

In this contribution we briefly discuss the recent extension of the three-loop
calculations for inclusive deep-inelastic scattering (DIS) \cite{Larin:1993vu,Larin:1996wd,Retey:2000nq,Moch:2004pa,Vogt:2004mw,Moch:2004xu,Vermaseren:2005qc,Vogt:2006bt} to the complete set of coefficient functions for
the charged-current (CC) case.
The new third-order expressions are too lengthy for this short overview. 
They can be found in Refs.~\cite{Moch:2007gx,Moch:2007rq} together with the 
calculational methods and a more detailed discussion. 
Furthermore the reader is referred to Refs.~\cite{Vogt:2008yw,MRVVprep} for our
first results on the three-loop splitting functions for the evolution of 
helicity-dependent parton distributions.

Structure functions in inclusive deep-inelastic scattering  are among the most 
extensively measured observables. The combined data from fixed-target 
experiments and the HERA collider spans about four orders of magnitude in both
Bjorken-$x$ variable and the scale $Q^2 = -q^2$ given by the momentum $q$ of 
the exchanged electroweak gauge boson \cite{Yao:2006px}. 
Here we consider the $W\!$-exchange charged-current case, see Refs.~\cite{Yang:2000ju,Tzanov:2005kr,Onengut:2005kv,Adloff:2003uh,Chekanov:2003vw,%
Aktas:2005ju,Chekanov:2006da} for recent data from neutrino DIS and HERA. 
With six structure functions, $F_2^{\,W^\pm}\!$, $F_3^{\, W^\pm}$ and 
$F_L^{\,W^\pm}\!$, this case has a far richer structure than, for example, 
electromagnetic DIS with only two independent observables, $F_2$ and~$F_L$.
 
Even taking into account a forthcoming combined H1/ZEUS final high-$Q^2$ data
set from HERA,  
more detailed measurements are required to fully exploit the resulting 
potential, for instance at a future neutrino factory, see Ref.~\cite{Mangano:2001mj}, and the LHeC, the proposed high-luminosity electron-proton
collider at the LHC~\cite{Dainton:2006wd}.
Already now, however, CC DIS provides important information on the
parton structure of the proton, e.g., its flavour decomposition and the 
valence-quark distributions. Moreover, present results are also sensitive to 
electroweak parameters of the Standard Model such as $\sin^2 \theta_W$, see 
Ref.~\cite{Zeller:2001hh}, and the space-like $W\!$-boson propagator
\cite{Aktas:2005iv}. As discussed, for example, in Refs.~\cite{Davidson:2001ji,McFarland:2003jw,Dobrescu:2003ta,Kretzer:2003wy}, a reliable 
determination of $\sin^2 \theta_W$ from neutrino DIS requires a detailed 
understanding of non-perturbative and perturbative QCD effects.    

Previous complete results on unpolarized DIS include the three-loop splitting 
functions \cite{Moch:2004pa,Vogt:2004mw} as well as the 3-loop coefficient 
functions for the photon-exchange structure functions $F_{\,2,L}$
\cite{Moch:2004xu,Vermaseren:2005qc}. However, most coefficient functions for 
CC DIS were not fully computed to three loops so far.

For this case it is convenient to consider linear combinations of the 
structure functions $F_a^{\,W^\pm}$ with simple properties under crossing, such 
as $F_a^{\,\nu p \pm \bar \nu p}$ ($a = 2,\: 3,\: L$) for neutrino DIS. For all
these combinations either the even or odd moments can be calculated in 
Mellin-$N$ space in the framework of the operator product expansion (OPE), 
see Ref.~\cite{Buras:1979yt}.
The results for the third-order coefficient functions for the even-$N$ 
combinations $\, F_{2,L}^{\,\nu p + \bar\nu p}$ can be taken over from 
electromagnetic DIS \cite{Moch:2004xu,Vermaseren:2005qc}. Also the coefficient 
function for the odd-$N$ based charged-current structure function 
$F_3^{\,\nu p +\bar\nu p}$ is completely known at three-loop accuracy, with the
results only published via compact parameterizations so far \cite{Vogt:2006bt}.
For the remaining combinations $\,F_{2,L}^{\,\nu p - \bar\nu p}$ and
$\,F_3^{\,\nu p - \bar\nu p\!}$, on the other hand, only recently the first six odd or even integer moments of the respective coefficient functions have 
been calculated to third order in Ref.~\cite{Moch:2007gx} following the 
approach of Refs.~\cite{Larin:1993vu,Larin:1996wd,Retey:2000nq} based on the
{\sc Mincer} program \cite{Gorishnii:1989gt,Larin:1991fz}.

The complete results of Refs.~\cite{Moch:2004xu,Vermaseren:2005qc,Vogt:2006bt} 
fix all even and odd moments $N$. Hence already the present knowledge of fixed 
Mellin moments for $\,F_{2,L}^{\,\nu p - \bar\nu p}$ and 
$\,F_3^{\,\nu p - \bar\nu p\!}$ is 
sufficient to determine also the lowest six moments of the differences of 
corresponding even-$N$ and odd-$N$ coefficient functions and to address a 
theoretical conjecture \cite{Broadhurst:2004jx} for these quantities, 
see Ref.~\cite{Moch:2007rq}.
Furthermore these moments facilitate $x$-space approximations in the style of, 
e.g, Ref.~\cite{vanNeerven:2001pe} which are sufficient for most 
phenomenological purposes, including the determination of the third-order QCD 
corrections to the Paschos-Wolfenstein relation~\cite{Paschos:1972kj} used for 
the extraction of $\sin^2 \theta_W$ from neutrino DIS.

\vspace{1mm}
The even-odd differences of the CC coefficient functions $\,C_a\,$ for 
$\,a = 2,\: 3,\: L\,$ can be defined by
\begin{eqnarray}
  \label{eq:cdiff}
  \delta\, C_{2,L} \; =\; C_{2,L}^{\,\nu p + {\bar \nu} p} 
    - C_{2,L}^{\,\nu p - {\bar \nu} p} \:\: , \qquad
  \delta\, C_3 \; =\;  C_3^{\,\nu p - {\bar \nu} p}
    - C_3^{\,\nu p + {\bar \nu} p} \:\: .
\end{eqnarray}
The signs are chosen such that the differences are always `even -- odd' in the 
moments $\, N$ accessible by the OPE \cite{Buras:1979yt}, and it is understood 
that the $d^{\:\!abc}d_{abc}$ part of $\,C_3^{\,\nu p + \bar\nu p}$ 
\cite{Retey:2000nq,Vogt:2006bt} is removed before the difference is formed. 
With $a_{\rm s} = \as /(4 \pi)$ these non-singlet quantities can be expanded 
as  
\begin{eqnarray}
\label{eq:cf-exp}
  \delta\, C_a \; = \; 
  \sum_{l=2} \: a_{\rm s}^{\, l}\: \delta\:\! c_{a}^{(l)}. 
\end{eqnarray}
There are no first-order contributions to these differences, hence the above 
sums start at $l = 2\,$.

\begin{figure}[ht]
\begin{center}
\includegraphics[width=12cm,angle=0]{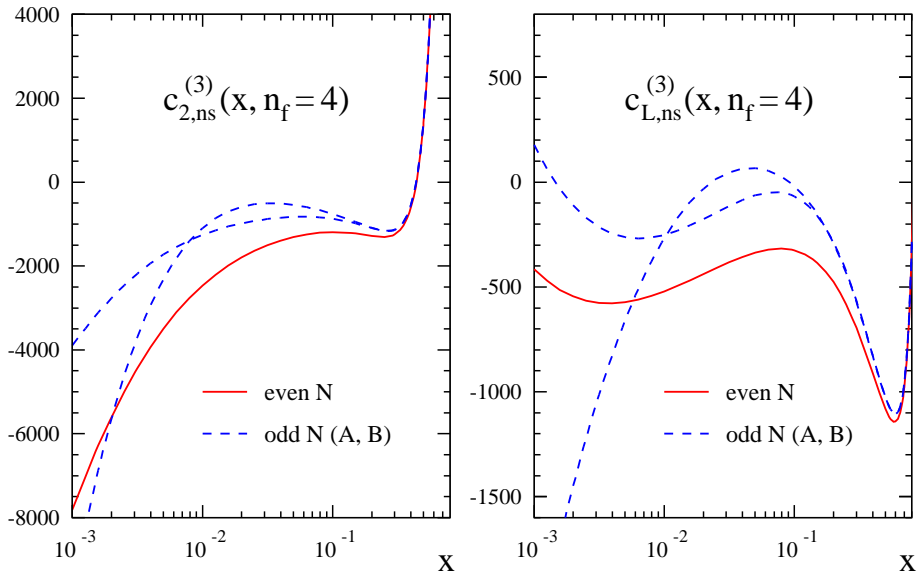}
\end{center}
\vspace*{-5mm}
\caption{The exact third-order coefficient functions of the even-$N$ structure 
   functions $\, F_{2,L}^{\,\nu p + \bar\nu p}$ for four massless flavours, and
   the approximate odd-moment quantities for ${\nu p - \bar\nu p}$ combination. 
\label{fig:c3diff}
}
\end{figure}

We start the illustration of these recent results by looking at the 
approximations for the $\,\nu p -\bar\nu p\,$ odd-$N$ coefficient functions 
$c_{2,L}^{(3)}(x)$ (see Ref.~\cite{Moch:2007rq} for a detailed discussion). 
These are compared in Fig.~\ref{fig:c3diff} to their exact counterparts 
\cite{Moch:2004xu,Vermaseren:2005qc} for the even-$N$ non-singlet structure 
functions. The dashed lines represent the uncertainty band 
due to the limited number of known moments.
The third-order even-odd differences remain
noticeable to larger values of $x$ than at two loops, e.g., up to $x \simeq
0.3$ for $F_2$ and $x \simeq 0.6$ for $F_L$ for the four-flavour
case shown in the figure. The moments $N = 1,\:3,\:\ldots,\: 9\,$ constrain
$\,\delta\:\! c_{2,L}^{(3)}(x)\,$ very well at $\,x \gsim 0.1$, and
approximately down to $\,x \approx 10^{-2}$.

\begin{figure}[th]
\centerline{\epsfig{file=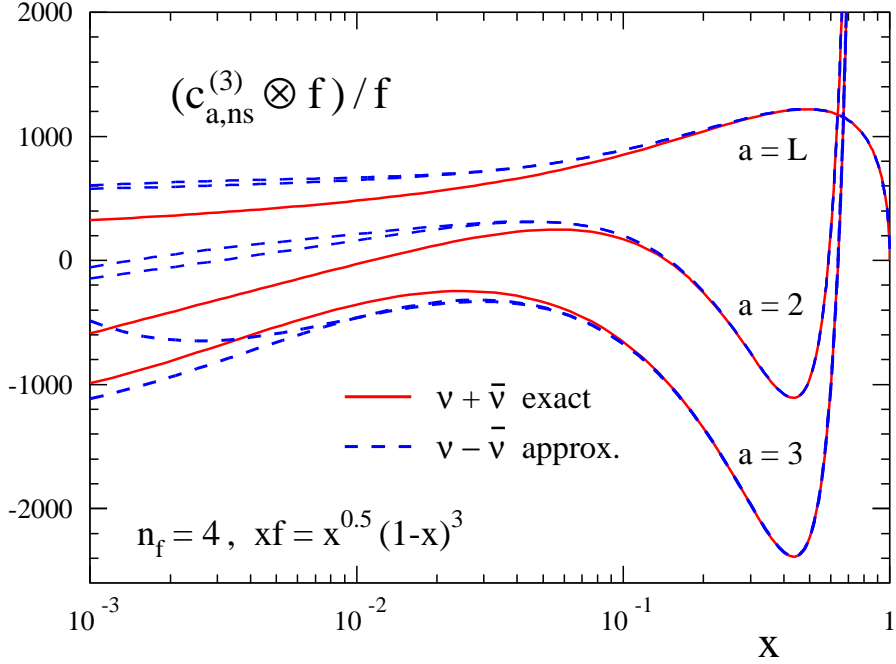,width=12cm,angle=0}\qquad}
\vspace*{-1mm}
\caption{Convolution of the six third-order CC coefficient functions for 
 $F_{\:\!2,\,3,\,L}$ in $\,\nu p + \bar\nu p\,$ 
and $\,\nu p - \bar\nu p\,$ DIS with a schematic but 
 typical non-singlet distribution $\!f$. All results have been normalized to 
 $\!f(x)$, suppressing the large but trivial variation of the absolute 
 convolutions.  \label{fig:c3dcnv} }
\vspace*{-2mm}
\end{figure}

\vspace{1mm}
Concerning low values of Bjorken-$x$ one should recall that the uncertainty 
bands shown by the dashed lines in Fig.~\ref{fig:c3diff} do not directly 
indicate the
range of applicability of these approximations, since the coefficient functions
enter observables only via smoothening Mellin convolutions with 
non-perturbative initial distributions. In Fig.~\ref{fig:c3dcnv} we therefore
present the convolutions of all six third-order CC coefficient functions with
a characteristic reference distribution. It turns out that the approximations
of the previous figure can be sufficient down to values even below $x=10^{-3}$,
which is amply sufficient for foreseeable applications to data. 
The uncertainty 
of $\,\delta\:\! c_{3}^{(3)}(x)$, on the other hand, becomes relevant already
at larger values, $\,x \lsim 10^{-2}$, as the lowest calculated moment of this
quantity, $\,N=2$, has far less sensitivity to the behaviour at low $x$.

\begin{figure}[hbt]
\centerline{\epsfig{file=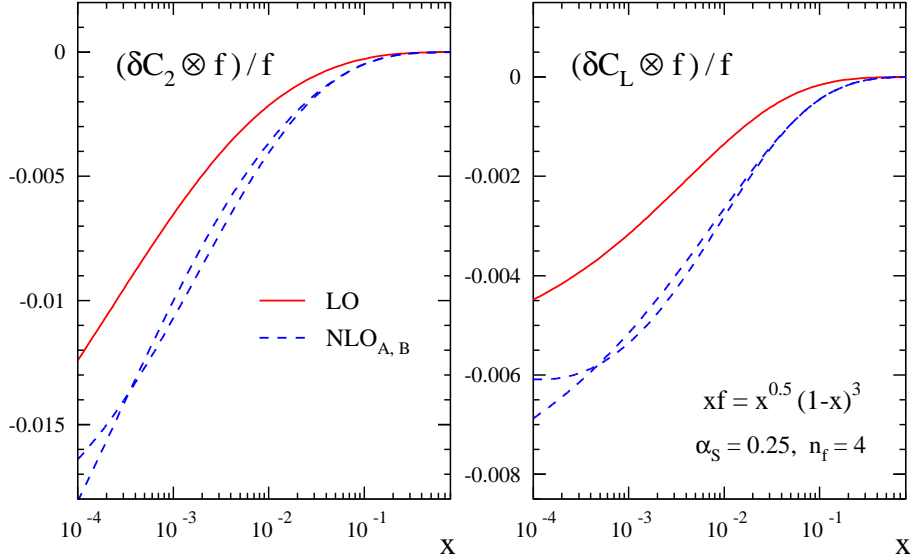,width=12.0cm,angle=0}\hspace{3mm}}
\vspace{-1mm}
\caption{The first two approximations, denoted by LO and NLO, of the 
 differences (\ref{eq:cdiff}) for $F_{\:\!2}$ and $F_{\:\!L}$ in 
 charged-current DIS. The results are shown for representative values of 
 $\alpha_{\rm s}$ and $n_f$ after convolution with the reference distribution 
 $\! f(x)$ also employed in Fig.~\ref{fig:c3dcnv}. The dashed curves correspond
 to the two approximation uncertainties for the new 
 $\alpha_{\rm s}^{\,3}$ contributions. \label{fig:c2lexp} }
\vspace*{-2mm}
\end{figure}

The three-loop corrections to the non-singlet structure functions are rather
small even well below the $x$-values shown in the figure~~--~~recall our small 
expansion parameter $a_{\rm s}\,$: the third-order coefficient are smaller by a 
factor $2.0\cdot 10^{-3}$ if the expansion is written in powers of 
$\alpha_{\rm s}$.  Their sharp rise for $\,x \rightarrow 1\,$ is understood in 
terms of soft-gluon effects
which can be effectively resummed, if required, to next-to-next-to-next-to-%
leading logarithmic accuracy \cite{Moch:2005ba}. Our even-odd differences 
$\,\delta\:\! c_{a}^{(3)}(x)$, on the other hand, are irrelevant at $x > 0.1$ 
but have a sizeable impact at smaller $x$ in particular on the corrections for 
$F_{\,2}$ and $F_{L}$.
The approximate results for $\,\delta\:\! c_{a}^{(3)}(x)$ facilitate a first 
assessment of the perturbative stability of the even-odd differences 
(\ref{eq:cdiff}). In Fig.~\ref{fig:c2lexp} we illustrate the known two orders 
for $F_{\:\!2}$ and $F_{\:\!L}$ for \mbox{$\alpha_{\rm s} = 0.25$} and 
$n_{\rm f}^{} = 4$ massless quark flavours, employing the same reference quark 
distribution as in Fig.~\ref{fig:c3dcnv}. 

Obviously our new $\alpha_{\rm s}^{\,3}$ corrections are important wherever 
these coefficient-function differences are non-negligible. On the other hand, 
our results confirm that these quantities are very small, and thus relevant 
only when a high accuracy is required. 
%
%
These conditions are fulfilled for the calculation of QCD corrections for 
the so-called Paschos-Wolfenstein relation. This relation is  
defined in terms of a ratio of neutral-current and charged-current cross 
sections for neutrino-nucleon DIS~\cite{Paschos:1972kj},
\beq
\label{eq:rminus}
R^{-} \; = \:\:  
\frac{\sigma(\nu_{\mu\,}N\rightarrow\nu_{\mu\,}X) \: - \: 
  \sigma(\bar \nu_{\mu\,}N\rightarrow\bar \nu_{\mu\,}X)}
{\sigma(\nu_{\mu\,}N\rightarrow\mu^-X) \: - \: 
  \sigma(\bar \nu_{\mu\,}N\rightarrow\mu^+X)} 
\:\: .
\eeq
The asymmetry $R^{-}$ directly measures $\,\sin^2 \theta_W$ if the up and down 
valence quarks in the target carry equal momenta, and if the strange and 
heavy-quark sea distributions are charge symmetric. Beyond the leading order 
this asymmetry can be presented as an expansion in $\alpha_s$ and inverse 
powers of the dominant isoscalar combination $u^- + d^-$, where $ q^- =\int_0^1
dx\; x \left( q(x) - \bar{q}(x) \right)$ is the second Mellin moment of the 
valence quark distributions.
Using the results for differences $\,\delta\:\! c_{a}^{(3)}(x), a=2,L,3$ one 
can present it in a numeric form,
\begin{eqnarray}
\label{eq:rminus-numbers}
R^{-} &\! = &
 \frac{1}{2} - \sin^2\theta_W 
  \:\: + \:\: \frac{u^- - d^- + c^- - s^-}{u^- + d^-} \: \Bigg\{
  1 - \frac{7}{3}\:\sin^2\theta_W  
  \; + \; \left( \frac{1}{2} - \sin^2\theta_W \right) \cdot
\nonumber \\ & & \mbox{} 
  \frac{8}{9} \frac{\alpha_s}{\pi} \left[ \,
    1
  + 1.689\,\as
  + (3.661 \pm 0.002)\,\as^2 \,
  \right]
  \Biggr\}
  \; + \; {\cal{O}} \left( (u^- + d^-)^{-2\,} \right)
  \; + \; {\cal{O}}\bigl(\alpha_s^4\bigr) 
\:\: , \quad
\end{eqnarray}
where the third term in the square brackets is determined by the $\alpha_s^3$ 
corrections $\,\delta\:\! c_{a}^{(3)}(x), a=2,L,3$.
The perturbation series in the square brackets appears reasonably well 
convergent for relevant values of the strong coupling constant, with the known
terms reading, e.g., 1 + 0.42 + 0.23 for $\as = 0.25$. Thus the $\as^2$ and 
$\as^3$ contributions correct the NLO estimate by 65\% in this case. On the 
other hand, due to the small prefactor of this expansion, the new third-order 
term increases the complete curly bracket in Eq.~(\ref{eq:rminus-numbers}) by 
only about 1\%, which can therefore by considered as the new uncertainty of 
this quantity due to the truncation of the perturbative expansion. Consequently
previous NLO estimates of the effect of, for instance, the (presumably mainly 
non-perturbative, see Refs.~\cite{Catani:2004nc,Lai:2007dq,Thorne:2007bt}) 
charge asymmetry of the strange sea remain practically unaffected by 
higher-order corrections to the coefficient functions.

To summarize, we have extended the fixed-$N$ three-loop calculations of 
inclusive DIS \cite{Larin:1993vu,Larin:1996wd,Retey:2000nq} to all 
charged-current cases not covered by the full (all-$N$) computations of 
Refs.~\cite{Moch:2004xu,Vermaseren:2005qc,Vogt:2006bt}. 
The region of applicability of these new results is restricted to Bjorken-$x$ 
values above about $10^{-3}$, a range amply sufficiently for any fixed-target
or collider measurements 
of those charged-current structure functions in the foreseeable future. 
Except for the longitudinal structure function $F_L$, the present coefficient 
functions are part of the next-to-next-to-next-to-leading order (N$^3$LO) 
approximation of massless perturbative QCD. Analyses at this order are possible
outside the small-$x$ region since the corresponding four-loop splitting 
functions will have a very small impact here, cf.~Ref.~\cite{Vogt:2007vv}.

\subsection{Small $x$ resummation
\protect\footnote{Contributing authors: G.~Altarelli, R.~D.~Ball,
  M.~Ciafaloni, D.~Colferai, G.~P.~Salam, 
A.~Sta\'sto, R.~S.~Thorne, C.~D.~White}}
\label{smallx-intro}
The splitting functions which govern the evolution of the parton
distributions (PDFs), together with the hard cross sections 
which relate those partons to hadronic physical observables, are
potentially unstable at high energy due to logarithmically enhanced
contributions. In particular, parametrizing 
observables such as deep-inelastic
structure (DIS) functions or  Drell-Yan (DY) or Higgs production cross section
in hadronic collisions in terms of a dimensionful scale $Q^2$ (photon
virtuality or invariant mass of the final state in DIS and DY
respectively) and
a dimensionless ratio $x$ (the Bjorken variable or $\frac{Q^2}{s}$ in
DIS and DY 
respectively), when $x\to0$ there are logarithmically enhanced
contributions to the  perturbation expansion of the form
$x^{-1}\alpha_S^n(Q^2)\log^m(1/x)$ ($n\geq m-1$). When $x$ is sufficiently
small, one must  resum such terms, reordering the
perturbation expansion in terms of leading logarithmic (LL) terms
followed by next-to-leading logarithmic (NLL) terms and so on. 

The problem can be traced to ladders of $t$-channel gluon exchanges at
LL order, with some quark mixing at NLL order and beyond. The
underlying framework for the resummation procedure is the BFKL
equation~\cite{Fadin:1975cb,Balitsky:1978ic}, an integral equation for
the unintegrated gluon $f(k^2,Q_0^2)$ that is currently known up to
full NLL order~\cite{Fadin:1998py,Camici:1997ij,Ciafaloni:1998gs},  
and approximate NNLL
order~\cite{Marzani:2007gk}.   This has the schematic form (up to NLL):
\begin{equation}
Nf(k^2,Q_0^2)=Nf_I(Q_0^2)+\abar(k^2)\int dk'^2\left[{\cal K}_0(k^2,{k'}^2,Q_0^2)+\abar(k^2){\cal K}_1(k^2,{k'}^2,Q_0^2)\right]f(k'^2),
\label{NLLBFKL} 
\end{equation}
where $f_I(Q_0^2)$ is a non-perturbative initial condition at some
initial scale $Q_0$, $\bar{\alpha}_S=3\alpha_S/\pi$ and ${\cal
  K}_{0,1}$ are the LL and NLL BFKL kernels. Different choices for the
argument of the running coupling are possible, leading to accordingly
modified  ${\cal K}_1$~\cite{Ciafaloni:2003ek,Ciafaloni:2003rd}.  

The solution of the BFKL equation can be used to extract leading and
subleading singular contributions to singlet DGLAP  splitting functions.
 The BFKL equation can either be solved numerically in its form
 given by Eq.~(\ref{NLLBFKL}), or else analytically by performing a 
double Mellin transform with respect to $x$ and $k^2$:
\begin{equation}
f(\gamma,N)=\int_0^\infty (k^2)^{-\gamma-1}\int_0^1 dx x^N f(x,k^2),
\label{mel}
\end{equation}
whereby the BFKL equation becomes a differential equation, with
kernels $\chi_{0,1}(\gamma)$ defined respectively as 
the Mellin transforms of ${\cal
  K}_{0,1}$.
 Furthermore, 
 by using the $k_t$-factorisation
theorem~\cite{Catani:1990eg}, one may determine leading small $x$
contributions to all orders to 
hard partonic cross sections for physical processes such as heavy
quark electroproduction~\cite{Catani:1990eg} and 
deep-inelastic scattering~\cite{Catani:1994sq}. Approximate subleading
results are also available~\cite{Bialas:2001ks,White:2006aj}. 

These results for splitting functions and hard partonic cross sections can then
be combined with fixed-order results to obtain resummed predictions
for physical observables. However,
it has now been known for some time that the LL BFKL equation is unable to
describe scattering data well, even when matched to a fixed order
expansion. Any viable resummation procedure must then, at the very least,
satisfy the following requirements: 
\begin{enumerate}
\item Include a stable solution to 
the BFKL equation with running coupling up to NLL order. 
\item Match to the standard DGLAP description at moderate and high $x$
  values (where this is known to describe data well).
\item Provide the complete set of splitting and coefficient functions for $F_2$ and $F_L$ in a well defined factorisation scheme.
\end{enumerate} 

Over the past few years, three approaches have emerged which, to some
extent, aim at fulfilling these conditions. Here we call these the ABF
\cite{Altarelli:1999vw,Altarelli:2001ji,Altarelli:2003hk,Altarelli:2005ni,Ball:2005mj,Ball:2007ra,Altarelli:2008aj,Altarelli:2008xp},
CCSS
\cite{Salam:1998tj,Ciafaloni:1998iv,Ciafaloni:1999yw,Ciafaloni:2000cb,Ciafaloni:2003rd,Ciafaloni:2005cg,Ciafaloni:2006yk,Ciafaloni:2007gf}
and TW
\cite{Thorne:1999rb,Thorne:1999fq,Thorne:1999sg,Thorne:2001nr,White:2006xv,White:2006yh}
approaches. 
In the ABF scheme all three requirements are met, and resummed
splitting functions  in the singlet sector have been
determined. Furthermore,
a complete control of the scheme dependence  at the
resummed level
has been
achieved, thereby allowing for a consistent determination of resummed
deep-inelastic coefficient functions, and thus of resummed structure functions.
However, the results obtained thus have not been fit  to the
data yet. In the CCSS formalism, resummed  splitting functions have
also been determined. However, results are given in a scheme which
differs from the \MS\ scheme at the resummed level; furthermore, resummed
coefficient functions and physical observables haven't been
constructed yet. The TW approach, instead, has already been compared to the
data in a global fit. However, this approach makes a number of
simplifying assumptions and the ensuing resummation is thus not as
complete as that which obtains in other approaches:
for example, this approach does not  
include the full collinear resummation of the BFKL kernel.

A comparison of resummed splitting functions and solution of evolution
equations determined in 
the ABF and CCSS approaches with $n_f=0$ was presented in
Ref.~\cite{Dittmar:2005ed}; the main features and differences of these
approaches were also discussed. Here, we extend this comparison to
the case of $n_f\not=0$ resummation, and also to the TW
approach. First, we will briefly summarize the main features of each
approach, and in particular we display the matrix of splitting functions
determined in the ABF and CCSS approaches. Then, we will compare
$K$-factors for physical observables determined using the ABF and TW
approach.

Note that there are some difference in notations between various
groups, which are retained here in order to simplify comparison to
the original literature. 
 In particular, the variable $N$ in Eq.~(\ref{mel}) will be referred to
 as $\omega$ in the CCS approach of Section~\ref{CCSS}, and the
 variable $\gamma$ in the same equation will be referred to as $M$ in
 the ABF approach of   Section~\ref{ABF}.
 
\subsubsection{The Altarelli-Ball-Forte (ABF) Approach}
\label{ABF}

In the ABF
approach~\cite{Ball:1999sh,Altarelli:1999vw,Altarelli:2000mh,Altarelli:2001ji,Altarelli:2003hk,Altarelli:2003kt,
  Altarelli:2004dq,Altarelli:2005ni,Ball:2005mj,Ball:2007ra,Altarelli:2008aj,
Altarelli:2008xp} 
one concentrates on the problem of obtaining an improved anomalous
dimension (splitting function) 
for DIS which reduces to the ordinary perturbative result at large 
$N$ (large $x$), thereby automatically satisfying renormalization
group constraints, 
while including resummed BFKL corrections at small $N$ (small $x$),
determined through the renormalization-group improved (i.e. running
coupling) version of the BFKL kernel. The ordinary
perturbative result for the singlet anomalous dimension is given by: 
\begin{equation}
\gamma(N,\alpha_s)=\alpha_s \gamma_0(N)~+~\alpha_s^2
\gamma_1(N)~+~\alpha_s^3
\gamma_2(N)~~\dots .
\label{gammadef}
\end{equation}
The BFKL corrections at small $N$ (small $x$)
are determined  by the BFKL
kernel $\chi(M,\alpha_s)$:
\begin{equation}
\chi(M,\alpha_s)=\alpha_s \chi_0(M)~+~\alpha_s^2 \chi_1(M)~+~\dots ,
\label{chidef}
\end{equation} 
which is the Mellin transform, with respect to $t=\ln \frac{k^2}{k_0^2}$, of the $N\to0$ angular averaged 
BFKL kernel.

The ABF construction is based on three ingredients.
\begin{enumerate}
\item {\it The duality relation} between the kernels $\chi$ and $\gamma$
\begin{equation}
\chi(\gamma(N,\alpha_s),\alpha_s)=N,
\label{dualdef}
\end{equation}
which is a consequence of the fact that at fixed coupling the
solutions of the BFKL and DGLAP 
equations should coincide at leading 
twist~\cite{Ball:1997vf,Ball:1999sh,Altarelli:1999vw}.
By using duality, one can use the perturbative expansions of
$\gamma$ and $\chi$ in powers of $\alpha_s$ to improve (resum) each
other: by combining them, one obtains a 
"double leading" (DL) expansion which includes all leading (and
subleading, at NLO) logs of $x$ and $Q^2$.
In particular, the DL expansion automatically
resums the collinear poles of $\chi$ at $M=0$. This eliminates the alternating
sign poles $+1/M, -1/M^2,.....$ that appear in $\chi_0$,
$\chi_1$,\dots, and  make the perturbative expansion of $\chi$
unreliable. This result is a model independent consequence
of momentum conservation
$\gamma(1,\alpha_s)=0$, whence, by duality: 
\begin{equation}
\chi(0,\alpha_s)=1.
\label{mom}
\end{equation}

\item {\it The symmetry of the BFKL kernel} upon gluon interchange.
In Mellin space, this symmetry
implies that at the fixed-coupling level the kernel $\chi$
for evolution in  $\ln \frac{s}{k k_0}$
 must satisfy
$\chi(M)=\chi(1-M)$. 
By exploiting this symmetry, one can use the collinear resummation of
the region $M\sim0$ which was obtained using the double-leading expansion
to also improve the BFKL kernel in the anti--collinear $M\simeq1$
region. This leads to a symmetric kernel which is an entire function for 
all  $M$, and has a minimum at $M=\frac{1}{2}$. The 
symmetry is broken by the DIS choice
 of variables $\ln\frac{1}{ x}=\ln\frac {s}{Q^2}$ and by the running of the
 coupling; however these symmetry breaking contribution can be
 determined exactly. This then leads to a stable resummed expansion of
 the resummed anomalous dimension at the fixed coupling level.

\item {\it The running-coupling resummation} of the BFKL solution. 
Whereas running coupling
  corrections to evolution equations are automatically included when
  solving the DGLAP evolution equation with resummed anomalous
  dimensions, the duality relation Eq.~(\ref{dualdef}) itself undergoes
  corrections when the running coupling is included in the BFKL
  equation~(\ref{NLLBFKL}). Running coupling corrections can then be
  derived order by order, and turn out to be affected by singularities
  in Mellin $M$ space. This implies that after Mellin inversion 
the associate splitting
  functions is enhanced as $x\to0$:
  their contribution grows as
  $\left(\alpha_s\beta_0\ln\frac{1}{x}\right)^n$ with the perturbative
  order. However the series of leading enhanced contribution can be
  summed at all orders in closed form, because it corresponds to the
  asymptotic expansion in powers of $\alpha_s$ of the solution to the
  running coupling BFKL equation~(\ref{NLLBFKL}) when the kernel $\chi$
  is approximated quadratically about its minimum. This exact solution
  can be expressed in terms of Airy functions~\cite{LipatovAiry,Altarelli:2001ji} when the kernel is
  linear in $\alpha_s$ and in terms of Bateman~\cite{Altarelli:2005ni} functions for generic
  kernels. Because both the exact solution and its asymptotic
  expansion are known, this BFKL running coupling resummation can be
  combined with the DGLAP anomalous dimension, already resummed at the
  BFKL fixed coupling level, with full control of overlap (double
  counting terms). Schematically, the result has the following form:
\begin{eqnarray}&&\gamma^{rc}_{\Sigma\,NLO}(\as(t),N)
=\gamma^{rc,\,pert}_{\Sigma\,NLO}(\as(t),N)+
\gamma^B(\as(t),N)- 
\gamma^B_s(\as(t),N)- 
\gamma^B_{ss}(\as(t),N)\nonumber\\&&\quad
-\gamma^B_{ss,0}(\as(t),N)+\gamma_{\rm match}(\as(t),N)
+\gamma_{\rm mom}(\as(t),N),\label{resumabfad}
\end{eqnarray}
where $\gamma^{rc,\,pert}_{\Sigma\,NLO}(\as(t),N)$ contains
all terms
  which are up to NLO in the double-leading expansion of point 1, 
symmetrized as
  discussed in point 2 above so that
  its dual $\chi$ has a minimum; $\gamma^B(\as(t),N)$ 
resums the series of singular running
  coupling corrections using the aforementioned 
exact BFKL solution in terms of a
  Bateman function; $\gamma^B_s(\as(t),N)$, 
$\gamma^B_{ss}(\as(t),N)$ $\gamma^B_{ss,0}(\as(t),N)$ are double
counting subtractions between the previous two contributions;
$\gamma_{\rm mom}$ subtracts subleading terms which spoil
  exact momentum conservation; $\gamma_{\rm match}$ subtracts any contribution which deviates
  from NLO DGLAP and at large $N$ doesn't drop at least as
  $\frac{1}{N}$. 
\end{enumerate}

\begin{figure}
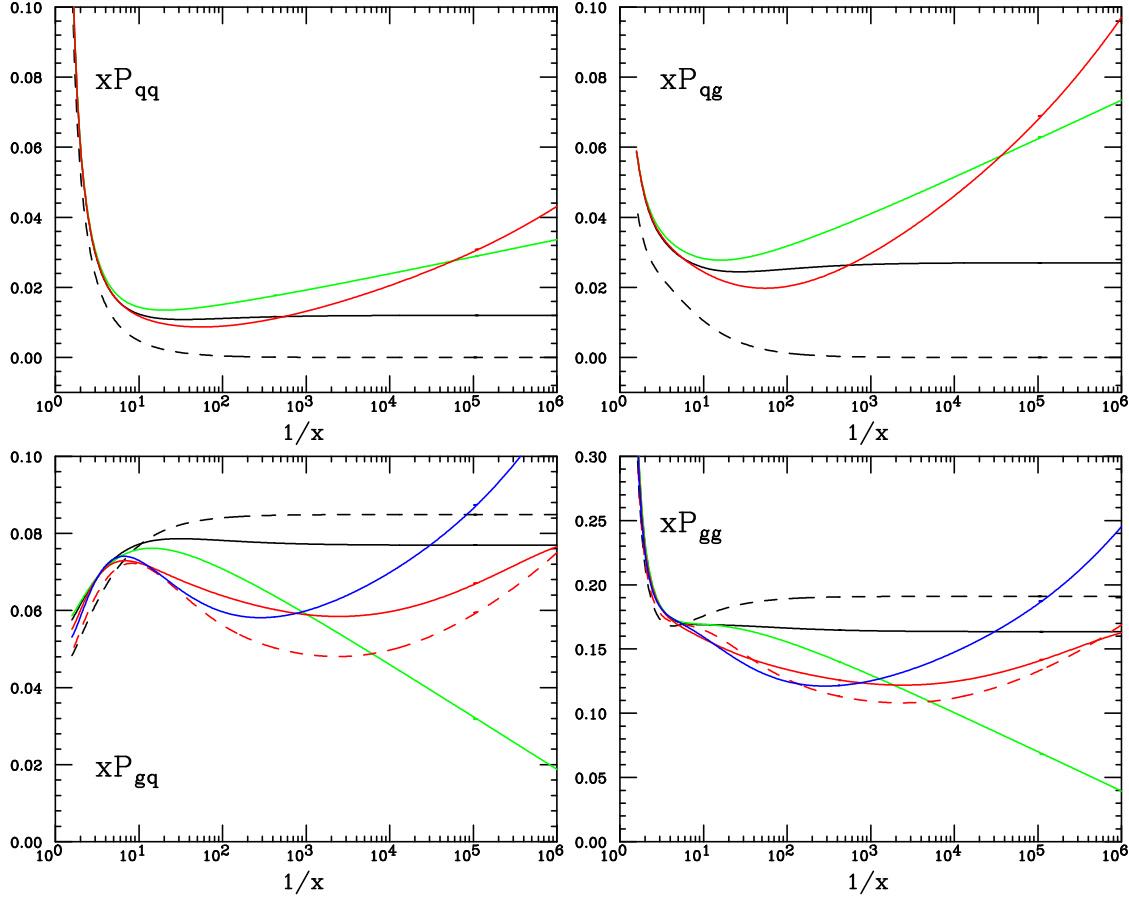

\begin{center}
\hbox{
\scalebox{0.5}{\includegraphics{pqq.ps}}
\scalebox{0.5}{\includegraphics{pqg.ps}}
}
\hbox{
\scalebox{0.5}{\includegraphics{pgq.ps}}
\scalebox{0.5}{\includegraphics{pgg.ps}}
}
\caption{The resummed splittings functions $P_{qq}$, $P_{qg}$, $P_{gq}$ and 
$P_{gg}$ in the ABF approach, all for $n_f=4$ and $\alpha_s=0.2$: LO
  DGLAP (dashed black), NLO DGLAP (solid black), NNLO DGLAP (solid green), LO
  resummed (red dashed), NLO resummed in the ${\rm Q}_0$\MS\ scheme (red) and
  in the \MS\ scheme (blue).} 
\label{Pfig}
\end{center}
\end{figure}

The anomalous dimension obtained through this procedure has a simple
pole as a leading small-$N$ (i.e. small $x$) singularity, like the LO
DGLAP anomalous dimension. The location of the pole is to the
right of the DGLAP pole, and it depends on the value of
$\alpha_s$. Thanks to the softening due to running of the coupling,
this value is however rather smaller than that which corresponds to
the leading BFKL singularity: for example, for $\alpha_s=0.2$, when
$n_f=0$ the pole is at $N=0.17$.

The splitting function obtained by Mellin inversion of the anomalous
dimension eq.~(\ref{resumabfad}) turns out to agree at the percent
level to that obtained by the CCSS group by
numerical resolution of the BFKL equation for all $x\lsim
10^{-2}$; for larger values of $x$ (i.e. in the matching  region)
the ABF result is closer to the NLO DGLAP result.

\begin{figure}[!t]
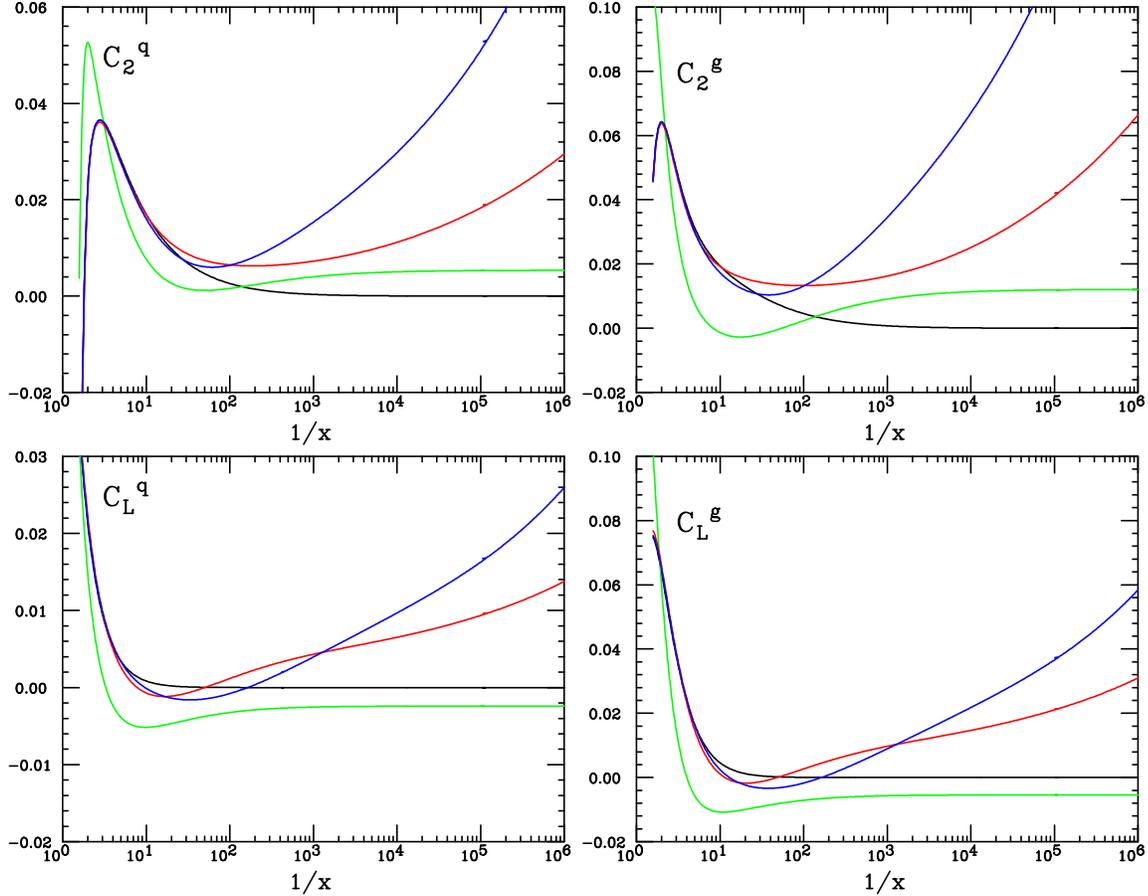

\begin{center}
\hbox{
\scalebox{0.5}{\includegraphics{c2q.ps}}
\scalebox{0.5}{\includegraphics{c2g.ps}}
}
\hbox{
\scalebox{0.5}{\includegraphics{clq.ps}}
\scalebox{0.5}{\includegraphics{clg.ps}}
}
\caption{The resummed DIS coefficient functions 
$C_{2q}$, $C_{2g}$, $C_{Lq}$ and 
$C_{Lg}$ in the ABF approach, all for $n_f=4$ and $\alpha_s=0.2$. The curves are labelled as in the previous figure.}
\label{Cfig}
\end{center}
\end{figure}

In order to obtain a full resummation of physical
observables, 
specifically for deep-inelastic scattering, the resummation discussed
so far has to be extended to the quark sector and to hard partonic
coefficients. This, on top of various technical complications,
requires two main conceptual steps:
\begin{itemize}
\item A factorization scheme must be defined at a resummed
  level. Because only one of the two eigenvectors of the matrix of
  anomalous dimensions is affected by resummation, once a scheme is
  chosen, the resummation discussed above determines entirely the
  two-by-two matrix of splitting functions in the singlet sector. The
  only important requirement is that the relation of this small $x$
  scheme choice to standard large $x$ schemes be known exactly, since
  this enables one to combine resummed results with known fixed order
  results.
\item PDFs evolved using resummed evolution equations
  must be combined with resummed coefficient functions. These are
  known, specifically for DIS~\cite{Catani:1994sq}, 
but are also known~\cite{Ball:2001pq} to be affected by singularities,
analogous to the running coupling singularities of the resummed
anomalous dimension discussed above, which likewise must be resummed
to all orders~\cite{Ball:2007ra}. 
This running coupling resummation of the coefficient
function significantly softens the small $x$ growth of the coefficient
function and substantially reduces its scheme dependence~\cite{Altarelli:2008aj}.
\end{itemize}

These steps have been accomplished in
Ref.~\cite{Altarelli:2008aj}, where resummed
anomalous dimensions (see fig.~\ref{Pfig}), coefficient functions
(see fig.\ref{Cfig})
 and structure functions (see section~\ref{comp} below)
have been determined. 
The scheme dependence of these results can be studied in detail:
results have been produced and compared in
both the  \MS\ and ${\rm Q}_0$\MS\ schemes, and furthermore
the variation of results upon variation of factorization and
renormalization scales has been studied.

Calculations of resummation corrections not only of deep inelastic
processes, but also of benchmark hadronic processes such as Drell-Yan,
vector boson, heavy quark and Higgs production are now possible and
should be explored.

\subsubsection{The Ciafaloni-Colferai-Salam-Stasto (CCSS) Approach}
\label{CCSS}

The Ciafaloni-Colferai-Salam-Stasto (CCSS) resummation approach
proposed in a series a papers
\cite{Salam:1998tj,Ciafaloni:1998iv,Ciafaloni:1999yw,Ciafaloni:2000cb,Ciafaloni:2003rd,Ciafaloni:2005cg,Ciafaloni:2006yk,Ciafaloni:2007gf}
is based on the  few general principles: 

\begin{itemize}
\item We impose the so-called kinematical
  constraint~\cite{Andersson:1995jt,Kwiecinski:1996td,Kwiecinski:1997ee}
 onto the real gluon emission terms in the BFKL kernel. The effect of
 this constraint is to cut out the regions of the phase space for
 which $k_T'^2 \ge k_T^2/z$ 
where $k_T,k_T'$ are the transverse momenta of the exchanged gluons
and $z$ is the fraction of the longitudinal momentum. 

\item The matching with the DGLAP anomalous dimension is done up to
  the next-to-leading order. 
\item  We impose the momentum sum rule onto the resummed anomalous dimensions.
\item Running coupling is included with the appropriate choice of
  scale. We take the argument of the running coupling to be the
  transverse momentum squared of the emitted gluon in the BFKL ladder
  in the BFKL part.  For the part which multiplies the DGLAP terms in
  the eigenvalue equation we choose the scale to be the maximal
  between $k_T^2$ and $k_T^{'2}$. 
\item All the calculations are performed directly in momentum space. This in particular enables easy implementation of the
running of the coupling with the choice of the arguments as described
above.  
\end{itemize}

The implementation at the leading logarithmic level in BFKL and DGLAP
(and in the single gluon channel case) works as follows. It is
convenient to go to the Mellin space representation 
where we denote by $\gamma$ and $\omega$ the Mellin variables
conjugated to  $\ln k_T$ and $\ln 1/x$ respectively. The full
evolution kernel 
can be represented as a series ${\cal K } = \sum_n \alpha_s^{n+1}
{\cal K }_n(\gamma,\omega)$. We take the resummed kernel at the lowest
order level 
to be 
\begin{equation}
{\cal K}_0(\gamma,\omega) \, = \, \frac{2 C_A}{\omega} \, \chi_0^{\omega}(\gamma) \, + \, [\gamma_0^{gg}(\omega)-\frac{2C_A}{\omega}]\chi_c^{\omega}(\gamma) \;  .
\label{eq:llxllo} 
\end{equation}
The terms in (\ref{eq:llxllo}) are the following
$$
\chi_0^{\omega}(\gamma)=2 \psi(1)-\psi(\gamma)-\psi(1-\gamma+\omega) \; ,
$$
is the leading logarithmic BFKL kernel eigenvalue with the kinematical
constraint imposed. This is reflected by the fact that the
singularities in the $\gamma$ plane at $\gamma=1$ are shifted by the
$\omega$. This ensures  the compatibility with the DGLAP collinear
poles, in the sense that 
we have only single  poles in $\gamma$. 
The function $\chi_c(\gamma)$ is the collinear part of the kernel
$$
\chi_c^{\omega}(\gamma) = \frac{1}{\gamma}+\frac{1}{1-\gamma+\omega} \; ,
$$
which includes only the leading collinear poles at $\gamma=0$ or
$1$. All the higher twist poles are neglected for this part of the
kernel. 
This kernel eigenvalue  is multiplied by the non-singular (in
$\omega$) part of the DGLAP anomalous dimension 
$\gamma_0^{gg}(\omega)-2C_A/\omega$
where $\gamma_0^{gg}(\omega)$ is the full anomalous dimension at the
leading order. 
The next-to-leading parts both in BFKL and DGLAP are included in  the
second term in the expansion, i.e. kernel ${\cal K}_1$ 
\begin{equation}
{\cal K}_1(\gamma,\omega) = \frac{(2 C_A)^2}{\omega} \tilde{\chi}_1^{\omega}(\gamma)+\tilde{\gamma}_1^{gg}(\omega)\chi_c^{\omega}(\omega)
\label{eq:nllxnllo}
\end{equation}
where $\tilde{\chi}_1^{\omega}(\gamma)$ is the NLL in x part of the
BFKL kernel eigenvalue with subtractions. These subtractions are
necessary 
to avoid double counting: we need to subtract the double and triple
collinear poles in $\gamma$ which are already included in the resummed
expression (\ref{eq:llxllo}) and which can be easily identified by
expanding this expression in powers of $\omega$ and using the LO
relation $\omega = \bar{\alpha}_s \chi_0(\gamma)$. The term
$\tilde{\gamma}_1^{gg}(\omega)$ in Eq.~(\ref{eq:nllxnllo}) is chosen
so that one obtains the correct DGLAP anomalous dimension at a fixed 
next-to-leading logarithmic level. The formalism described above has
been proven to work successfully in the single channel case, that is
for evolution of gluons only.  The solution was shown to be very
stable with respect to the changes of the resummation scheme.  

The quarks are included in the CCSS
approach by a matrix formalism. The basic assumptions in this construction are:
\begin{itemize}
\item Consistency with the collinear matrix factorization of the
  PDFs in the singlet evolution. 
\item Requirement that only single pole singularities in both in $\gamma$
  and $\omega$ are present in the  kernel eigenvalues. 
This assumption allows for the natural consistency with DGLAP and BFKL
respectively. Higher order singularities can be generated at higher
orders only through the subleading dependencies on these two
variables. 
\item Ability to compute all the anomalous  dimensions which can be
  directly compared with the DGLAP approach. 
This can be done by using set of recursive equations which allow to
calculate the anomalous dimensions order by order 
from the kernel eigenvalues.
\item Impose the collinear-anticollinear symmetry of the kernel matrix
  via the similarity transformation. 
\item Incorporate NLLx BFKL and DGLAP up to NLO (and possibly NNLO).
\end{itemize} 

\begin{figure}
  \centering
  \includegraphics[width=0.6\textwidth]{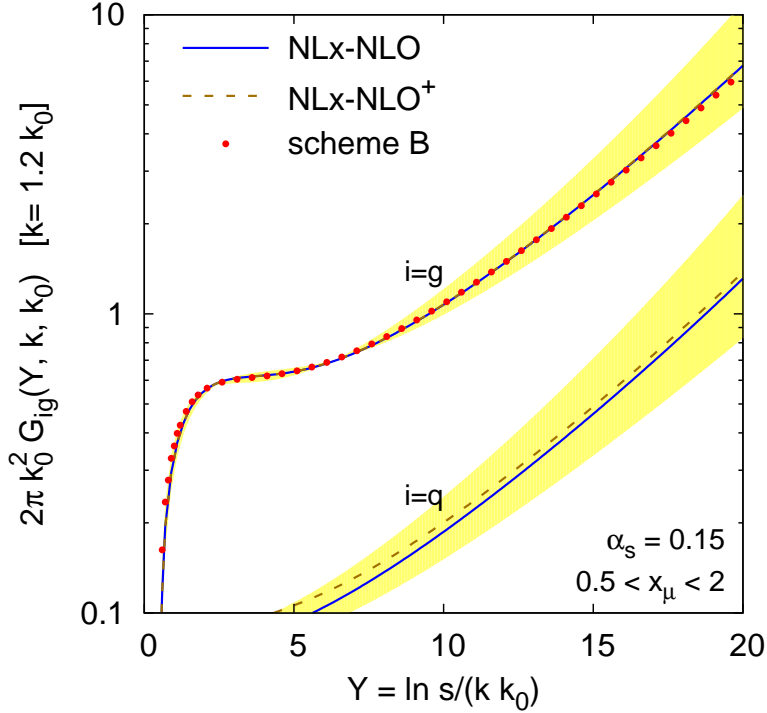}
  \caption{\it Gluon-induced part of the Green function for the NL$x$-NLO and
    NL$x$-NLO$^+$ models, compared to the results the single channel approach.
    For the models of this paper both gluon-gluon  and quark-gluon Green's function are shown.
    The value chosen for the coupling, $\as=0.15$, corresponds to
    $k_0\simeq 20 \GeV$. The band indicates the spread in the result for
    the NL$x$-NLO model when varying the renormalization scale in the
    range $0.5 < x_\mu < 2$.}
  \label{fig:green_ccss}
\end{figure}

The direct solutions to the  matrix equations  are the quark and gluon
Green's functions. These are presented in Fig.~\ref{fig:green_ccss} 
for the case of the gluon-gluon and quark-gluon part. The  resulting
gluon-gluon part is increasing exponentially with the logarithm of energy
$\ln s$  
with an effective intercept of about $\sim 0.25$. It is much
suppressed with respect to the leading logarithmic order.  
We also note that the  single channel results and the matrix results
for the gluon-gluon Green's function are very similar to each other. 
In Fig.~\ref{fig:green_ccss} we also present the quark-gluon channel
which is naturally suppressed in normalization with respect to the
gluon-gluon one by a factor of the strong coupling constant. This
can be intuitively understood as the (singlet) quarks are radiatively
generated from the gluons, 
and therefore this component follows the gluon density very
closely. The yellow bands indicate the change of the Green's functions
with respect to the change of the scale.  

\begin{figure}[ht]
  \centering
  \includegraphics[width=1.0\textwidth]{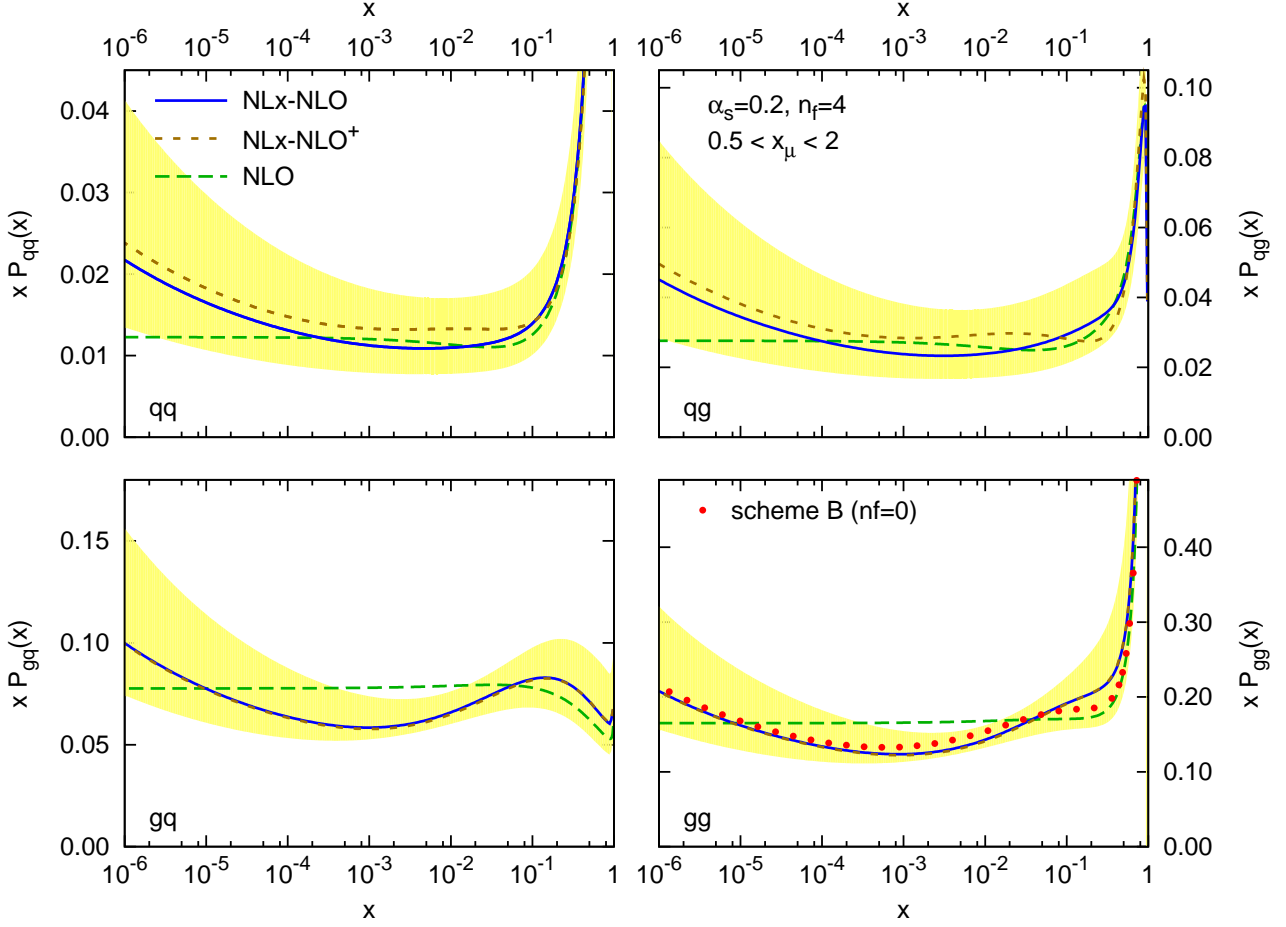}
  \caption{\it The matrix of NL$x$-NLO (and NL$x$-NLO$^+$) splitting functions
    together with their scale uncertainty and the NLO splitting
    functions for comparison. In the $gg$ channel, we also show the
    old scheme~B result ($n_f=0$, no NLO contributions, 1-loop
    coupling) . The band corresponds to the span of
    results (NL$x$-NLO) obtained if one chooses $x_\mu=0.5$ and
    $x_\mu=2.0$.}
  \label{fig:splitting_ccss}
\end{figure}

In
Fig.~\ref{fig:splitting_ccss} we present all four splitting functions
for fixed value of scale $Q^2$. Here, again the results are very close to the previous single channel
approach in the case of the gluon-gluon splitting function. 
The gluon-quark channel is very close to the gluon-gluon one, with the
characteristic dip of this function at about $x\sim 10^{-3}$. 
The dip delays the onset of rise of the splitting function only to
values of x of about $10^{-4}$. The scale dependence growths with 
decreasing $x$ but it is not larger than in the fixed NLO case. The
quark-gluon and quark-quark splitting functions tend to have slightly
larger uncertainty due to the scale change but are also slightly
closer to the plain NLO calculation. They also tend to have a less
pronounced dip structure. 
\subsubsection{The Thorne-White (TW) Approach}
\label{TW}

Substituting the LO running coupling $\abar(k^2)$ into equation
(\ref{NLLBFKL}) and performing a double Mellin transform according to
equation~(\ref{mel}), the BFKL equation~\ref{NLLBFKL}, as mentioned in
Section~\ref{smallx-intro}, becomes a differential
equation: 
\begin{equation}
\frac{d^2f(\gamma,N)}{d\gamma^2}=\frac{d^2f_I(\gamma,Q_0^2)}{d\gamma^2}-\frac{1}{\bb N}\frac{d(\chi_0(\gamma)f(\gamma,N))}{d\gamma}+\frac{\pi}{3\bb^2 N}\chi_1(\gamma)f(\gamma,N)
\label{NLLBFKL2},
\end{equation}
where $\chi_{0,1}(\gamma)$ are the Mellin transforms of ${\cal
  K}_{0,1}$. 
The solution for $f(N,\gamma)$ of Eq.~(\ref{NLLBFKL2})
has the following form~\cite{Collins:1988ze,Ciafaloni:1998iv}:
\begin{equation}
f(N,\gamma)=\exp\left(-\frac{X_1(\gamma)}{\bb N}\right)\int_\gamma^\infty A(\tilde{\gamma})\exp\left(\frac{X_1(\tilde{\gamma})}{\bb N}\right)d\tilde{\gamma}.
\label{sol1}
\end{equation}
Up to power-suppressed corrections, one may shift the lower limit of the 
integral $\gamma\rightarrow0$, so that the gluon distribution factorises into 
the product of a perturbative and a non-perturbative piece. 
The nonperturbative piece depends on the bare input gluon distribution and 
an in principle calculable hard contribution. However, this latter part is 
rendered ambiguous by diffusion into the infrared, and in this approach is 
contaminated by infrared renormalon-type contributions. The perturbative 
piece is safe from this and is sensitive to diffusion into the ultraviolet 
region of weaker coupling. Substituting equation~(\ref{sol1}) 
into (\ref{NLLBFKL2}), one finds that the perturbative piece is given 
(after transforming back to momentum space):
\begin{equation}
{\cal G}_E^1(N,t)=\frac{1}{2\pi\imath}\int_{1/2-\imath\infty}^{1/2+\imath\infty}\frac{f^{\beta_0}}{\gamma}\exp\left[\gamma t-X_1(\gamma,N)/(\bb N)\right]d\gamma,
\label{sol2}
\end{equation}
where:
\begin{equation}
X_1(\gamma,N)=\int_{\frac{1}{2}}^\gamma\left[\chi_0(\tilde{\gamma})+N\frac{\chi_1(\tilde{\gamma})}{\chi_0(\tilde{\gamma})}\right]d\tilde{\gamma}.
\label{X1}
\end{equation}
Structure functions $F_i$ also factorize, and the perturbative factors
have a similar form to Eq.~(\ref{sol2}), but involve an
additional impact factor $h_i(\gamma,N)$ in the integrand according to
the $k_t$-factorisation theorem \cite{Catani:1994sq}. Crucially,
coefficient functions and anomalous dimensions involve ratios of the
above quantities, such that the non-perturbative factor cancels. Thus,
once all the impact factors are known, the complete set of coefficient
and splitting functions can be disentangled. Finally they can be
combined with the standard NLO DGLAP results (which are known to
describe data well at higher $x$ values) using the simple
prescription: 
\begin{equation}
P^{tot.}=P^{NLL}+P^{NLO}-\left[P^{NLL(0)}+P^{NLL(1)}\right],
\label{add}
\end{equation}
where $P$ is a splitting or coefficient function, and $P^{NLL(i)}$ the ${\cal O}(\alpha_s^i)$ contribution to the resummed result which is subtracted to avoid double-counting. 
It should be noted that the method of subtraction of the 
resummed contribution in the matching is different to that for the ABF 
approach outlined after Eq.~(\ref{resumabfad}). For example, at NLO in the 
resummation the BFKL equation provides both the $\alpha_S/N$ part of 
$P_{gg}$ and the part at ${\cal O}(\alpha_S)$ constant as $N \to \infty$. 
Hence we choose to keep all terms constant as $N \to \infty$ generated 
by Eq.~(\ref{sol2}), with similar considerations for other splitting 
functions and coefficient functions, though these can contain terms 
$\propto N$. Hence, we include terms which will 
have some influence out to much higher $x$ than in the ABF approach.   
\begin{figure}
\begin{center}
\scalebox{0.8}{\includegraphics{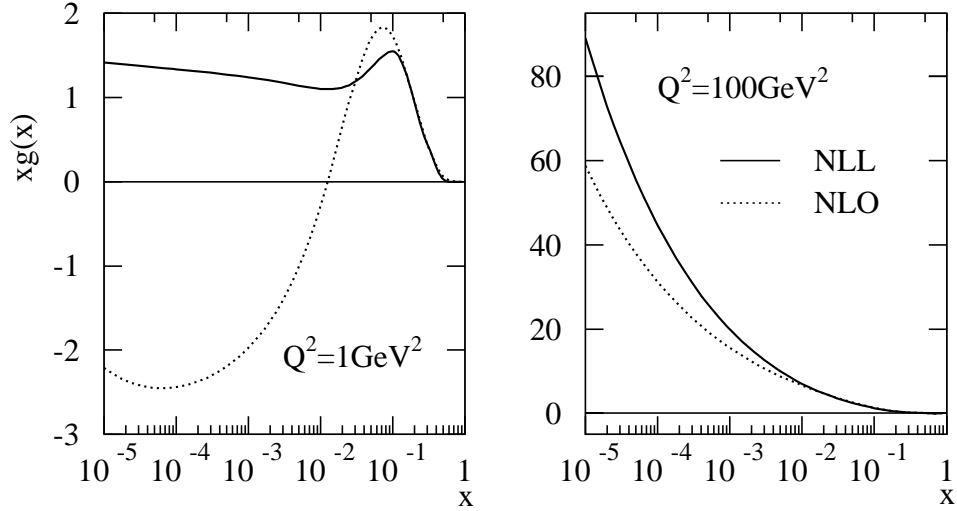}}
\caption{Gluons arising from a global fit to scattering data including NLL small x resummations in the DIS($\chi$) factorisation scheme (solid). Also shown is the result from an NLO DGLAP fit in the same scheme.}
\label{gluons}
\end{center}
\end{figure}

In the TW manner of counting orders LL is defined as the first order at which contributions appear, so while for the gluon splitting function this is 
for $\abar^n\ln^m(1/x)$ for $m=n-1$ for impact factors this is for $m=n-2$.
A potential problem therefore arises in that the NLL impact factors are not 
known exactly. However, the LL impact factors with conservation of energy of 
the gluon imposed are known in cases of both massless and massive quarks 
\cite{Bialas:2001ks,White:2006aj}, and are known to provide a very good 
approximation to the full ${\cal O}(\alpha_S^2)$ and ${\cal O}(\alpha_S^3)$ 
quark-gluon splitting functions and coefficient functions\cite{White:2005wm},
implying that they must contain much of the important higher-order 
information. These can then be used to 
calculate NLL coefficient and splitting functions within a particular
factorisation scheme. One must also specify a general mass variable
number scheme for consistent implementation of heavy quark mass
effects. Such a scheme (called the DIS($\chi$) scheme) has been given
in \cite{White:2006xv,White:2006yh} up to NLL order in the high energy
expansion, and NLO order in the fixed order expansion.  

\begin{figure}
  \centering
  \includegraphics[width=1.0\textwidth]{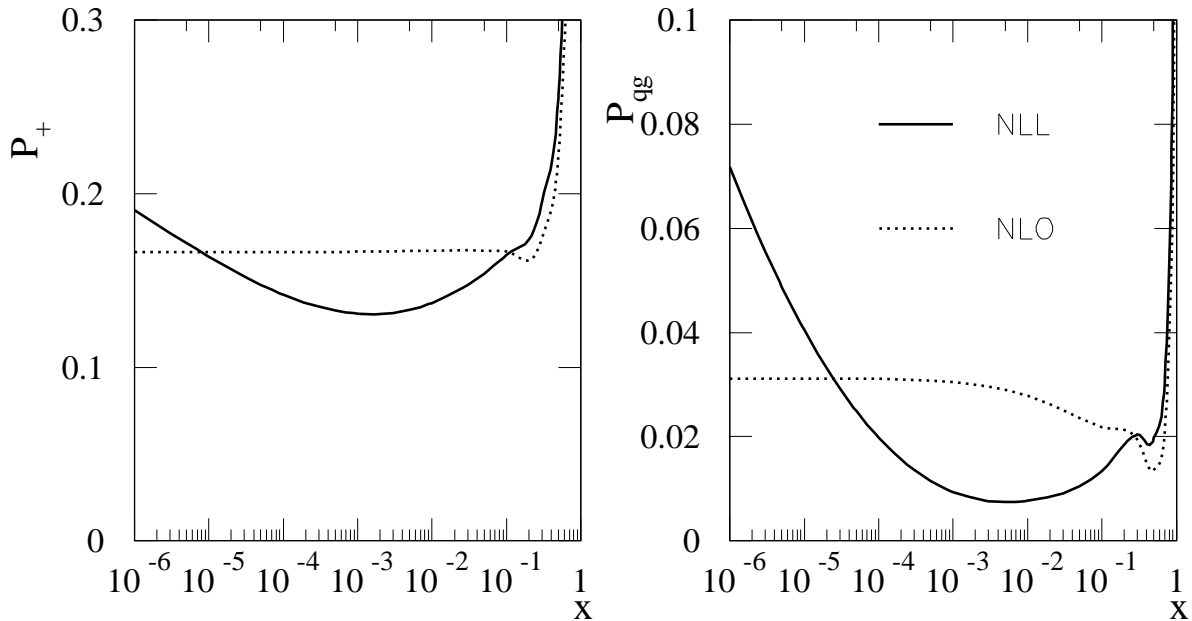}
  \caption{\it The resummed splitting functions (solid) 
   $P_+ \approx P_{gg}$ and $P_{qg}$ in the TW approach, both for 
   $n_f=4$ and $\alpha_S=0.16$, compared to the corresponding NLO
   forms (dotted).}
  \label{fig:splitting_tw}
\end{figure}

The form of the resummed splitting functions shown in 
fig.~\ref{fig:splitting_tw} are qualitatively consistent with 
those from the ABF approach, fig.~\ref{Pfig}, and CCSS 
approach fig.~\ref{fig:splitting_ccss} (note however that in these 
plots the value of $\alpha_s$ is a little larger, and the scheme is different).
This is despite the fact that the approach does not 
include the explicit collinear resummation of the BFKL kernel adopted 
in the other two approaches. It was 
maintained in \cite{Thorne:1999sg,Thorne:2001nr} that the diffusion into 
the ultraviolet, effectively making the coupling weaker, hastens the 
perturbative convergence for splitting functions, and the kernel near 
$\gamma=0$, making this additional resummation less necessary.
There is no particular obstruction to including this resummation in the 
approach, it is simply cumbersome. Indeed, in Ref.~\cite{Thorne:2001nr}
the effect was checked, and modifications found to be no greater than 
generic NNLO corrections to the resummation, so it was omitted. 
(Note that any process
where there are two hard scales, sensitive to $\gamma \approx 0.5$, or
attempted calculation of the  
hard input for the gluon distribution, sensitive to $\gamma=1$, would
find this resummation essential.) 
The main feature of the resummed
splitting functions is a significant dip below the NLO DGLAP results,
followed by an eventual rise at very low $x\simeq 10^{-5}$. This
behaviour drives a qualitative change in the gluon distribution, when
implemented in a fit to data.

The combined NLO+NLL splitting and coefficient functions (in the TW
approach) have been implemented in a global fit to DIS and related
data in the DIS($\chi)$ scheme, thus including small $x$ resummations
in both the massless and massive quark sectors
\cite{White:2006yh}. The overall fit quality was better than a
standard NLO fit in the same factorisation scheme, and a similar NLO
fit in the more conventional \MS\ factorisation scheme. The
principal reason for this is the dip in the resummed evolution
kernels, which allows the gluon distribution to increase at both high
and low values of $x$. This reduces a tension that exists between the
high $x$ jet data of \cite{Tevjet1,Tevjet2} and the low $x$ HERA data
\cite{Adloff:2000qj,Adloff:2000qk,Adloff:2003uh,Breitweg:1998dz,Chekanov:2001qu}. The gluon distributions arising from
the NLL and NLO fits are shown in figure \ref{gluons}, for the
starting scale $Q^2=1$GeV$^2$ and also for a higher value of
$Q^2$. One sees that whilst the NLO gluon wants to be negative at low
$x$ and $Q^2$, the resummed gluon is positive definite and indeed
growing slightly as $x\rightarrow 0$. The gluons agree well for higher
$x$ values (where the DGLAP description is expected to dominate), but
deviate for $x\leq10^{-2}$. This can therefore be thought of as the
value of $x$ below which resummation starts to become relevant. 
\begin{figure}
\begin{center}
\scalebox{0.6}{\includegraphics{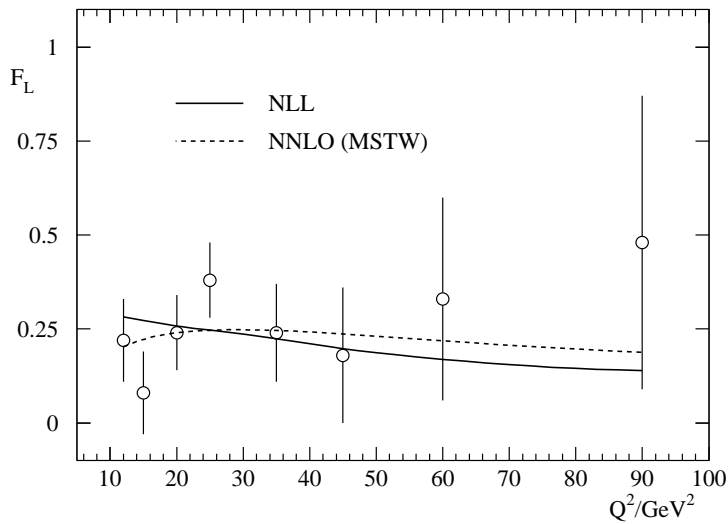}}
\caption{Recent H1 data on the longitudinal structure function $F_L$,
  together with the NLL resummed prediction from the TW approach, and
  a recent NNLO result from the MSTW group.} 
\label{flfig}
\end{center}
\end{figure}

The qualitatively different gluon from the resummed fit (together with
the decreased evolution kernels w.r.t. the fixed order description)
has a number of phenomenological implications: 
\begin{enumerate}
\item The longitudinal structure function $F_L$ is sensible at small $x$ and $Q^2$ values, where the standard DGLAP description shows a marked instability \cite{Martin:2006qv}.
\item As a result of the predicted growth of $F_L$ at small $x$ the
  resummed result for the DIS reduced cross-section shows a turnover
  at high inelasticity $y$, in agreement with the HERA data. This
  behaviour is not correctly predicted by some fixed order fits. 
\item The heavy flavour contribution (from charm and bottom) to $F_2$
  is reduced at higher $Q^2$ in the resummed approach, due mainly to
  the decreased evolution, as already noted in a full analysis in the
  fixed-order expansion at  
NNLO \cite{Thorne:2006qt}. Nevertheless, it remains a significant
fraction of the total structure function at small $x$. 
\end{enumerate}

Other resummation approaches should see similar results when
confronted with data, given the qualitative (and indeed quantitative)
similarities between the splitting functions. It is the decreased
evolution with respect to the DGLAP description that drives the
qualitative change in the gluon distribution. This is then the source
of any quantitative improvement in the description of data, and also
the enhanced description of the longitudinal structure function and
reduced cross-section. 

The resummed prediction for $F_L$ is shown alongside the recent H1
data \cite{:2008tx} in figure \ref{flfig}, and compared with an
up-to-date NNLO fixed order result \cite{Martin:2007bv}. One sees that
the data cannot yet tell apart the predictions, but that they are
starting to diverge at low $x$ and $Q^2$, such that data in this range
may indeed be sensitive to the differences between resummed and fixed
order approaches. 

\begin{figure}[t!]
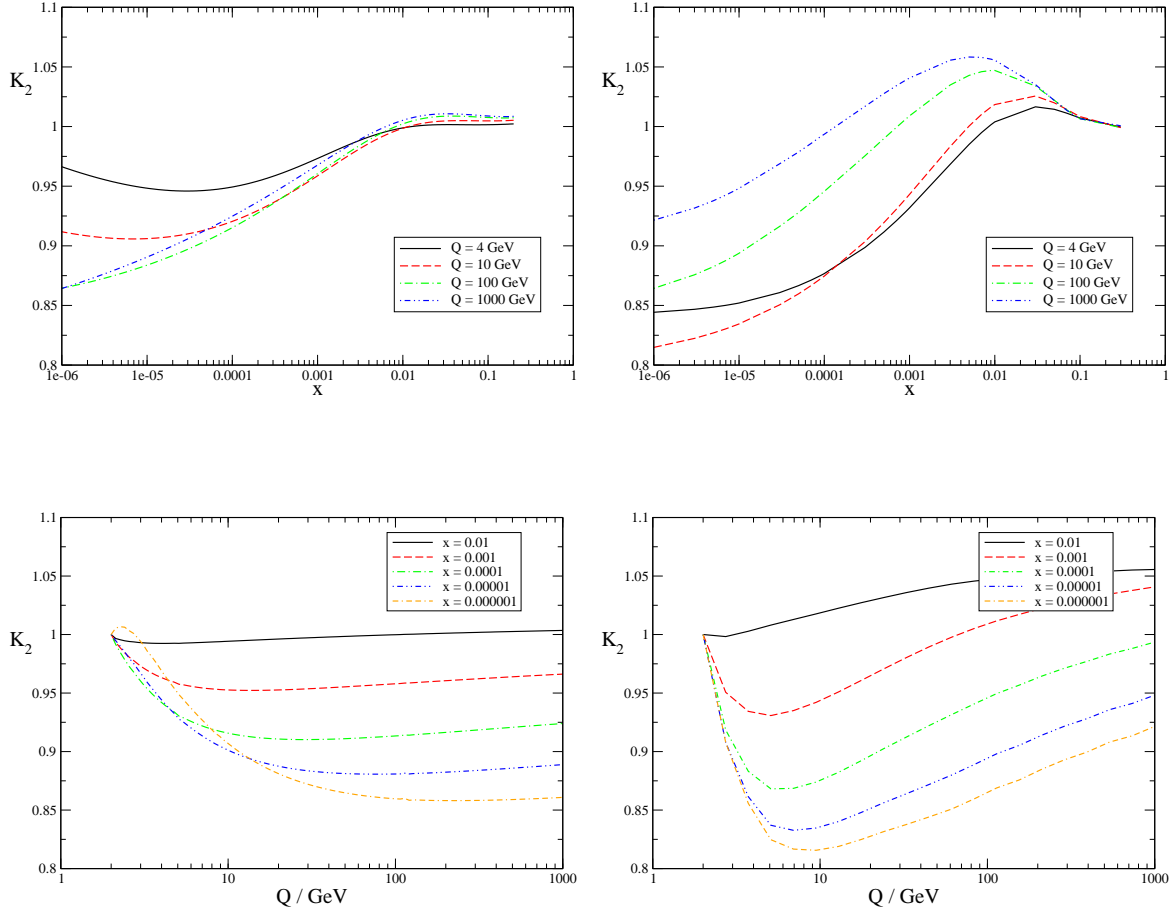

\begin{center}
\includegraphics[width=.45\linewidth]{KFSABFf.eps}\hspace{.2cm}
\includegraphics[width=.45\linewidth]{KFSTWf.eps}\\\vspace{1.5cm}
\includegraphics[width=.45\linewidth]{KFSABFQf.eps}\hspace{.2cm}
\includegraphics[width=.45\linewidth]{KFSTWQf.eps}\\
\caption{The ratio $F_2^{NLL}/F_2^{NLO}$  in the ABF approach (left)
  and the TW approach (right), 
using toy PDFs, given in eq.~\ref{parts},  calculated as function
of $x$ at fixed 
for $Q^2$ (upper ), and as a function of
$Q^2$ at fixed $x$ (lower).}
\label{K2}
\end{center}
\end{figure}

\subsubsection{Resummed structure functions: comparison of the ABF and TW approaches}
\label{comp}
In this section, we present an application of the ABF and TW
approaches to the resummed determination of the 
$F_2$ and $F_L$ deep-inelastic structure functions. The
corresponding exercise for the CCSS approach has not yet been
finalised. 
A direct comparison of the two approaches is complicated by issues of 
factorisation scheme dependence: whereas in the ABF approach 
results may be obtained in any scheme, and in particular the 
\MS\ and closely related $Q_0$-\MS\ scheme,  in
the TW formalism 
splitting functions and coefficient functions beyond NLO in 
$\alpha_S$ are resummed in the
${\rm Q}_0$-DIS scheme\cite{Ciafaloni:1995bn,Ciafaloni:2006yk}, 
which coincides with the standard DIS scheme at large $x$ but differs 
from it at the resummed level; the scheme change needed in
order to obtain the coefficient functions from the DIS-scheme ones is
performed exactly up to NLO and approximately beyond it.
Thus, without a more precise definition of the relation of this scheme 
to \MS, one cannot 
compare splitting and coefficient functions, which are factorisation
scheme dependent.  

\begin{figure}[t!]
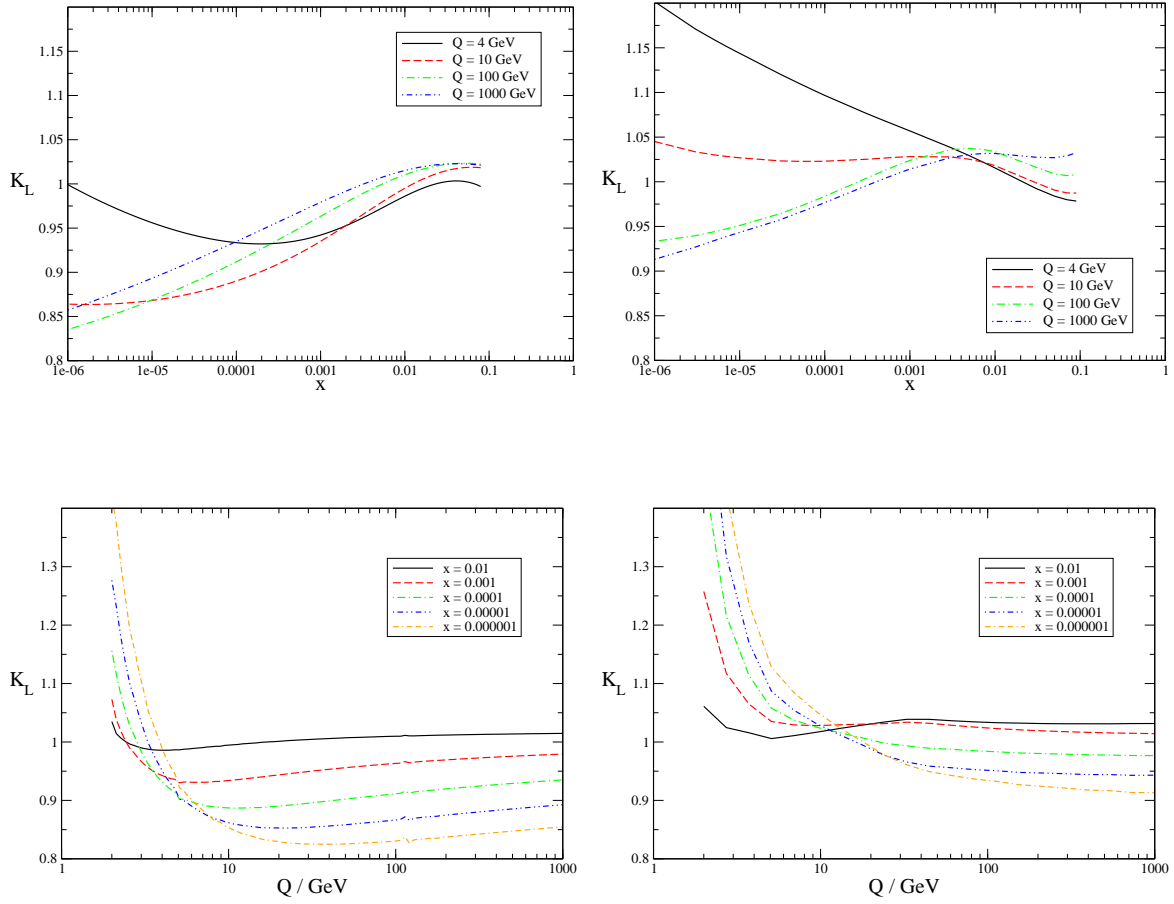

\begin{center}
\includegraphics[width=.45\linewidth]{KFLABFf.eps}\hspace{.2cm}
\includegraphics[width=.45\linewidth]{KFLTWf.eps}\\\vspace{1.5cm}
\includegraphics[width=.45\linewidth]{KFLABFQf.eps}\hspace{.2cm}
\includegraphics[width=.45\linewidth]{KFLTWQf.eps}\\
\caption{The ratio $F_L^{NLL}/F_L^{NLO}$  in the ABF approach (left)
  and the TW approach (right), 
using toy PDFs, given in eq.~\ref{parts},  calculated as function
of $x$ at fixed 
for $Q^2$ (upper ), and as a function of
$Q^2$ at fixed $x$ (lower).}
\label{KL}
\end{center}
\end{figure}

A useful compromise is to present the respective results for the ratio
of structure function predictions: 
\begin{equation}
K_i=\frac{F_i^{NLL}(x,Q^2)}{F_i^{NLO}(x,Q^2)},
\label{ridef}
\end{equation}
where $i\in{2,L}$, 
and the $F_i$ are calculated by convoluting the relevant coefficients
with PDFs obtained by perturbative evolution of a
common set of  of partons, defined at a starting scale of
$Q_0^2=4$GeV$^2$. The  
number of flavors is fixed to three, to avoid ambiguities due to  
heavy quark effects.
The initial PDFs are assumed to be fixed
(i.e., the same at the unresummed and unresummed level) in the DIS
factorization scheme at the scale $Q_0$. 
Of course, in a realistic situation the
data are fixed and the PDFs are determined by a fit to
the data: hence they are not the same at the resummed and unresummed
level (compare Fig.~\ref{gluons} above). 
However, in the DIS factorization scheme the structure function
$F_2$ is simply proportional to the quark distribution, hence by fixing
the PDFs in this scheme one ensures that $F_2$ is fixed at the
starting scale.

 This starting  PDFs are constructed as follows: the quark and gluon
 distributions are chosen to have the representative form also used in
Ref.~\cite{Altarelli:2008aj}
\begin{equation}
xg(x)=k_s x S(x)=k_gx^{-0.18}(1-x)^5;\quad xq_v=k_qx^{0.5}(1-x)^4,
\label{parts}
\end{equation}
in the \MS\ scheme,
where $g(x)$ is the gluon, $S(x)$ the sea quark distribution, and $x
q_v(x)$ denotes a valence quark distribution.   We choose $k_s=3$, and
then all other parameters are fixed by momentum and number sum
rules. Note that  the
gluon is the same as that used in the previous comparison of
Ref.~\cite{Dittmar:2005ed}. The PDFs eq.~(\ref{parts}) 
are then transformed to the DIS
factorization scheme~\cite{Diemoz:1987xu} using the NLO (unresummed) scheme 
change at the scale $Q_0$. The result is then used as a fixed boundary
condition for all (unresummed and resummed, ABF and TW) calculations.
In the TW approach, the DIS scheme for unresummed quantities and 
Q$_0$DIS scheme as discussed above is then 
used throughout. 
In the ABF approach, the
fixed DIS-scheme boundary condition is transformed to the ${\rm Q}_0$\MS\
scheme~\cite{Ball:1995tn,Altarelli:2008aj} (which at the unresummed
level coincides with standard \MS) by using the unresummed or
resummed scheme change function as appropriate, and then all
calculations are performed in ${\rm Q}_0$\MS.
One might hope that most of the residual scheme dependence 
cancels upon taking the ratio of the NLL and NLO results, at least for 
schemes that are well defined and without unphysical singularities.  

The results for $K_2$ and $K_L$ are shown in figures~\ref{K2} for
$F_2$ in the ABF and TW procedures respectively and similarly in
figures~\ref{KL} for $F_L$. One sees that for $x$ sufficiently small,
and for $Q$ not too large, the resummed $F_2$  is
consistently lower than its fixed order counterpart in both
approaches, due to the decreased evolution of the gluon, and also (in
the \MS\ scheme) due
to the fact that resummed coefficient functions are much larger than
the NLO ones at
small $x$ and low $Q^2$. Similarly the resummed $F_L$ is larger than the 
fixed order at low $Q$ and small enough $x$, but falls rapidly as $Q$ 
increases. However despite these superficial similarities, the two 
approaches differ quantitatively in several respects:
\begin{itemize}
\item the ABF resummed $F_2$ matches well to the NLO for 
$x \gsim 10^{-2}$ at all scales, while the TW $F_2$ shows a  rise
  around
$x\simeq 10^{-2}$, which is largest at low $Q$. This may be due to the 
significant differences between resummed and NLO splitting functions 
at very high $x$ in 
fig.~\ref{fig:splitting_tw}. A similar mismatch may be seen at $x\sim 0.1$ 
in the $F_L$ K-factor.
\item at large scales the ABF resummation stabilises, due to the 
running of the coupling, so the K-factors becomes rather flat: they grow 
only logarithmically in $\ln Q$. By contrast the TW $F_2$ K-factor
still shows a marked $Q^2$ dependence. This may be 
 related to the fact that
the TW resummation does not resum the collinear singularities 
in the BFKL kernel,
and to
the TW choice (see Sect.~\ref{TW}) not to
include subtraction of terms induced by the resummation
which do not drop at large $x$. This choice  
induces a change in the PDFs at
higher $x$ in the TW approach, 
which results in effects which persist to higher $Q^2$ at
smaller $x$. 
\item at the initial scale $Q_0$ the TW resummed $F_L$ grows much more 
strongly as $x$ decreases than the ABF resummed $F_L$. This is likely 
to be due to the different treatment of the coefficient functions: in
this respect, the fully consistent treatment of the factorization
scheme, 
the effect of collinear resummation, and the different 
definitions of what is called resummed 
NLO used by the two groups all play a part. 
\end{itemize}

\subsubsection{Conclusion}

The problem of understanding the small $x$ evolution of structure  
 functions in the domain of $x$ and $Q^2$ values of relevance for HERA  
 and LHC physics has by now reached a status where all relevant  
 physical ingredients have been identified, even though not all groups
 have quite reached the stage  at which the formalism can  
 be transformed into a practical tool for a direct connection with the  
 data. 

 In this report we summarised the status of the three  
 independent approaches to this problem by ABF, CCSS and TW, we  
 discussed the differences in the  adopted procedures and finally we  
 gave some recent results. The most complete formalisms are those by  
 ABF and CCSS while the TW approach is less comprehensive
 but simpler to handle, and thus has been 
 used in fit to data. We recall that, at the level of  
 splitting functions the ABF and CCSS have been compared in
 ref.~\cite{Dittmar:2005ed}  and  
 found to be in very good agreement. The singlet splitting function  
 obtained by TW was also compared with ABF and CCSS in
 ref.~\cite{White:2006yh}  and also found to be  
 in reasonable agreement, at least at small $x$. 

 Here we have shown the  
 results of an application to the structure functions $F_2$ and $F_L$  
 of the ABF and TW methods. The same input parton densities at  
 the starting scale $Q_0$ were adopted by these two groups and the
 $K$-factors  for resummed versus fixed NLO perturbative structure  
 functions were calculated using the respective methods. The results  
 obtained are in reasonable qualitative agreement  for $F_2$, less so for  
 $F_L$. Discrepancies may in part be due to the choice of
 factorization scheme, but our study suggests that the
 following are  also likely to make a quantitative difference:  
whether or not a  resummation of collinear singularities in the BFKL 
 kernel is performed,  whether 
 contributions from the resummation which persist at large $x$ are
 subtracted and whether the factorization scheme is consistently
  defined in the same way 
at resummed and NLO levels.


\subsection{Parton saturation and geometric
  scaling\protect\footnote{Contributing authors: G.~Beuf, F.~Caola, F.~Gelis, 
L.~Motyka, 
C.~Royon, D.~\v S\'alek, A.~M.~Sta\'sto} }
\label{sec:satgs}
\subsubsection{Introduction\protect\footnote{Contributing authors: F.~Gelis, A. M. Sta\'sto}}

The degrees of freedom involved in hadronic collisions at sufficiently
high energy are partons, whose density grows as the energy increases
(i.e., when $x$, their momentum fraction, decreases).  This growth of
the number of gluons in the hadronic wave functions is a phenomenon
which has been well established at HERA. One expects however that it
should eventually ``saturate'' when non linear QCD effects start to
play a role.

An important feature of partonic interactions is that they involve
only partons with comparable rapidities. Consider the interaction
between a hadron and some external probe (e.g. a virtual photon in Deep Inelastic Scattering) and consider what happens when one boosts the hadron, increasing
its rapidity in successive steps.  In the first step, the valence
constituents become Lorentz contracted in the longitudinal direction
while the time scale of their internal motions is Lorentz dilated.  In
addition, the boost reveals new vacuum fluctuations coupled to the
boosted valence partons.  Such fluctuations are not Lorentz contracted
in the longitudinal direction, and represent the dynamical degrees of
freedom; they are the partons that can interact with the probe.
Making an additional step in rapidity would freeze these fluctuations,
while making them Lorentz contracted as well. But the additional boost
also produces new quantum fluctuations, which become the new dynamical
variables. This argument can be repeated, and one arrives at the
picture of a high-energy projectile containing a large number of
frozen, Lorentz contracted partons (the valence partons, plus all the
quantum fluctuations produced in the previous boosts), and partons
which have a small rapidity, are not Lorentz contracted and can
interact with the probe.  This space-time description was developed
before the advent of QCD (see for instance \cite{feyn}; in Bjorken's
lectures \cite{Bjork2}, one can actually foresee the modern
interpretation of parton evolution as a renormalization group
evolution).

\begin{wrapfigure}{l}{0.4\columnwidth}
\begin{center}
\resizebox*{!}{6cm}{\includegraphics{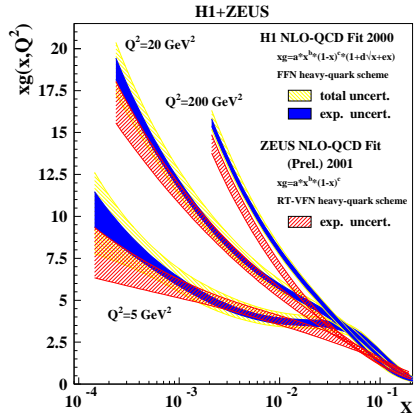}}
\end{center}
\caption{\label{fig:HERA-small-x} The gluon structure function in a
proton measured at HERA.}
\end{wrapfigure}
This space-time picture, which was deduced from rather general
considerations, can now be understood in terms of QCD. In fact,
shortly after QCD was established as the theory of strong interaction,
quantitative equations were established, describing the phenomenon
outlined above \cite{KuraeLF1,Balitsky:1978ic,Gribov:1972ri,Gribov:1972rt,Altarelli:1977zs,Dokshitzer:1977sg}.
In particular, the equation derived by Balitsky, Fadin, Kuraev and
Lipatov \cite{KuraeLF1,Balitsky:1978ic} describes the growth of the
non-integrated gluon distribution in a hadron as it is boosted towards
higher rapidities.
Experimentally, an important increase of the number of gluons at small
$x$ has indeed been observed in the DIS experiments performed at HERA
(see Fig.~\ref{fig:HERA-small-x}), down to $x\sim 10^{-4}$. Such a
growth raises a problem: if it were to continue to arbitrarily small
$x$, it would induce an increase of hadronic cross-sections as a power
of the center of mass energy, in violation of known unitarity bounds.

However, as noticed by Gribov, Levin and Ryskin in \cite{GriboLR1},
the BFKL equation includes only branching processes that increase the
number of gluons ($g\to gg$ for instance), but not the recombination
processes that could reduce the number of gluons (like $gg\to g$).
While it may be legitimate to neglect the recombination process when
the gluon density is small, this cannot remain so at arbitrarily high
density: a saturation mechanism of some kind must set in.  Treating
the partons as ordinary particles, one can get a crude estimate of the
onset of saturation, which occurs at:
\begin{equation}\label{Qsaturation}
Q^2 = Q_s^2\; ,\quad{\rm with\ } Q_s^2
\sim \alpha_s(Q_s^2)\frac{xG(x,Q_s^2)}{\pi R^2}\; .
\end{equation}
The momentum scale that characterizes this new regime, $Q_s$, is
called the saturation momentum \cite{Muell3}. Partons with transverse
momentum $Q> Q_s$ are in a dilute regime; those with $Q<Q_s$ are in
the saturated regime. The saturation momentum increases as the gluon
density increases. This comes from an increase of the gluon structure
function as $x$ decreases. The increase of the density may also come
from the coherent contributions of several nucleons in a nucleus. In
large nuclei, one expects $Q_s^2\propto A^{1/3}$, where $A$ is
the number of nucleons in the nucleus.

Note that at saturation, naive perturbation theory breaks down, even
though $\alpha_s(Q_s)$ may be small if $Q_s$ is large: the saturation
regime is a regime of weak coupling, but large density. At saturation,
the gluon occupation number is proportional to $ 1/\alpha_s$.  In such
conditions of large numbers of quanta, classical field approximations
become relevant to describe the nuclear wave-functions.

Once one enters the saturated regime, the evolution of the parton
distributions can no longer be described by a linear equation such as
the BFKL equation.  The color glass condensate formalism (for a
review, see \cite{Iancu:2003xm}), which relies on the separation of
the degrees of freedom in a high-energy hadron into frozen partons and
dynamical fields, as discussed above, provides the non linear
equations that allow us to follow the evolution of the partonic
systems form the dilute regime to the dense, saturated, regime. For
instance, the correlator ${\rm
  tr}\big<U^\dagger(\x_\perp)U(\y_\perp)\big>$ of two Wilson lines
--which enters in the discussion of DIS-- evolves according to the
Balitsky-Kovchegov \cite{Balit1,Kovch3} equation:
\begin{eqnarray}
&&\frac{\partial {\rm tr}\big<U^\dagger(\x_\perp)U(\y_\perp)\big>_{x}}
{\partial\ln(1/x)}
=
-\frac{\alpha_s}{2\pi^2}\int_{\z_\perp}
\frac{(\x_\perp-\y_\perp)^2}{(\x_\perp-\z_\perp)^2(\y_\perp-\z_\perp)^2}
\nonumber\\
&&\times\Big[
N_c {\rm tr}\big<U^\dagger(\x_\perp)U(\y_\perp)\big>_{x}
-{\rm tr}\big<U^\dagger(\x_\perp)U(\z_\perp)\big>_{x}
{\rm tr}\big<U^\dagger(\z_\perp)U(\y_\perp)\big>_{x}
\Big]\; .
\label{eq:BK}
\end{eqnarray}
(This equation reduces to the BFKL equation in the low density limit.)


The geometric scaling phenomenon was first introduced in the context of the dipole picture of the deep inelastic electron-proton scattering \cite{Stasto:2000er}. The process of the scattering of the virtual photon on a proton at very small values of $x$ can be conveniently formulated in the dipole model. In this picture the photon fluctuates into the quark-antiquark pair (dipole) and subsequently
interacts with the target.  In the small $x$ regimes these two processes factorize and they  can be encoded into the dipole formula for the total $\gamma^* p $ cross section
\begin{equation}
\sigma_{T,L}(x,Q^2) \, = \, \int d^2 {\bf r} \int dz |\Psi_{T,L}(r,z,Q^2)|^2\, 
\hat{\sigma}(x,r)\label{DipoleFact}
\end{equation}
where $\Psi_{T,L}$ is the wave function for the photon and
$\hat{\sigma}$ is the dipole cross section. $r$ is the dipole size and
$z$ is the light-cone fraction of the longitudinal momentum carried by
the quark (or antiquark).  The photon wave functions $\Psi$ are known,
the dipole cross section can be expressed in terms of the correlator
of Wilson lines whose evolution is driven by Eq.~(\ref{eq:BK})~:
\begin{equation}
\hat{\sigma}(x,r)
=
\frac{2}{N_c}\int d^2\X \;
{\rm tr}\Big<1-U(\X+\frac{\bf r}{2})U^\dagger(\X-\frac{\bf r}{2})\Big>\; .
\end{equation}
Alternatively, it can be modeled or extracted from the data.  In the
GBW model it was assumed that the dipole cross section has a form
\begin{equation}
\label{eq:gbw_dip}
\hat{\sigma} = \sigma_0 \left[ 1-\exp(-r^2/R_0(x)^2)\right]
\end{equation}
where $R_0(x)=(x/x_0)^{-\lambda}$ is a saturation radius (its inverse
is usually called the saturation scale $Q_s(x)$) and $\sigma_0$ a
normalisation constant. One of the key properties of the model was the
dependence on the dipole size and the Bjorken $x$ through only one
combined variable $r^2 Q_s^2(x)$.  This fact, combined with the
property of the dipole formula, allows to reformulate the total cross
section as a function of $Q^2/Q_s^2(x)$ only. This feature is known as
the geometric scaling of the total $\gamma^* p $ cross section.
Initially postulated as a property of the GBW model, it was then shown
that the experimental data do indeed exhibit the aforementioned
regularity in a rather wide range of $Q^2$ and for small values of
Bjorken $x$. 

Although it is a postulate in the GBW model, this property can be
derived from the small-$x$ behavior of the solutions of
Eq.~(\ref{eq:BK}) \cite{Munier:2003sj} : for a wide class of initial
conditions, the BK equation drives its solution towards a function
that obeys this scaling. Note also that the saturation scale,
introduced by hand in the GBW model, is dynamically generated by the
non linear evolution described by Eq.~(\ref{eq:BK}).  This suggested
that the regularity seen in the data could be explained by the scaling
property of the solutions to the nonlinear equations in the saturated
regime - and thus may provide some indirect evidence for gluon
saturation.

Nevertheless, several important questions remained. One of them, is
the problem of the compatibility of the DGLAP evolution with the
property of the geometric scaling. It is known from the global fits
that the standard DGLAP evolution works quite well for the description
of the of the deep inelastic data even in the very low x and $Q^2$
regime. That suggests that the saturation should be confined to the
very tight kinematic regime, and it is therefore questionable whether
the observed regularity could be attributed to the saturation at all.
In the present contribution we discuss several approaches to this
problem.
\label{sec:xxx}

\subsubsection{Phenomenology\protect\footnote{Contributing authors: C. Royon, D. \v S\'alek}}\label{phenomenology}

In order to compare the quality of different scaling laws,
it is useful to use a quantity called {\it quality factor} (QF).
It is also used to find the best parameters
for a given scaling.
In the following, this method is used to compare the
scaling results for the proton structure function $F_2$ and $F_2^c$,
the deeply virtual Compton scattering, the diffractive structure function,
and the vector meson cross section data measured at HERA.

\paragraph{Quality Factor}

Given a set of data points $(Q^2, x, \sigma=\sigma(Q^2, x))$ and a parametric
scaling variable $\tau = \tau(Q^2,Y,\lambda)$ (with $Y=\ln 1/x$) we want to know
whether the cross-section can be parametrised as a function of the variable $\tau$ only.
Since the function of $\tau$ that describes the data is not known,
the $QF$ has to be defined independently of the form of that function.

For a set of points $(u_i, v_i)$, where $u_i$'s are ordered and normalised
between 0 and 1,
we introduce $QF$ as follows~\cite{Gelis:2006bs}
\begin{equation}
QF(\lambda) = \biggl[ \sum_{i} \frac{(v_i-v_{i-1})^2}{(u_i-u_{i-1})^2+\epsilon^2} \biggr]^{-1}
\label{QF},
\end{equation}
where $\epsilon$ is a small constant that prevents the sum from being infinite in case
of two points have the same value of $u$.
According to this definition, the contribution to the sum in~(\ref{QF}) is large
when two successive points are close in $u$ and far in $v$.
Therefore, a set of points lying close to a unique curve
is expected to have larger $QF$ (smaller sum in~(\ref{QF}))
compared to a situation where the points are more scattered.


Since the cross-section in data differs by orders of magnitude and $\tau$ is
more or less linear in $log(Q^2)$,
we decided to take $u_i = \tau_i(\lambda)$ and $v_i = log(\sigma_i)$.
This ensures that low $Q^2$ data points contribute to the $QF$
with a similar weight as higher $Q^2$ data points.


\paragraph{Fits to $F_2$ and DVCS Data}

\begin{figure}[t]
\hfill
\begin{minipage}[t]{.45\textwidth}

\epsfig{file=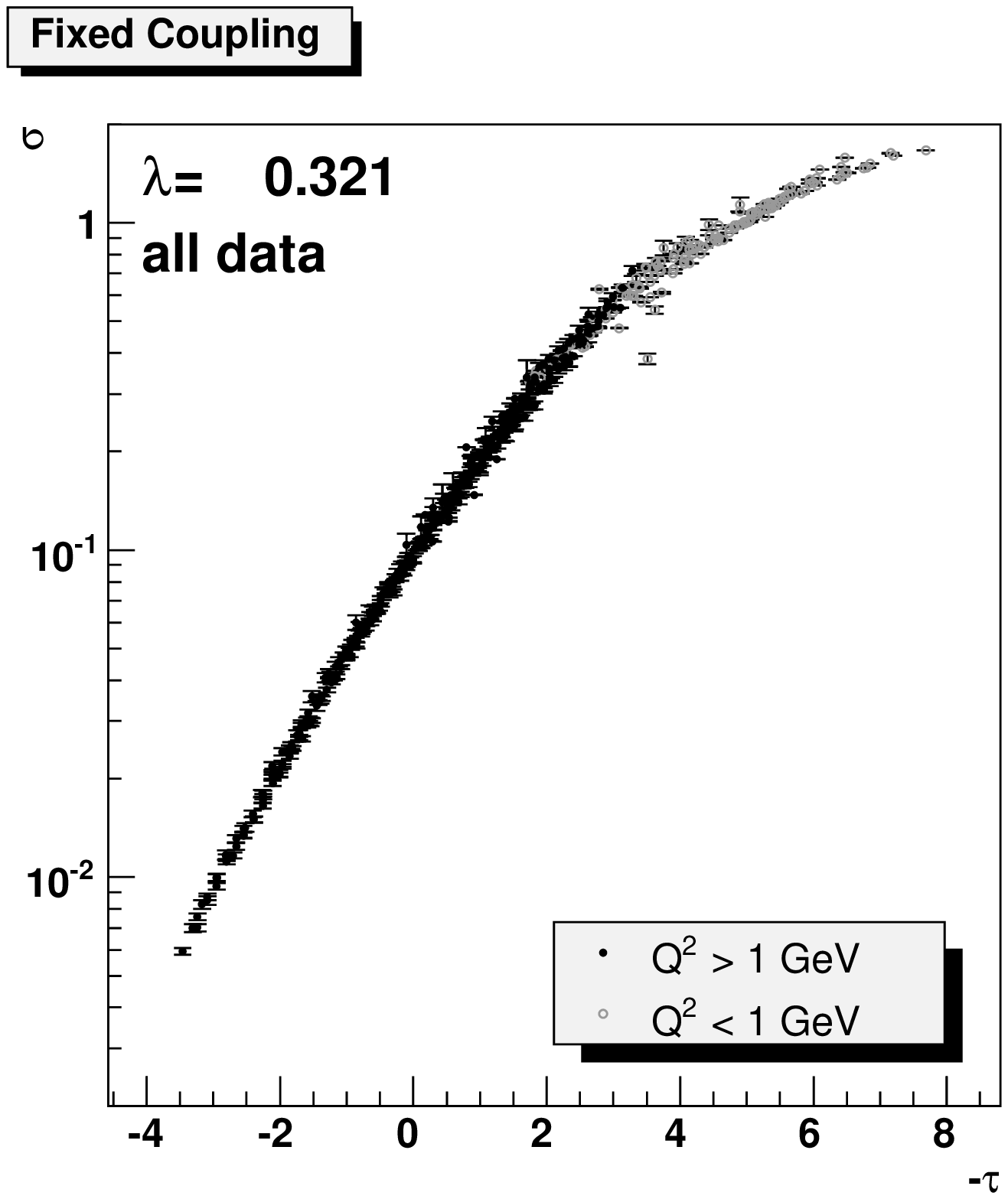,width=6.cm} 
\caption{{\bf $F_2$ data:} 
Scaling curve $\sigma = \sigma(\tau)$ for ``Fixed Coupling''.
A $Q^2>1$ GeV$^2$ cut was applied to the data.}
\label{F2_QF_1}

\end{minipage}
\hfill
\begin{minipage}[t]{.45\textwidth}

\epsfig{file=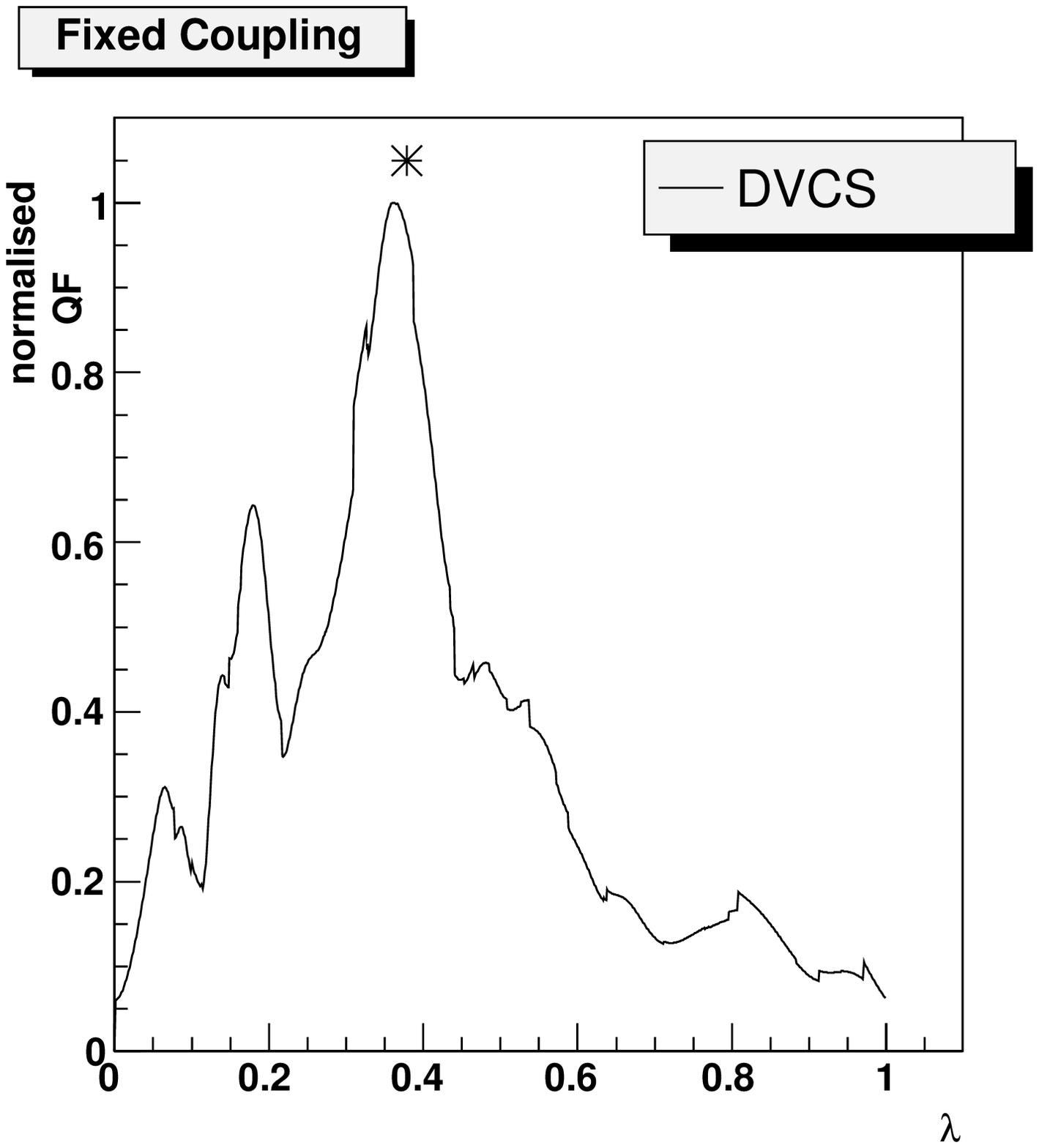,width=6.cm} 
\caption{{\bf DVCS data:}
Quality factor normalised to 1 plotted against the parameter $\lambda$.
Star denotes the fit result for $F_2$ data.}
\label{DVCS}

\end{minipage}
\hfill
\end{figure}

We choose to consider all available data from H1, ZEUS, NMC and E665
experiments~\cite{Adloff:2000qk, Adloff:2003uh, Breitweg:1998dz, Breitweg:2000mu, Chekanov:2001qu, Chekanov:2003yv, Arneodo:1996qe, Adams:1996gu}
with $Q^2$ in the range $[1;150]$~GeV$^2$ and $x<0.01$\begin{footnote}{The data in the last ZEUS paper include contributions for $F_L$ 
and $xF_3$ but those can be neglected within the kinematical domain we 
consider.}\end{footnote}.
We exclude the data with $x>10^{-2}$ since they are dominated by the valence
quark densities, and the formalism of saturation does not apply in this kinematical
region. In the same  way, the upper $Q^2$ cut is introduced while the lower $Q^2$ cut
ensures that we stay away from the soft QCD domain. We will show
in the following that the data points with $Q^2<1$~GeV$^2$ spoil the fit stability.
Two kinds of fits to the scaling laws are performed, either 
in the full mentioned $Q^2$ range, or in a tighter $Q^2$ range $[3;150]$~GeV$^2$
to ensure that we are in the domain where perturbative QCD applies.


Figure~\ref{F2_QF_1} shows the scaling plot for ``Fixed Coupling" in the
$Q^2$ range $[1;150]$~GeV$^2$, which shows that the lowest $Q^2$ points in grey 
have a tendency to lead to worse scaling.
The QF values are similar for the ``Fixed Coupling", ``Running Coupling I",
and ``Running Coupling IIbis" --- with a tendency to be slightly better for ``Running Coupling
IIbis" --- and worse for diffusive scaling~\cite{Beuf:2008mf}.

The amount of the DVCS data~\cite{:2007cz, Aktas:2005ty} measured by H1 and ZEUS is smaller
(34 points for H1 and ZEUS requiring $x \le 0.01$ as for $F_2$ data), therefore the precision
on the $\lambda$ parameter is weaker.
The kinematic coverage of the DVCS data covers smaller region in $x$ and $Q^2$ than $F_2$:
$4<Q^2<25$ GeV$^2$ and $5\cdot10^{-4}<x<5\cdot10^{-3}$.
The DVCS data lead to similar
$\lambda$ values as in the $F_2$ data (see Fig.~\ref{DVCS}), showing the consistency of the
scalings. The values of the QF show a tendency to favour ``Fixed Coupling'', but all different
scalings (even ``Diffusive Scaling") lead to reasonable values of QF.

\begin{figure}[t]
\hfill
\begin{minipage}[t]{.45\textwidth}

\epsfig{file=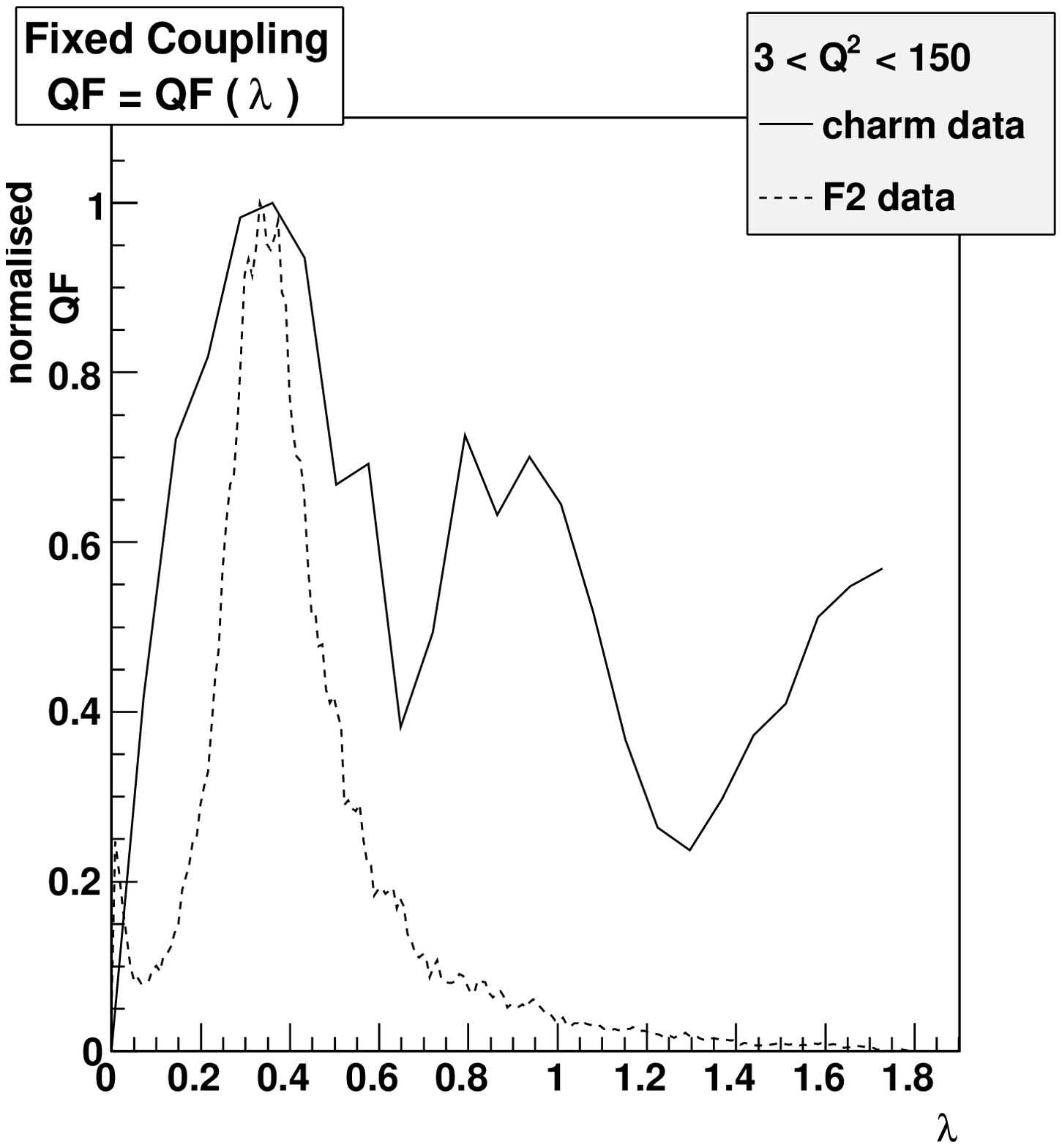,width=6.cm}
\caption{{\bf $F_2^{c}$ data:}
Comparison of the $\lambda$ parameter
for $F_2$ and $F_2^c$ data for $Q^2>3$ GeV$^2$.
}
\label{F2c_data}

\end{minipage}
\hfill
\begin{minipage}[t]{.45\textwidth}

\epsfig{file=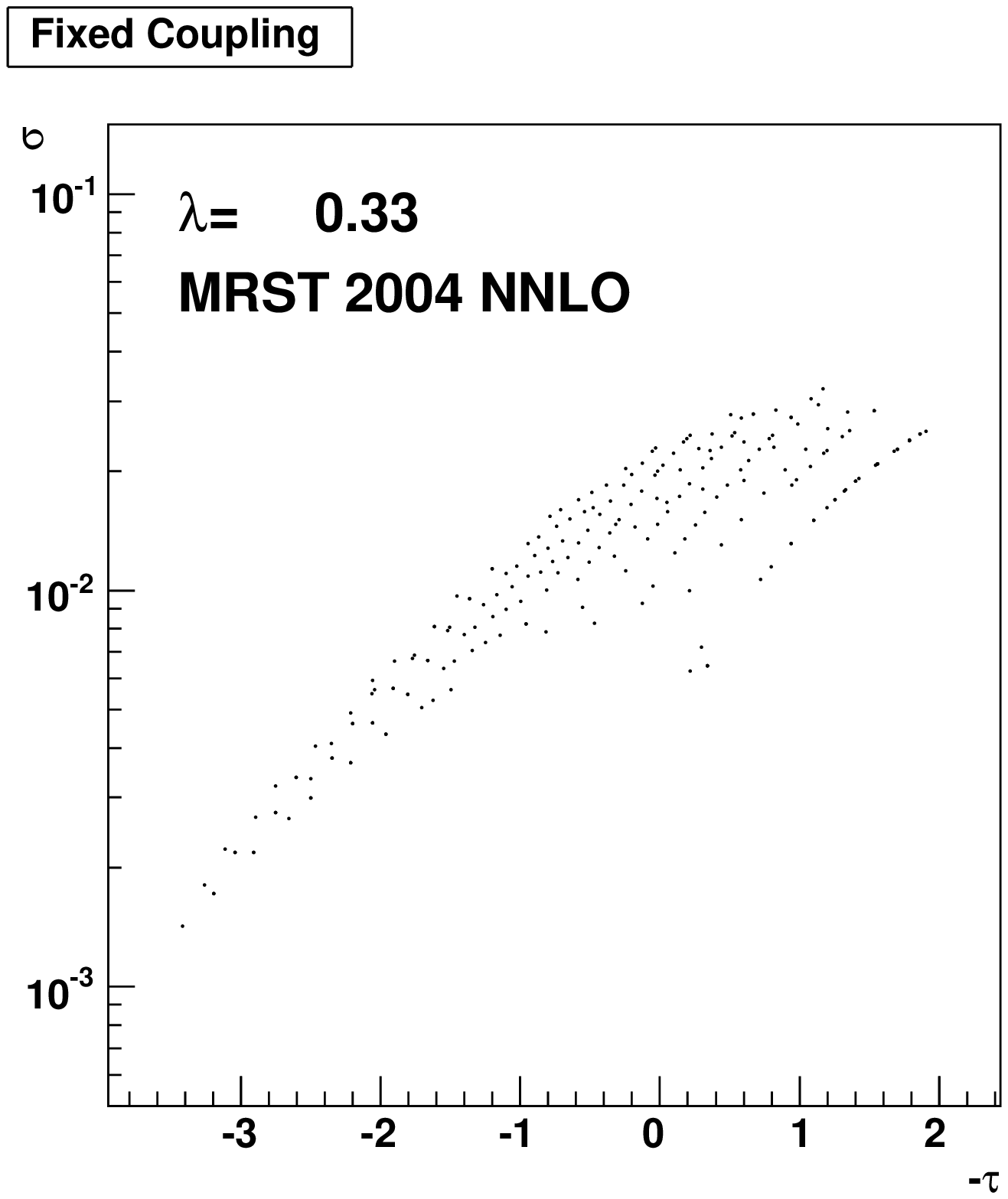,width=6.cm}
\caption{{\bf $F_2^c$ parametrisation:}
Scaling curve $\sigma=\sigma(\tau)$ for fixed coupling using the MRST 2004 NNLO
parametrisation for $\lambda=0.33$ as obtained in the fit to experimental data.
No scaling is observed for $Q^2>3$ GeV$^2$.}
\label{MRST033}

\end{minipage}
\hfill
\end{figure}


\paragraph{Implications for Diffraction and Vector Mesons}

We used the values of the parameters obtained from the fit to $F_2$ data to test the
various scaling variables on the diffractive cross section and vector meson
data~\cite{Aktas:2006hy, Chekanov:2005vv, Chekanov:2004hy}. We tested both the fixed
$\beta$ scaling behaviour in $x_{I\!\!P}$ and the fixed $x_{I\!\!P}$ scaling behaviour in
$\beta$. At fixed $\beta$, we find a scaling behaviour up to $\beta=0.65$.
At fixed $x_{I\!\!P}$, the scaling behaviour of the diffractive cross section as a
function of $\beta$ and $Q^2$ is far less obvious. This is not a surprise, as
not enough data is available in the genuine small $\beta$ region. A sign
of scaling is however observed for the $x_{I\!\!P}=0.03$ bin.

Concerning $\rho$, $J/\Psi$, and $\phi$ production~\cite{Chekanov:2005cqa, Aktas:2005xu, Adloff:1999kg}, we found a
reasonable scaling behaviour for all tested scaling variables, with the hard
scale $Q^2+M_V^2$, borrowed from vector mesons wave function studies.
Surprisingly, the best scaling is for all three vector mesons the ``Diffusive
scaling''.

\paragraph{Fits to $F_2$ and $F_2^c$ in QCD Parametrisations}

First we test the scaling properties using experimental $F_2^c$ data.
The requirements on the kinematical domain remain the same as in the case of $F_2$ studies.
The lower $Q^2>3$~GeV$^2$ cut also allows to remove eventual charm mass effects.
We use the charm $F_2^c$ measurements from the H1 and ZEUS experiments\cite{Adloff:1996xq,Adloff:2001zj,Breitweg:1997mj,Aubert:1982tt}.
Only 25 data points lie in the desired kinematical region.


Since the statistics in the data is low, the fit results are not precise.
Nevertheless, they still lead to clear results that are comparable to $F_2$ fits.
The results are found similar between $F_2$ and $F_2^c$ (see Fig.~\ref{F2c_data}).
All $\lambda$ parameters are similar for $F_2$ and $F_2^c$ except for ``Diffusive Scaling''.
As in the case of the $F_2$ scaling analysis, ``Fixed Coupling'', ``Running Coupling I''
and ``Running Coupling II'' give similar values of $QF$, and
``Diffusive Scaling'' is disfavoured.

The QCD parametrisations~\cite{Nadolsky:2008zw, Martin:2004ir, Gluck:1998xa} of the structure function have been tested using
CTEQ, MRST, GRV. The same $Q^2$ and $x$ points as in the experimental data
were taken into account. Parametrisations of $F_2$ are able to
reproduce the scaling results seen in the experimental data.
However, they are not successful in describing the scaling properties
in case of $F_2^c$. Fig.~\ref{MRST033} shows the scaling curve
of ``Fixed Coupling'' in the MRST NNLO 2004 parametrisation of $F_2^c$
where the value of $\lambda=0.33$ is imposed (as seen in the experimental data).
The scaling curve is plotted with all the points used in the $F_2$ study.
Therefore the fact that there is not just a single scaling curve
in $F_2^c$ parametrisation is not in direct disagreement with the data
--- with 25 point only, the curves in parametrisation and data look similar.
However the fit values of $\lambda$ are different.

The CTEQ, MRST or GRV parametrisations are unable to
reproduce the scaling properties in $F_2^c$. It seems a sea-like intrinsic charm
component like the one used in CTEQ 6.6 C4 helps to get results closer to
a single scaling curve~\cite{uscharm}.
Scaling is not present at all in the MRST or GRV parametrisations at low $Q^2$.


\subsubsection{Geometric scaling and evolution equations with
  saturation\protect\footnote{Contributing author: G. Beuf}}

Let us now recall how scaling properties arise from saturation, as shown in \cite{Munier:2003sj}, using methods and results from non-linear physics (see \cite{Iancu:2002tr,Mueller:2002zm} for alternative demonstrations). Our discussion, independent of the precise saturation formalism, is valid \emph{e.g.} for the JIMWLK and BK equations (see \cite{Iancu:2003xm} and references therein), at LL, NLL or even higher order in $\log (1/x)$. We will discuss separately the fixed and the running $\alpha_s$ cases, as running coupling is the main effect which can modify the discussion.

Saturation amounts to add a non-linear damping contribution to the BFKL evolution.
One writes formally the evolution equation at LL for the dipole-proton cross section $\hat{\sigma}$ (\ref{DipoleFact}) 
\begin{equation}
\partial_Y \hat{\sigma}(Y,L)=\bar{\alpha} \chi(-\partial_L) \hat{\sigma}(Y,L) - \textrm{non-linear terms in }\hat{\sigma}(Y,L)\label{eqBFKLsat}\ ,
\end{equation}
where $Y\equiv \log (1/x)$, $L\equiv-\log (r^2 \Lambda_{QCD}^2)$ and $\chi(\gamma)$ is the characteristic function of the BFKL kernel. The nonlinear damping ensures that, for any $Y$, $\hat{\sigma}(Y,L)$ grows at most as a power of $|L|$ for $L\rightarrow -\infty$ (\emph{i.e.} $r\rightarrow +\infty$). The color transparency property of the dipole cross section implies $\hat{\sigma}(Y,L)\propto e^{-L}$ for $L\rightarrow +\infty$.
Using a double Laplace transform with partial waves $e^{-\gamma L +\omega Y}$, the linear part of (\ref{eqBFKLsat}) reduces to the BFKL dispersion relation $\omega = \bar{\alpha} \chi(\gamma)$, which gives the partial waves solutions $e^{-\gamma [L -\bar{\alpha} \chi(\gamma) Y/\gamma]}$. In the relevant interval $0\!<\!\gamma\!<\!1$, the phase velocity $\lambda(\gamma) =  \bar{\alpha} \chi(\gamma)/\gamma$ has one minimum, for the critical value $\gamma=\gamma_c\simeq 0.63$ which is the solution of $\chi(\gamma_c)= \gamma_c \chi'(\gamma_c)$. In the presence of saturation terms in the evolution equation, the wave with $\gamma=\gamma_c$ is selected dynamically.

In order to understand the dynamics of the problem, let us consider an arbitrary initial condition, at some rapidity $Y=Y_0$. With the definition $\gamma_{eff}(L,Y)\equiv -\partial_L \log(\hat{\sigma}(Y,L))$, $\gamma_{eff}(L,Y_0)$ gives the exponential slope of the initial condition in the vicinity of $L$. That vicinity will then propagates for $Y\geq Y_0$ at a velocity $\lambda(\gamma_{eff}(L,Y))=\bar{\alpha} \chi(\gamma_{eff}(L,Y))/\gamma_{eff}(L,Y)$. One finds easily that, if $\gamma_{eff}(L,Y_0)$ is a growing function of $L$, the regions of smaller velocity will spread during the $Y$ evolution, and invade the regions of larger velocity. Restricting ourselves to initial conditions verifying the saturation at $L\rightarrow -\infty$ and the color transparency at $L\rightarrow +\infty$ as discussed previously, one obtains that $\gamma_{eff}(L,Y_0)$ goes from $0$ at low $L$ to $1$ at large $L$. At intermediate $L$, $\gamma_{eff}(L,Y_0)$ will cross the value $\gamma_c$, corresponding to the minimal velocity $\lambda_c=\lambda(\gamma_c)$. Hence, one conclude that, as $Y$ grows, there is a larger and larger domain in $L$ where $\gamma_{eff}(L,Y)=\gamma_c$ and thus $\lambda=\lambda_c$. In that domain, one has $\hat{\sigma}(Y,L)\propto e^{-\gamma_c (L-\lambda_c Y)}$, and hence the geometric scaling $\hat{\sigma}(Y,L)\equiv f(L\!-\!\lambda_c Y)=f(-\log(r^2 Q_s^2(x)))$, with a saturation scale $Q_s^2(x)= e^{\lambda_c Y} \Lambda_{QCD}^2=x^{-\lambda_c} \Lambda_{QCD}^2 $.
One finds that the geometric scaling window is limited to $L< \lambda_c Y + \sqrt{\bar{\alpha} \chi''(\gamma_c) Y/2}$, and separated from the region still influenced by the initial condition by a cross-over driven by BFKL diffusion.
So far, we discussed only scaling properties of the dipole cross section $\hat{\sigma}$. As explained in the introduction, they imply similar scaling properties of the virtual photon-proton cross section, with the replacement $r\mapsto 1/Q$.

The mechanism of wave selection explained above happens mainly in the linear regime\footnote{We call linear (non-linear ) regime the (Y,L) domain where the explicit value of the non-linear terms in (\ref{eqBFKLsat}) is (is not) negligible compared to the value of the linear terms.}, \emph{i.e.} for small $\hat{\sigma}$, or equivalently $r$ smaller than $Q_s^2(x)$. However, the geometric scaling property stays also valid in the non-linear regime, \emph{i.e.} for $r$ larger than $Q_s^2(x)$, which is reached after a large enough evolution in $Y$. The only, but decisive, role of saturation in the linear domain is to provide the following dynamical boundary condition in the IR to the linear BFKL evolution: when $\hat{\sigma}$ is large, it should be quite flat ($\gamma_{eff}(L)\simeq 0$). Indeed, one can simulate successfully the impact of saturation on the solution in the linear regime by studying the BFKL evolution in the presence of an absorptive wall \cite{Mueller:2002zm}, set at a $Y$-dependent and selfconsistently determined position near the saturation scale.

At NLL and higher order level, the terms different from running coupling ones do not affect the previous discussion. They just change the kernel eigenvalues $\chi(\gamma)$ and thus shift the selected parameters $\gamma_c$ and $\lambda_c$. On the contrary, going from fixed to running coupling brings important changes. As the mechanism of spreading of smaller velocity regions of the solution towards larger velocity ones is local, one expect that it holds in the running coupling case. But it selects coupling-dependent velocity and shape of the front, the coupling itself being $L$-dependent. Hence, the picture is the following. We still have the formation of a specific traveling wave front solution, which progressively loses memory of its initial condition. However, the selected values of the velocity and shape of the front drift as the front propagate towards larger $L$ (smaller $r$), due to asymptotic freedom.
So far, this running coupling case has been solved analytically \cite{Mueller:2002zm,Munier:2003sj} only at large $L$ and large $Y$, keeping the relevant geometric scaling variable $-\log(r^2 Q_s^2(x))$ finite. One finds that the evolution is slower than in the fixed coupling case, as the large $Y$ behavior of the saturation scale is now $Q_s^2(x)\sim e^{\sqrt{v_c Y/b}}\Lambda_{QCD}^2$, with $b\equiv(33-2 N_f)/36$ and $v_c\equiv 2 \chi(\gamma_c)/\gamma_c$. In addition, the geometric scaling window is narrower: asymptotically in $Y$, it is expected to hold only for\footnote{$\xi_1\simeq-2.34$ is the rightmost zero of the Airy function.} $L< \sqrt{v_c Y/b} +  (|\xi_1|/4)\  (\chi''(\gamma_c))^{1/3} Y^{1/6}/(2 b \gamma_c \chi(\gamma_c))^{1/6}$.
The convergence of the selected front towards this asymptotic solution seems rather slow, which may weaken its phenomenological relevance. The whole theoretical picture is nevertheless consistent with numerical simulations \cite{Gardi:2006rp,Albacete:2007yr}. Both leads to a universal traveling wave front structure of the solution, implying scaling properties also subasymptotically.

In order to do phenomenological studies, one can try to extrapolate to finite $L$ and $Y$ the scaling behavior found asymptotically. However, this extrapolation is not unique \cite{Beuf:2008mb}. There is indeed an infinite family of scaling variables
\begin{equation}
\tau_\delta \equiv \left[1 - \left(\frac{v_c Y}{b L^2}\right)^\delta\right]L, 
\end{equation}
parameterized by $\delta$, which are different from each other at finite $L$ and $Y$ but all converge to the same asymptotic scaling previously mentioned. The parameter $\delta$ seems quite unconstrained, both from the theory and from the DIS data, as shown in the phenomenological section of the present contribution. We considered as benchmark points in that family two specific choices of $\delta$. The choice $\delta=1/2$ leads to the only scaling variable of the family which is a genuine geometric scaling variable, \emph{i.e.} is equivalent to a scaling with $r^2 Q_s^2(x)$. It is named \emph{running coupling I} in the phenomenological section. The choice $\delta=1$ leads to the scaling variable obtained by substitution of the fixed coupling by the running coupling directly in the original fixed coupling geometric scaling variable. It is called \emph{running coupling II}.

Finally, one expects scaling properties in any case from evolution equations with saturation, both in the non-linear regime, and in a scaling window in the linear regime. In the linear regime, the solution still obey the linearized equation, and saturation play only the role of a dynamically generated boundary condition. Hence, geometric scaling there, although generated by saturation, is not a hint against the validity of PDF fits. However, geometric scaling occurs also in the non-linear regime, where the scaling function is no more a solution of the linear BFKL or DGLAP equations.

\subsubsection{DGLAP evolution and the saturation boundary
  conditions\protect\footnote{Contributing author: A. M. Sta\'sto}}
One of the issues  that could be studied in the context of the geometric scaling versus DGLAP evolution is the possibility of the different boundary conditions for the DGLAP evolution equations. These boundary conditions   would incorporate the saturation effects and posses the scaling property. Typically, in the standard approach, to obtain the solution to the linear DGLAP evolution equations, one imposes the initial conditions onto the parton densities at fixed value of $Q_0^2$ and then performs the evolution into the region of larger values of  $Q^2$. However, in the presence of saturation these might not be the correct boundary conditions for DGLAP equations.
 As mentioned earlier the saturation regime is specified by the critical line, the saturation scale $Q_s(x)$
which is a function of $x$ Bjorken and its value increases as the Bjorken $x$ decreases (or as we go to yet higher energies).
 In that case it seems legitimate to ask, what is the behavior of the DGLAP solutions when evolved from the saturation boundary $Q^2=Q_s^2(x)$ rather then from the fixed  scale $Q^2=Q_0^2$.
To answer this question we imposed \cite{Kwiecinski:2002ep} the boundary condition for the gluon density at the saturation scale $Q^2=Q_s^2$ which possesses the scaling property namely $\frac{\alpha_s}{2\pi} xg(x,Q^2=Q^2_s(x))=\frac{\alpha_s}{2\pi}r^0 x^{-\lambda}$ (in the fixed coupling case). The solution for the gluon density at small $x$ (at fixed coupling) which can be derived from solving the DGLAP equations with this boundary is given by
\begin{equation}
\frac{\alpha_s}{2\pi}\frac{xg(x,Q^2)}{Q^2} \sim \frac{\alpha_s}{2\pi} \bigg( \frac{Q^2}{Q_s^2(x)}\bigg)^{(\alpha_s/2\pi) \gamma_{gg} (\omega_0)-1}
\end{equation}
where $\gamma_{gg}$ is the gluon-gluon DGLAP anomalous dimension. This  solution clearly has the geometrical scaling property as it is only a function of $Q^2/Q_s^2(x)$.
It is interesting to note that there exists a critical value of the  exponent $\lambda$ of the saturation scale
which determines the  existence of scaling.
For example in the double leading logarithmic approximation the scaling is present for rather large values of the exponent $\lambda \ge 4 \alpha_s \pi /3$ whereas there is no scaling for smaller values of $\lambda$.  The formula shown above is however only approximate, as in the derivation we included only the leading behavior which should be dominant at asymptotically small values of $x$. At any finite value of $x$ the scaling will be mildly violated by the nonleading terms. We checked numerically that this is indeed the case, though the violation was very small.
 This analysis was extended for the case of the more realistic DGLAP evolution with the running coupling. As expected the presence of the scale violation due to the running coupling will lead to the violation of the scaling.
In this case the geometric scaling is only approximate with the solution for the gluon density given by
$$
\frac{\alpha_s(Q^2)}{2\pi}\frac{xg(x,Q^2)}{Q^2} \; \sim \; \frac{Q_s^2(x)}{Q^2} \bigg[ 1+\frac{\alpha_s(Q_s^2(x))}{2\pi b}\ln [Q^2/Q_s^2(x)]\bigg]^{b \gamma_{gg}(\lambda)-1} \; ,
$$
with $b$ being the beta function of the QCD running coupling.
The scaling here is present provided we have  $\alpha_s(Q_s(x)) \ln [Q^2/Q_s^2(x)]/(2\pi b) \ll 1$.  Thus the geometric scaling violating term can be factored out.

In summary, this  analysis shows that the geometric scaling property can be build into the DGLAP initial conditions, and that
the solution to the linear evolution equation which do not include the parton saturation effects can preserve the scaling even in the regime of high $Q^2$ values,
outside the saturation region.

\subsubsection{Geometric scaling from DGLAP
  evolution\protect\footnote{Contributing author: F. Caola}}

From the DGLAP point of view there is another possible explanation
for geometric scaling: the scaling behaviour can be generated by the evolution
itself, rather than being a preserved boundary condition. In fact, it is
possible to show \cite{Caola:2008xr} both analytically and numerically that
in the relevant HERA region approximate geometric scaling is a feature
of the DGLAP evolution. In order to see this, one has first to rewrite the
DGLAP solution as a function
of $t-\lambda(t,x) \log 1/x$ (``fixed-coupling scaling'') or
$t-\lambda(t,x) \sqrt{\log 1/x}$ (``running-coupling scaling'')\begin{footnote}
{The labels ``fixed-coupling'' or ``running-coupling'' are here
a bit misleading. In fact, all the results shown here
are obtained with the full running-coupling DGLAP solution.
We kept this notation only for comparison with saturation-based approaches.}
\end{footnote}. Then from the explicit form of the DGLAP solution it
follows that in the relevant kinematic region $\lambda(t,x)$ is approximatively
constant, leading to  $\sigma_{DGLAP}(t,x)\approx \sigma_{DGLAP}\left(t-t_s(x)
\right)$. Hence  approximate
geometric scaling in the HERA region is a feature of the DGLAP evolution.
Interestingly enough,
this DGLAP-generated geometric scaling is expected to hold also at large
$Q^2$ and relatively large $x$ (say $x\lsim 0.1$), in contrast with the
saturation-based geometric scaling which should be a
small~$x$, small (or at least moderate)~$Q^2$ effect.

\begin{figure}
\vskip-0.5cm
\centering
\epsfig{file=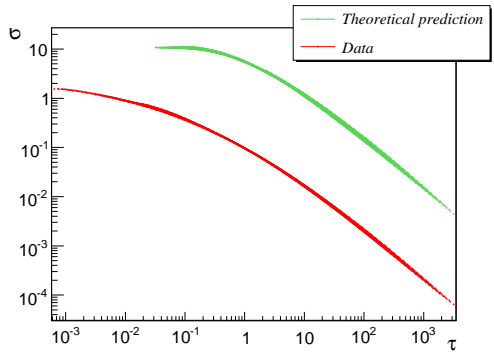,width=0.4\columnwidth}
\caption{Scaling plot with $x<0.1$. For the theoretical DGLAP curve,
only points with
$Q^2>1$~GeV$^2$ were kept. Curves are offset for clarity.
\label{phenoFix}}
\end{figure}

In order to make more quantitative statements, one can use the quality
factor method introduced in Sec. \ref{phenomenology}. As a
starting point, one can consider the leading-order small $x$
 DGLAP evolution of a flat boundary condition. At the level of accuracy
of geometric scaling, this approximation should be accurate enough in
a wide kinematic region, say
$Q^2\gsim 10$~GeV$^2$, $x\lsim 0.1$ at HERA. Now, a quality-factor analysis
shows
that in this region
the leading-order small $x$ DGLAP solution has an excellent scaling behaviour,
even better than the scaling behaviour observed in HERA data. Also the
DGLAP predictions for the geometric slope $\lambda$ perfectly agree with the
phenomenological values: from the DGLAP solution we obtain
$\lambda_{fix}^{DGLAP}=0.32\pm0.05$ (''fixed-
coupling'' scaling) and $\lambda_{run}^{DGLAP}=1.66\pm0.34$
(''running-coupling'' scaling),
to be compared with $\lambda_{fix}^{exp}=0.32\pm0.06$,
$\lambda_{run}^{exp}=1.62\pm0.25$. Moreover, data exhibit geometric scaling
also for larger $x$, larger $Q^2$ (say $x\lsim 0.1$ at HERA), as predicted
by the DGLAP evolution. All these results are summarized in Fig.~\ref{phenoFix},
 where
we plot the theoretical and phenomenological\begin{footnote}{In fact, in order
to make a more flexible analysis, we didn't use the actual HERA data but
a neural network interpolation of world DIS data \cite{DelDebbio:2004qj}. As long as one stays in the
HERA region the output of the net is totally reliable.} \end{footnote}
reduced cross sections in
function of the ''fixed-coupling'' scaling variable $\ln \tau=t-\lambda \ln 1/x$, with $\lambda=0.32$, in the HERA
region with the cut $x<0.1$. An analogous plot can be obtained for the
''running-coupling'' scaling \cite{Caola:2008xr}. We interpret these results
as striking evidence that for $Q^2>10$~GeV$^2$ the geometric scaling seen
at HERA is generated by the DGLAP evolution itself, without need of a
peculiar saturation ansatz or of a suitable scaling boundary condition.

For $Q^2<10$~GeV$^2$ the leading-order DGLAP solution exhibits violations
of geometric scaling at small $x$. However, in this region any
fixed-order DGLAP calculation fails because it does not resum small $x$
logarithms. If one consider the
DGLAP evolution at the resummed level, geometric scaling reappears quite
naturally, both in the ''fixed-coupling'' and ''running-coupling'' forms
\cite{Caola:2008xr}. Hence, small $x$ resummation
extends the region where geometric scaling is expected to values of $Q^2$
lower than  10~GeV$^2$. However at low $Q^2$ sizeable
higher twist and non perturbative effects can spoil the
universal behaviour of the DGLAP solution. In this region hence the HERA
scaling could still be generated by some DGLAP evolution, but, differently
 from the
$Q^2>10$~GeV$^2$ region, here there is no
 strong evidence that this is in fact the case.

\subsubsection{Saturation model and higher 
twists\protect\footnote{Contributing author: L.~Motyka}}

The QCD description of hard scattering processes within the Operator 
Product Expansion (OPE) approach leads to the twist expansion of matrix 
elements of process-dependent composite operators. Contributions of emerging 
local operators with the increasing twists, $\tau$, are suppressed by 
increasing inverse powers of the hard scale, $Q^2$.
In DIS, at the lowest order (i.e.\ when the anomalous dimensions vanish), 
the twist-$\tau$ contribution to the DIS cross section scales as 
$Q^{-\tau}$. Therefore, at sufficiently large $Q^2$ it is justified to 
neglect higher twist effects, and retain only the leading twist-2 
contribution. This leads to the standard collinear factorisation approach
with universal parton density functions evolving according to the DGLAP 
evolution equation. It should be kept in mind, however, that the higher twist 
effects do not vanish completely and that they introduce corrections to 
theoretical predictions based on the DGLAP approach. Thus, the 
higher twist corrections may affect the determination of parton density 
functions. The importance of these corrections depends
on the level of precision required and on the kinematic domain. In particular, 
in the region of very small~$x$ the higher twist effects are expected to 
be enhanced, so that they may become significant at moderate $Q^2$.
Thus, it should be useful to obtain reliable estimates of higher twist
effects at small~$x$. In this section we shall present higher twist 
corrections to $F_T$, $F_L$ and $F_2$ structure functions following from 
the DGLAP improved saturation model~\cite{Bartels:2002cj}. 
The results presented in this section have been obtained in the course of 
an ongoing study~\cite{Bartels:2008,Motyka:2008}. The method applied to perform the 
twist decomposition of the DGLAP improved saturation model is a generalisation
of the Mellin space approach proposed in Ref.~\cite{Bartels:2000hv}.

A rigorous QCD analysis of the higher twist contributions to DIS at high 
energies is a complex task. So far it has been performed for the $q\bar q gg$ 
operators~\cite{Ellis:1982cd}, but the evolution of twist~4 purely gluonic 
operators has not been resolved, --- even the proper complete basis of the 
operators has not been found yet. The collinear evolution is known at all 
twists, however, for so called {\em quasi-partonic operators}, for which the 
twist index is equal to the number of partons in the 
$t$-channel~\cite{Bukhvostov:1985rn}.
Such operators should receive the strongest enhancement from the QCD 
evolution. At the leading logarithmic approximation the collinear evolution 
of quasi-partonic operators is relatively simple --- it is given by 
pair-wise interactions between the partons in the $t$-channel.
The interactions are described by the non-forward DGLAP 
kernel~\cite{Bukhvostov:1985rn}.
Within this formalism, the evolution of four-gluon quasi-partonic 
operators was investigated in Ref.~\cite{Bartels:1993ke,Bartels:1993it} 
in the double logarithmic  approximation. 
At small~$x$ the scattering amplitudes are 
driven by exchange of gluons in the $t$-channel, and the quark exchanges
are suppressed by powers of~$x$. Thus we shall focus on the dominant 
contribution of the multi-gluon exchanges in the $t$-channel. 
In the large $N_c$-limit, the dominant singularities of the four 
gluon operator are those corresponding to states in which gluons get 
paired into colour singlet states. In other words,
the four-gluon operator evolves like a product of two independent gluon
densities. In general, for $1/N_c \to 0$, the $2n$-gluon (twist-$2n$) 
operator factorizes into the product of $n$ twist-2 gluon densities. 
After suitable inclusion of the AGK cutting rules and the 
symmetry factors of $1/n!$, one arrives at the eikonal picture of 
$n$-ladder exchange between the probe and the target. This is to be 
contrasted with the Balitsky-Kovchegov picture of Pomeron fan diagrams, 
which was obtained as a result of resummation of the terms enhanced by 
powers of large $\ln(1/x)$ rather than by powers of $\ln Q^2$.

The eikonal form of the multiple scattering was assumed in the saturation 
model proposed by Golec-Biernat and W\"{u}sthoff (GBW) 
\cite{GolecBiernat:1998js,GolecBiernat:1999qd}. 
The dipole cross-section given by Eq.~\ref{eq:gbw_dip} has a natural 
interpretation in terms of a resummation of multiple scattering amplitudes. 
The scatters are assumed to be independent of each other, and the 
contribution of $n$ scatterings is proportional to $\,[r^2 / R^2 _0(x)]^n\,$. 
The connection of the saturation model to the QCD evolution of quasi-partonic 
operators  is further strengthened by the DGLAP improvement of the dipole 
cross section~\cite{Bartels:2002cj}. In the DGLAP improved saturation model 
the dipole cross section depends on the collinear gluon density,
\begin{equation} 
\label{eq:dipimpr}
  \hat\sigma(x,r) = \sigma_0\left[
1-\exp\left(-\frac{\pi^2 r^2}{N_c \sigma_0}
\alpha_s(\mu^2)\,xg(x,\mu^2)\,\right) \right],
\end{equation}
where the scale $\mu^2$ depends on the dipole size, $\mu^2 = C/r^2$ for
$C/r^2 > \mu_0 ^2$, and  $\mu^2 = \mu_0^2$ for $C/r^2 < \mu_0 ^2$.
The gluon density applied has been obtained from the LO~DGLAP evolution
without quarks, with the input assumed at the scale $\mu_0 ^2$
\footnote{
In the original DGLAP-improved model\cite{Bartels:2002cj} 
a different definition of the scale was adopted, 
$\mu^2 =  C/r^2 + \mu_0 ^2$, but this choice 
is less convenient for the QCD analysis.}. 
Clearly, in Eq.~(\ref{eq:dipimpr}) one sees an exact matching between the 
power of $r^2$ and the power of $xg(x,\mu^2)$ suggesting a correspondence 
between the term $\sim [r^2 \alpha_s(\mu^2)\,xg(x,\mu^2)]^n$ in the 
expansion of $\hat\sigma(x,r)$ and the twist-$2n$ contribution to the 
dipole cross section. Thus, we expect that the saturation model 
approximately represents higher twist contributions in the deep inelastic
scattering generated by the gluonic quasi-partonic operators.

The twist analysis of the DIS cross-section must include a treatment of 
the quark box that mediates the coupling of the virtual photon, $\gamma^*$, 
to the $t$-channel gluons. In the dipole model the $\gamma^* g \to q\bar q$ 
amplitude, computed within QCD, 
is Fourier transformed (w.r.t.\ the transverse  momentum of the quark) 
to the coordinate representation and appears as the photon wave function, 
compare Eq.~(\ref{eq:gbw_dip}). In more detail, one uses the $\gamma^* g$ 
amplitude computed within the $k_T$-factorisation framework. 
This amplitude receives contributions from all twists. The twist structure
of the quark box is transparent in the space of Mellin moments, and the same 
is true for the dipole cross-section. Thus we define,
\begin{equation} 
\label{eq:Mellin}
\tilde H_{T,L} (\gamma,Q^2) =  \int_0 ^1 dz \int_0 ^\infty {dr^2} \; 
r^2 \, \left| \Psi_{T,L} (r,z,Q^2) \right| ^2 \; r^{2(\gamma-1)}\, ,
\end{equation}
\begin{equation} 
\tilde{\hat\sigma}(x,\gamma) =  \int_0 ^\infty {dr^2} \, \hat\sigma(x,r^2) \,
r^{2(\gamma-1)}\, .
\end{equation}
It then follows from the Parsival formula that,
\begin{equation} 
\label{eq:parsival}
\sigma_{T,L}(x,Q^2) = \int_{\cal C} {d\gamma \over 2\pi i} \, 
\tilde H_{T,L}(-\gamma,Q^2) \, \tilde{\hat \sigma}(x,\gamma).
\end{equation}
For the massless quark case one has 
$\tilde H_{T,L}(\gamma,Q^2) = \tilde H_{T,L} (\gamma)\,  Q^{-2\gamma}$. 
The contour of integration, ${\cal C}$, in Eq.~\ref{eq:parsival} belongs to 
the fundamental Mellin strip, $-1<\mathrm{Re}\, \gamma <0$.

In order to obtain the twist expansion of $\sigma$, one extends the 
contour ${\cal C}$ in the complex $\gamma$-plane into a 
contour ${\cal C}'$ closed from the left-hand side. 
The Mellin integral in Eq.~\ref{eq:parsival} may be then decomposed into
contributions coming from singularities of 
$\tilde H_{T,L}(-\gamma,Q^2) \, \tilde{\hat \sigma}(x,\gamma)$.
The function $\tilde H_{T} (-\gamma)$ ($\tilde H_{L} (-\gamma)$) 
has simple poles at all negative integer values of $\gamma$, except 
of $\gamma = -2$ ($\gamma = -1$), where $\tilde H_{T}$  ($\tilde H_{L}$) 
is regular. The singularity structure of the dipole cross section, 
$\tilde{\hat\sigma}(\gamma)$, depends on the specific form of 
$\hat\sigma(x,r^2)$. For $\hat\sigma(x,r^2)$ used in the GBW model, 
the  $\tilde{\hat\sigma}(x,\gamma)$ has simple poles at all negative 
integers $\gamma$'s. For the DGLAP improved form of $\hat \sigma$ given 
by (\ref{eq:Mellin}), $\tilde{\hat \sigma}(x,\gamma)$ has cut singularities 
that extend to the left from $\gamma=k$ where $k=-1,-2,\;$~etc. 
The leading behaviour of  $\tilde{\hat \sigma}$ around a branch
point at $\gamma=k$ is given by $\sim (\gamma-k)^{p(k)}$, where the 
exponent~$p(k)$ is generated by the DGLAP evolution. 
As the cuts extend to the left from the branch points, the dominant
contribution to the cross section at the given twist comes from the 
vicinity of the corresponding branch point.

The singularity structure of the quark box part $\tilde H_{T,L}(\gamma)$ 
plays the crucial role in understanding the strength of the subleading
twist effects. To see that one expands $\tilde H_{T,L}(\gamma)$ 
around the singular points, $\gamma = 1$ and $\gamma = 2$ 
(recall that the argument of $\tilde H_{T,L}$ is $-\gamma$
in the Parsival formula (\ref{eq:parsival})):
\begin{equation} 
\label{eq:htlexp2}
\tilde H_{T}(\gamma) = {a_T ^{(2)} \over \gamma-1} +  b_T ^{(2)} 
+ {\cal O}(\gamma-1), \qquad
H_{L}(\gamma) = b_L ^{(2)} + {\cal O}(\gamma-1),
\end{equation}
for twist-2, and
\begin{equation} 
\label{eq:htlexp4}
\tilde H_{T}(\gamma) = b_T ^{(4)} + {\cal O}(\gamma-2), \qquad
H_{L}(\gamma) = {a_L ^{(4)} \over \gamma-2} + b_L ^{(4)} + {\cal O}(\gamma-2),
\end{equation}
for twist-4.
The singular $1/(\gamma-1)$ and  $1/(\gamma-2)$ terms in (\ref{eq:htlexp2})
and (\ref{eq:htlexp4}) generate an additional enhancement, 
$\sim\ln(Q^2)$, of the corresponding twist-2 and twist-4 contributions to 
the DIS cross-section. The constant pieces, proportional to 
$b_{T,L} ^{(2)}$ and $b_{T,L} ^{(4)}$, produce no new 
logarithms (thus they are interpreted as the next-to-leading order 
(NLO) QCD corrections) and the higher terms in the Laurent expansion give 
yet higher orders in the perturbative expansion of the $g\to q$ splitting 
functions and to the coefficient functions. 
We summarize this discussion by displaying 
below the most leading contributions to $\sigma_{T,L}$ at twist-2
($\sigma^{(2)}_{T,L}$) and at twist-4 ($\sigma^{(4)}_{T,L}$) obtained
in the DGLAP improved saturation model:
\begin{equation} 
\sigma^{(2)} _T \; \sim  \; 
{a^{(2)} _T \over Q^2} 
\int_{\mu_0^2} ^{Q^2} {dQ'^2 \over Q'^2} \alpha_s(Q'^2)xg(x,Q'^2) \, ,
\qquad
\sigma^{(2)} _L \; \sim \;
{b^{(2)} _L\over Q^2} \, \alpha_s(Q^2)xg(x,Q^2)\, ,
\end{equation}
for twist-2, and
\begin{equation} 
 \sigma^{(4)} _T \; \sim  \; 
{b^{(4)} _T \over Q^4} [\alpha_s(Q^2)xg(x,Q^2)]^2 \, ,
\qquad
\sigma^{(4)} _L \; \sim \;
{a^{(4)} _L\over Q^4} \,
\int_{\mu_0^2} ^{Q^2} {dQ'^2 \over Q'^2} [\alpha_s(Q'^2)xg(x,Q'^2)]^2\, ,
\end{equation}
for twist-4. These results imply that the 
the relative twist-4 correction to $F_T$ is strongly suppressed w.r.t.\ the 
twist-2 contribution, as the subleading twist-4 term in $F_T$ appears
only at the NLO. On the contrary, for $F_L$, the leading twist term
enters only at the NLO, and the the twist-4 correction enters 
at the leading order. So, the relative twist-4 effects in $F_L$ are expected to
be enhanced. Note, that both in the case of  $F_T$ and $F_L$ the  
twist-4 effects are enhanced w.r.t.\ the twist-2 contribution by an
additional power of the gluon density, $xg(x,Q^2)$.
For the structure function $F_2 = F_T + F_L$ we expect small relative 
corrections from the higher twists because of the opposite
sign of coefficients $a^{(4)}_{L}$ and $b^{(4)}_{T}$, that leads to
cancellations between the twist-4 contributions from $F_T$ and $F_L$
at moderate $Q^2$. These conclusions about the importance of the higher
twist corrections are expected to be quite general, because they 
follow directly from the twist structure of the quark box and do not 
depend on the detailed form of the twist-4 gluon distribution.

\begin{wrapfigure}{l}{0.4\columnwidth}
\label{fig:twist}
\centerline{\epsfig{file=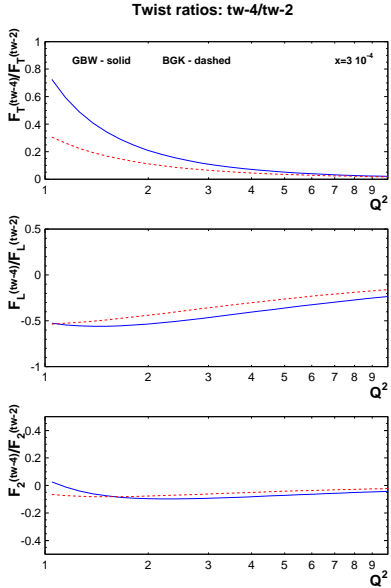,width=0.3\columnwidth}}
\caption{The ratio of twist-4 to twist-2 components of 
$F_T$, $F_L$ and $F_2$ at $x=3\cdot 10^{-4}$ in the GBW model
(continuous lines) and in the DGLAP improved saturation model
(dashed lines).
}
\end{wrapfigure}

We performed\cite{Bartels:2008,Motyka:2008} 
an explicit numerical evaluation of the twist-4  
corrections to $F_T$, $F_L$ and $F_2$ in the DGLAP improved saturation
model, and compared the results to results obtained~\cite{Bartels:2000hv}  
within the GBW model without the DGLAP evolution. 
The parameters of the DGLAP model were fitted
to describe all $F_2$ data at small~$x$. In the model we took into account
three massless quark flavours and the massive charm quark. The twist analysis,
however, has been, so far, performed only for the massless quark contribution.
The obtained relative twist-4 corrections to $F_T$, $F_L$ and $F_2$
are displayed in Fig.~\ref{fig:twist}, as a function of $Q^2$, 
for $x=3\cdot 10^{-4}$. 
The continuous curves correspond to the GBW model~\cite{Bartels:2000hv},
and the dashed ones have been obtained~\cite{Bartels:2008,Motyka:2008} 
in the DGLAP improved saturation model.
Although there are some quantitative differences between the models,
the qualitative picture is quite consistent and confirms
the results of the analytic analysis outlined above. 
Thus, the higher twist corrections are strongest in $F_L$, and much 
weaker in $F_T$. In $F_2$ there occurs a rather fine cancellation between 
the twist-4 contributions to $F_T$ and $F_L$, at all $Q^2$, 
down to 1~GeV$^2$. Although an effect of this kind was expected, it still
remains somewhat surprising that this cancellation works so well.
We estimate that, for $x=3\cdot 10^{-4}$,  the twist-4 relative 
correction to $F_2$ is $2\,$--$\,4$\% at $Q^2=10$~GeV$^2$, and 
smaller than 10\% for all $Q^2$ down to 1~GeV$^2$.
For $F_L$, the relative correction is $\sim 20\%$ at $Q^2=10$~GeV$^2$, and
strongly increases with the decreasing scale, reaching $\sim 50\%$ at
$Q^2=1$~GeV$^2$. It implies that the determination of parton
densities from twist-2 $F_2$ data is safe even at small~$x$ and moderate
$Q^2$. On the other hand $F_L$ at small~$x$ may provide a sensitive 
probe of higher twist effects and parton saturation.

\subsubsection{Conclusions}
There are many possible explanations for the scaling properties of HERA data, 
some of them based on saturation effects and some others based on pure 
linear evolution. In order to separate between these different explanations, 
it is fundamental to specify a kinematic window. 

In particular, for large 
enough $Q^2$ and not too small $x$ (say $Q^2\gsim 10$~GeV$^2$ in the HERA 
region) the observed geometric scaling is determined by the DGLAP evolution,
irrespective of the boundary condition. For smaller values of $Q^2$, the evolution 
of parton densities is still linear, but is sensitive to a boundary condition. 
In an evolution toward smaller $x$, like BFKL, this boundary condition is dynamically 
generated by saturation, and it leads to the geometric scaling window. 
It is possible to take these effects 
into account also in a  $Q^2$ evolution, like DGLAP, by imposing as initial 
condition the same boundary condition. We have seen that, in this case, even 
the LO DGLAP equation is able to propagate geometric scaling towards larger $Q^2$. 
In that domain, although geometric scaling may arise as saturation effect, 
the evolution is still linear, and thus compatible with standard PDFs analysis.
However, at yet lower $Q^2$ and $x$ standard linear evolution is no longer 
reliable. In particular, for $Q^2$ smaller than a $x$ dependent saturation
scale $Q_s(x)$,  the evolution of parton densities becomes fully nonlinear, 
and this spoils the actual determination of the PDFs. Results from 
inclusive diffraction and vector meson exclusive production at HERA, and from 
dA collisions at RHIC all suggest that in the kinematic accessible $x$ region 
$Q_s\sim1-2$~GeV. 

In conclusion, we can say that for large enough $Q^2\gsim 10$~GeV$^2$ geometric
scaling is fully compatible with linear DGLAP evolution. For smaller 
$Q^2$ the situation becomes more involved. For $Q^2\gsim 5$~GeV$^2$ 
 the HERA scaling is still compatible with DGLAP, maybe with some small $x$
resummation or some suitable boundary condition. However, other effects may
be relevant in this region. For yet lower $Q^2$ and $x$ the linear theory 
becomes unreliable and saturation could be the right explanation for 
geometric scaling. Unfortunately at HERA we have too few data for a definitive 
explanation of geometric scaling in the very small $x$ region, since 
many different approaches lead approximatively to the same results and 
it is very difficult to separate among them. For example, in the low $x$ region 
both saturation and perturbative resummations lead to a decrease of the 
gluon and to geometric scaling. 
At the LHC, where higher center-of-mass energy is available, the $x$ region is significantly 
extended down to very small values.  Especially in the fragmentation region the typical 
values of $x$ which can be probed can reach down to $10^{-6}$ for partons with transverse 
momenta of about few  GeV.  This fact combined with the very wide rapidity coverage of 
the main LHC detectors opens up a completely new window for the study of 
parton saturation, and its relations with geometric scaling and linear evolution
will possibly be clarified.

\pagebreak

\section{BENCHMARKING OF PARTON DISTRIBUTIONS AND THEIR
  UNCERTAINTIES\protect\footnote{ Contributing
authors:  
R.~D.~Ball, L.~Del~Debbio, J.~Feltesse, 
S.~Forte, A.~Glazov, A.~Guffanti, J.~I.~Latorre,
A.~Piccione, V.~Radescu, J.~Rojo, R.~S.~Thorne, M.~Ubiali, G.~Watt
}}
\label{sec:pdfunc}
\subsection{Introduction}

The proper treatment of  uncertainties associated to the fit of Parton
Distribution Functions (PDF) has become a subject of great interest
in the last few years. A simple way of understanding differences between
available approaches to parton fits is to fix some hypothesis (say,
experimental data, QCD parameters, input parameterizations, error
treatment), and check what is the effect of the remaining assumptions. 
Such studies were previously done in the framework of the first
HERA--LHC workshop~\cite{Dittmar:2005ed}.

In the following we will discuss three benchmark fits. The first one is
presented in Sect.~\ref{sec:h1appr}. It is 
based on the  H12000 parton fit~\cite{Adloff:2003uh},
and it compares a  new version of this fit, in which  uncertainty bands are
determined~\cite{Giele:1998gw,Giele:2001mr}  using a Monte
Carlo method, to the reference fit, where uncertainty bands are obtained using 
the standard Hessian 
method. The main motivation of this
benchmark is to study the impact of  possible non-Gaussian behaviour of
the data and, more generally, the dependence on the error treatment. 

The second benchmark is presented in Sect.~\ref{sec:heralhc}. It
is based on the study performed 
by S.~Alekhin
and R.~Thorne in Ref.~\cite{Dittmar:2005ed}, which compared the 
fits by their respective groups to a common reduced set of data with
common assumptions, and also to
their respective reference (global) fits. 
This comparison is extended here in two ways. First, the comparison is
extended to include an  
NNPDF fit to the same reduced set of data with the same assumptions, 
and the NNPDF1.0
reference fit~\cite{Ball:2008by}. Second, results are
also compared to
a fit based on the recent MSTW 2008~\cite{Watt:2008hi,Thorne:2007bt}
analysis. As in the Thorne benchmark fit, this uses  slightly different 
data sets and assumptions; it is furthermore
modified to use the same  input   
parameterization and improved treatment of uncertainties as  MSTW.
The main purpose of these comparisons  is to answer the
questions (a) to which extent fit results from various groups
obtained using different methodologies still
differ from each other when  
common or similar assumptions and a common or similar reduced dataset are  used and (b) how the fits to the reduced
dataset by each group compare to the fit to the full dataset. 
 
The third benchmark, discussed  in Sect.~\ref{sec:h1bench},
is a further elaboration on the benchmark presented in
Sect.~\ref{sec:h1appr}, extended to 
include the NNPDF fit, which also uses a Monte Carlo
approach. The main purpose of this benchmark is  to compare two
fits (H1 and NNPDF) which have the same error treatment but different
parton parameterizations. The 
inclusion in this benchmark of  the NNPDF fit  is also interesting because
it allows a comparison of a fit
based on a very consistent  set of data coming
from the H1 collaboration only, to fits which include all 
DIS data sets,
which are less compatible than the H1 sets alone.

\subsubsection{Settings for the H1 benchmark}
\label{sec:h1set}

\begin{figure}[ht]
\begin{center}
\includegraphics[scale=.4]{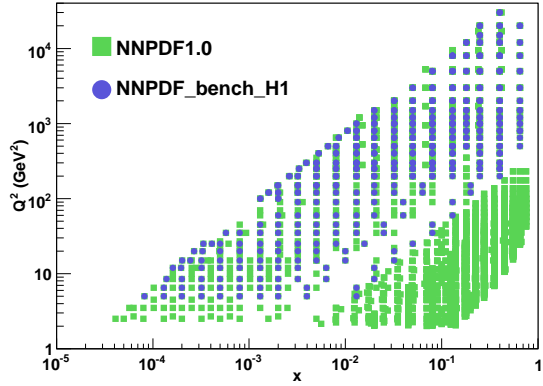}
\caption{The data used in the H1 benchmark and in the NNPDF
  reference fit.\label{fig:bench2_data}
}
\end{center}
\end{figure}

This analysis is based on all the DIS inclusive data by the
H1 collaboration from the HERA-I run. A kinematic cut of $Q^2 >
3.5~{\rm GeV}^2$ is applied to avoid any higher twist effect. The data
points used in the analysis are summarized in
Table~\ref{tab:bench2_data} and Fig.~\ref{fig:bench2_data}. 

\begin{table}[ht]
  \begin{center}
    \begin{tabular}{|l|r|l|l|}
      \hline
      Data Set & Data points & Observable & Ref. \\
      \hline
      H197mb & 35   & $\tilde{\sigma}^{NC,+}$ &  \cite{Adloff:2000qk}\\
      H197lowQ2 & 80 & $\tilde{\sigma}^{NC,+}$ &  \cite{Adloff:2000qk}\\
      H197NC    & 130  & $\tilde{\sigma}^{NC,+}$ & \cite{Adloff:1999ah}\\
      H197CC    & 25  & $\tilde{\sigma}^{CC,+}$ & \cite{Adloff:1999ah}\\
      H199NC    & 126  & $\tilde{\sigma}^{NC,-}$ & \cite{Adloff:2000qj}\\
      H199CC    & 28  & $\tilde{\sigma}^{CC,-}$ & \cite{Adloff:2000qj}\\
      H199NChy   & 13  & $\tilde{\sigma}^{NC,-}$ & \cite{Adloff:2000qj}\\
      H100NC    & 147 & $\tilde{\sigma}^{NC,+}$ & \cite{Adloff:2003uh}\\
      H100CC    & 28  & $\tilde{\sigma}^{CC,+}$ & \cite{Adloff:2003uh}\\
      \hline
      Total & 612 & \multicolumn{2}{l}{}\\
      \cline{1-2}
    \end{tabular}
  \end{center}
\caption{Data points used in the H1 benchmark after kinematic cuts 
of $Q^2> 3.5~{\rm GeV}^2$.\label{tab:bench2_data}
}
\end{table}

The theoretical assumptions are:
\begin{itemize}
\item NLO perturbative QCD in the $\overline{\rm MS}$ renormalization
and factorization scheme;
\item zero-mass variable flavour number scheme with quark masses
$m_c=1.4~{\rm GeV}$ and $m_b=4.5~{\rm GeV}$;
\item the strong coupling  fixed to
$\alpha_s (M_Z)=0.1185$; 
\item momentum and valence sum rules enforced;
\item starting scale for the evolution at $Q_0^2= 4~{\rm GeV}^2$;
\item  
strange contribution fixed as
\begin{eqnarray}
s(x, Q_0^2) = \bar{s}(x, Q_0^2)=f_s \bar{D}(x, Q_0^2)
=\frac{f_s}{1-f_s}\bar{d}(x, Q_0^2),
\end{eqnarray}
with $U=u+c$ and $D=d+s+b$ and
with $f_s=0.33$; 
\item  charm contribution fixed as
\begin{eqnarray}
c(x, Q_0^2) = \bar{c}(x, Q_0^2)=f_c \bar{U}(x, Q_0^2)
=\frac{f_c}{1-f_c}\bar{u}(x, Q_0^2),
\end{eqnarray}
with $f_c=0.15$;
\item five independent PDFs: gluon and $U$, $D$, $\bar U$, $\bar D$
  (see definition above);
\item iterated solution for  evolution (see,
e.g.~\cite{Giele:2002hx}, Sect. 1.3).
\end{itemize}

Both the H1 and NNPDF methodologies are based on
\begin{itemize}
\item Monte Carlo method to determine uncertainties.
This method will be discussed in detail in Sect.~\ref{sec:mcmethod} below.

\end{itemize}
They differ in the way PDFs are parameterized: 
\begin{itemize}
\item H1 parameterizes PDFs as 
\begin{eqnarray}
x g(x, Q_0^2) &=& A_g x^{B_g}(1-x)^{C_g} [1+D_g x]\,,\nonumber\\
x U(x, Q_0^2) &=& A_U x^{B_U}(1-x)^{C_U} [1+D_U x+F_U x^3]\,,\nonumber\\
x D(x, Q_0^2) &=& A_D x^{B_D}(1-x)^{C_D} [1+D_D x]\,,\\
x \bar{U}(x, Q_0^2) &=& A_{\bar{U}} x^{B_{\bar{U}}}(1-x)^{C_{\bar{U}}}\,,\nonumber\\
x \bar{D}(x, Q_0^2) &=& A_{\bar{U}} x^{B_{\bar{D}}}(1-x)^{C_{\bar{D}}}\,,\nonumber\\
\end{eqnarray}
which yields 10 free parameters after sum rules are imposed;
\item NNPDF parameterizes PDFs with a 2-5-3-1 neural network, 
which implies 185 free parameters to be fitted.
\end{itemize}
Because of the large number of parameters, the minimum of the NNPDF
fit is determined using the stopping criterion discussed in
Sect.~\ref{sec:bench_intro} below, 
while the minimum of the H1 fit is determined as the
standard minimum $\chi^2$ (or maximum likelihood) point of parameter space.

\subsubsection{Settings for the HERA--LHC benchmark}
\label{sec:settings}

\begin{figure}[ht]
\begin{center}
\includegraphics[scale=0.4]{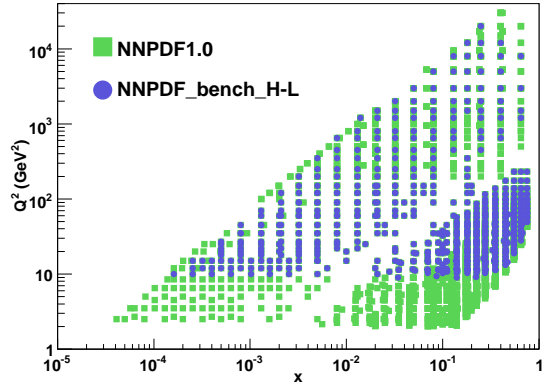}
\caption{The data used in the HERA--LHC benchmark and in the NNPDF
  reference fit.\label{fig:bench1_data}
}
\end{center}
\end{figure}

This benchmark was first presented in Ref.~\cite{Dittmar:2005ed},
where its settings were defined.
In order to have a conservative ensemble of experimental
data and observables, only structure function DIS data
are used. Large kinematic cuts are applied to avoid any higher 
twist effect. The data points used in the Alekhin analysis 
are summarized in Table~\ref{tab:bench1_data} and Fig.~\ref{fig:bench1_data}. 

\begin{table}[ht]
  \begin{center}
    \begin{tabular}{|l|r|l|l|}
      \hline
      Data Set & Data points & Observable & Ref.\\
      \hline
      ZEUS97  & 206       & $F_2^{p}$ & \cite{Chekanov:2001qu} \\
      H1lowx97  & 77       & $F_2^{p}$ & \cite{Adloff:2000qk} \\
      NMC     & 95       & $F_2^p$      & \cite{Arneodo:1996qe} \\
      NMC\_pd & 73       & $F_2^d/F_2^p$  & \cite{Arneodo:1996kd}\\
      BCDMS   & 322       & $F_2^{p}$   &  \cite{Benvenuti:1989rh}  \\
      \hline
      Total & 773 & \multicolumn{2}{l}{}\\
      \cline{1-2}
    \end{tabular}
  \end{center}
\caption{Data points used in the HERA--LHC benchmark after kinematic cuts 
of $Q^2> 9~{\rm GeV}^2$ and $W^2> 15~{\rm GeV}^2$ are applied. \label{tab:bench1_data}
}
\end{table}

The theoretical assumptions are:
\begin{itemize}
\item NLO perturbative QCD in the $\overline{\rm MS}$ renormalization
and factorization scheme;
\item zero-mass variable flavour number scheme with quark masses
$m_c=1.5~{\rm GeV}$ and $m_b=4.5~{\rm GeV}$;
\item $\alpha_s (M_Z)$ fitted: the best-fit values are
$0.1110\pm0.0012$ (Alekhin) and $0.1132\pm0.0015$ (Thorne);
\item momentum and valence sum rules imposed;
\item starting scale for  evolution $Q_0^2= 1~{\rm GeV}^2$;
\item four independent input PDFs ($u$ and $d$ valence, the sea and
  the gluon);
\item no light sea asymmetry: $\bar{u}=\bar{d}$;
 \item no independent  strange PDF:
\begin{eqnarray}
s(x, Q_0^2) +\bar{s}(x, Q_0^2)=0.5 (\bar{u}(x, Q_0^2)+\bar{d}(x, Q_0^2))\,;
\end{eqnarray}
\item iterated solution of evolution equations;
\end{itemize}

The NNPDF analysis presented here is based on the same data set and 
theoretical
assumptions, the only difference being that the
strong coupling is fixed to $\alpha_s (M_Z)=0.112$,
i.e.~the average of the fitted values of S.~Alekhin and R.~Thorne.

\begin{table}[ht]
  \begin{center}
    \begin{tabular}{|l|r|l|l|}
      \hline
      Data Set & Data points & Observable & Ref.\\
      \hline
      ZEUS97  & 206       & $\tilde{\sigma}^{{\rm NC},+}$ & \cite{Chekanov:2001qu} \\
      H1lowx97  & 86       &$\tilde{\sigma}^{{\rm NC},+}$ & \cite{Adloff:2000qk} \\
      NMC     & 67     & $F_2^p$      & \cite{Arneodo:1996qe} \\
      NMC\_pd & 73       & $F_2^d/F_2^p$  & \cite{Arneodo:1996kd}\\
      BCDMS   & 157       & $F_2^{p}$   &  \cite{Benvenuti:1989rh}  \\
      \hline
      Total & 589 & \multicolumn{2}{l}{}\\
      \cline{1-2}
    \end{tabular}
  \end{center}
\caption{Data points used in the MSTW benchmark fit after kinematic cuts 
of $Q^2> 9~{\rm GeV}^2$ and $W^2> 15~{\rm GeV}^2$ are
applied. \label{tab:benchmstw_data} 
}
\end{table}

The Thorne benchmark used somewhat different data sets and assumptions.
Namely:
\begin{itemize}
\item  A somewhat different dataset is used, as displayed in
  Table~\ref{tab:benchmstw_data}. This differs from the dataset of
  Table~\ref{tab:bench1_data} and Figure~\ref{tab:bench1_data} because
  the NMC and BCDMS fixed-target data on $F_2^p$ 
used are averaged over different beam energies, and also,
 HERA reduced cross sections rather than structure function data 
are used,
resulting in an additional nine H1 points. Note that the Thorne
benchmark in Ref.~\cite{Dittmar:2005ed} also included the $F_2^d$ BCDMS deuterium data. 
\item All correlations between systematics are neglected, and
  statistical and systematic errors are added in quadrature.
\item Normalizations of individual data sets are fitted 
with
a rescaling of uncertainties to avoid systematic bias. 
\item The $F_2^d/F_2^p$ data are corrected for nuclear 
shadowing effects~\cite{Badelek:1994qg}. 

\end{itemize}

The MSTW analysis presented here makes the same choices as the Thorne
benchmark, but with $\alpha_s (M_Z)=0.112$, and additionally
\begin{itemize}
\item  a  global correction of $-3.4\%$ is applied to the luminosity of the 
published H1 MB 97 data~\cite{Adloff:2000qk} following a luminosity 
reanalysis~\cite{VargasTrevino:2007zz}.
\item a quartic penalty term in the $\chi^2$ definition is given to
  normalizations which deviate from the central value.
\end{itemize}

\subsection{Experimental Error Propagation\protect\footnote{Contributing
authors: J.~Feltesse, A.~Glazov, V.~Radescu}}
\label{sec:h1appr}

\subsubsection{Introduction}
\begin{figure}[t]
\begin{center}
\includegraphics[width=0.6\linewidth]{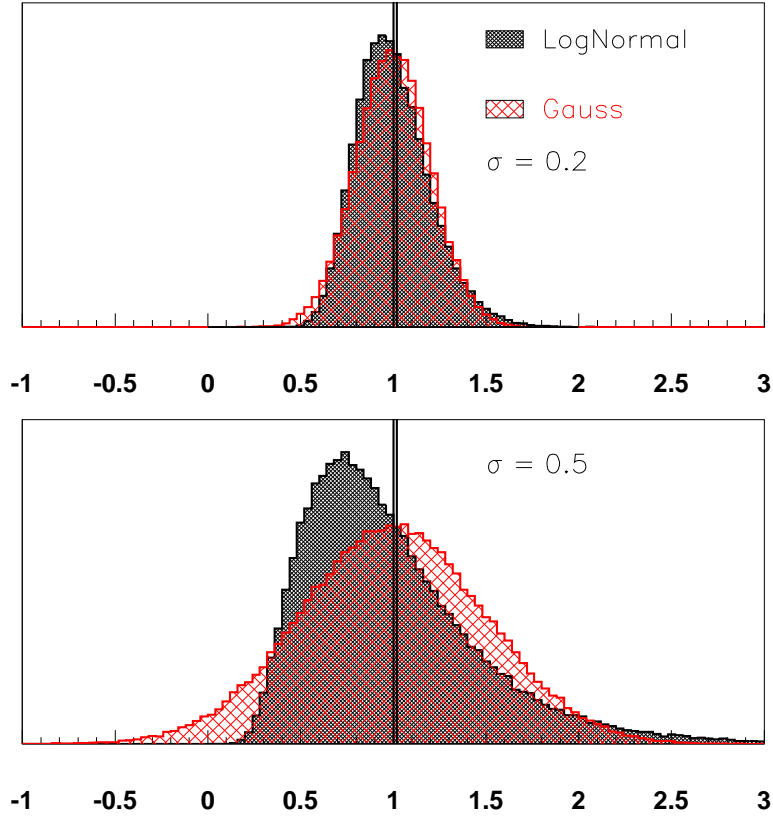}
\caption {Comparison of the  lognormal  (black, darker
  hatching) and Gaussian (red, lighter hatching) probability distributions.
The distributions are shown with mean equal to one, and two different
choices for the RMS (for both distribution): $\sigma=0.2$ (top) and
$\sigma=0.5$ .\label{Fig:gauss}
}
\end{center}
\end{figure}

Standard error estimation of proton parton distribution functions (PDFs) relies on the assumption that all
errors follow Gaussian (or normal) statistics. 
However, this assumption may not always be correct.
Some systematic uncertainties such as luminosity and detector
acceptance follow rather a log-normal distribution (see
Section~\ref{sec:combfit}).
Compared to the Gaussian case, the lognormal distribution which has
the same mean and root mean square (RMS), 
is asymmetric and has a shifted peak, as shown illustratively in
Figure~\ref{Fig:gauss}. 
Therefore, the non-Gaussian behaviour of the experimental uncertainties could lead to an additional uncertainty of the resulting PDFs.
An alternative to the standard error propagation is a toy Monte Carlo (MC) method.
Here,  an implementation of the  MC method is presented for estimation
of the PDF uncertainties with various assumptions for the error
distribution.
In addition, this MC method  provides an independent cross check of
the standard error propagation when assuming the Gaussian error distributions.

\subsubsection{Method}
\label{sec:mcmethod}

The Monte Carlo technique consists firstly in preparing replicas of
the initial data sets which have the central value of the cross
sections, $\sigma_i$, fluctuating within its systematic and
statistical uncertainties taking into account all point to point
correlations.
Various assumptions can be considered for the error distributions.
When dealing with the statistical and point to point uncorrelated
errors, one could allow each data point to randomly fluctuate within
its uncorrelated uncertainty assuming either Gauss, lognormal, or any
other desired form of the error distribution. For example, for
Gaussian errors
\begin{equation}
  \sigma_i \longrightarrow \sigma_i\left(1+\delta_i^{uncorr}\cdot R_i
\right),
\end{equation}
where $\delta_i^{uncorr}$ corresponds to the uncorrelated
uncertainties  and
$R_i$ is a random number chosen from a normal distribution with a
mean of $0$ and a standard deviation of $1$.
 Hence, the central value of  each cross section point $i$ is
shifted by $\delta_{i}^{uncorr}\cdot R_i$.

For the systematic errors, the treatment is a bit more complicated
than above.
This is  due to the correlation between data points and
that,  in general,  the data points are sensitive to the systematic
sources  with a different strength $\delta_{ij}$, where index $i$ ($j$)
runs over all the cross section points (all systematic sources). 
In order to take this into account, for each systematic source $j$ 
a uniformly distributed  {\em fluctuation probability} \rm $P_j$ is
selected. 
Then, for each data point $i$  
the central value of cross section is shifted such that probability of
this shift, which depends on $\delta_{ij}$ and the exact
form of the probability distribution function,  
is equal $P_j$  (for positive $\delta_{ij}$) or ($1-P_j$) (for
negative $\delta_{ij}$). 
In other words, each central value of the cross
section is shifted with the same probability of the corresponding
systematic shift.
For example for the Gaussian errors, this procedure is equivalent to

\begin{equation}
\sigma_i \longrightarrow \sigma_i\left(1+\delta_i^{uncorr}\cdot R_i
+\sum_j^{N_{sys}}\delta_{ij}^{corr} \cdot R_j 
\right),
\end{equation}
where in addition to the shifts for the uncorrelated errors previously
explained,   $R_j$ corresponds to another random number chosen from a 
normal distribution with mean of $0$ and standard deviation of
$1$ as a fluctuation for the systematic source $j$.  Hence, the 
central values of the cross sections are shifted  in addition
by $\delta_{ij}^{corr}\cdot R_j$ for each systematic shift.

This preparation of the data is repeated for $N$ times, where high 
statistics is desirable for more accurate results.
For this study  we used $N>100$ which proved to suffice.
For each 
replica, a next to leading  order (NLO) QCD fit is performed to extract the PDFs.  
The errors on the PDFs are estimated from the RMS of the spread of the
$N$ lines corresponding to the  $N$ individual fits to extract PDF.

A fit to the published H1 HERA-I data of neutral and charged current
$e^\pm p$ scattering cross sections~\cite{Adloff:2003uh}
using the settings discussed in Sect.~\ref{sec:h1set} has been performed,
using the  QCDNUM program~\cite{qcdnum}. 
%


\subsubsection{Validation of the Method}

\begin{figure}[ht]
\begin{center}
\includegraphics[width=0.4\columnwidth]{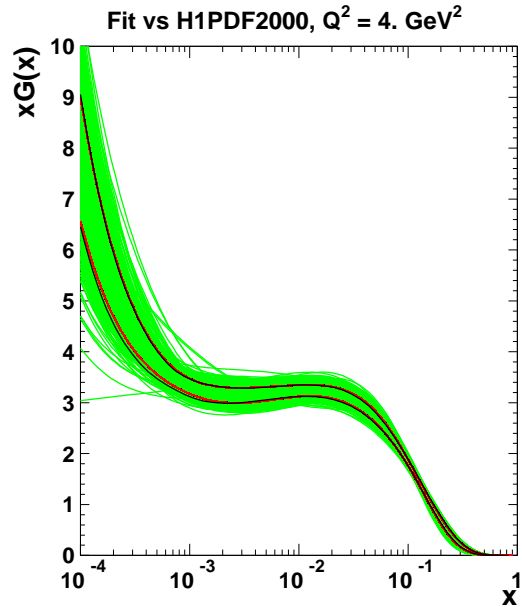}
\caption{Comparison between the standard error calculations and the
  Gauss error distribution is shown for  the gluon PDF. Green lines
  represent the spread of Monte Carlo generated allowances for the
  errors, and the red lines are the RMS of this spread. The black
  lines correspond to the standard error calculations of the PDF
  errors. \label{Fig:mv}
 }
\end{center}
\end{figure}

The MC method is tested by comparing the standard error estimation of
the PDF uncertainties with the MC techniques by assuming that all the
errors (statistical and systematic) follow Gaussian (normal)
distribution.
Figure \ref{Fig:mv} shows good agreement between the methods.

\subsubsection{Test of various assumptions for the error distributions}

\begin{figure}[ht]
\begin{center}
\includegraphics[width=0.4\columnwidth]{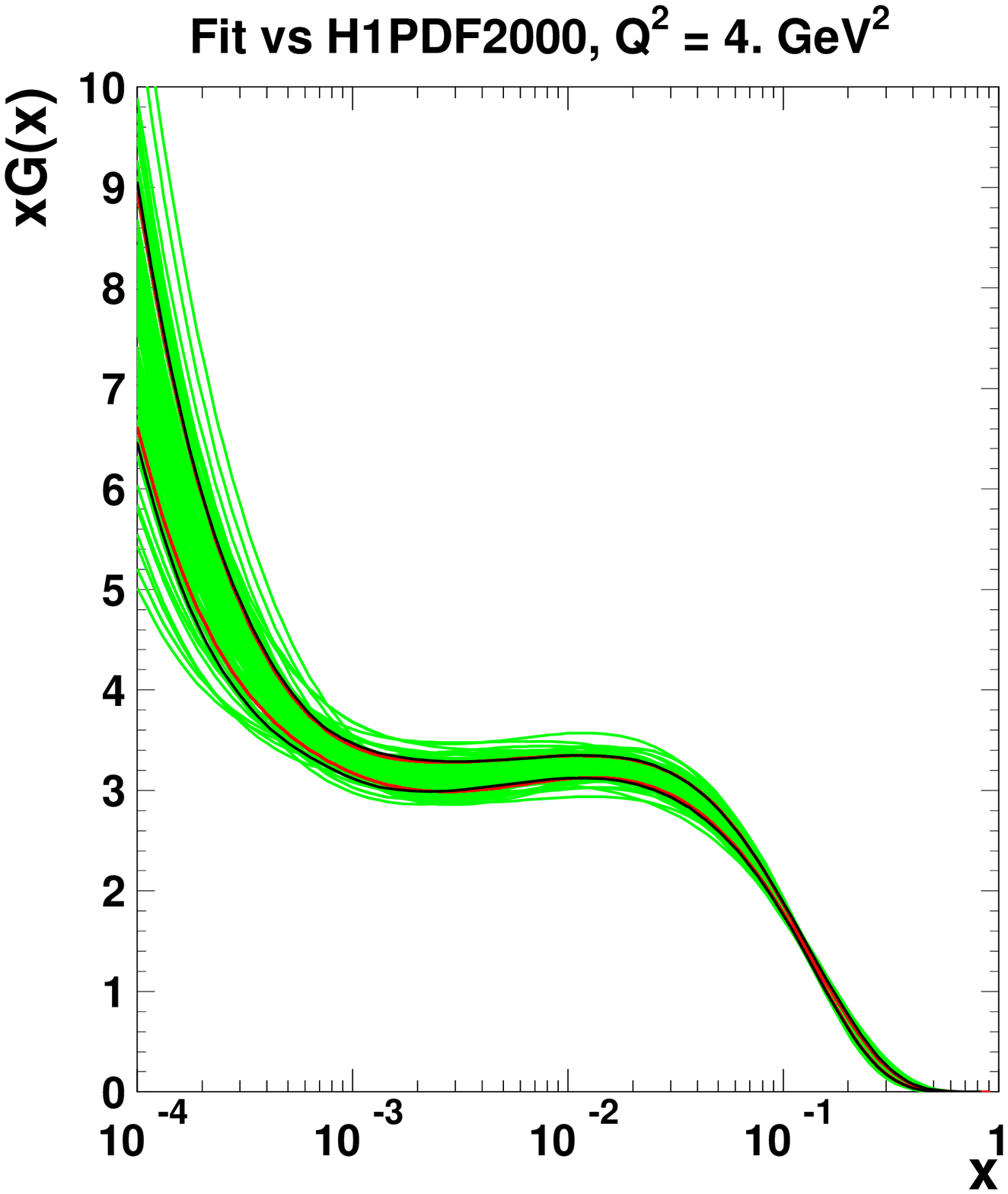}
\includegraphics[width=0.4\columnwidth]{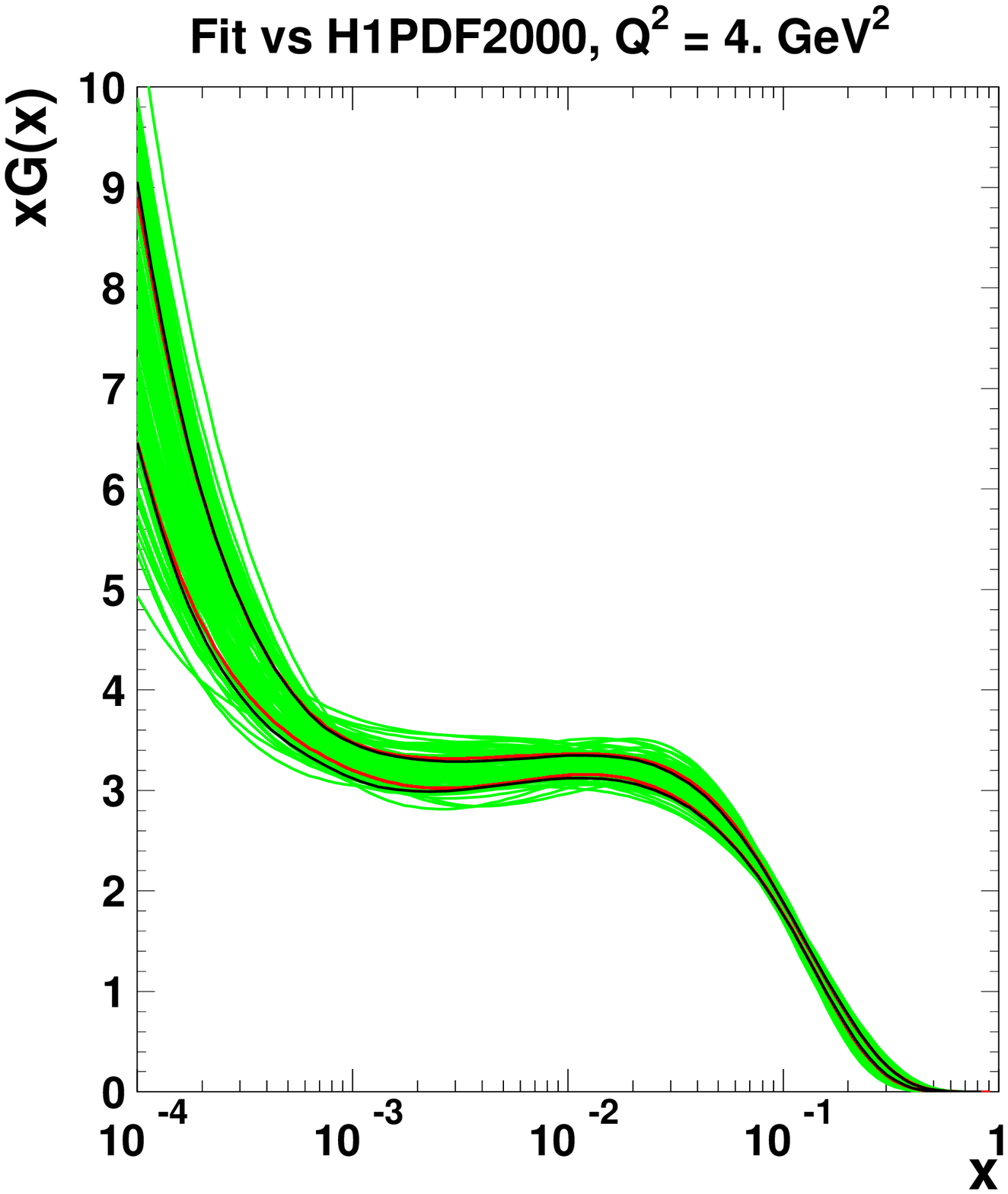}
\caption {Comparison between errors on PDFs obtained via standard
  error calculation (black) where Gauss assumption is used, and errors
obtained via Monte Carlo method (red) where luminosity uncertainty is
allowed to fluctuate according to lognormal distributions and all the
other uncertainties follow the Gaussian distribution (left), and where
 all the systematic uncertainties are
allowed to fluctuate according to lognormal distributions (right). Only the
gluon PDF is shown, where the errors are larger. The green lines show
the spread of the $N$ individual fits.\label{Fig:lumilog}
}
\end{center}
\end{figure}

Two cases are considered which may represent most likely the error
distributions:
(1) the lognormal distribution for the luminosity uncertainty and the rest of
the errors are set to follow the Gauss shape, (2) the lognormal
distributions for all the systematic errors and the statistical errors
are set to follow the Gauss distributions.
The results for the first case (1) are 
shown in Figure~\ref{Fig:lumilog}.
The results of the  tests for the case when  lognormal distributions
for all the systematic uncertainties are assumed is shown in
Figure \ref{Fig:lumilog}. 
We observe that for the precise H1 HERA-1 data the effect of using
lognormal distribution,  which is considered for some systematic
uncertainties more physical, is similar to the pure gauss distribution case.

\subsubsection{Conclusions}

A simple method to estimate PDF uncertainties has
been built within QCD Fit framework.
 Assuming only gauss distribution of all errors, the
results agree well with the standard error estimation. 
This method allows to check the effect of non-
gauss assumptions for distributions of the
experimental uncertainties.{}
For the H1 data, results are similar to the gauss case
when using lognormal.
 The method could be extended for other physical
variables (i.e. cross sections) for cross checks with the
standard error evaluation.

\subsection{HERA--LHC Benchmark}
\label{sec:heralhc}

This benchmark is based on the Alekhin/Thorne benchmark of
Ref.~\cite{Dittmar:2005ed}, whose settings has been given in
Sect.~\ref{sec:settings}. 
Both the Alekhin and Thorne fits had the following features:
\begin{itemize}
\item uncertainties determined using the
Hessian method with $\Delta\chi^2=1$;
\item input PDFs are parameterized using  the following functional form:
\begin{eqnarray}
x\,f_i(x,Q_0^2)=A_i(1-x)^{b_i}(1+\epsilon_i x^{0.5}+\gamma_i x)x^{a_i}\,.
\end{eqnarray}
with $\epsilon_i$ and $\gamma_i$ set to zero for the 
sea and gluon distributions.
Hence, there were a total of 13 free PDF parameters
plus $\alpha_s(M_Z)$ after imposing sum rules.
\end{itemize}

Here, we  reanalyze it within the MSTW and NNPDF approaches.
First, we summarize the respective MSTW and NNPDF approaches, and
especially their differences when compared to the previous HERALHC
benchmark fits of Ref.~\cite{Dittmar:2005ed}. 
Then,  results for benchmark
fits obtained with the various different approaches are compared to
each other. Finally,
we compare each benchmark fit to its counterpart
based on a wider range of data, i.e.~the NNPDF1.0~\cite{Ball:2008by}
reference and the MRST01~\cite{Martin:2002aw} and
 MSTW08~\cite{Watt:2008hi,Thorne:2007bt} PDFs.

\subsubsection{MSTW approach\protect\footnote{Contributing authors: 
R.~S.~Thorne, G.~Watt}}
\label{sec:bench_intro_MSTW}

The benchmark analysis is now much more closely aligned to the global 
analysis than was the case for the Thorne benchmark compared to the 
MRST global analysis. It follows 
the general approach taken by the MRST (or more recently,
MSTW) group, and is similar to that described in Ref.~\cite{Martin:2002aw}.
There are some new features which are explained below.

\begin{itemize}
\item[-]{\it Input parameterization.} We take the input PDF parameterization at 
$Q_0^2 = 1$ GeV$^2$ to be:
\begin{eqnarray}
  xu_v(x,Q_0^2) &=& A_u\,x^{\eta_1} (1-x)^{\eta_2} (1 + \epsilon_u\,\sqrt{x} + \gamma_u\,x)\,,\\
  xd_v(x,Q_0^2) &=& A_d\,x^{\eta_3} (1-x)^{\eta_4} (1 + \epsilon_d\,\sqrt{x} + \gamma_d\,x)\,,\\
  xS(x,Q_0^2) &=& A_S\,x^{\delta_S} (1-x)^{\eta_S} (1 + \epsilon_S\,\sqrt{x} + \gamma_S\,x)\,,\\
  xg(x,Q_0^2) &=& A_g\,x^{\delta_g} (1-x)^{\eta_g} (1 + \epsilon_g\,\sqrt{x} + \gamma_g\,x) + A_{g^\prime}\,x^{\delta_{g^\prime}} (1-x)^{\eta_{g^\prime}}\,,
\end{eqnarray}
where $S = 2(\bar{u}+\bar{d}+\bar{s})$, $s=\bar{s}=0.1\,S$ and
$\bar{d}=\bar{u}$. The parameters $A_u$, $A_d$ and $A_g$ 
are fixed by sum rules, leaving potentially 19 free parameters.  In practice, 
to reduce the
number of highly correlated parameters, making linear error propagation
unreliable, 
we determine the central value of the benchmark fit by freeing all 
19 parameters, then fix 6 of those at the best-fit values when calculating 
the Hessian matrix used to determine the PDF uncertainties, giving a total 
of 13 eigenvectors.  This is the same procedure as used in the MSTW 2008 
global fit \cite{Watt:2008hi,Thorne:2007bt}, where there are an additional 3 free
parameters
associated with $\bar{d}-\bar{u}$ and an additional 4 free parameters 
associated with strangeness, giving a total of 20 eigenvectors.  Note that 
the parameterization used in the previous Alekhin/Thorne benchmark fits 
was considerably more restrictive, where the $\epsilon_S$, $\gamma_S$, 
$\epsilon_g$ and $\gamma_g$ parameters were set to zero, and 
the second (negative) gluon term was omitted entirely.  In addition, 
$\epsilon_u$ was held fixed for the Thorne benchmark fit, 
leaving a total of 12 eigenvectors.  We find that the more flexible gluon 
parameterization, allowing it to go negative at very small $x$, is very highly
correlated with the value obtained for $\alpha_s$, and a value of 
$\alpha_s(M_Z) = 0.105$ is obtained if it is allowed to go free at the 
same time as the other parameters, therefore we instead choose to fix it at 
$\alpha_s(M_Z) = 0.112$ as in the NNPDF benchmark fit.

\item[-]{\it Error propagation.} Apart from the more flexible input 
parameterization, the other major difference in the new MSTW version of the 
HERA--LHC benchmark fit, with respect to the previous Thorne (MRST) version,
is the choice of tolerance, $T = \sqrt{\Delta\chi^2}$.  The MRST
benchmark fit used the standard choice $T=1$ for one-sigma uncertainties.  More
precisely, the 
distance $t$ along each normalized eigenvector direction was taken to be 1, 
and ideal quadratic behaviour about the minimum was assumed, giving 
$T\approx t=1$.  The MRST global fit used $T=\sqrt{50}$ for a 90\% confidence 
level (C.L.) uncertainty band; however, this is not appropriate when fitting 
a smaller number of data sets.  Recently, a new procedure has been developed~\cite{Watt:2008hi,Thorne:2007bt} which enables a \emph{dynamic} determination
of the tolerance 
for each eigenvector direction, by demanding that each data set must be 
described within its one-sigma (or 90\%) C.L.~limits according to a 
hypothesis-testing criterion, after rescaling the $\chi^2$ for each data set 
so that the value at the global minimum corresponds to the most probable 
value.  Application of this procedure to the MSTW benchmark fit gives 
$T\sim 3$ for one-sigma uncertainties and $T\sim 5$ for 
90\% C.L.~uncertainties.  For the MSTW global fit, the typical values of $T$ 
required are slightly larger, with more variation between different 
eigenvector directions. The increase in $T$ in the global fit is mainly
due to the inclusion of 
some less compatible data sets, while the greater variation in $T$ between
eigenvectors is due to the fact that some parameters, particularly those
associated with $s$ and $\bar{s}$, are 
constrained by far fewer data sets than others.  In the MSTW fits, the data
set normalizations are allowed to vary, with the aforementioned penalty term,
when determining the PDF uncertainties. For global fits this 
automatically leads to a small increase in uncertainty 
compared to the MRST determinations, where data set normalisations
were held fixed when calculating the Hessian matrix used for error propagation.
In the MRST benchmark fit the data
set normalizations were allowed to vary. To calculate the
uncertainty bands from the eigenvector PDF sets, we use the formula for
asymmetric errors given, for example, in Eq.~(13) of
Ref.~\cite{Martin:2002aw}.

\end{itemize}

\subsubsection{NNPDF approach\protect\footnote{Contributing authors: 
R.~D.~Ball, L.~Del~Debbio, S.~Forte, A.~Guffanti, J.~I.~Latorre,
A.~Piccione, J.~Rojo, M.~Ubiali}}
\label{sec:bench_intro}

The NNPDF approach was proposed in Ref.~\cite{Forte:2002fg}, and it was
applied there and in  Ref.~\cite{DelDebbio:2004qj} to the parameterization 
of the structure function $F_2 (x, Q^2)$ with only two or more
experimental data sets respectively. In Ref.~\cite{DelDebbio:2007ee} it was first
used for the determination of a single PDF (the isotriplet quark
distribution), and in Ref.~\cite{Ball:2008by} a full set of  PDFs fit based
on DIS data (NNPDF1.0) was presented. Because the method has been
discussed extensively in these references, here  we only  summarize
briefly its main features.

\begin{itemize}
\item[-]{\it Error propagation}. We make a Monte Carlo sample of the
  probability distribution of the experimental data by generating an
  ensemble of $N$ replicas of artificial data following a
  multi-gaussian distribution centered on each data point with 
  full inclusion of the experimental covariance matrix. Each replica
  is used to construct a set of PDFs, thereby propagating the
  statistical properties of the data Monte Carlo sample to a final
  Monte Carlo sample of PDFs. Here we shall take $N=100$. The method
  is the same as discussed in Sect.~\ref{sec:mcmethod}, the only
  difference being the treatment of normalization errors: relative
  normalizations are fitted in the H1 approach, while they are included
  among the systematic errors in the Monte Carlo data generation in
  the NNPDF approach (see Refs.~\cite{Adloff:2003uh,Ball:2008by} for
  details of the respective procedures) .
\item[-]{\it Input parameterization}. Each PDF is parameterized with a
  functional form provided by a neural network. 
  The architecture for the neural network
  is the same for all PDFs, and yields a parameterization with
   37 free parameters
   for each PDF. This is a very redundant parameterization, it is
   chosen
 in order to
   avoid parameterization bias; neural networks are a particularly
   convenient way of dealing with redundant parameterizations. Note
   that sum rules are also imposed.
\item[-]{\it Minimization}. A redundant parameterization allows for
  fitting not only the underlying physical behaviour, but also
  statistical noise. Therefore, the
  minimization is stopped not at the absolute minimum of the $\chi^2$, 
  but rather before one starts fitting  noise. This optimal stopping
  point is determined as follows: the data in each replica are randomly
  assigned either to  a training or to a validation set. The fit is performed
  on data of the training set only, while the
  validation set is used as a monitor. The fit is stopped when the
  quality of the fit to the training set keeps improving, but the
  quality of the fit to the validation set deteriorates.
\end{itemize}

\subsubsection{Comparison between the Benchmark Parton Distributions}
\label{sec:mrstvsnnpdf}

\begin{table}[ht]
  \begin{center}
    \begin{tabular}{|l|c|c|}
      \hline
      Data Set & $\chi_{\rm bench}^2/N_{\rm data}$& $\chi_{\rm global}^2/N_{\rm data}$ \\
      \hline
      ZEUS97  &  1.09 &1.18\\
      H1lowx97  &  1.03& 1.00\\
      NMC     &  1.40 & 1.45 \\
      NMC\_pd &  1.24 & 1.32 \\
      BCDMS   &  1.21 & 1.98\\
      \hline
      Total &  1.19& 1.53 \\
      \hline
    \end{tabular}
  \end{center}
\caption{NNPDF $\chi^2$ for the total and each single data set,  both
  for the benchmark and global fit. \label{tab:nnpdf_chi2_bench1}
}
\end{table}

\begin{table}[ht]
  \begin{center}
    \begin{tabular}{|l|c|c|}
      \hline
      Data set & ${\chi^{\rm diag}_{\rm bench}}^2 / N_{\rm data}$ & ${\chi^{\rm diag}_{\rm global}}^2 / N_{\rm data}$ \\ \hline
      ZEUS97  & 0.76 & 0.79 \\
      H1lowx97 & 0.53 & 0.54 \\
      NMC  & 1.08 & 1.11 \\
      NMC\_pd & 0.78 & 0.89 \\
      BCDMS & 0.74 & 1.13 \\
      \hline
      Total & 0.76 & 0.89 \\
      \hline
    \end{tabular}
  \end{center}
\caption{ MSTW $\chi^2$ for the total and each single data set, both
  for the benchmark and global fit. Notice that statistical and
  systematic errors are added in quadrature and that relative data set
  normalizations are  
fitted.}
\label{tab:mstwchisq}
\end{table}

The $\chi^2$ per data point for the NNPDF and MSTW fits are shown in
Table~\ref{tab:nnpdf_chi2_bench1} and~\ref{tab:mstwchisq}
respectively. Note that in the MSTW fit statistical and systematic
errors are added in quadrature, so the quantity shown is the diagonal
contribution to the $\chi^2$. The quality of the NNPDF is seen to be
uniformly good. The quality of the MSTW is also uniform, though it
cannot be compared directly because of the different way systematics
are treated. The comparison of each benchmark fit to the corresponding
global fit will be discussed in Sect.~\ref{sec:nnpdf_mrst_global} below.

\begin{figure}[ht]
\begin{center}
\includegraphics[width=0.3\linewidth]{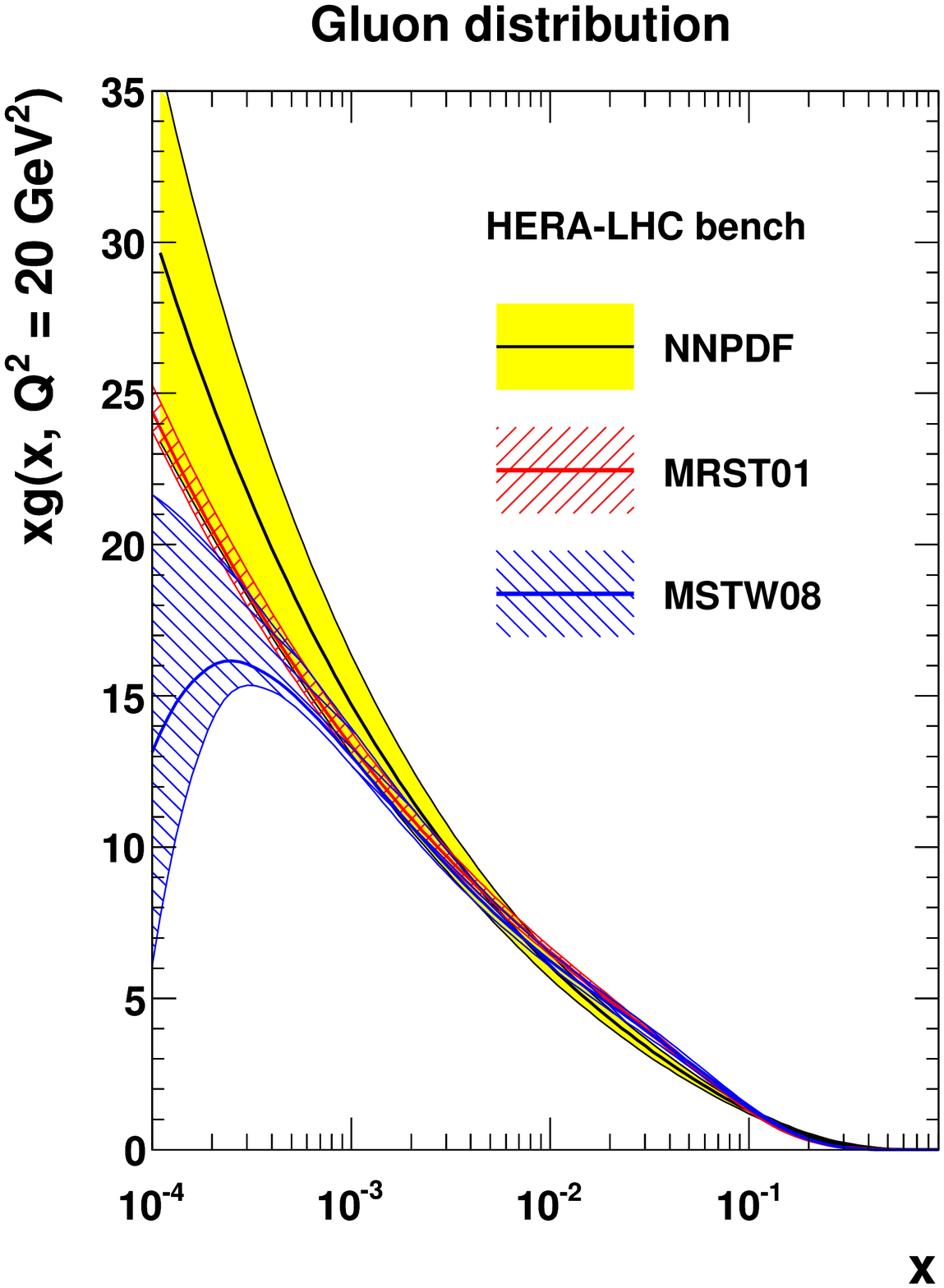}%
\includegraphics[width=0.3\linewidth]{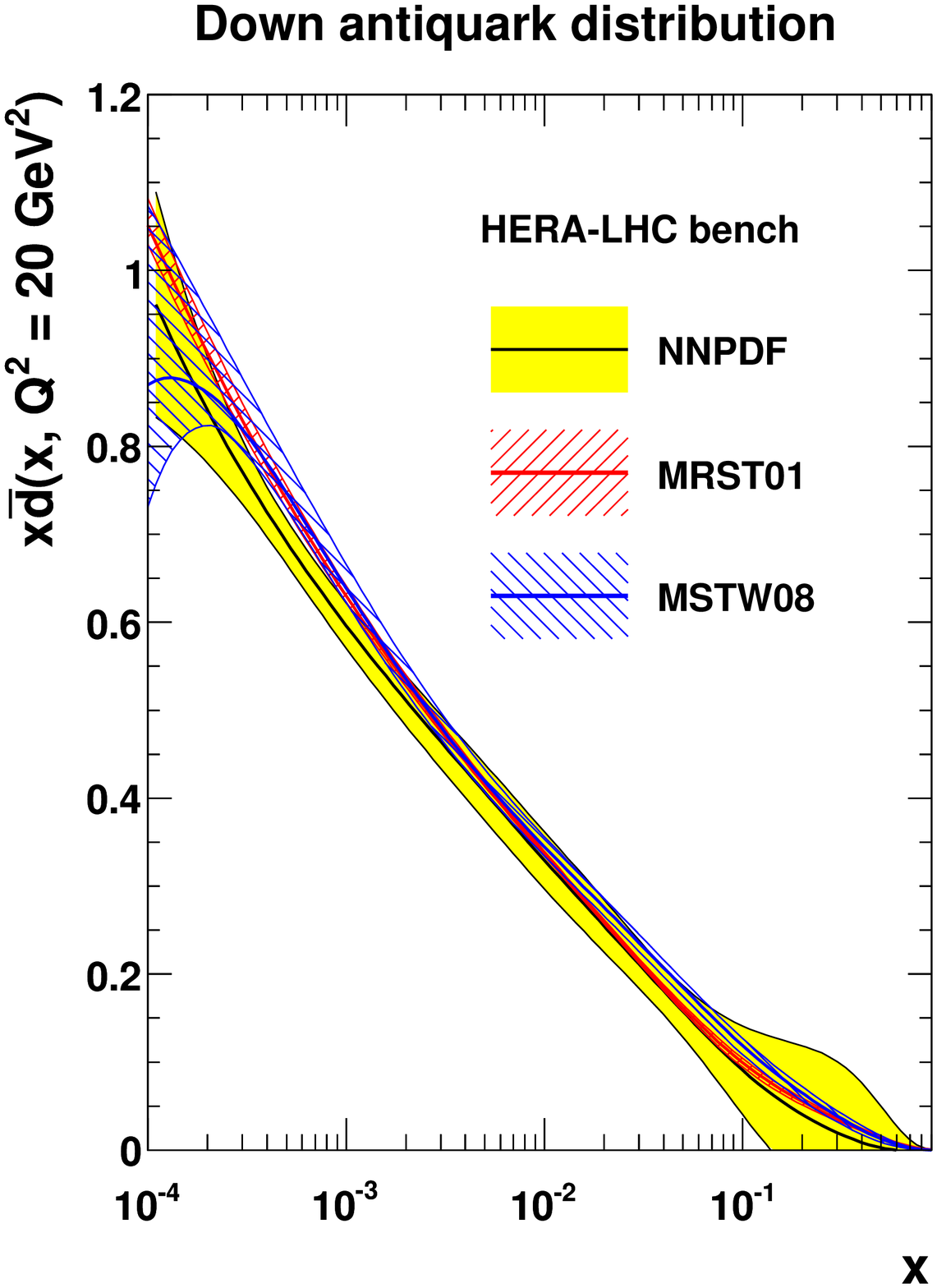}\\
\includegraphics[width=0.3\linewidth]{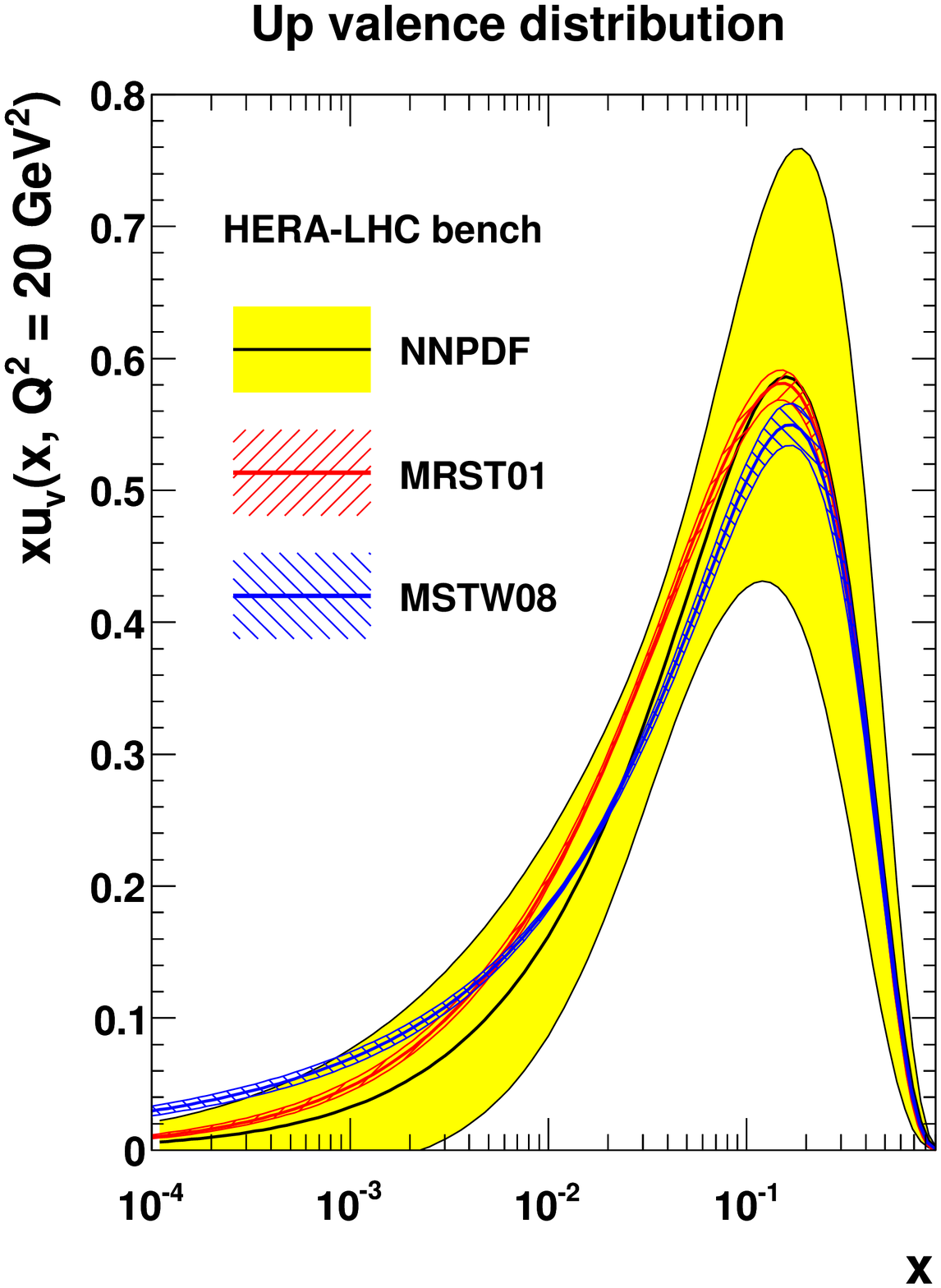}%
\includegraphics[width=0.3\linewidth]{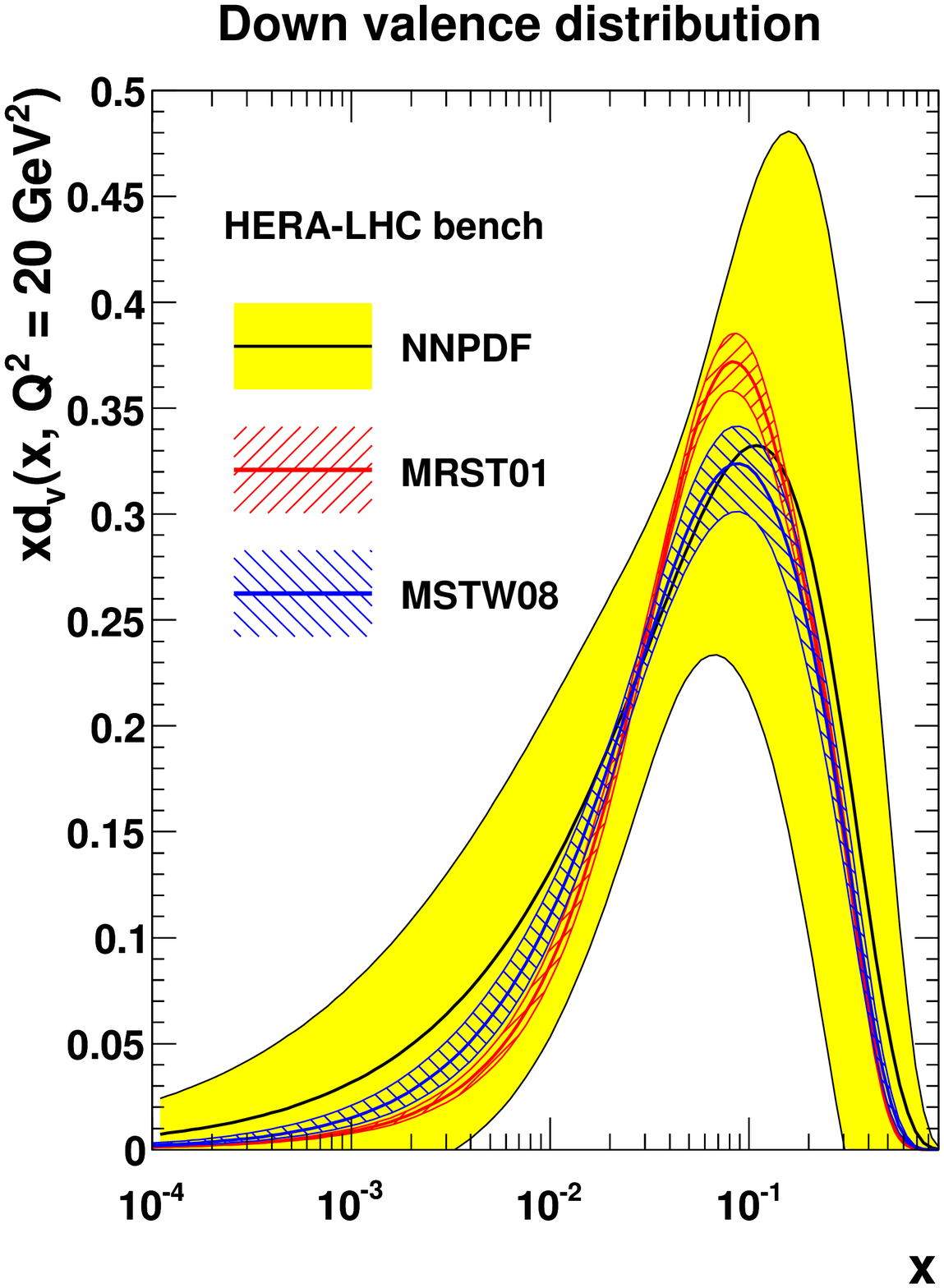}
\vspace{-.7cm}
\caption{Comparison of the NNPDF,  MRST  and MSTW benchmark fits for the
gluon, $d$-sea, $u$-valence and $d$-valence at $Q^2=20~{\rm GeV}^2$.
All uncertainties shown correspond to one--$\sigma$ bands.
\label{fig:nnpdf_mrst_bench}}
\end{center}
\end{figure}

In Fig.~\ref{fig:nnpdf_mrst_bench} the PDFs from the NNPDF and MSTW  benchmark
fits presented here are  
compared to those by Thorne from Ref.~\cite{Dittmar:2005ed} at the
same reference scale of $Q^2=20~{\rm GeV}^2$ used there (denoted as
MRST01 in the figure).
The benchmark fit
by  Alekhin \cite{Dittmar:2005ed} is not shown as the PDFs are very close to
the those by Thorne displayed in Fig.~\ref{fig:nnpdf_mrst_bench}.

For PDFs and kinematical regions where data are available,
namely the small-$x$ gluon and sea
quark and the large-$x$ $u_v$ distributions,
the central values of the NNPDF fit are quite
close to those of the MRST and MSTW  fits, 
despite the differences
in methodology. The central 
values of the PDFs are slightly different for the MRST and MSTW benchmark
fits due to the use of BCDMS $F_2^d$ data in the former, which affects
mainly valence quarks.
Where extrapolation is
needed, such as for the $d_v$ distribution, which is constrained only
by the small amount of data on the ratio $F_2^d/F_2^p$, or the
large-$x$ sea quark, central values are rather more different (though
the Alekhin/MRST/MSTW benchmark central values  are within  the NNPDF 
error band). The exception is the smallest-$x$ gluon, where the form of the 
MSTW parameterization results in a very sharp turn-over. However, even here 
the uncertainty bands are close to overlapping.

Differences are sizeable in the estimation of uncertainties.
Firstly, uncertainty bands for NNPDF benchmark are significantly larger
than for the  MSTW benchmark, which in turn are in general somewhat larger
than those for the MRST benchmark. The difference between MRST and MSTW,
which are based on similar methodology, is due to use of a dynamic
tolerance and a  more flexible gluon
parameterization in MSTW (see Sect.~\ref{sec:bench_intro_MSTW}).
Secondly, 
the width of the uncertainty band for NNPDF benchmark  varies rather more
than that of the MRST benchmark
according to the PDF and the kinematic region, though this is not quite so 
much the case comparing to MSTW benchmark. Indeed, the NNPDF
uncertainties are quite small
in the region between $x=0.01$ and $x=0.1$ (where there is the bulk of
HERA and fixed-target data), while they blow up
in the large-$x$ region for the sea quark or the 
small-$x$ gluon, where there is less or no experimental information.
The smallness of the uncertainty band for MSTW for the small-$x$ valence 
quarks  may be partially due to the lack of flexibility in the
parameterization: note that because of sum rules, the size of 
uncertainties in the data
and extrapolation region are correlated.

Finally, the MRST/MSTW central value generally falls within the NNPDF
uncertainty band, but the NNPDF central value tends to fall outside
the MRST/MSTW uncertainty band whenever the central values differ 
significantly.

\subsubsection{Comparison of the Benchmark Parton Distributions and Global Fits}
\label{sec:nnpdf_mrst_global}

\begin{figure}[ht]
\begin{center}
\includegraphics[width=0.3\linewidth]{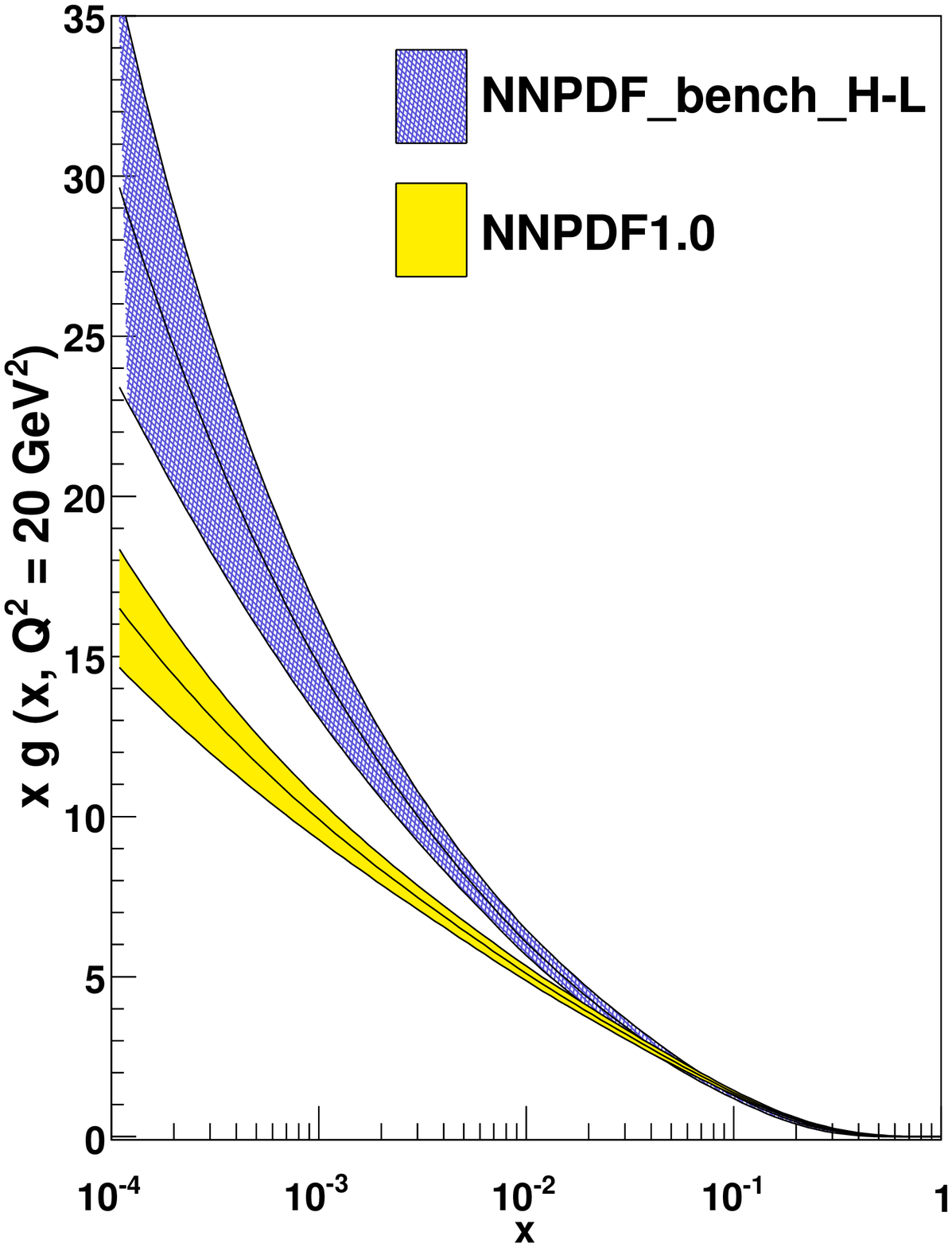}
\includegraphics[width=0.3\linewidth]{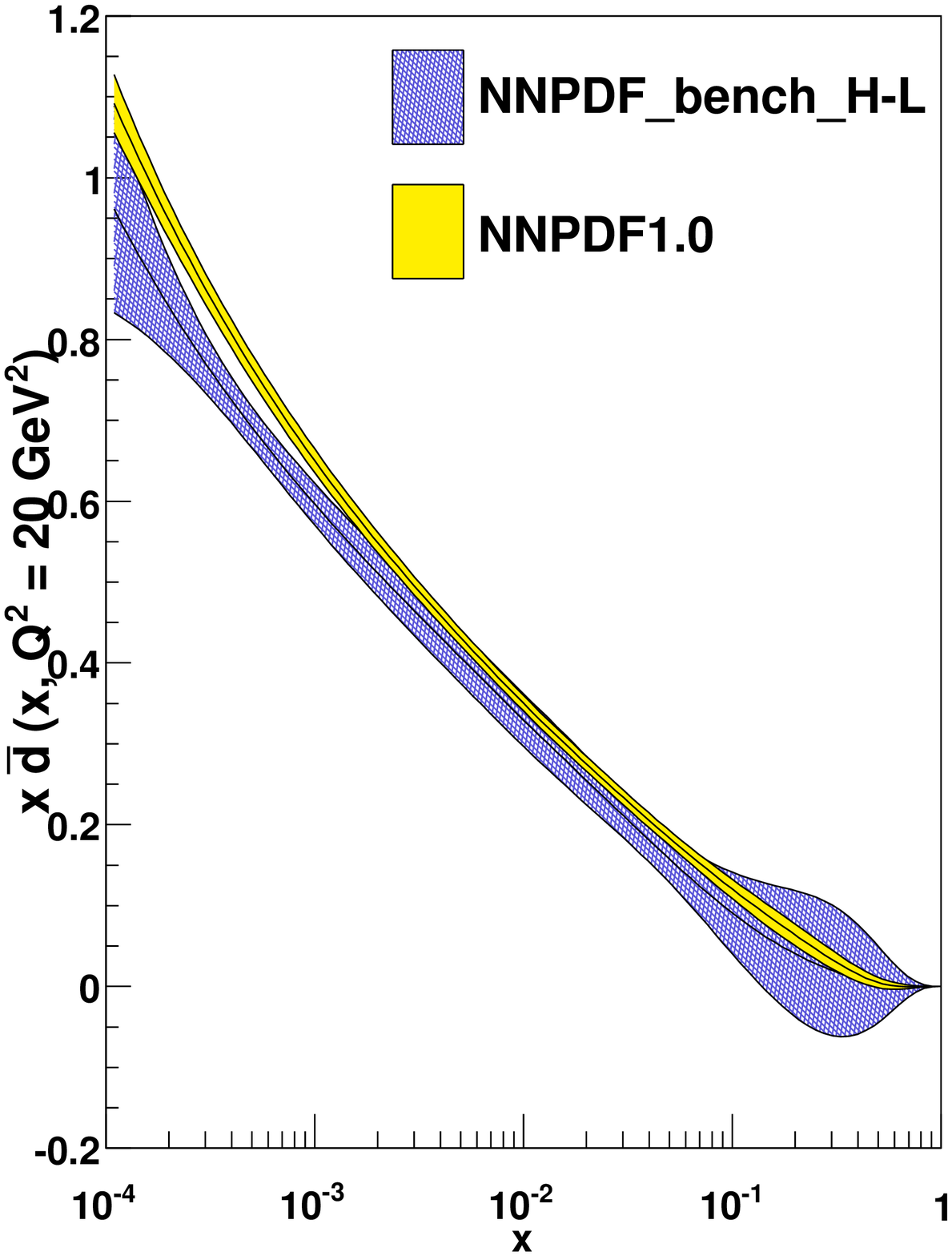}\\
\includegraphics[width=0.3\linewidth]{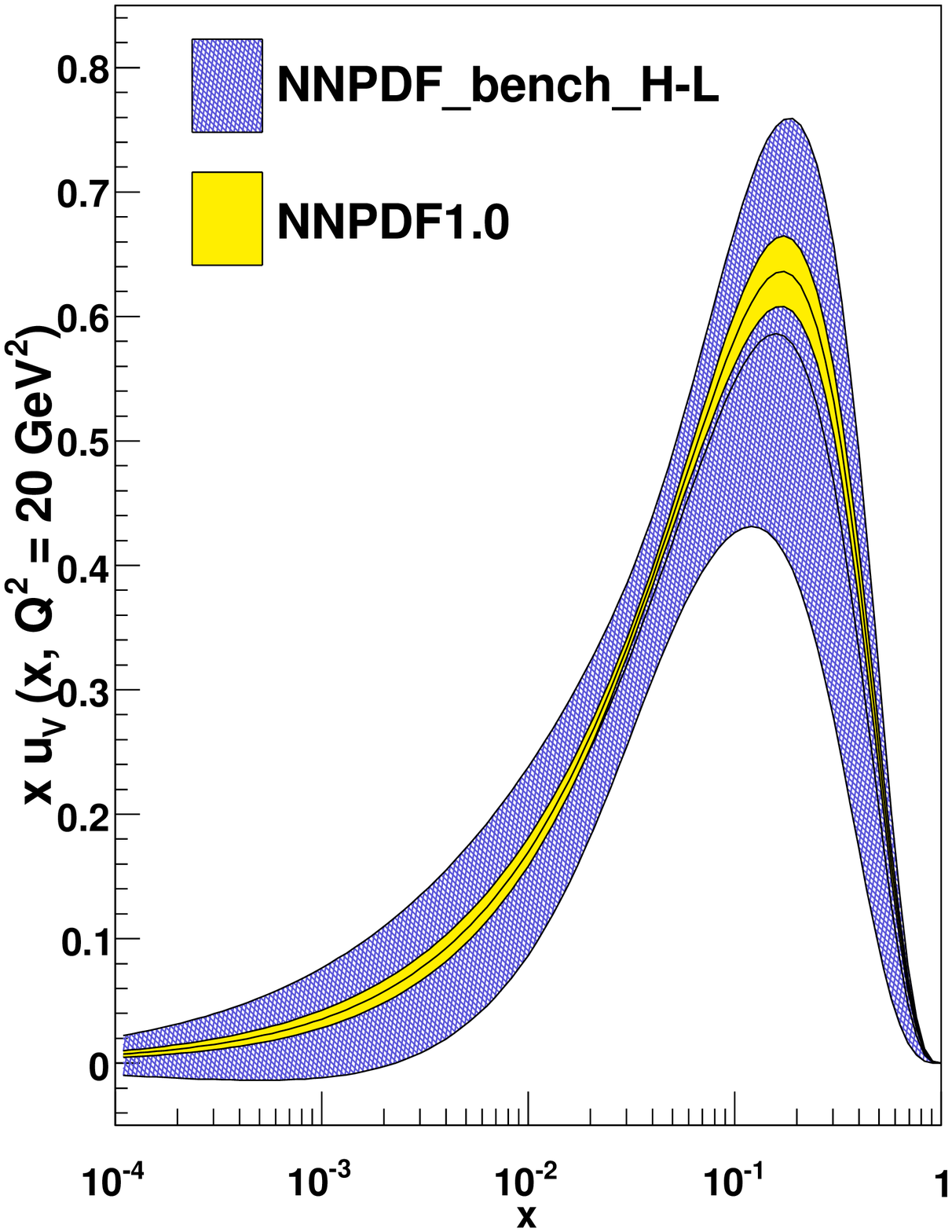}
\includegraphics[width=0.3\linewidth]{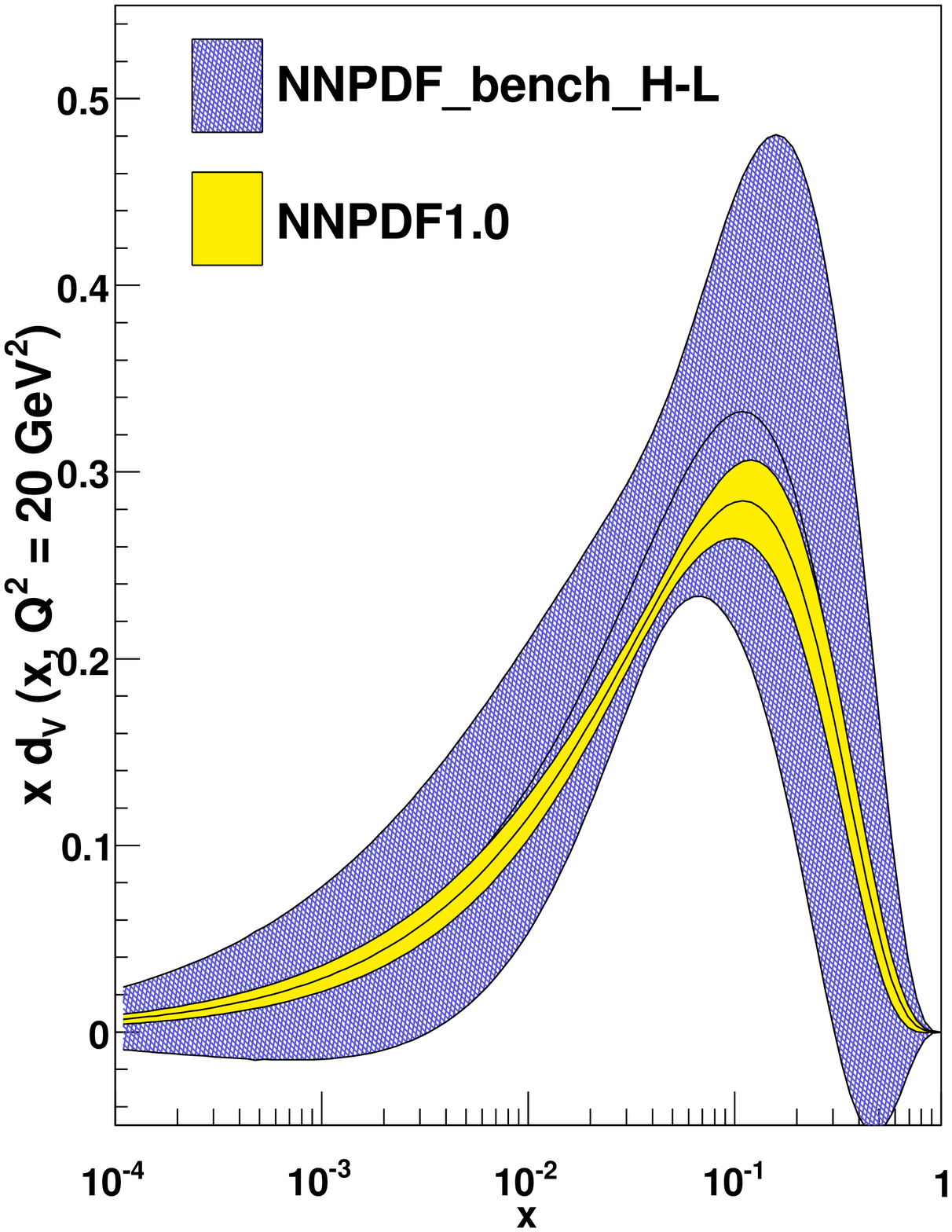}
\vspace{-.5cm}
\caption{Comparison of the NNPDF benchmark and reference fits for the
gluon, $d$-sea, $u$-valence and $d$-valence at $Q^2=20~{\rm GeV}^2$.
\label{fig:nnpdf_mrst_global}
}
\end{center}
\end{figure}

\begin{figure}[ht]
  \begin{center}
    \includegraphics[width=0.3\linewidth]{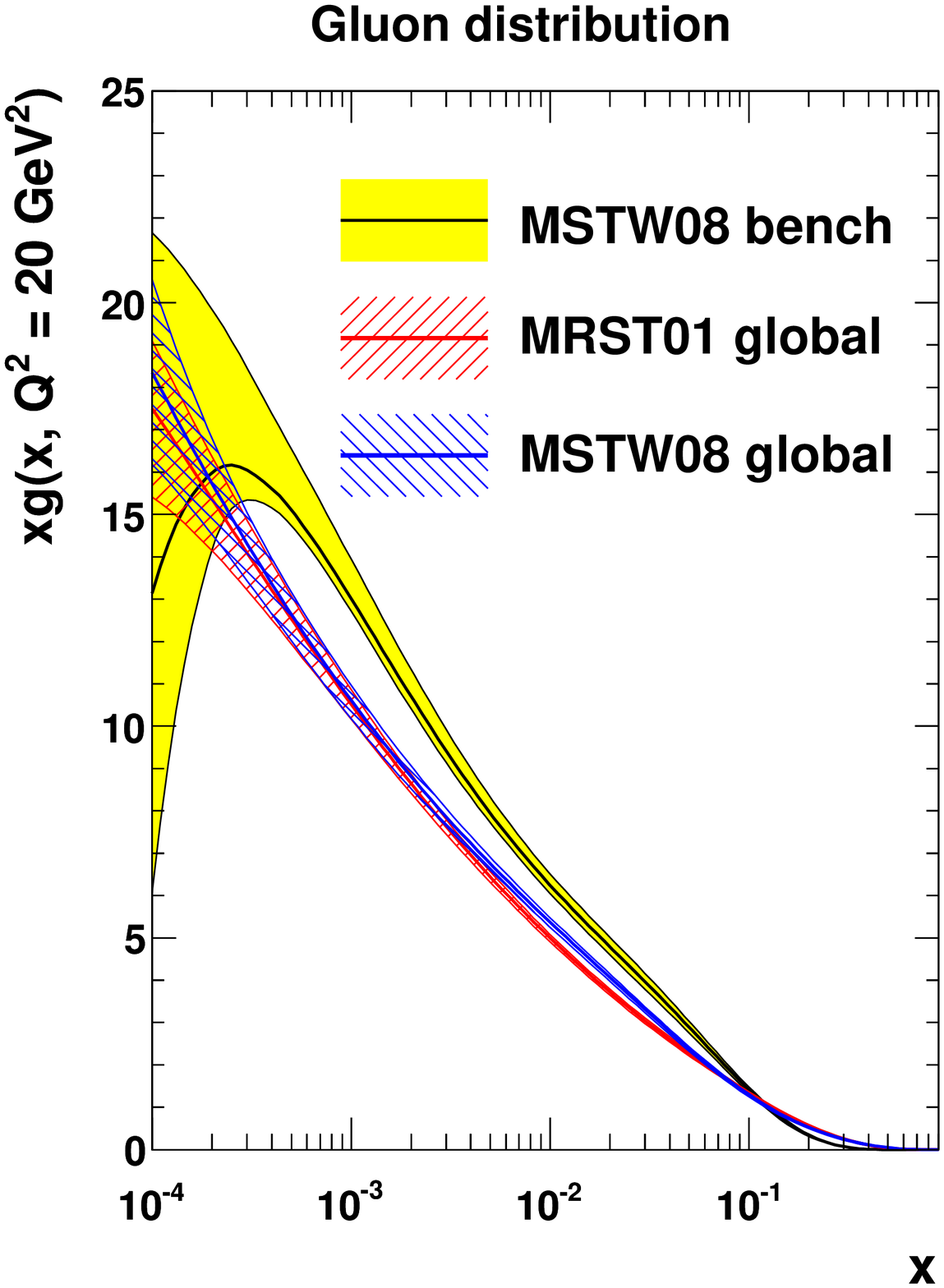}%
    \includegraphics[width=0.3\linewidth]{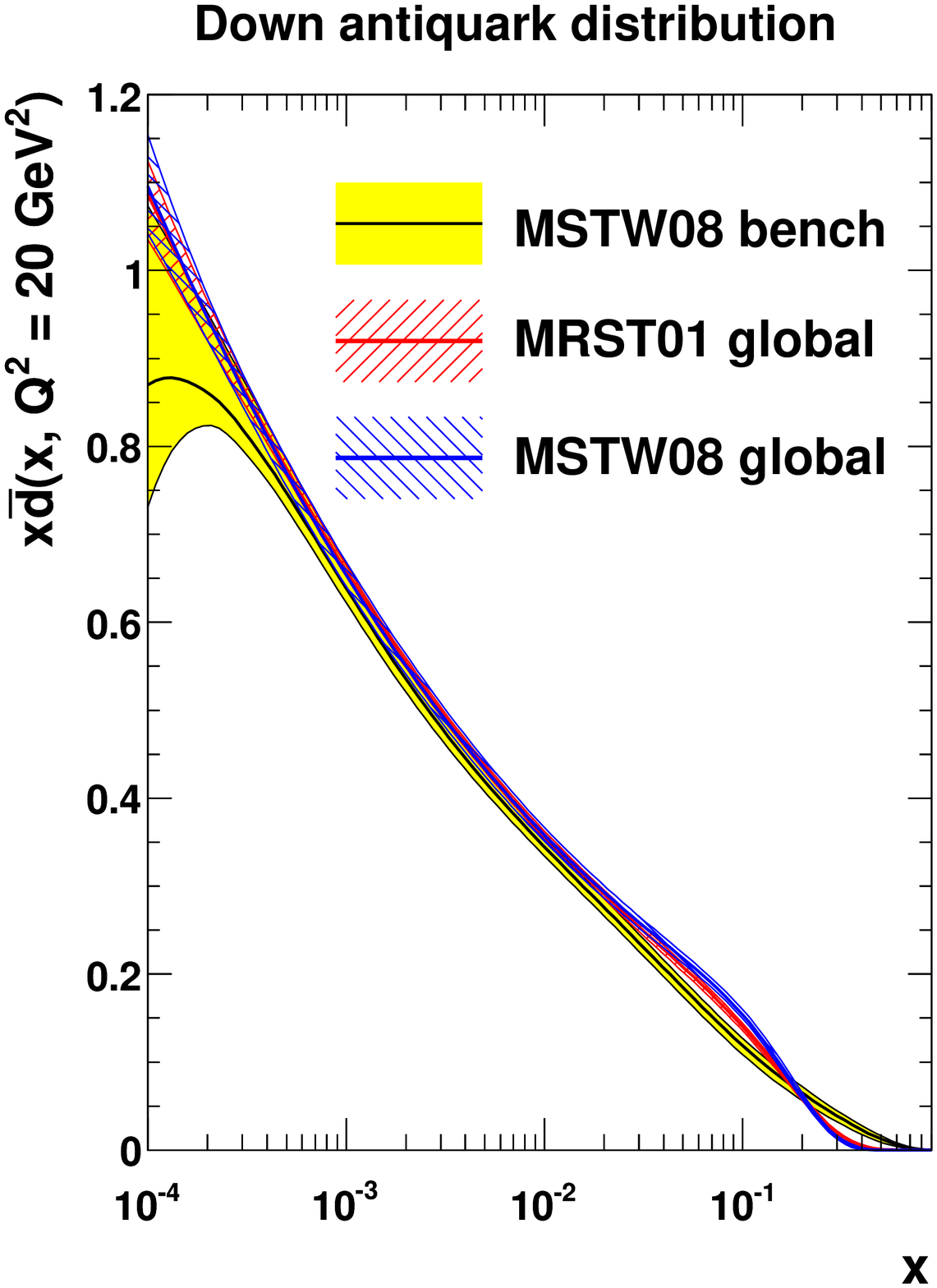}\\
    \includegraphics[width=0.3\linewidth]{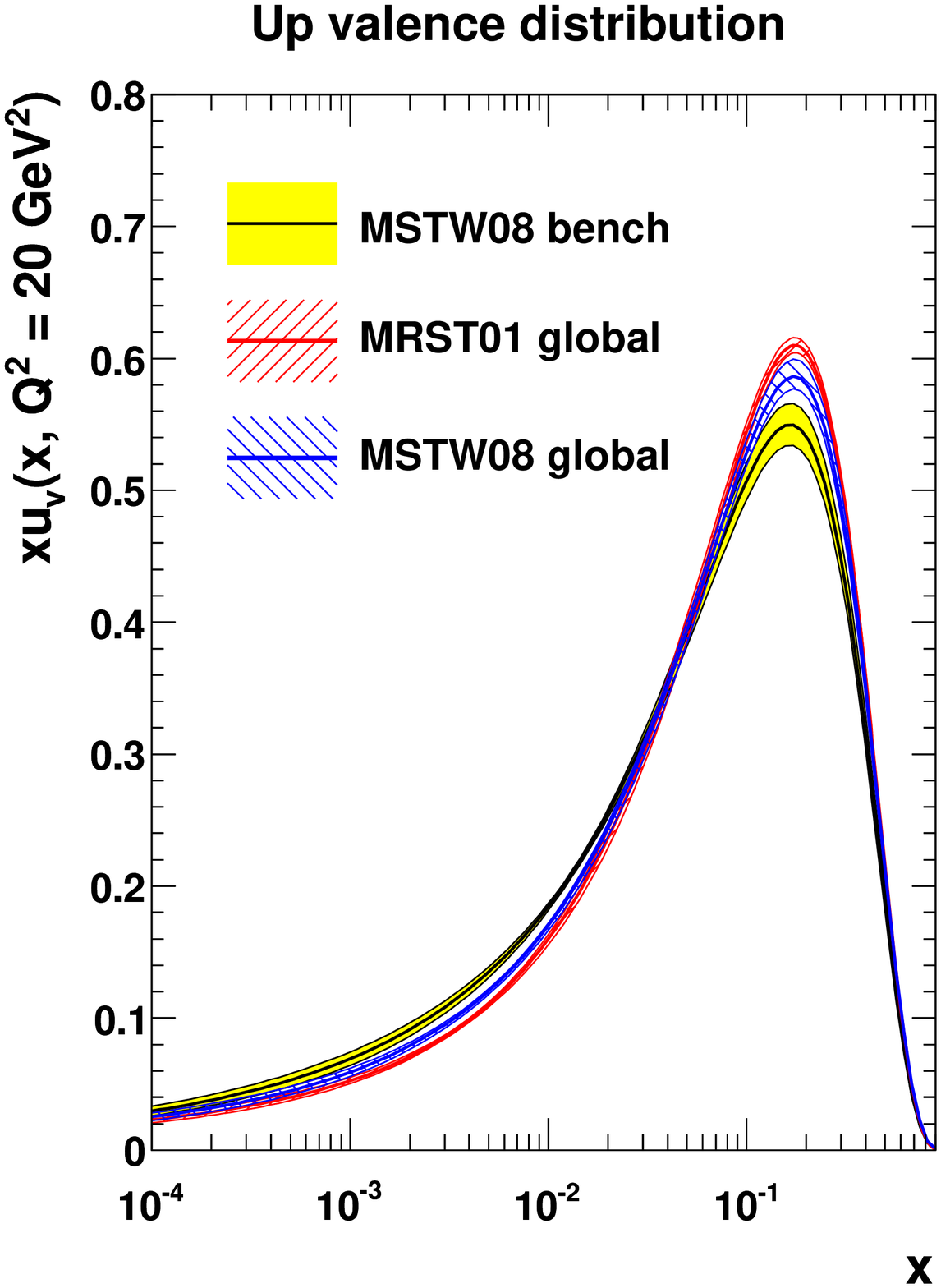}%
    \includegraphics[width=0.3\linewidth]{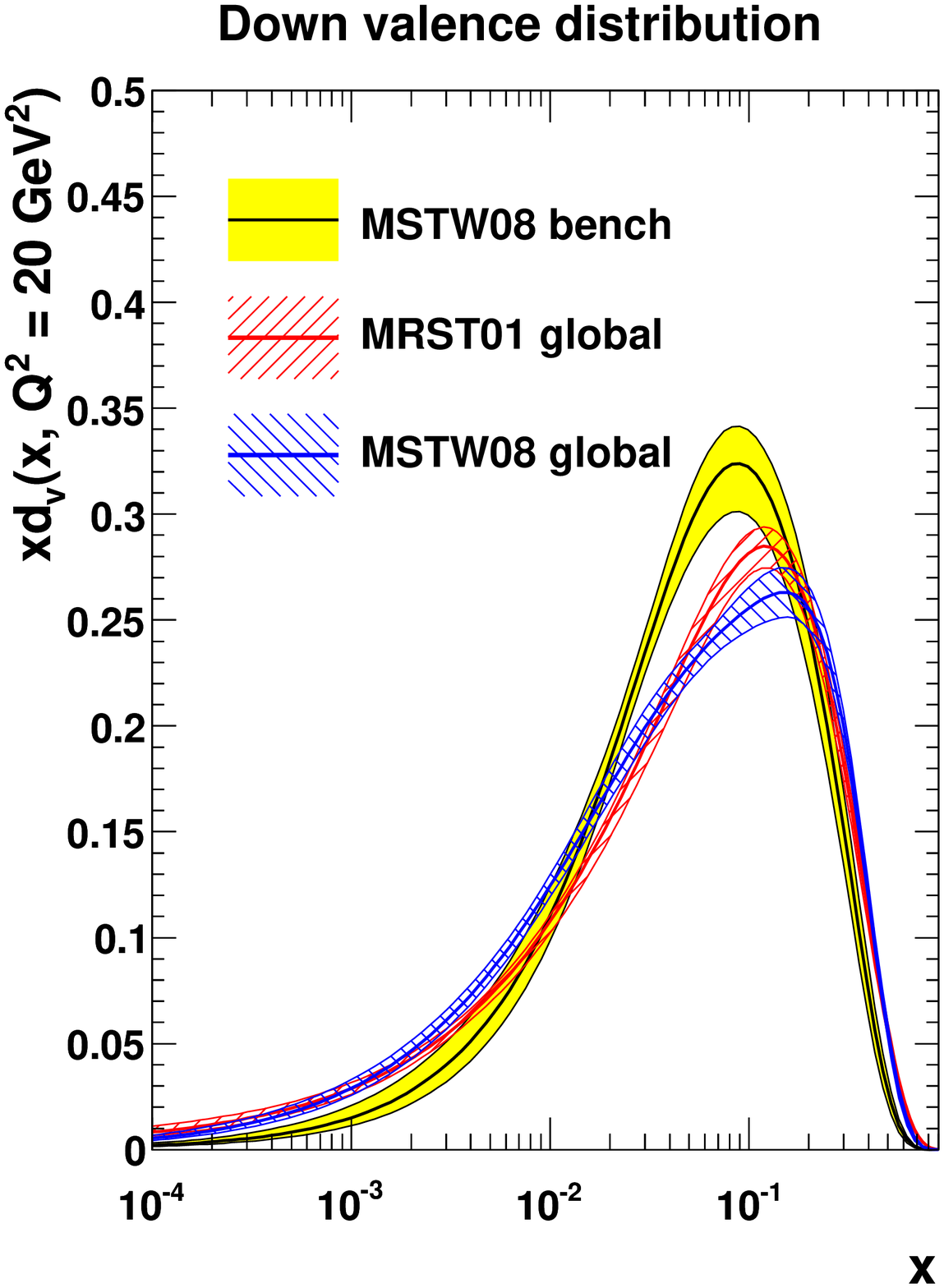}
\vspace{-.5cm}
    \caption{Comparison of the MSTW benchmark and MRST/MSTW global fits for the
      gluon, $d$-sea, $u$-valence and $d$-valence at $Q^2=20~{\rm GeV}^2$.
      All uncertainties shown correspond to one--$\sigma$ bands.
      \label{fig:mstw08bench_mstw08global}
    }
  \end{center}
\end{figure}

In Fig.~\ref{fig:nnpdf_mrst_global}  we compare the NNPDF benchmark
fit to the NNPDF1.0 reference fit of Ref.~\cite{Ball:2008by} (NNPDF
global, henceforth), while
in
 Fig.~\ref{fig:mstw08bench_mstw08global} 
we compare the MSTW benchmark fit 
to the MRST01~\cite{Martin:2002aw} (MRST global, henceforth) and
MSTW08~\cite{Watt:2008hi,Thorne:2007bt} 
global fits (MSTW global, henceforth).

The $\chi^2$  of the NNPDF benchmark and global fits are compared
in Table~\ref{tab:nnpdf_chi2_bench1}, while those of the
MSTW
benchmark and global fits are compared in
Table~\ref{tab:mstwchisq}. Note that for the NNPDF fits the $\chi^2$
is computed using the full covariance matrix, while for the MSTW fits
systematic and statistical uncertainties are added in
quadrature. Note also that
the MRST and MSTW
global fits are carried out in a general-mass variable flavour number scheme
rather than the zero-mass variable flavour number scheme used in the
corresponding benchmark fits, whereas for NNPDF both global and
benchmark fits are done with a  zero-mass variable flavour number scheme.
Comparison of the quality of each  benchmark to the corresponding global fit 
to the same points in Table~\ref{tab:mstwchisq}  shows a significant 
deterioration in
the quality of the fit (total $\Delta\chi^2 \gg 1$),
especially for the BCDMS $F_2^p$ data.  
All fits appear to be acceptable for
all data sets: for instance, even though the $\chi^2$ of the NNPDF
global fit for the benchmark subset of data is $1.98$, it is equal to
$1.59$~\cite{Ball:2008by}  for the full BCDMS set of data. However, the
increase in $\chi^2$ suggests that there might be data
inconsistencies.

Let us now compare each pair of benchmark and global fits.
For NNPDF, the difference in central value between benchmark and
reference
 is comparable to
that found between the MRST or Alekhin global fits and
their benchmark counterparts in Ref.~\cite{Dittmar:2005ed}.
However, the NNPDF global and benchmark fits remain compatible within their
respective error bands. Indeed, the NNPDF benchmark fit has a rather
larger error band than the reference, as one would expect from
a fit based on a rather smaller set of (compatible) data. Such a
behaviour was however not observed in the comparison between global
and benchmark MRST and Alekhin fits of Ref.~\cite{Dittmar:2005ed}.

It is interesting to observe that the 
gluon shape at low $x$ of the benchmark and global NNPDF disagree
at the one $\sigma$ level (though they agree at two $\sigma$). This 
can be understood as a consequence of the fact that the value of
$\alpha_s$ in the two fits is sizably different ($\alpha_s=0.112$
vs. $\alpha_s=0.119$). Theoretical uncertainties related to the value
of $\alpha_s$ were shown in Ref.~\cite{Ball:2008by} to be negligible and
thus not included in the NNPDF error band, but of course they become
relevant if $\alpha_s$ is varied by several standard deviations
(3.5~$\sigma$, in this case).

Coming now to MSTW, we first notice that,
as discussed in Sect.~\ref{sec:mrstvsnnpdf}, the
MSTW benchmark set has somewhat larger uncertainty bands than the 
MRST benchmark set and thus also 
than each of the sets obtained from global fits. 
Consequently, 
the MSTW benchmark PDFs are 
generally far more consistent with the MSTW global fit sets than the 
corresponding comparison between MRST benchmark PDFs and global fit PDFs shown
in Ref.~\cite{Dittmar:2005ed}, largely due to the more realistic uncertainties
in the MSTW benchmark. Comparing central values we see exactly the same 
feature in the gluon 
distribution as the NNPDF group, and the explanation is likewise the same, 
highlighting possible difficulties in comparing PDFs obtained with different 
values of $\alpha_s(M_Z)$. 

\begin{figure}[ht]
\begin{center}
\includegraphics[width=0.3\linewidth]{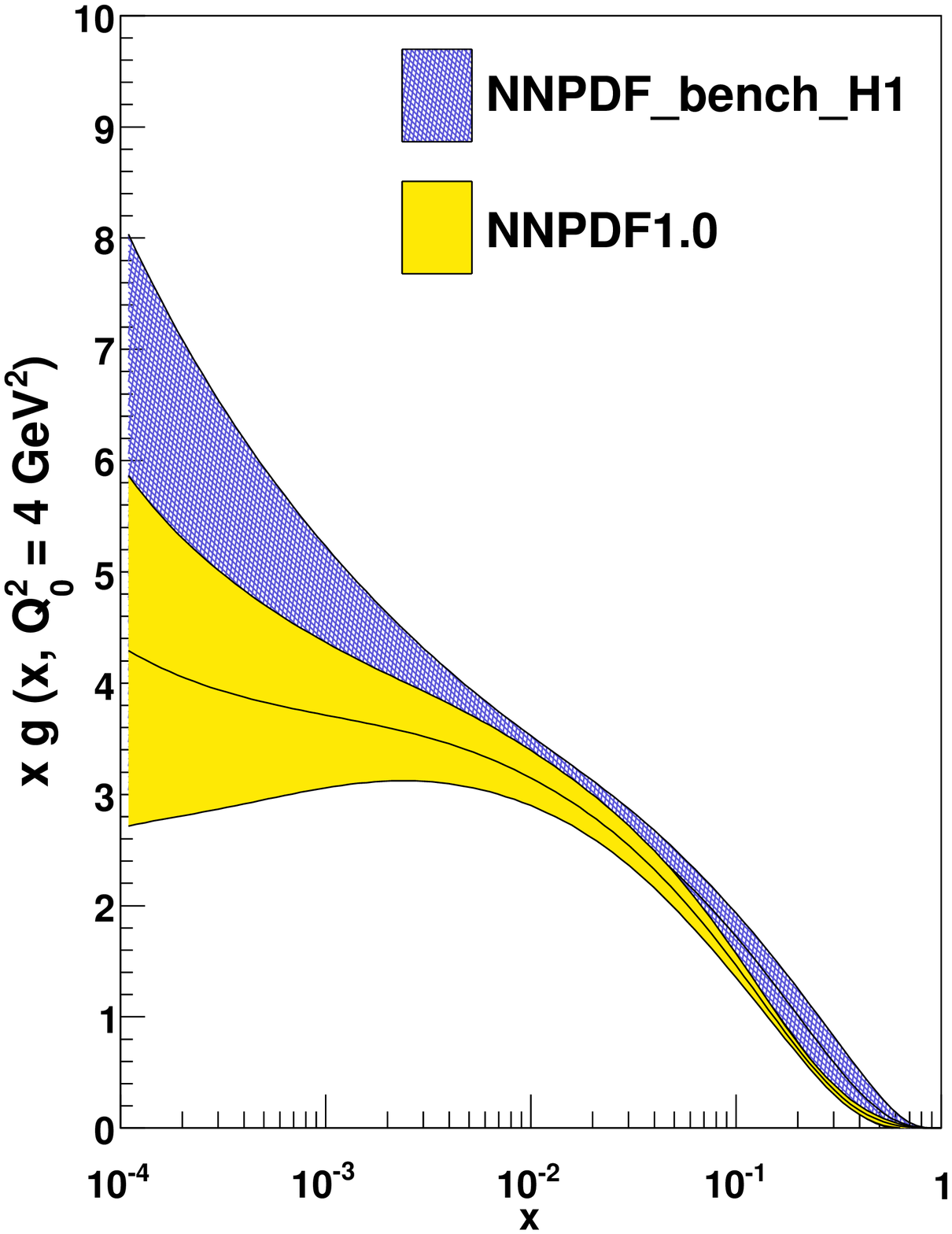}
\includegraphics[width=0.3\linewidth]{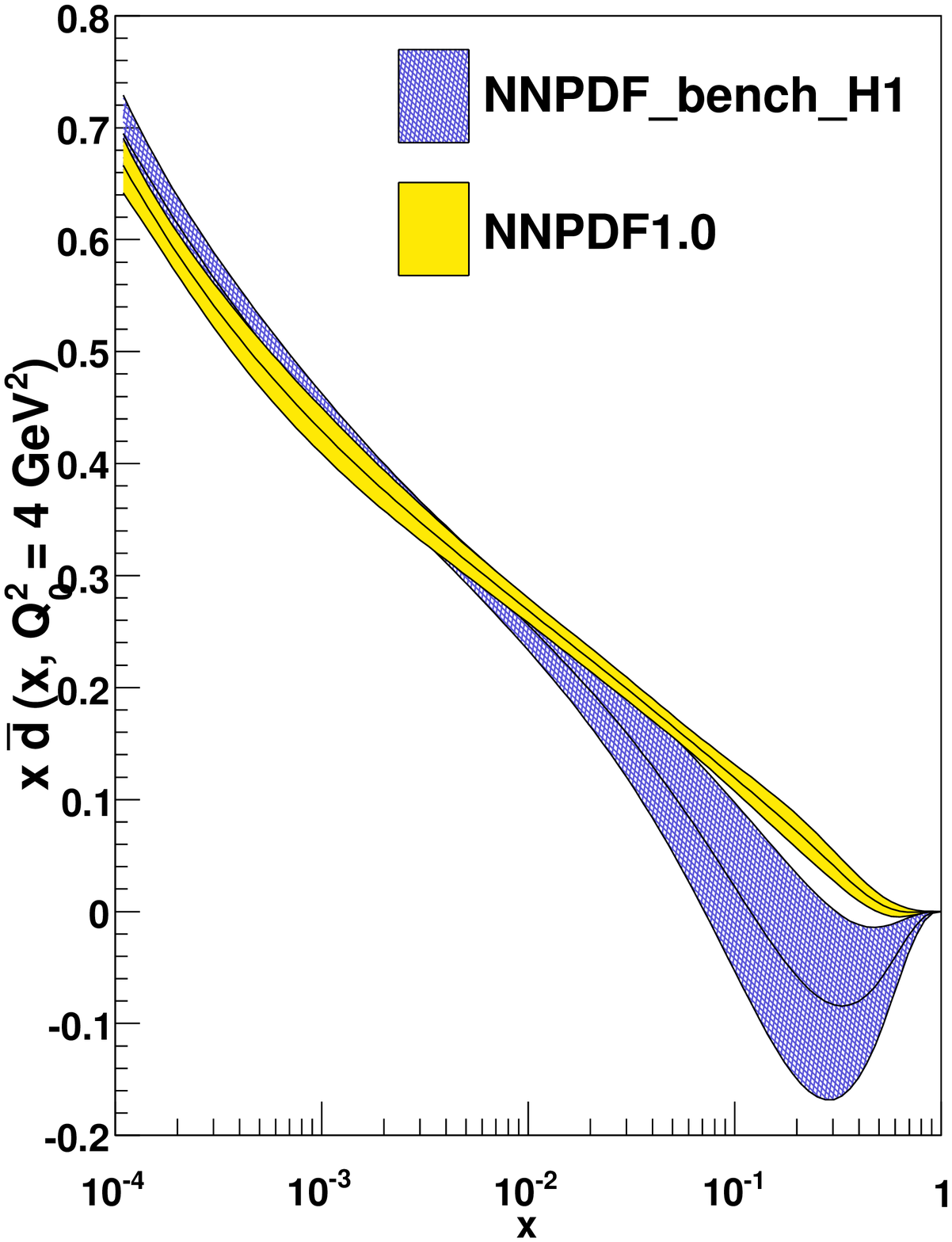}\\
\includegraphics[width=0.3\linewidth]{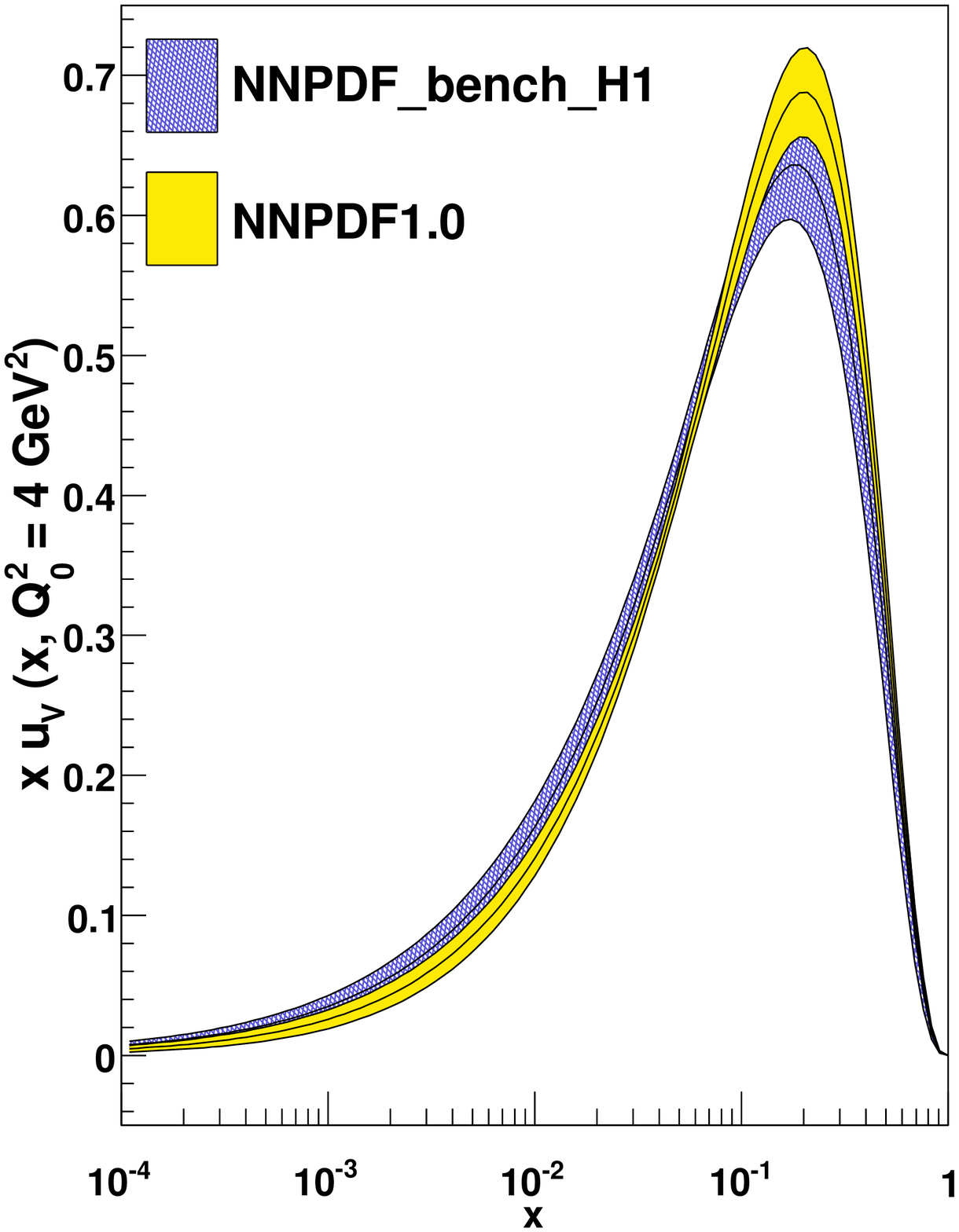}
\includegraphics[width=0.3\linewidth]{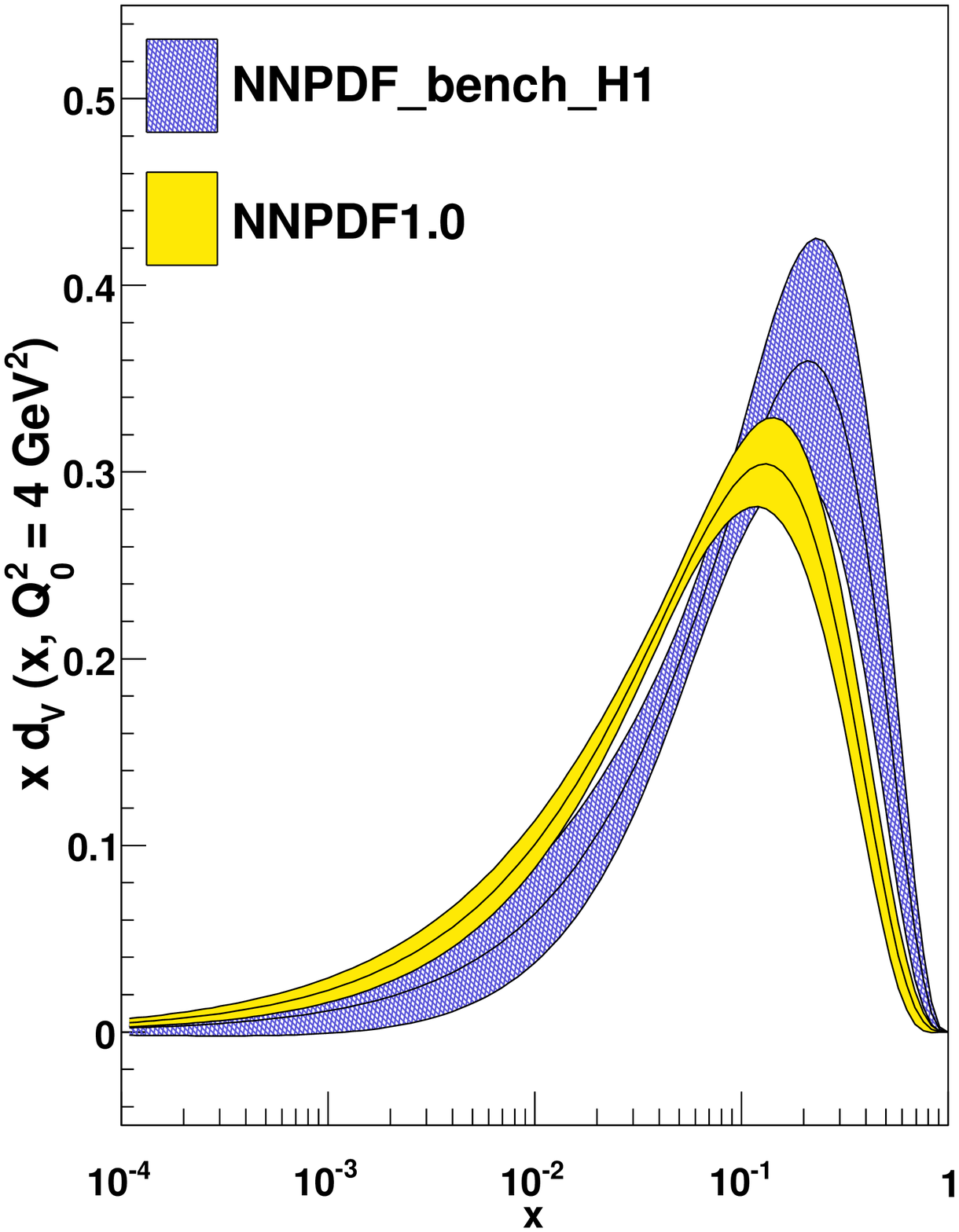}
\vspace{-.5cm}
\caption{Comparison of the NNPDF benchmark and reference fits for the
gluon, $d$-sea, $u_v$ and $d_v$ at $Q^2=4~{\rm GeV}^2$.\label{fig:nnpdf_h1_global}
}
\end{center}
\end{figure}

Unlike for the NNPDF group, the MSTW group sees some degree of 
incompatibility between the benchmark PDFs and the global fit PDFs for the 
valence quarks, particularly in the case of the down valence.
This may be related to the assumption\ $\bar u = \bar d$, which
constrains valence quarks and sea quarks in an artificial manner
since there is less flexibility to alter each
independently. Indeed, in the global fits there is an excess of\ 
$\bar d$ over\ $\bar u$ which maximizes at $x=0.1$. Forcing equivalence of
antiquark distributions might therefore lead to a deficit of down sea
quarks and a corresponding excess of up sea quarks, and also, for the
same reason, to an excess of down valence quarks. These are indeed
seen both in the NNPDF and MSTW benchmark fits when compared to the
respective global fits. The effect is however well within
the uncertainty bands for NNPDF, which indeed do not observe any
statistically significant difference between results of a fit to the
reduced benchmark data set with the\ $\bar u=\bar d$ assumption 
(as presented in Fig.~\ref{fig:nnpdf_mrst_global}) or without
it (as presented in Ref.~\cite{Ball:2008by}, Fig.~12).

As well as this important effect one sees that the main discrepancy at 
$x=0.1$ for down valence quarks is greater when comparing the benchmark fits
to the global MSTW fit than to the global MRST fit. This is because
recent new Tevatron data on $Z$ rapidity distributions
and lepton asymmetry from $W$ decays provide a strong constraint on
the down quark, and some of this new data shows considerable tension
with other data sets.

\subsection{H1 Benchmark} 
\label{sec:h1bench}

We now discuss the extension of the fit using the
settings of Sect.~\ref{sec:h1set} to also include the NNPDF
approach. Results are compared both to those of the NNPDF reference
fit, and to those obtained by the H1 fit of Sect.~\ref{sec:h1appr}
to the same data. 
We then compare the NNPDF benchmark and reference, with the specific
aim of addressing the issue of the dependence of the results 
on the size of the data set (H1 dataset vs. the HERA--LHC dataset of
Sect.~\ref{sec:heralhc}). Finally, the H1 and NNPDF benchmark fits are
compared to each other with the purpose of understanding the impact of
the respective methodologies.

\subsubsection{NNPDF analysis\protect\footnote{Contributing authors R.~D.~Ball, 
L.~Del~Debbio, S.~Forte, A.~Guffanti, J.~I.~Latorre, A.~Piccione, J.~Rojo,
M.~Ubiali}}
\label{sec:nnpdfh1b}

\begin{figure}[ht]
\begin{center}
\includegraphics[width=0.3\linewidth]{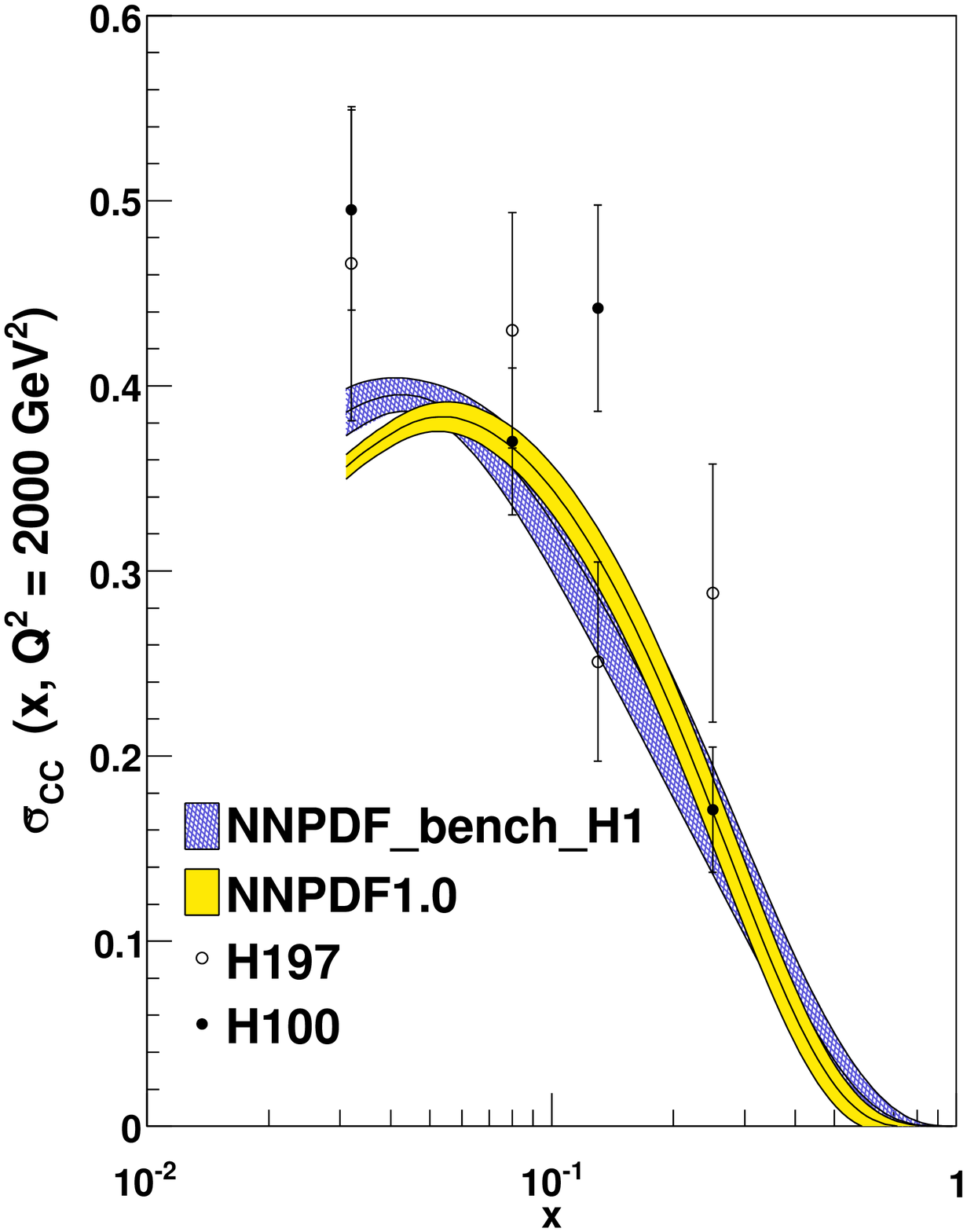}
\includegraphics[width=0.3\linewidth]{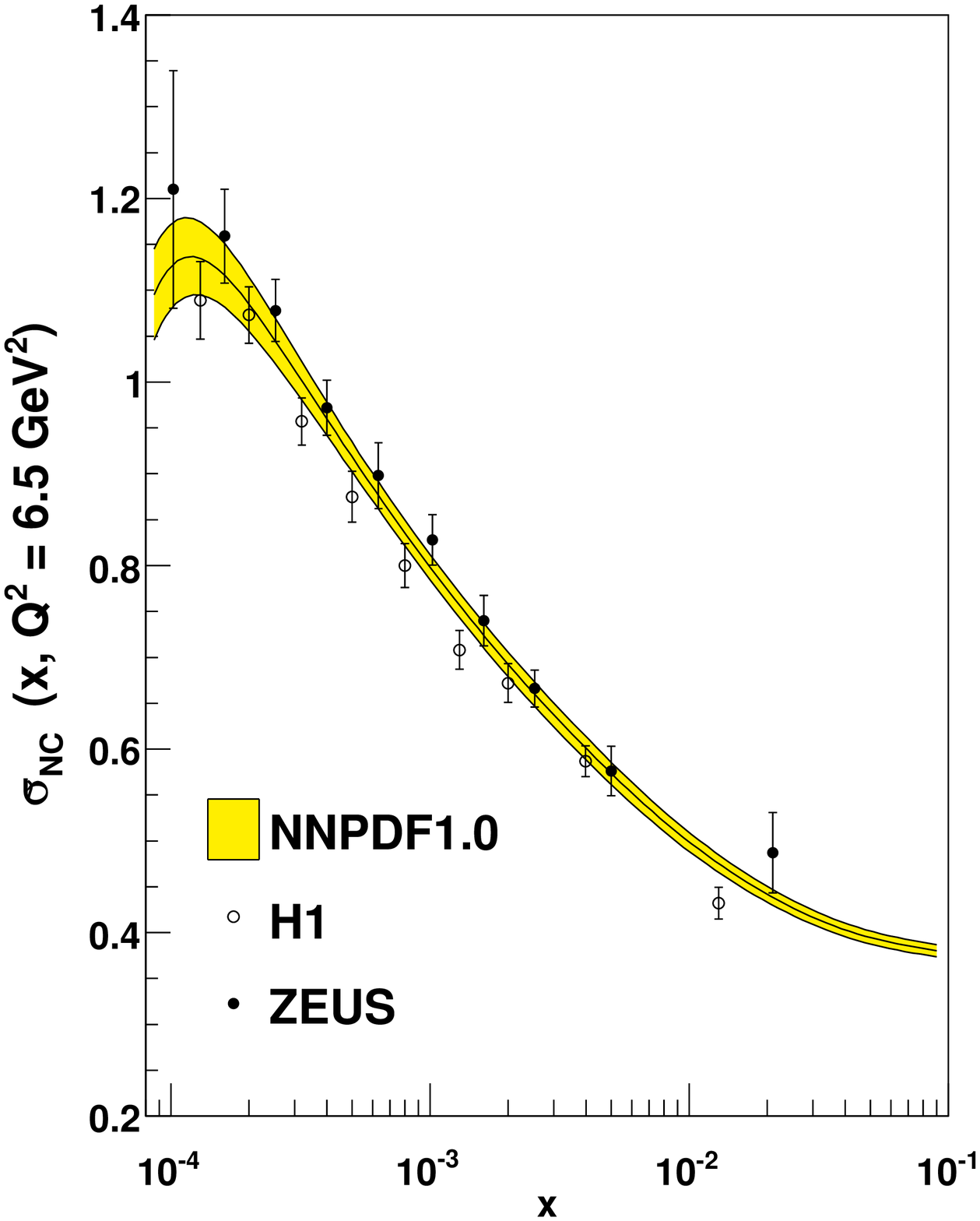}
\includegraphics[width=0.3\linewidth]{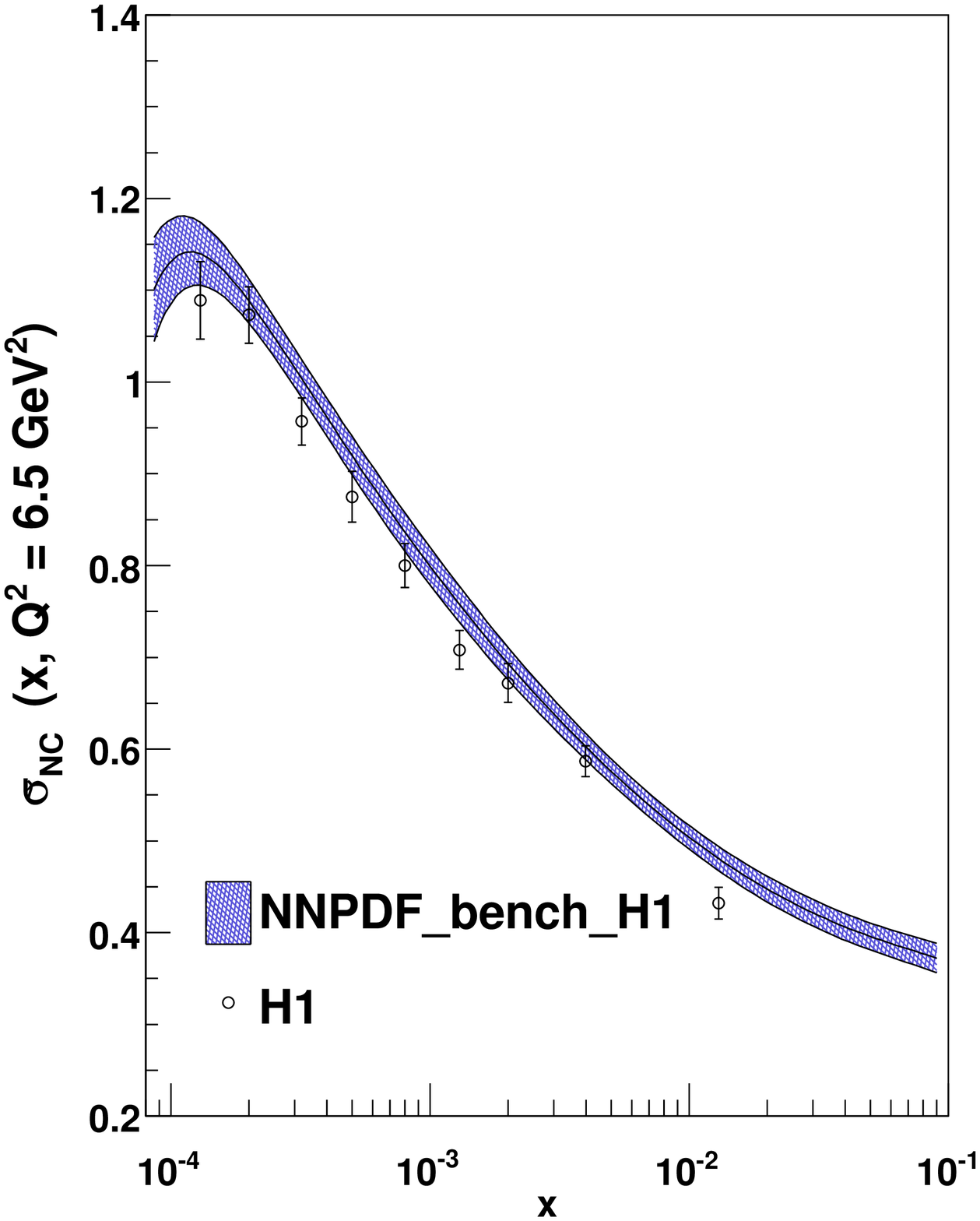}
\caption{Left: NNPDF benchmark and reference fits 
at $\sqrt{s}=301 {\rm GeV}$ compared to 
H1 charged current data. Center: NNPDF reference fit compared to H1 and ZEUS
neutral current data. Right: NNPDF benchmark fit compared to H1
neutral current data.\label{hera_cc_nc}
}
\end{center}
\end{figure}
The results of the NNPDF
benchmark are compared to the NNPDF reference fit results in
Fig.~\ref{fig:nnpdf_h1_global}.  
The general features of the benchmark are analogous to those of 
the HERA--LHC benchmark discussed in
Section~\ref{sec:nnpdf_mrst_global}, with some effects being more
pronounced because the benchmark dataset is now even smaller. Specifically, 
we observe that uncertainties bands blow up when data are removed:
this is very clear for instance  in the\ $\bar{d}$ distribution at
large-$x$, as a consequence of the fact that the benchmark dataset of
Table~\ref{tab:bench2_data} does not include deuterium data. The
negative value of this PDF at large $x$ is presumably unphysical and
it would disappear if positivity of charged current cross sections
were imposed, including also the (anti-)neutrino ones. 
The only positivity constraint in the NNPDF fit is
imposed on the $F_L$ structure function~\cite{Ball:2008by}, because
this is the only DIS observable whose positivity is not constrained
by the full data set.

It is interesting to note however that this effect is not observed for
the $u_v$ distribution, where instead the benchmark and the reference
fit show almost equal uncertainties. In order to understand this, in
Fig.~\ref{hera_cc_nc} we compare two situations with or without error
shrinking, by examining the predictions obtained using the benchmark
and reference fits for some observables to the corresponding data.  A
first plot (left) shows the shrinking of the uncertainty on the
prediction for the charged--current cross section in the reference
fit. This is mostly due to the CHORUS neutrino data, which are in the
reference and not in the benchmark. These data are clearly consistent
with the H1 data shown in the plot. The subsequent pair of plots
compares (center) the prediction for the neutral--current cross
section from the reference fit compared to H1 and ZEUS data (both of
which are used for the reference fit), and (right) from the benchmark
fit to the H1 data only (which are the only ones used in the benchmark
fit). The uncertainty bands in the two fits are similar size: indeed,
the ZEUS and H1 data display a systematic disagreement which is
approximately the size of this uncertainty band. Hence, the (small but
significant) systematic inconsistency between the ZEUS and H1 data
prevents reduction of the uncertainty band when the ZEUS data are
added to the fit, beyond the size of this discrepancy. Therefore, the
NNPDF methodology leads to combined uncertainties for inconsistent
data which are similar to those obtained with the so--called PDG (or
scale-factor) method~\cite{pdg}.

\begin{figure}[ht]
\begin{center}
\includegraphics[width=0.3\linewidth]{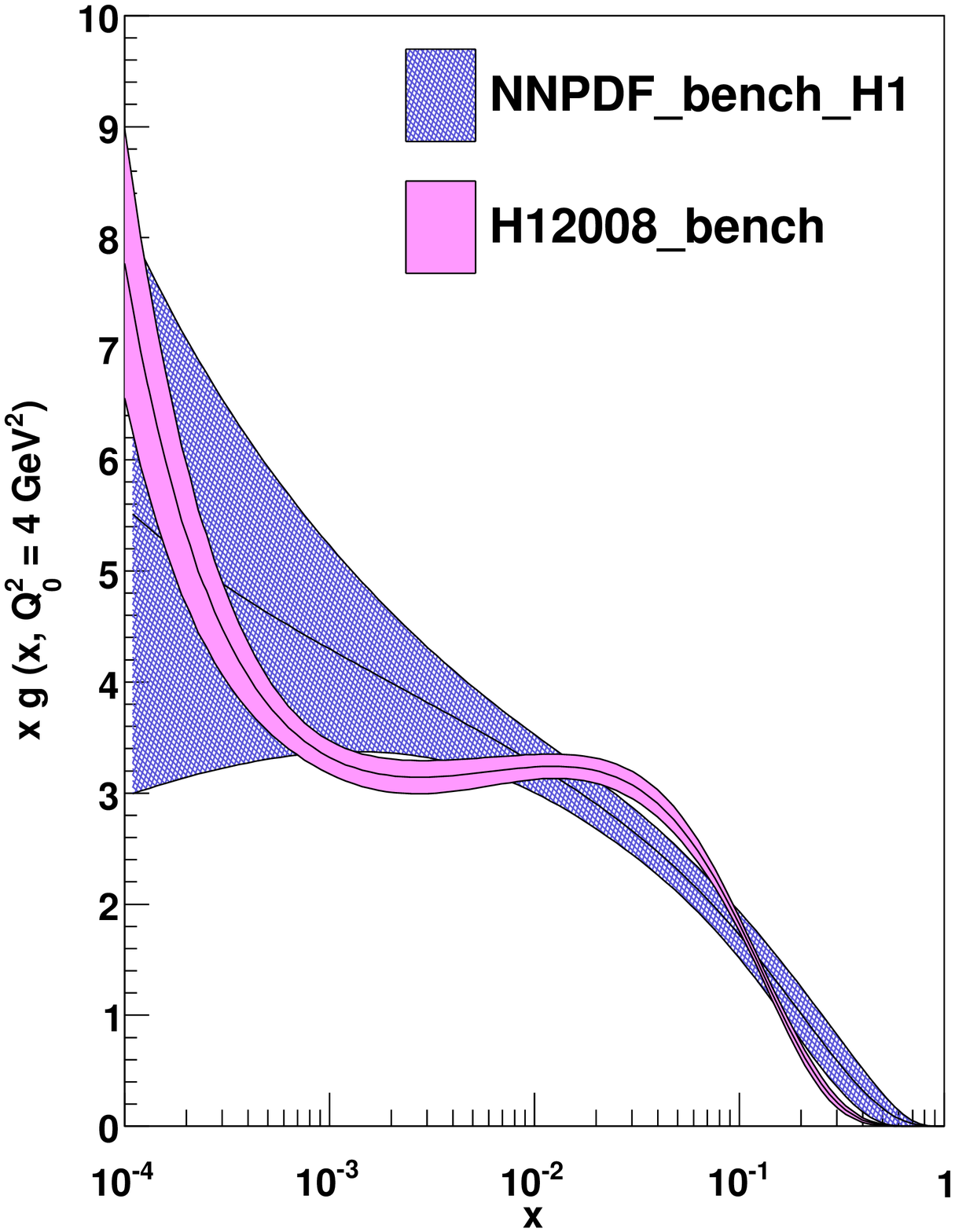}
\includegraphics[width=0.3\linewidth]{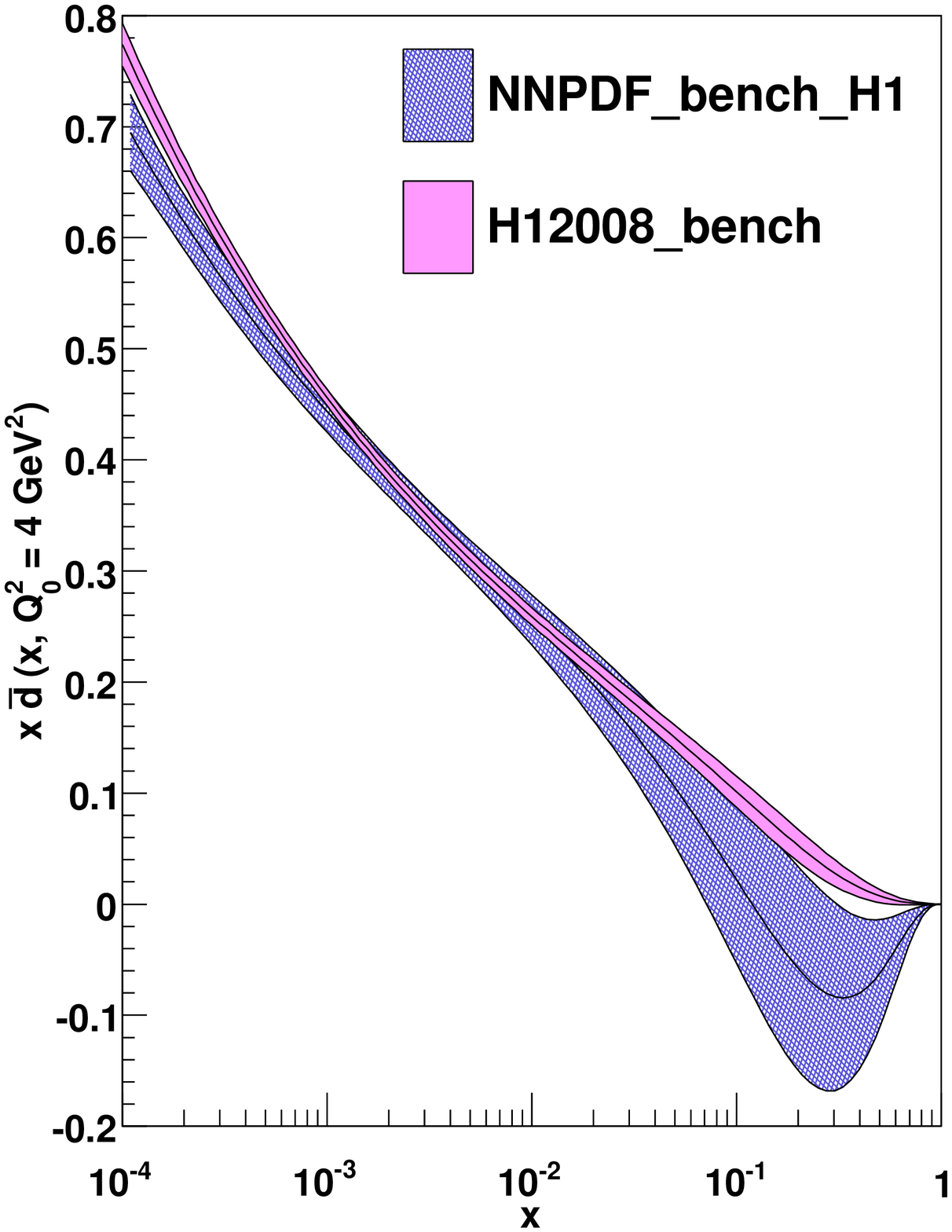}\\
\includegraphics[width=0.3\linewidth]{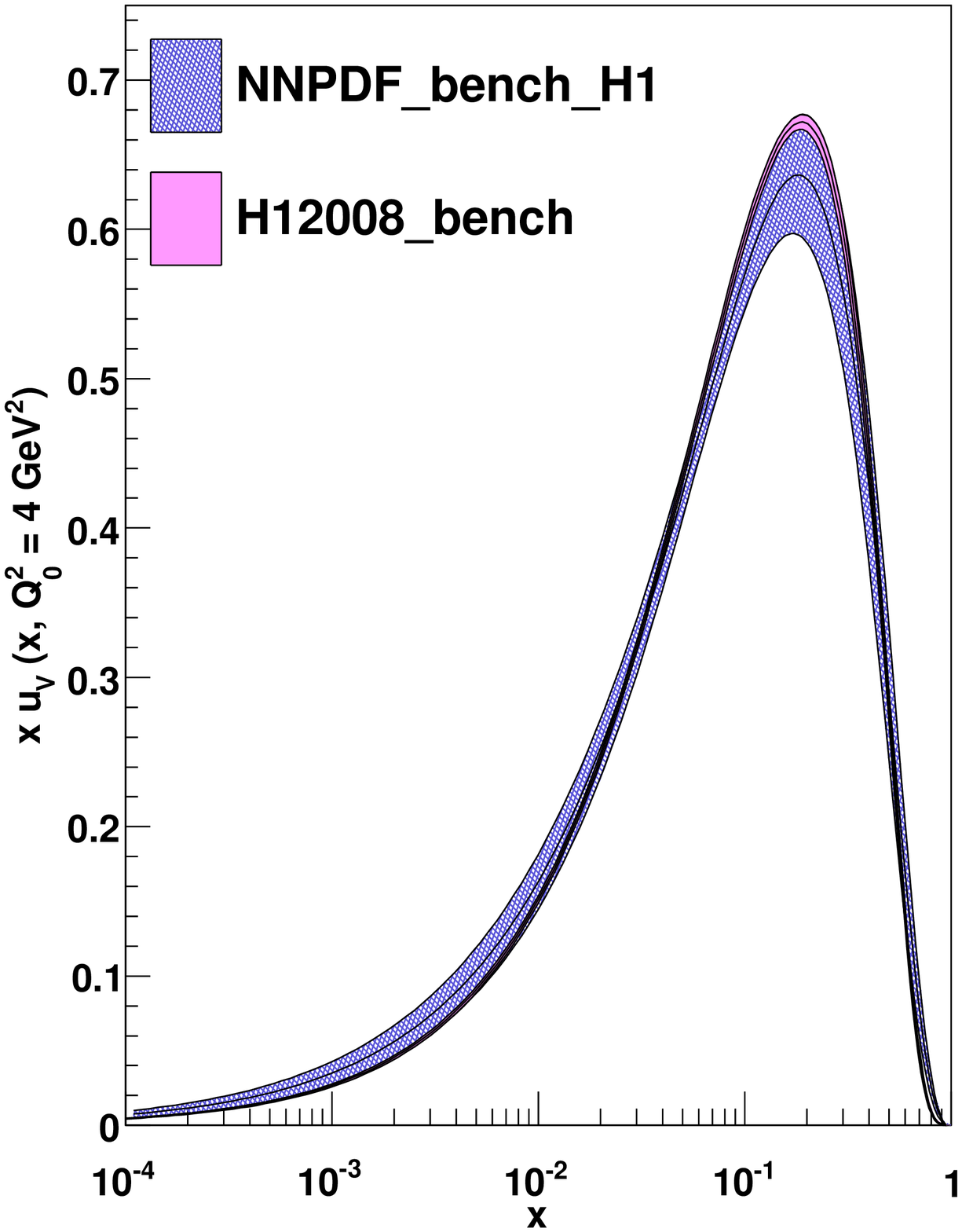}
\includegraphics[width=0.3\linewidth]{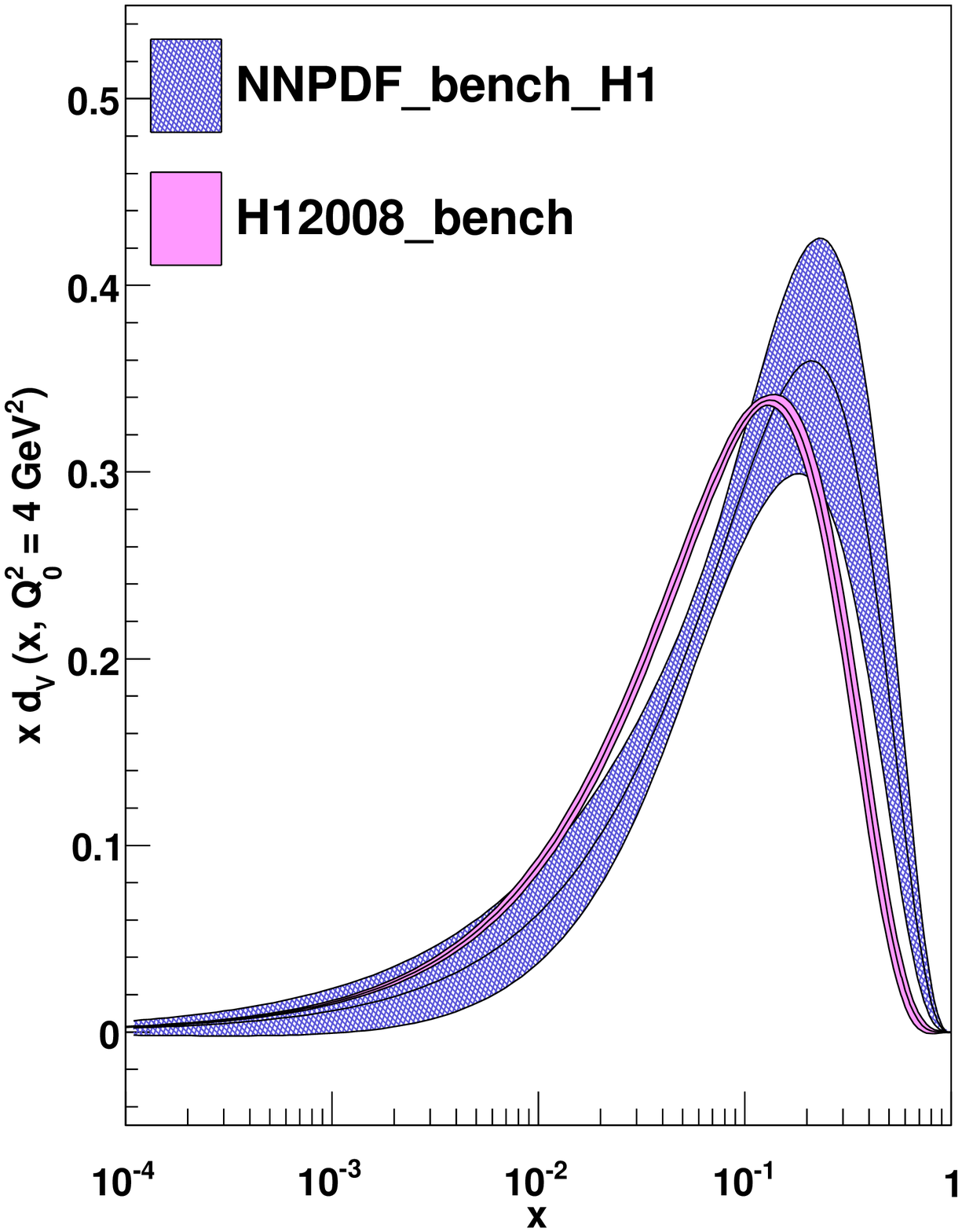}
\vspace{-.5cm}
\caption{Comparison of the NNPDF and H1 benchmark fit for the
gluon, $d$-sea, $u_v$ and $d_v$ at $Q^2=4~{\rm GeV}^2$.\label{fig:nnpdf_h1_bench}
}
\end{center}
\end{figure}

\begin{figure}[ht]
\begin{center}
\includegraphics[scale=0.45]{gauss.eps}
\includegraphics[scale=0.35]{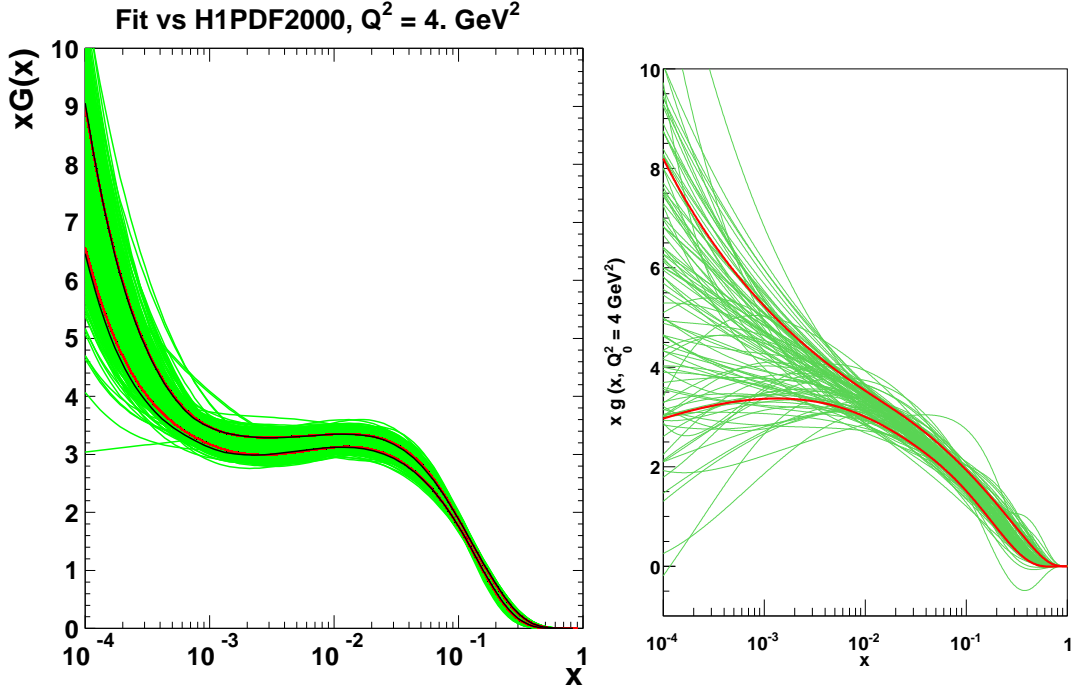}
\caption{The Monte Carlo set of gluon PDFs for the 
H1 benchmark
(left, same as Fig.~\ref{Fig:mv}) and the NNPDF benchmark. The red
lines show the one-sigma contour calculated from the Monte Carlo set,
and in the H1 case the black lines show the Hessian one-sigma contour.
\label{fig:nnpdfvsh1}
}
\end{center}
\end{figure}

\begin{table}[h!]
  \begin{center}
    \begin{tabular}{|l|r|r|}
      \hline
      Data Set & $\chi^2_{\rm H1}/N_{\rm data}$ & $\chi^2_{\rm NNPDF}/N_{\rm data}$  \\
      \hline
      H197mb    & 0.83 & 0.82 \\
      H197lowQ2 & 0.90 & 0.87 \\
      H197NC    & 0.69 & 0.80 \\
      H197CC    & 0.73 & 0.97 \\
      H199NC    & 0.88 & 1.01 \\
      H199CC    & 0.62 & 0.84 \\
      H199NChy  & 0.35 & 0.35 \\
      H100NC    & 0.97 & 1.00 \\
      H100CC    & 1.07 & 1.38 \\
      \hline
      Total & 0.88 & 0.96 \\
      \hline
    \end{tabular}
  \end{center}
\caption{H1 and NNPDF $\chi^2$ for the total and each single data set.
Cross correlations among data sets are neglected to evaluate
the $\chi^2$ of a single data set.\label{tab:chi2comp}
}
\end{table}

Notice that if relative normalization  are fitted (as done by
in the H1 approach of Sect.~\ref{sec:h1appr}) instead of being treated
simply as a source of systematics, this systematic inconsistency would
be significantly reduced in the best-fit. The associate uncertainty
however then appears as an addition source of systematics.
 This happens when H1 and
ZEUS data are combined in a single dataset (see
Section~\ref{sec:combfit} below). In the NNPDF
approach, instead, this systematics is produced by the Monte Carlo
procedure. 

\subsubsection{Comparison between the Benchmark Parton Distributions}
\label{sec:h1nnpdfcomp}

The $\chi^2$ of the H1 and NNPDF benchmarks are given in
Table~\ref{tab:chi2comp}, while the corresponding PDFs are compared in
Fig.~\ref{fig:nnpdf_h1_bench}. Furthermore, in Fig.~\ref{fig:nnpdfvsh1} we
show the respective full Monte Carlo PDF sets in the case of
the gluon distribution.  

The quality of the two fits is comparable, the differences in
$\chi^2$ being compatible with statistical fluctuations.
In the region where experimental information
is mostly concentrated, specifically for the $u_v$ distribution over all
the $x$-range and for the\ $\bar{d}$ and the $d_v$ distributions in the
small-$x$ range, the results of the two fits are in good agreement,
though the H1 uncertainty bands are generally somewhat smaller.

In the region where experimental information is scarce or missing,
sizable differences are found, similar to those observed when
comparing the MRST/MSTW bench and NNPDF bench to the HERA--LHC benchmark of
Sect.~\ref{sec:mrstvsnnpdf}. 
Specifically, in these regions NNPDF uncertainties are
generally larger than H1 bands: the width of the uncertainty band for
the H1 fit varies much less between the data and extrapolation regions
than that of the NNPDF bench. 
Also, the H1 central value always falls within the NNPDF
uncertainty band, but the NNPDF central value tends to fall outside
the H1 uncertainty band whenever the central values differ significantly.
Figure~\ref{fig:nnpdfvsh1} suggests that
this may be due to the greater flexibility of the functional form in
the NNPDF fit. Specifically, the\ $\bar d$ quark distribution at large
$x$ does not become negative in the H1 fit, because this behaviour is
not allowed by the parameterization.

\pagebreak
\section{DETERMINATION OF PARTON DISTRIBUTIONS}
\label{sec:pdfdet}
\subsection{Extraction of the proton PDFs from a combined fit
of H1 and ZEUS inclusive DIS cross sections 
\protect\footnote{Contributing authors: A.~Cooper-Sarkar,  A.~Glazov, G.~Li
for the H1-ZEUS combination group.}}
\label{sec:combfit}
\subsubsection{Introduction}
\label{sec:intro}
The kinematics
of lepton hadron scattering is described in terms of the variables $Q^2$, the
invariant mass of the exchanged vector boson, Bjorken $x$, the fraction
of the momentum of the incoming nucleon taken by the struck quark (in the 
quark-parton model), and $y$ which measures the energy transfer between the
lepton and hadron systems.
The differential cross-section for the neutral current (NC) process is given in 
terms of the structure functions by
\[
\frac {d^2\sigma(e^{\pm}p) } {dxdQ^2} =  \frac {2\pi\alpha^2} {Q^4 x}
\left[Y_+\,F_2(x,Q^2) - y^2 \,F_L(x,Q^2)
\mp Y_-\, xF_3(x,Q^2) \right],
\]
where $\displaystyle Y_\pm=1\pm(1-y)^2$. 
The structure functions $F_2$ and $xF_3$ are 
directly related to quark distributions, and their
$Q^2$ dependence, or scaling violation, 
is predicted by perturbative QCD. For low $x$, $x \leq 10^{-2}$, $F_2$ 
is sea quark dominated, but its $Q^2$ evolution is controlled by
the gluon contribution, such that HERA data provide 
crucial information on low-$x$ sea-quark and gluon distributions.
At high $Q^2$, the structure function $xF_3$ becomes increasingly 
important, and gives information on valence quark distributions. 
The charged current (CC) interactions also
enable us to separate the flavour of the valence distributions 
at high-$x$, since their (LO) cross-sections are given by, 
\[
\frac {d^2\sigma(e^+ p) } {dxdQ^2} = \frac {G_F^2 M_W^4} {(Q^2 +M_W^2)^2 2\pi x}
x\left[(\bar{u}+\bar{c}) + (1 - y)^2 (d + s) \right],
\]
\[
\frac {d^2\sigma(e^- p) } {dxdQ^2} = \frac {G_F^2 M_W^4} {(Q^2 +M_W^2)^2 2\pi x}
x\left[(u + c) + (1 - y)^2 (\bar{d} + \bar{s}) \right].
\]

Parton Density Function (PDF) determinations are usually obtained in global NLO 
QCD fits~\cite{mrst,cteq,Chekanov:2005nn}, which use fixed target 
DIS data as well as HERA data. In such analyses, the high statistics HERA NC 
$e^+p$ data have determined the low-$x$ sea and 
gluon distributions, whereas the fixed target data have determined 
the valence distributions. Now that high-$Q^2$ HERA data on NC and CC
 $e^+p$ and $e^-p$ inclusive double 
differential cross-sections are available, PDF fits can be made to HERA 
data alone, since the HERA high $Q^2$ cross-section 
data can be used to determine the valence distributions. This has the 
advantage that it eliminates the need for heavy target corrections, which 
must be applied to the $\nu$-Fe and $\mu D$ fixed target data. Furthermore
there is no need to assume isospin symmetry, i.e. that $d$ in the 
proton is the same as $u$ in the neutron, 
since the $d$ distribution can be obtained directly from CC $e^+p$ data. 

The H1 and ZEUS collaborations have both used their data to make PDF 
fits~\cite{Chekanov:2005nn},~\cite{Adloff:2003uh}. Both of these data sets
have very small statistical uncertainties, so that the contribution of 
systematic uncertainties becomes dominant and consideration of 
point to point correlations between systematic uncertainties is essential.
The ZEUS analysis takes account of correlated experimental 
systematic errors by the Offset Method, whereas H1 uses the Hessian 
method~\cite{durham}. 
Whereas the resulting ZEUS and H1 PDFs are compatible, the gluon PDFs 
have rather different shapes, see Fig~\ref{fig:summary}, and the 
uncertainty bands spanned by these analyses 
are comparable to those of the global fits.

It is possible to improve on this situation 
since ZEUS and H1 are measuring the same physics in the same 
kinematic region. These data have been combined using a 'theory-free' 
Hessian fit in which the only assumption is that there is a true 
value of the cross-section, for each process, at each $x,Q^2$ 
point~\cite{combination,*Feltesse}. 
Thus each experiment has been calibrated to the other. 
This works well because the 
sources of systematic uncertainty in each experiment are rather different, 
such that all the systematic uncertainties are re-evaluated. 
The resulting correlated systematic uncertainties 
on each of the combined data points are significantly smaller than the 
statistical errors. This combined data set has been used as the 
input to an NLO QCD PDF fit. The consistency of the input data set 
and its small systematic uncertainties enables us 
to calculate the experimental uncertainties on the PDFs using the 
$\chi^2$ tolerance, $\Delta\chi^2=1$. This represents a further advantage 
compared to the global fit analyses where increased tolerances of 
$\Delta\chi^2=50-100$ are used to account for data inconsistencies. 

\begin{figure}[tbp]
\centerline{
\epsfig{figure=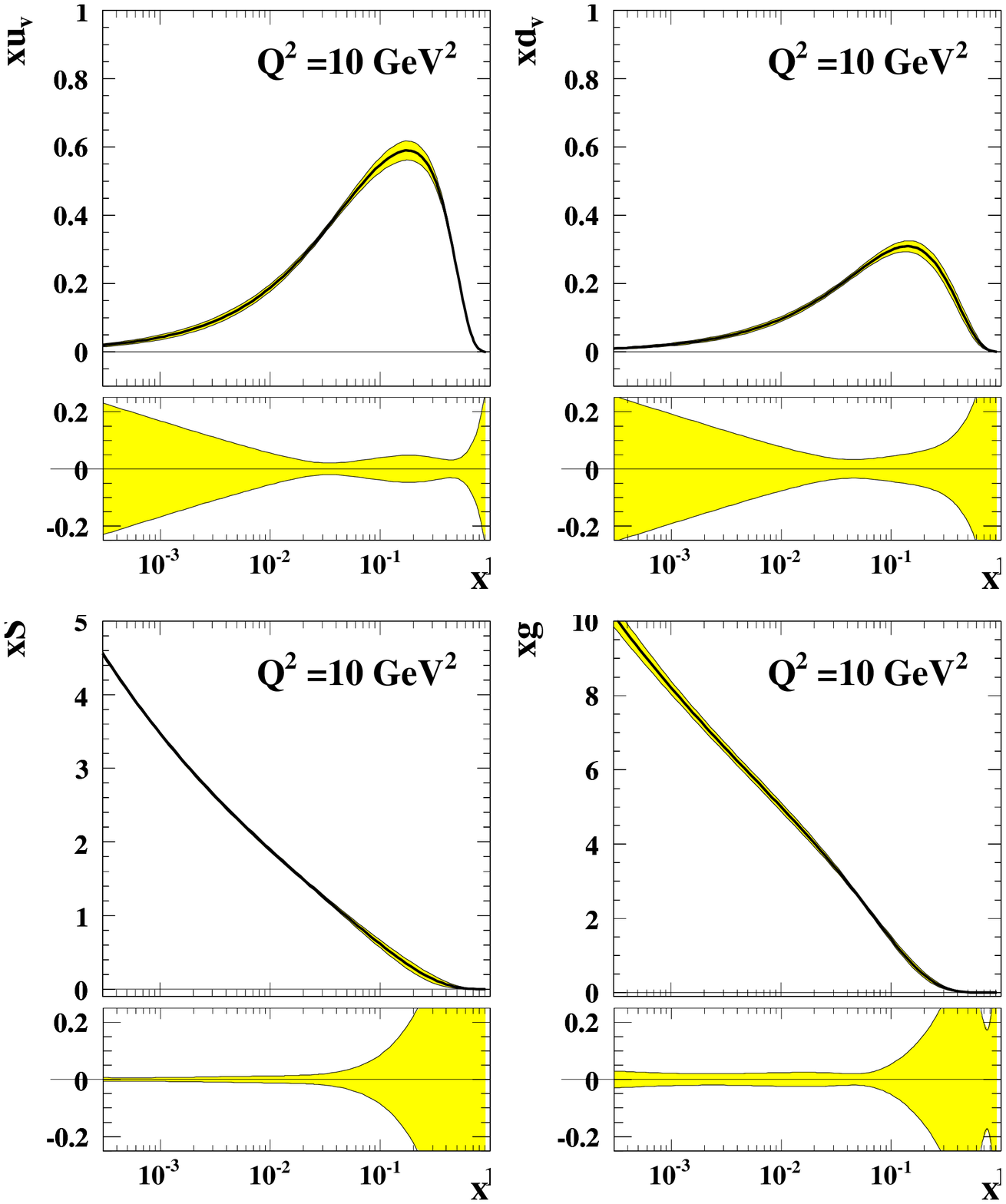,width=0.33\textwidth,height=5.5cm}
\epsfig{figure=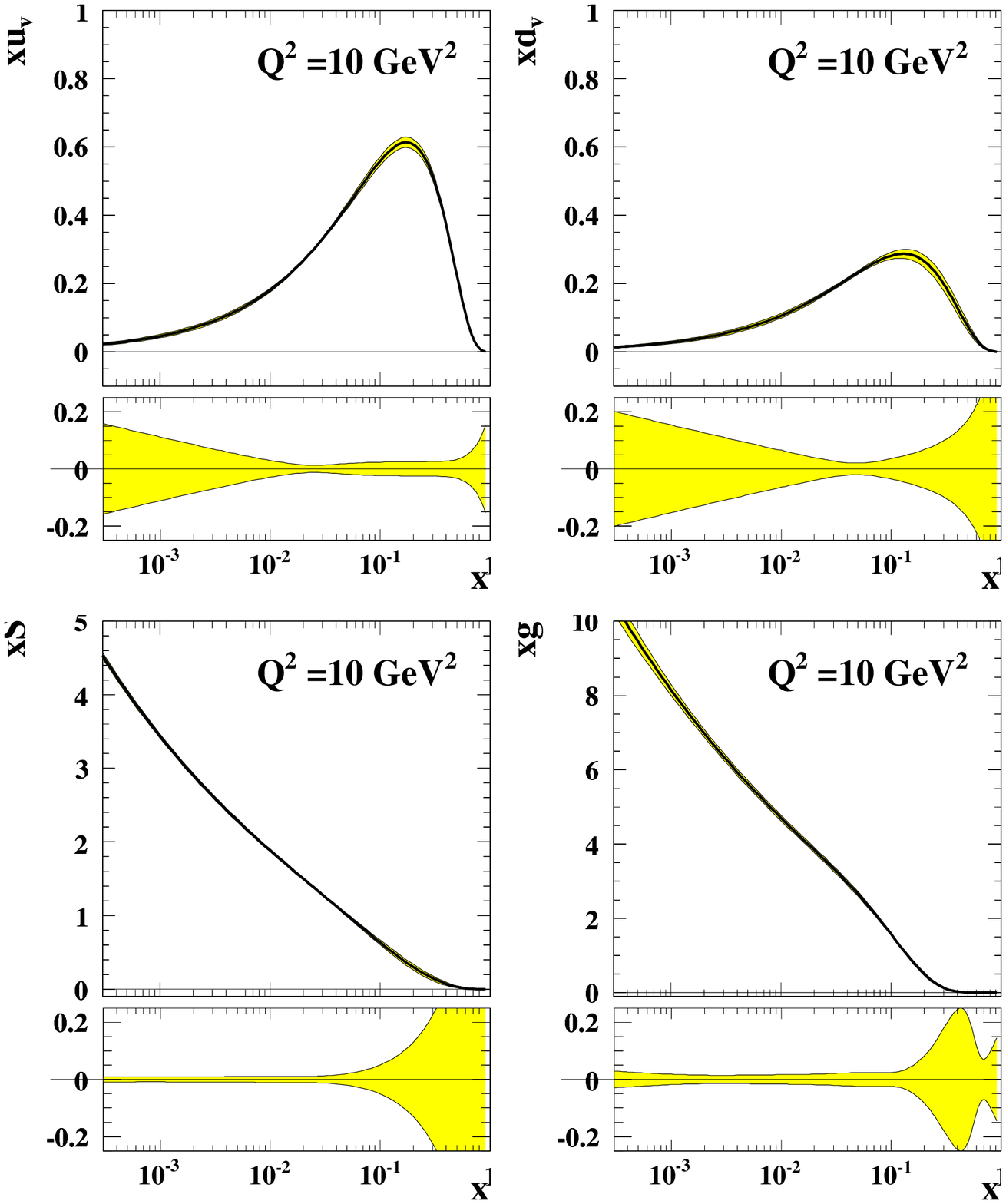,width=0.33\textwidth,height=5.5cm}}
\caption {HERAPDFs, $xu_v,xd_v,xS,xg$ at $Q^2=10$GeV$^2$. (Left) with 
experimental uncertainties evaluated as for the central fit (see text) and 
(right) with experimental uncertainties evaluated by accounting for the 
47 systematic errors by the Hessian method.}
\label{fig:hessian}
\end{figure}

For the HERAPDF0.1 fit presented here, the role of correlated 
systematic uncertainties is no longer crucial since these uncertainties are 
relatively small. This ensures that similar results are 
obtained using either Offset or Hessian methods, or by simply combining 
statistical and systematic uncertainties in quadrature. 
The $\chi^2$ per degree of freedom for a Hessian fit is $553/562$ and for a 
quadrature fit it is $428/562$.
For our central fit we have 
chosen to combine the 43 systematic uncertainties 
which result from the separate ZEUS and H1 data sets in quadrature, 
and to Offset the 4 sources of uncertainty which result from the combination 
procedure. The $\chi^2$ per degree of freedom for this fit is $477/562$.
This procedure results in the most conservative estimates on the resulting 
PDFs as illustrated in Fig.~\ref{fig:hessian} which compares the PDFs and 
their experimental uncertainties as evaluated by the procedure of our central 
fit and as evaluated by treating the 47 systematic uncertainties by the 
Hessian method.

Despite this conservative procedure,
 the experimental uncertainties on the resulting 
PDFs are impressively small and a thorough consideration of further 
uncertainties 
due to model assumptions is necessary. In Section~\ref{sec:datacomb}
we briefly describe the data combination procedure.
In Section~\ref{sec:anal} we describe 
the NLO QCD analysis and model assumptions. In Section~\ref{sec:results} 
we give results. In Section~\ref{sec:conc} we give a summary of the fit results
and specifications for release of the HERAPDF0.1 to LHAPDF. In 
Section~\ref{sec:WZ} we investigate the predictions of the HERAPDF0.1 for 
$W$ and $Z$ cross-sections at the LHC.
 
\subsubsection{Data Combination}
\label{sec:datacomb}
The data combination is based on assumption that the H1 and ZEUS experiments measure
the same cross section at the same kinematic points. The systematic uncertainties
of the measurements are separated, following the prescription given by the H1 and ZEUS,
into point to point correlated sources $\alpha_j$ and uncorrelated systematic uncertainty,
which is added to the statistical uncertainty in quadrature to  result in total uncorrelated
uncertainty $\sigma_{i}$ for each bin $i$. The correlated systematic sources are considered to be 
 uncorrelated between H1 and ZEUS. All uncertainties
are treated as multiplicative i.e. proportional to the central values,
which is a good approximation for the measurement of  the cross sections. 

A correlated probability distribution function for the physical cross sections $M^{i, {\rm true}}$ 
and systematic uncertainties $\alpha_{j,{\rm true}}$
for a single experiment  corresponds to a $\chi^2$ function:
\begin{equation}
 \chi^2_{\rm exp}\left(M^{i,{\rm true}},\alpha_{j,{\rm true}}\right) = 
 \sum_i
 \frac{\left[M^{i,{\rm true}}-\left(M^i 
+ \sum_j \frac{\textstyle \partial M^i}{\textstyle \partial \alpha_j} 
\frac{\textstyle M^{i,{\rm true}}}{\textstyle M^i} (\alpha_{j,{\rm true}})\right)\right]^2}
{\left(\sigma_{i} \frac{\textstyle M^{i,{\rm true}}}{\textstyle M^i}\right)^2}
 + \sum_j \frac{( \alpha_{j,{\rm true}})^2} {\sigma^2_{\alpha_j}},
\label{eq:ave}
\end{equation}
where $M^i$ are the central values measured by the experiment, $\partial M^i/\partial \alpha_j$
are the sensitivities to the correlated systematic uncertainties and $\sigma_{\alpha_j}$
are the uncertainties of the systematic sources. For more than one experiment, total  $\chi_{\rm tot}^2$
can be represented as a sum of $\chi^2_{\rm exp}$. The combination procedure 
allows to represent $\chi_{\rm tot}^2$ in the following form:
\begin{equation}
 \chi^2_{\rm tot}\left(M^{i,{\rm true}},\beta_{j,{\rm true}}\right) = 
\chi^2_0 +
 \sum_i
 \frac{\left[M^{i,{\rm true}}-\left(M^{i,{\rm ave}} 
+ \sum_j \frac{\textstyle \partial M^{i,{\rm ave}}}{\textstyle \partial \beta_j}
\frac{\textstyle M^{i,{\rm true}}}{\textstyle M^{i,{\rm ave}}} (\beta_{j,{\rm true}})\right)\right]^2}
{\left(\sigma_{i,{\rm ave}} \frac{\textstyle M^{i,{\rm true}}}{\textstyle M^{i,{\rm ave}}}\right)^2}
 + \sum_j \frac{( \beta_{j,{\rm true}})^2} {\sigma^2_{\beta_j}}.
\label{eq:ave2}
\end{equation}
Here the sum runs over a union set of the cross section bins.
The value of the $\chi^2_{tot}$ at the minimum, $\chi^2_0$, quantifies 
consistency of the experiments. $M^{i,{\rm ave}}$ are the average values
of the cross sections and $\beta_j$ correspond to the new systematic
sources which can be obtained from the original sources $\alpha_j$
through the action of an
orthogonal matrix.  In essence, the average of several data sets
allows one to represent the total $\chi^2$ in a form which is similar
to that corresponding to a 
single data set, Eq.~\ref{eq:ave}, but with modified
systematic sources. 

The combination is applied to NC and CC cross section data taken with $e^+$ and $e^-$ beams simultaneously
to take into account correlation of the systematic uncertainties. The data taken 
with proton beam energies of $E_p=820$~GeV and $E_p=920$~GeV are combined together
for inelasticity $y<0.35$, for this a small center of mass energy correction is applied.
For the combined data set there are 596 data points and 43 experimental systematic sources.
The $\chi^2_0/dof = 510/599$ is below $1$, which indicates conservative estimation
of the uncorrelated systematics.

Besides the experimental uncertainties,  four additional sources 
related to the assumptions made for the systematic uncertainties are considered. Two of the extra sources
deal with  correlation of the H1 and ZEUS data for estimation of the photoproduction background
and simulation of hadronic energy scale. These sources introduce additional $\sim 1\%$ uncertainty
for $y>0.6$ and $y<0.02$ data. The third source  covers uncertainty 
arising from the center of mass
correction by varying $F_L = F_L^{QCD}$ to $F_L=0$. The resulting uncertainty reaches few per mille level
for $y\sim 0.35$. Finally, some of the systematic uncertainties, for example background subtraction,
may not be necessary   multiplicative but rather additive, independent of the cross section central values.
The effect of additive assumption for the errors is evaluated by comparing the average obtained using
Eq.~\ref{eq:ave} and an average in which $M^{i,{\rm true}}/M^{i,{\rm ave}}$ scaling is removed
for all but global normalization errors.

\subsubsection{QCD Analysis}
\label{sec:anal}
 
The QCD predictions for the structure functions 
are obtained by solving the DGLAP evolution 
equations~\cite{Dokshitzer:1977sg,Gribov:1972ri,Altarelli:1977zs} 
at NLO in the \msbar scheme with the
renormalisation and factorization scales chosen to be $Q^2$~\footnote{
The programme QCDNUM~\cite{qcdnum} has been used and checked against the 
programme QCDfit~\cite{pz}.}.
The DGLAP equations yield the PDFs
 at all values of $Q^2$ provided they
are input as functions of $x$ at some input scale $Q^2_0$. This scale has been 
chosen to be $Q^2_0 = 4$GeV$^2$ and variation of this choice is considered 
as one of the model uncertainties.
The resulting PDFs are then convoluted with NLO coefficient functions to 
give the structure functions which enter into the expressions for the 
cross-sections. The choice of the heavy quark 
masses is, $m_c=1.4, m_b=4.75$GeV, and variation of these 
choices is included in 
the model uncertainties. For this preliminary analysis, the heavy quark 
coefficient functions have been calculated in the zero-mass variable flavour 
number scheme. The strong coupling constant was fixed to 
$\asmz =  0.1176$~\cite{pdg}, and variations in this value of $\pm 0.002$ 
have also been considered.

The fit is made at leading twist. 
The HERA data have a minimum invariant mass of the hadronic system, $W^2$, 
of $W^2_{min} = 300$~GeV$^2$ and a maximum $x$, $x_{max} = 0.65$, 
such that they are in a kinematic region where there is no
sensitivity to target mass and large-$x$ higher 
twist contributions. However a minimum $Q^2$ cut is imposed 
to remain in the kinematic region where
perturbative QCD should be applicable. This has been chosen to be
 $Q^2_{min} = 3.5$~GeV$^2$. Variation of this cut is included as one
 of the model uncertainties. 

 A further 
model uncertainty is the choice of the initial parameterization at 
$Q^2_0$. Three types of parameterization have been considered. For each of 
these choices 
the PDFs are parameterized by the generic form 
\begin{equation}
 xf(x) = A x^{B} (1-x)^{C} (1 + D x + E x^2 +F x^3),
\label{eqn:pdf}
\end{equation}
and the number of parameters is chosen by 'saturation of the $\chi^2$', 
such that parameters $D,E,F$ are only varied if this brings significant 
improvement to the $\chi^2$. Otherwise they are set to zero.

The first parameterization considered follows that used by the ZEUS 
collaboration. The PDFs for $u$ valence, $xu_v(x)$,  $d$ valence, $xd_v(x)$, 
total sea, $xS(x)$, the 
gluon, $xg(x)$, and the difference between the $d$ and $u$
contributions to the sea, $x\Delta(x)=x(\bar{d}-\bar{u})$, are parameterized. 
 
\[
xu_v(x) = A_{uv} x^{B_{uv}} (1-x)^{C_{uv}} (1 + D_{uv} x +E_{uv} x^2)
\]
\[
xd_v(x) = A_{dv} x^{B_{dv}} (1-x)^{C_{dv}}
\]
\[xS(x) = A_{S} x^{B_{S}} (1-x)^{C_{S}}\]
\[xg(x) = A_{g} x^{B_{g}} (1-x)^{C_{g}} (1 + D_g x)\]
\[x\Delta(x)= A_{\Delta} x^{B_\Delta} (1-x)^{C_\Delta}
\]
The total sea is given by, 
$xS=2x(\bar{u} +\bar{d} +\bar{s}+ \bar{c} +\bar{b})$, where 
$\bar{q}=q_{sea}$ for each flavour, $u=u_v+u_{sea}, d=d_v+d_{sea}$ and 
$q=q_{sea}$ for all other flavours. There is no 
information on the shape of the $x\Delta$ distribution in a fit 
to HERA data alone and so this distribution has its parameters 
fixed, such that its shape is consistent
with Drell-Yan data and its normalization is consistent 
with the size of the Gottfried sum-rule violation. 
A suppression of the strange sea with respect to the non-strange sea 
of a factor of 2 at $Q^2_0$, is imposed
consistent with neutrino induced dimuon data from NuTeV. 
The normalisation parameters, $A_{uv}, A_{dv}, A_g$, are constrained to 
impose the number sum-rules and momentum sum-rule. 
The $B$ parameters, $B_{uv}$ and $B_{dv}$  are set equal, 
since there is no information to constrain any difference. 
Finally this ZEUS-style parameterization has eleven free parameters.

The second parameterization considered follows that of the H1 Collaboration 
The choice of quark PDFs which are 
parameterized is different. The quarks are considered as $u$-type and $d$-type,
 $xU= x(u_v+u_{sea} + c)$, 
$xD= x(d_v +d_{sea} + s)$, $x\bar{U}=x(\bar{u}+\bar{c})$ and 
$x\bar{D}=x(\bar{d}+\bar{s})$, assuming $q_{sea}=\bar{q}$, as 
usual. These four (anti-)quark distributions are parameterized separately.
\[
xU(x) = A_{U} x^{B_{U}} (1-x)^{C_{U}} (1 + D_{U} x +E_{U} x^2 + F_U x^3)
\]
\[
xD(x) = A_{D} x^{B_{D}} (1-x)^{C_{D}} (1 + D_D x)
\]
\[
x\bar{U}(x) = A_{\bar{U}} x^{B_{\bar{U}}} (1-x)^{C_{\bar{U}}}
\]
\[
x\bar{D}(x)= A_{\bar{D}} x^{B_{\bar{D}}} (1-x)^{C_{\bar{D}}}
\]
\[xg(x) = A_{g} x^{B_{g}} (1-x)^{C_{g}} \]
 Since the valence distributions must vanish as $x \to 0$, 
the parameters, $A$ and $B$ are set equal for $xU$ and $x\bar{U}$; 
$A_U=A_{\bar{U}}$, $B_U=B_{\bar{U}}$;  and for $xD$ and 
$x\bar{D}$; $A_D=A_{\bar{D}}$, $B_D=B_{\bar{D}}$. 
Since there is no information on the flavour structure of the sea 
it is also necessary to set  $B_{\bar{U}}=B_{\bar{D}}$,
such that there is a single $B$ parameter for all four quark distributions. 
The normalisation, $A_g$, of the gluon is determined from the momentum 
sum-rule and the parameters $D_U$ and $D_D$ are determined by the number 
sum-rules.
Assuming that the strange and charm quark distributions can be expressed as 
$x$ independent fractions, $f_s=0.33$ and $f_c=0.15$, of the $d$ and $u$ 
type sea respectively, 
gives the further constraint $A_{\bar{U}}=A_{\bar{D}} (1-f_s)/(1-f_c)$, 
which ensures that $\bar{u}=\bar{d}$ at low $x$.  
Finally this H1-style parameterization has 10 free parameters.

The third parameterization we have considered combines the best features of the 
previous two. It has less model 
dependence than the ZEUS-style parameterization in that it makes fewer
assumptions on the form of sea quark asymmetry $x\Delta$, and it has less model dependence than the
 H1-style 
parameterization in that it does not assume equality of all $B$ parameters.
Furthermore, although all types of parameterization give acceptable $\chi^2$ 
values, the third parameterization has the best $\chi^2$ and 
it gives the most conservative experimental errors. This is the 
parameterization which we chose for our central fit.
The PDFs which are parameterized are $xu_v$, $xd_v$, $xg$ and 
$x\bar{U}$, $x\bar{D}$. 
\[
xu_v(x) = A_{uv} x^{B_{uv}} (1-x)^{C_{uv}} (1 + D_{uv} x +E_{uv} x^2)
\]
\[
xd_v(x) = A_{dv} x^{B_{dv}} (1-x)^{C_{dv}}
\]
\[
x\bar{U}(x) = A_{\bar{U}} x^{B_{\bar{U}}} (1-x)^{C_{\bar{U}}}
\]
\[
x\bar{D}(x)= A_{\bar{D}} x^{B_{\bar{D}}} (1-x)^{C_{\bar{D}}}
\]
\[xg(x) = A_{g} x^{B_{g}} (1-x)^{C_{g}} \]
The normalisation parameters, $A_{uv}, A_{dv}, A_g$, are constrained to 
impose the number sum-rules and momentum sum-rule. 
The $B$ parameters, $B_{uv}$ and $B_{dv}$  are set equal, $B_{uv}=B_{dv}$ and 
the $B$ parameters  $B_{\bar{U}}$ and $B_{\bar{D}}$ are also set equal,
 $B_{\bar{U}}=B_{\bar{D}}$, such that 
there is a single $B$ parameter for the valence and another different single 
$B$ parameter for the sea distributions. 
Assuming that the strange and charm quark distributions can be expressed as 
$x$ independent fractions, $f_s=0.33$ and $f_c=0.15$, of the $d$ and $u$ 
type sea, 
gives the further constraint $A_{\bar{U}}=A_{\bar{D}} (1-f_s)/(1-f_c)$.
The value of $f_s=0.33$ has been chosen to be consistent with determinations 
of this fraction using neutrino induced di-muon production. This value
 has been varied to evaluate model
uncertainties. The charm fraction has been set to be consistent with dynamic
 generation of charm
from the start point of $Q^2= m_c^2$, in a zero-mass-variable-flavour-number 
scheme. A small variation of the value of 
$f_c$ is included in the model uncertainties. Finally this parameterization 
has 11 free parameters.

\begin{figure}[tbp]
\centerline{
\epsfig{figure=herapdf_exp_all.eps,width=0.33\textwidth,height=5.5cm}
\epsfig{figure=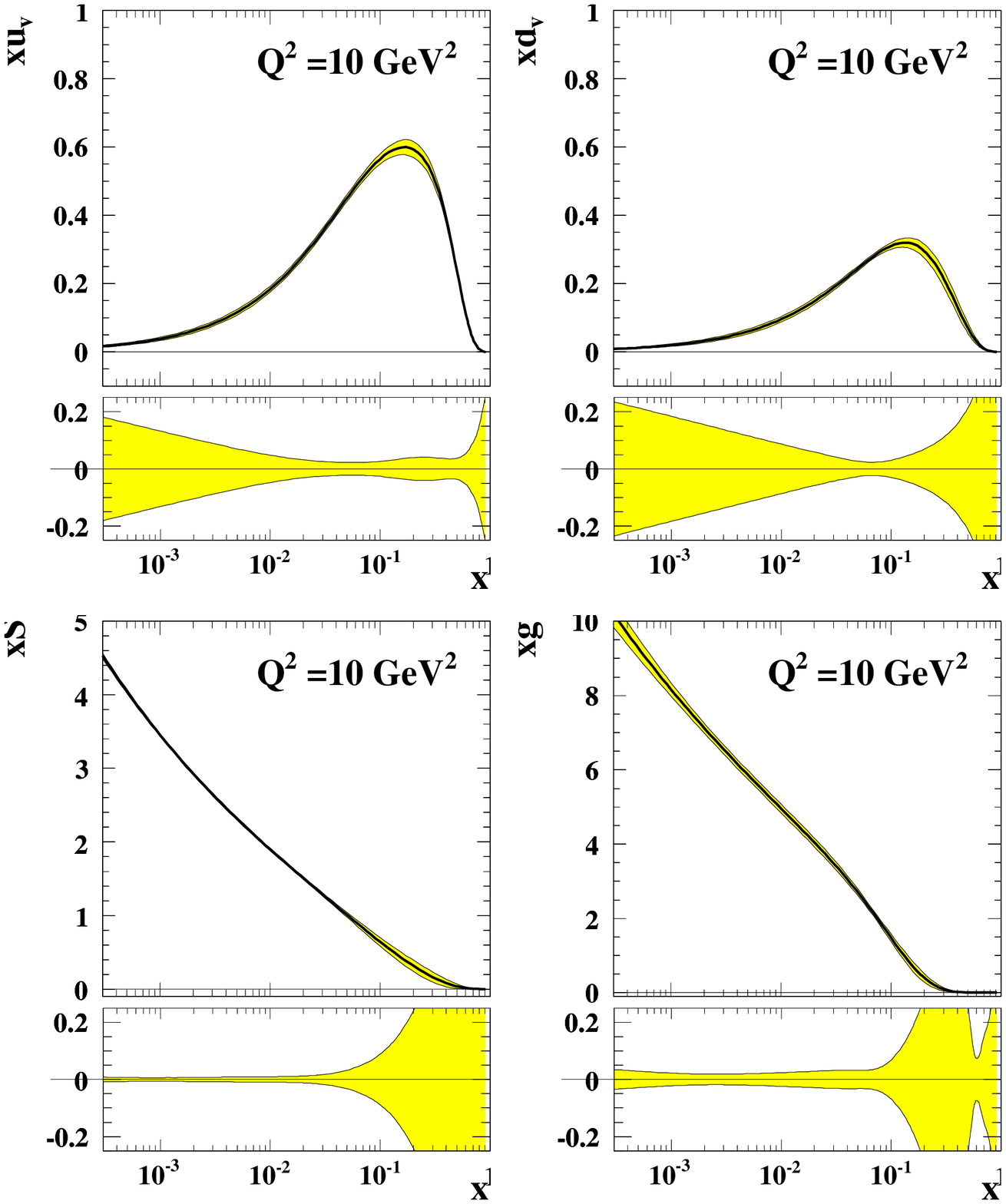,width=0.33\textwidth,height=5.5cm}
\epsfig{figure=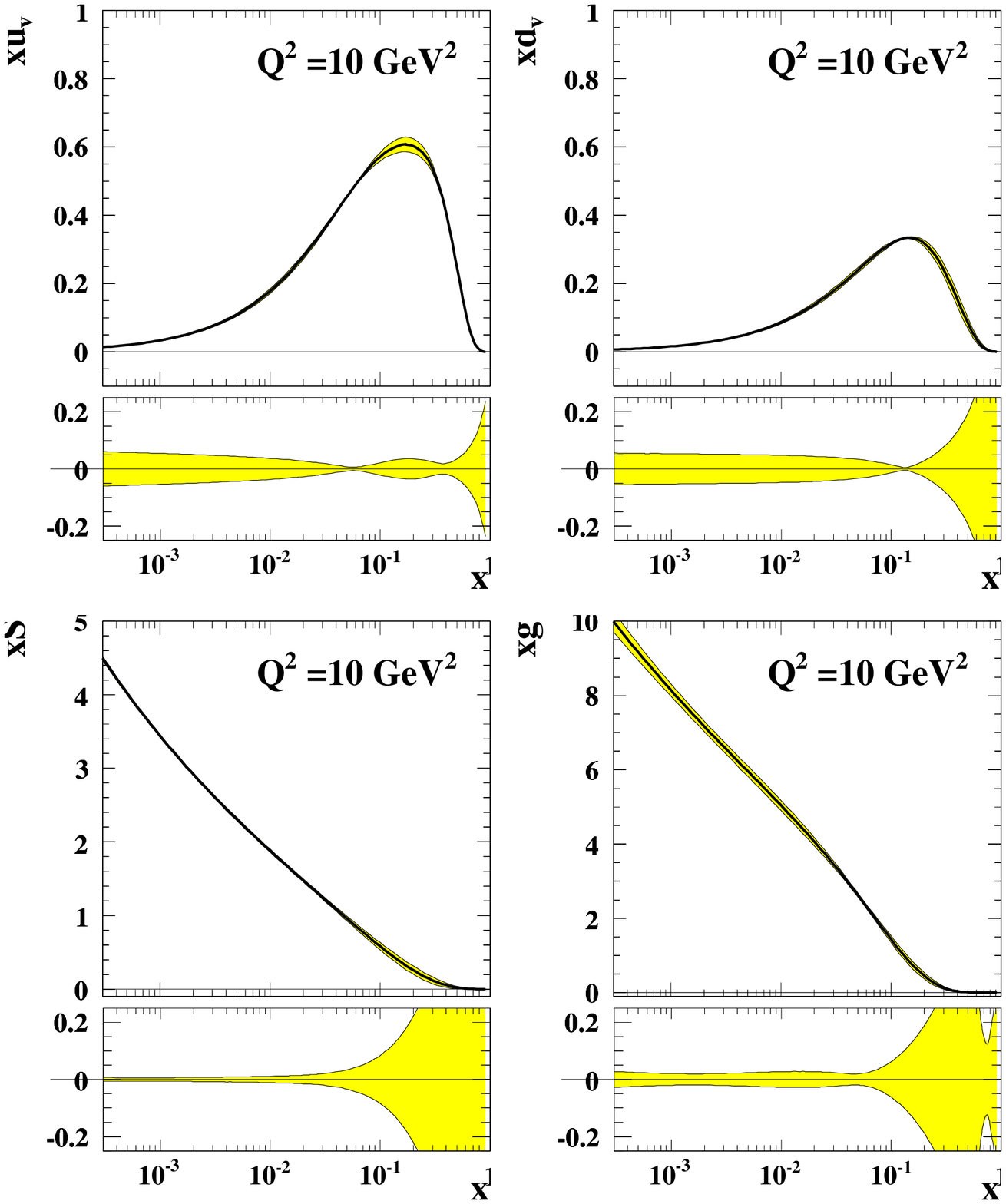,width=0.33\textwidth,height=5.5cm}}
\caption {HERAPDFs, $xu_v,xd_v,xS,xg$ and their uncertainties at 
$Q^2=10$GeV$^2$. (Left) for the central fit; (centre) for the ZEUS-style 
parameterization; (right) for the H1-style parameterization}
\label{fig:param}
\end{figure}

It is well known that the choice of parameterization can affect both PDF 
shapes and the size of the PDF uncertainties. Fig~\ref{fig:param} compares 
the PDFs and their uncertainties as evaluated using these three different 
parameterizations. As mentioned earlier, the third parameterization results 
in the most conservative uncertainties.

We present results for the HERA PDFs 
based on the third type of parameterization, including six 
sources of model uncertainty as specified in 
Table~\ref{tab:model}.  We also compare to results obtained by varying 
$\asmz$ and by varying the choice of parameterization to those of the ZEUS and 
the H1 styles of parameterization.
\begin{table}[tbp]
\centerline{
\begin{tabular}{|l|l|l|r|}
\hline
  Model variation& Standard value & Upper Limit & Lower limit  \\
\hline
$m_c$ & $1.4$  & $1.35$ & $1.5$   \\
$m_b$ & $4.75$  & $4.3$ & $5.0$  \\
$Q^2_{min}$ & $3.5$  & $2.5$ & $5.0$   \\
$Q^2_0$ & $4.0$  & $2.0$ & $6.0$  \\
$f_s$ & $0.33$  & $0.25$ & $0.40$   \\
$f_c$ & $0.15$  & $0.12$ & $0.18$  \\
\hline
\end{tabular}}
\caption{Standard values of input parameters and cuts, and the variations 
considered to evaluate model uncertainty
}
\label{tab:model}
\end{table}

\subsubsection{Results}
\label{sec:results} 
\begin{figure}[tbp]
\centerline{
\epsfig{figure=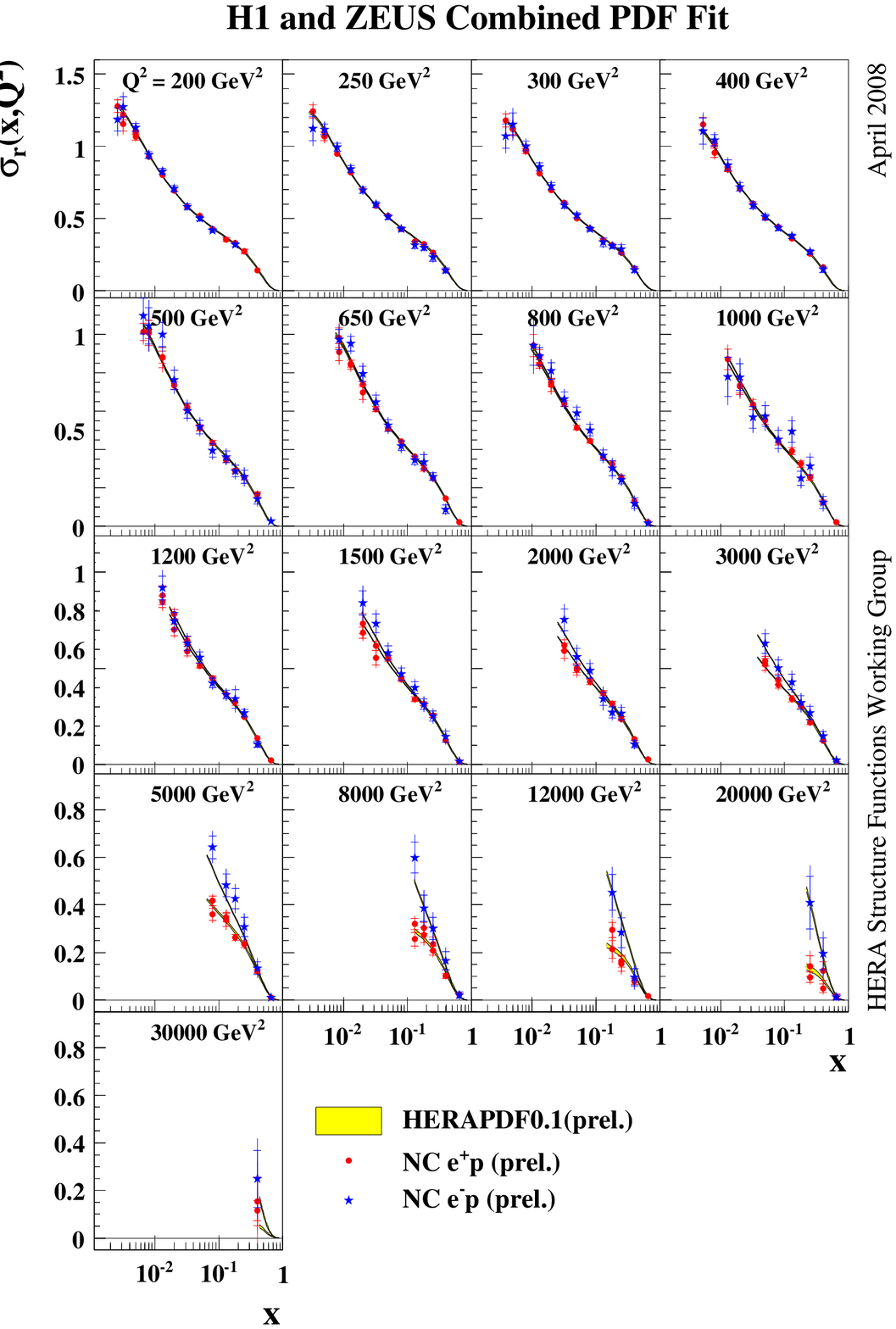,width=0.5\textwidth,height=6.5cm}
\epsfig{figure=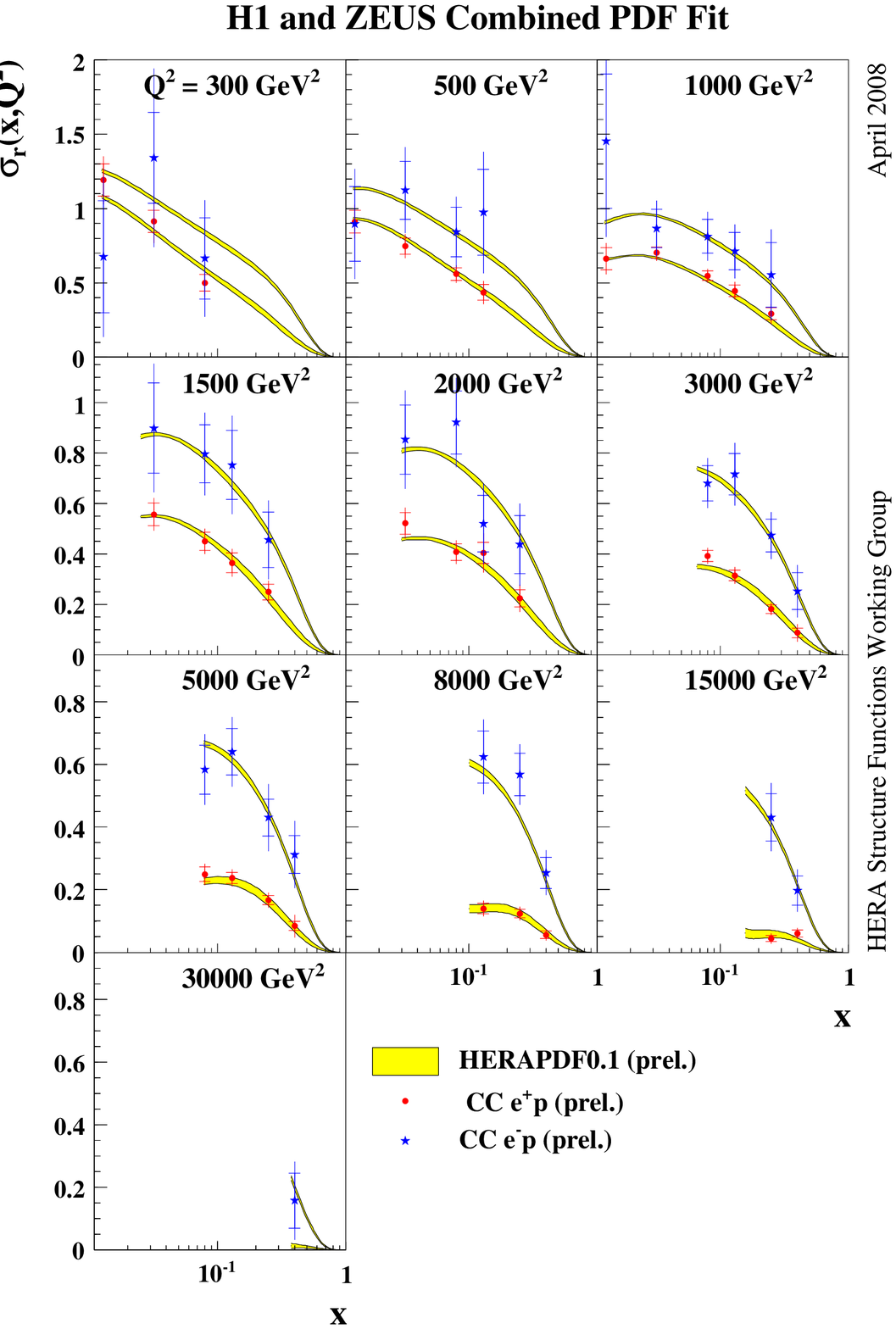,width=0.5\textwidth,height=6.5cm}}
\caption {HERA combined NC (left) and CC (right) data. 
The predictions of the HERAPDF0.1 fit are superimposed. The uncertainty bands
 illustrated derive from both experimental and model sources}
\label{fig:data}
\end{figure}

\begin{figure}[tbp]
\centerline{
\epsfig{figure=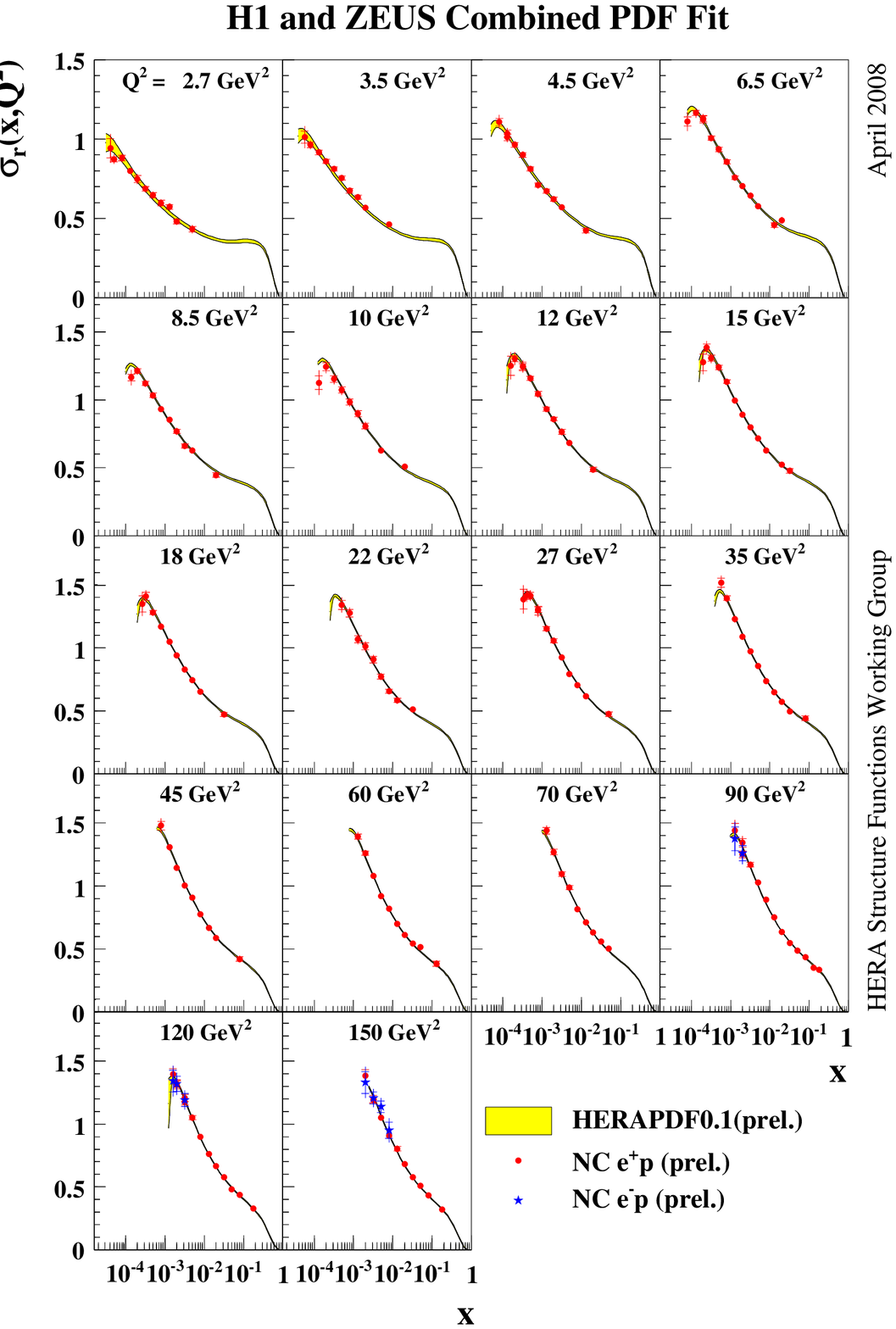,width=0.5\textwidth,height=6.5cm}
\epsfig{figure=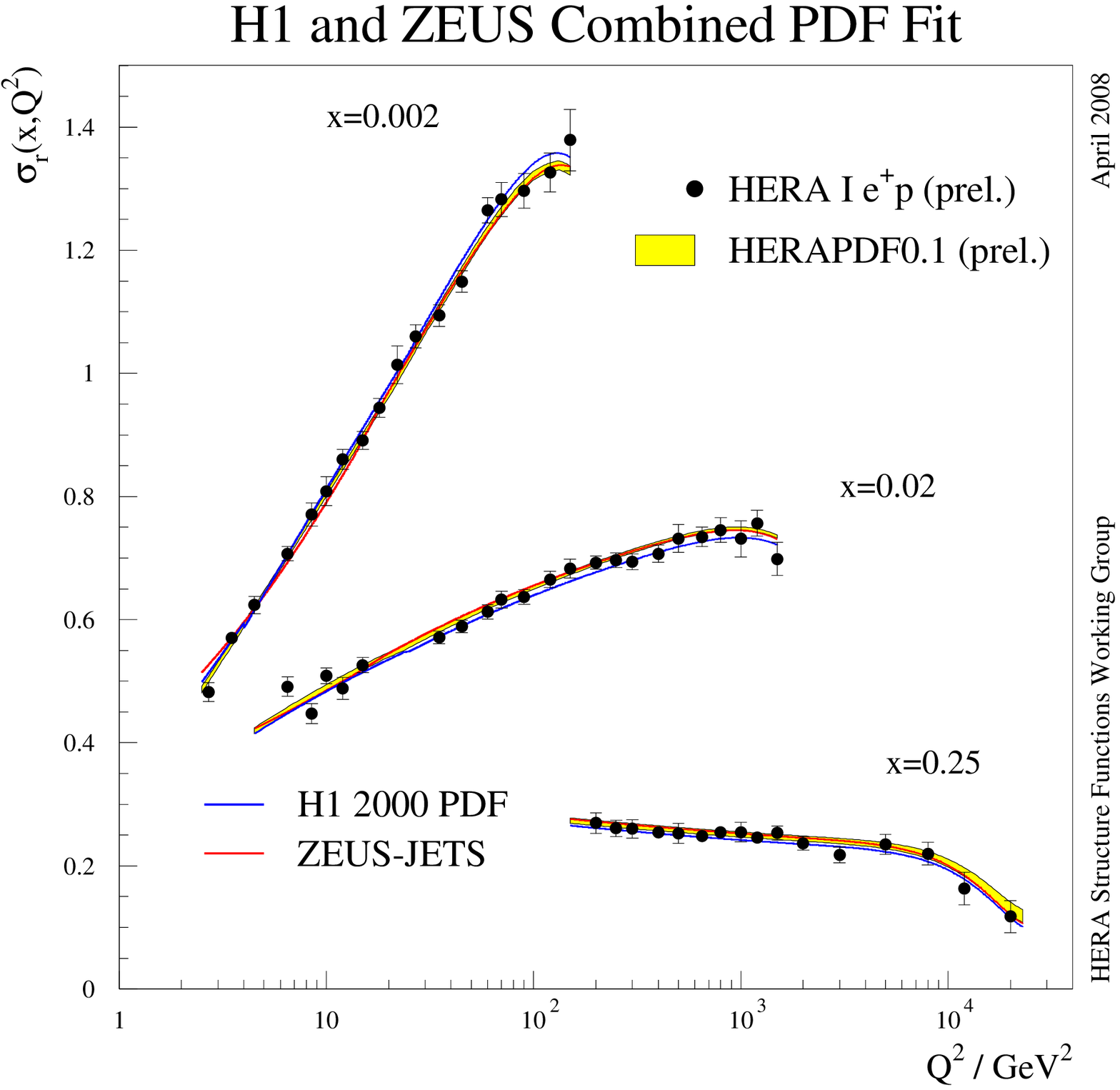,width=0.5\textwidth,height=6.5cm}}
\caption {Left: HERA combined NC data at low $Q^2$. Right: the NC reduced 
cross-section vs $Q^2$ for three $x$-bins.
The predictions of the HERAPDF0.1 fit are superimposed, together with
 the predictions of the ZEUS-JETS and H1PDF2000 PDFs}
\label{fig:vsQ2}
\end{figure}

In Fig.~\ref{fig:data} we show the HERAPDF0.1 superimposed on the 
combined data set for NC data and CC data.
In Fig~\ref{fig:vsQ2} we show the NC data at low $Q^2$, and we 
illustrate scaling violation by showing the reduced cross-section vs. $Q^2$  
for a few representative $x$ bins. The predictions of the HERAPDF0.1 fit are 
superimposed, together with
 the predictions of the ZEUS-JETS and H1PDF2000 PDFs. 

\begin{figure}[tbp]
\centerline{
\epsfig{figure=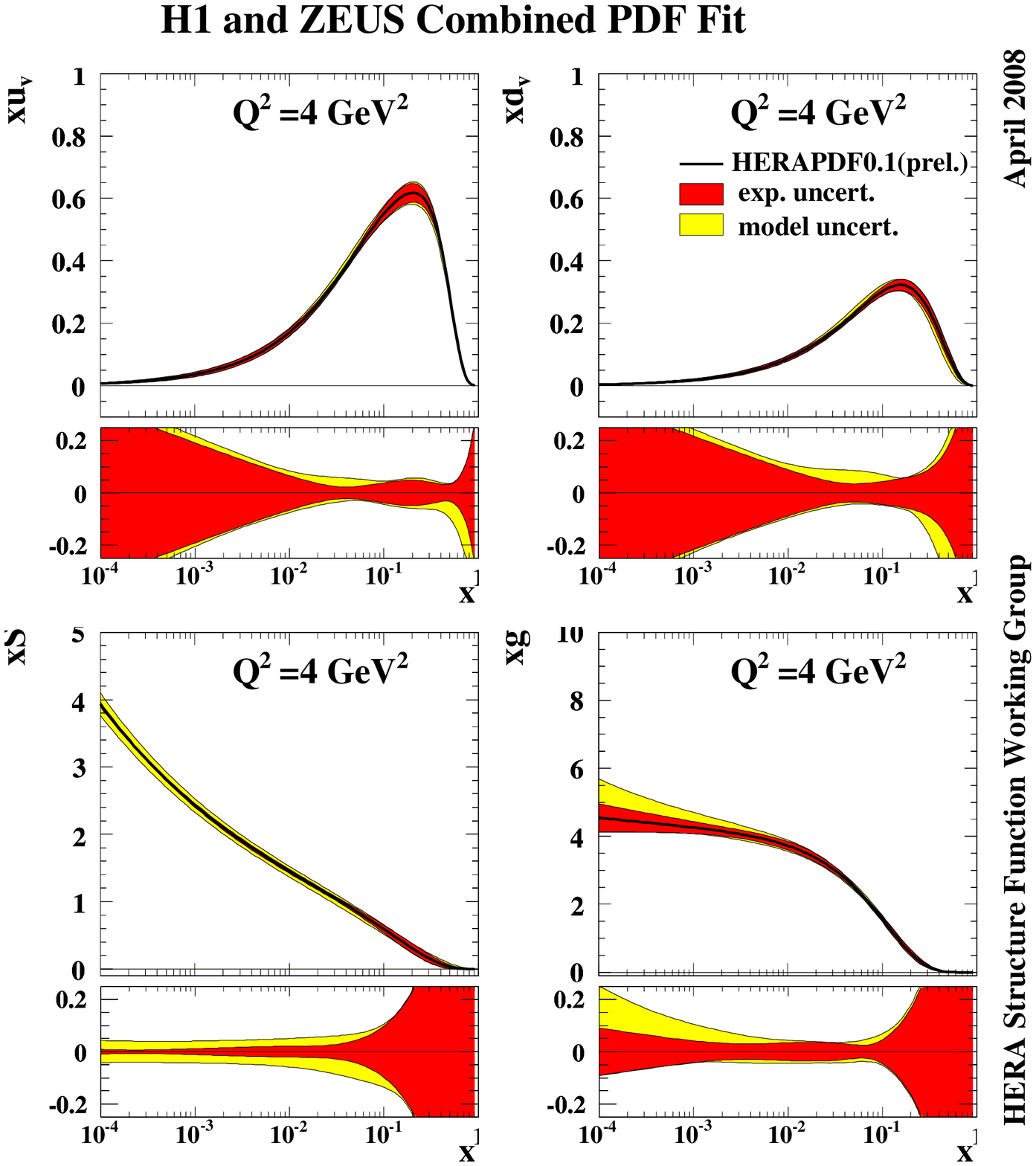 ,width=0.5\textwidth,height=6.5cm}
\epsfig{figure=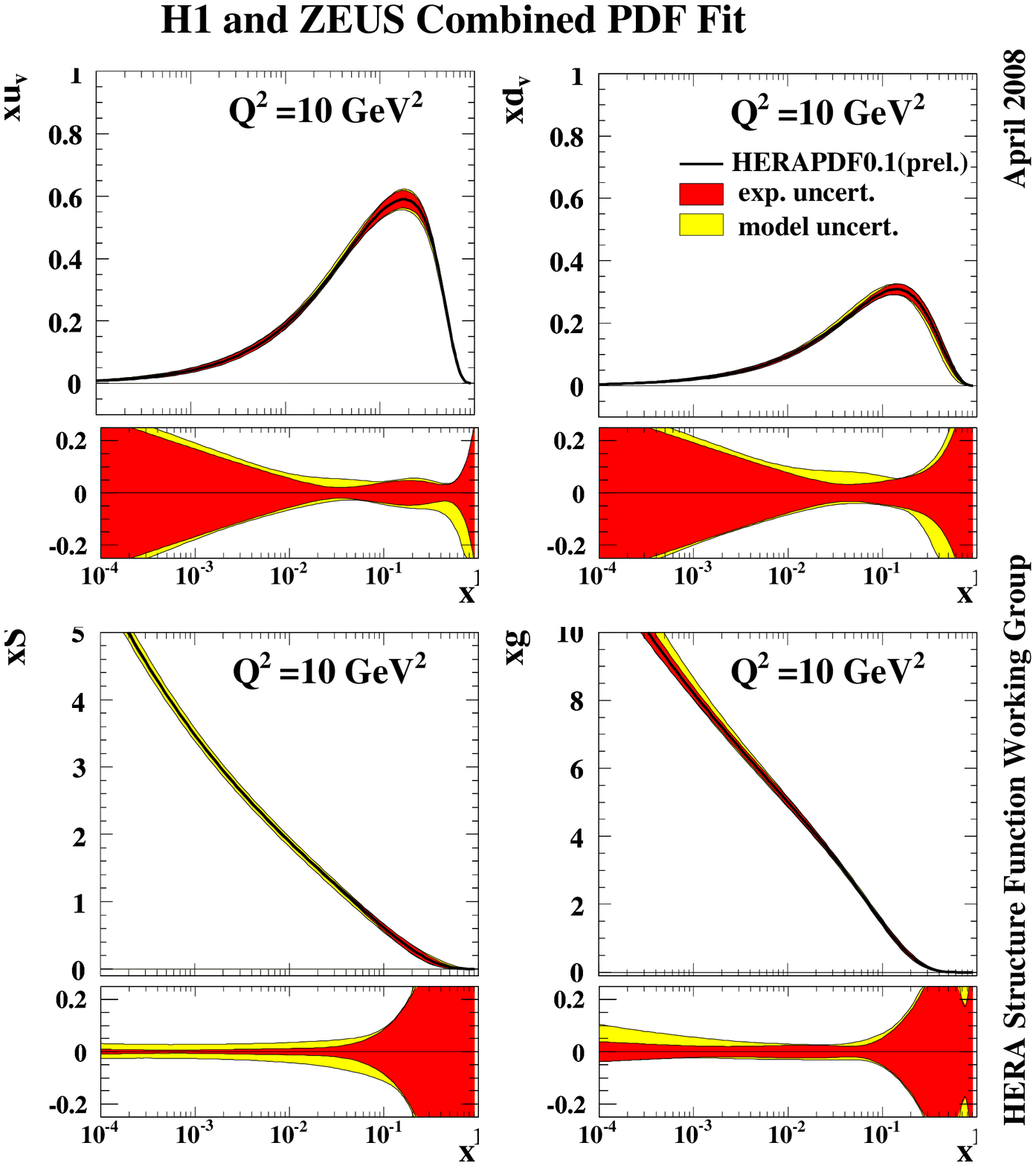 ,width=0.5\textwidth,height=6.5cm}}
\caption {HERAPDFs, $xu_v,xd_v,xS,xg$, at (left) $Q^2=4$~GeV$^2$ and (right)
$Q^2=10$~GeV$^2$. Fractional uncertainty bands are shown 
beneath each PDF. The experimental and model uncertainties are shown 
separately as the red and yellow bands respectively}
\label{fig:pdfs410}
\end{figure}

\begin{figure}[tbp]
\centerline{
\epsfig{figure=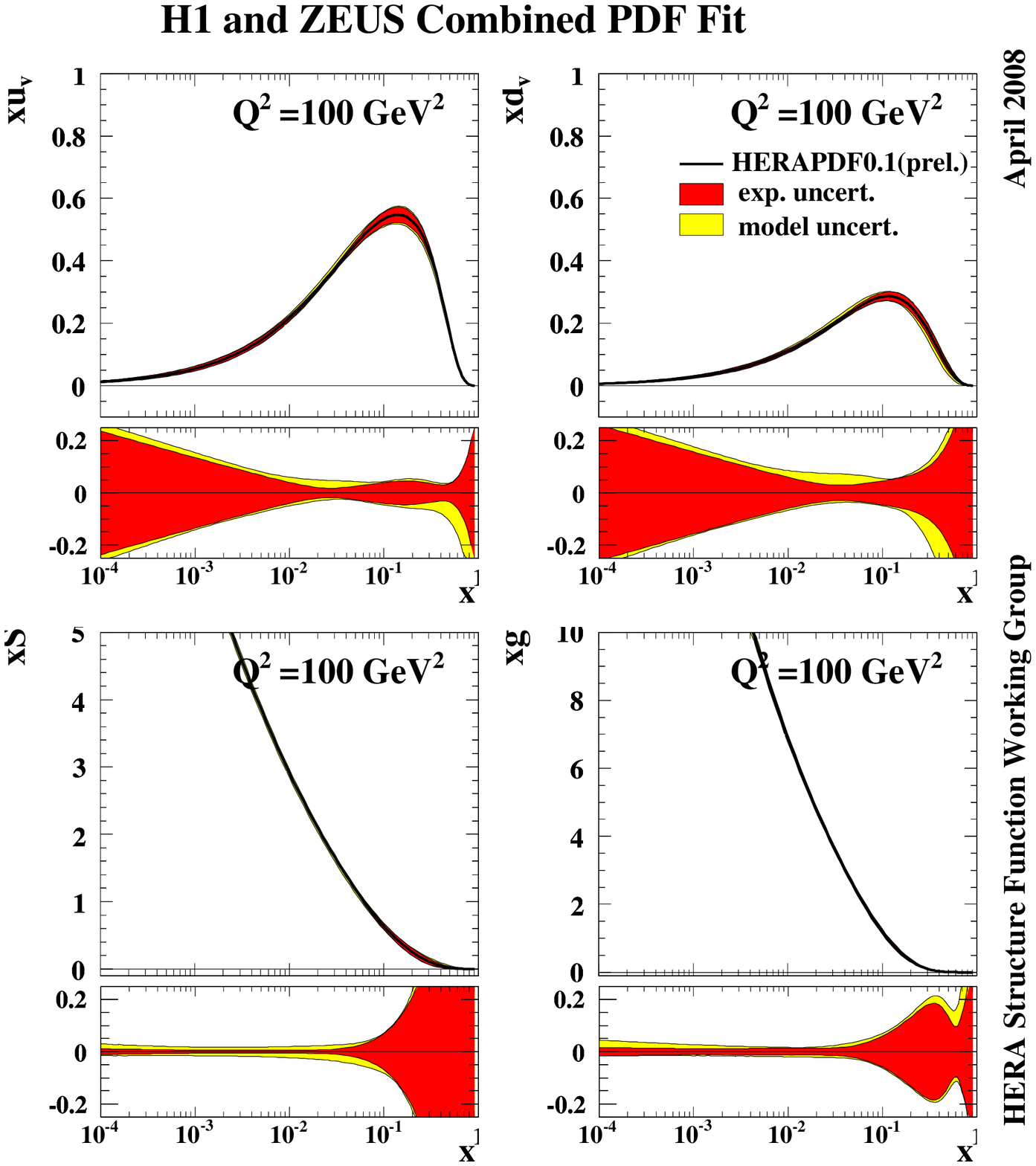 ,width=0.5\textwidth,height=6.5cm}
\epsfig{figure=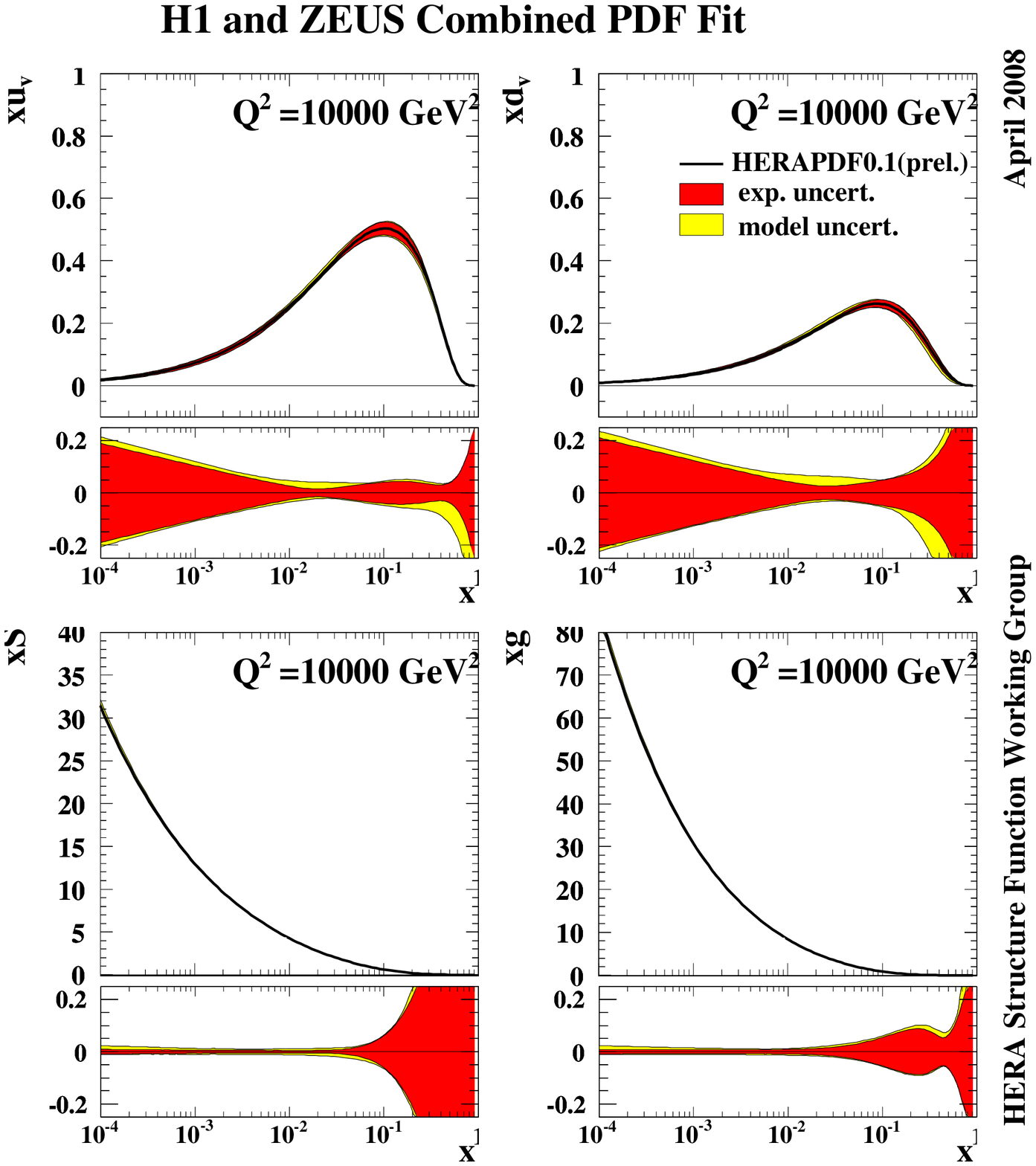 ,width=0.5\textwidth,height=6.5cm}}
\caption {HERAPDFs, $xu_v,xd_v,xS,xg$, at (left) $Q^2=100~$GeV$^2$ and (right)
$Q^2=10000~$GeV$^2$. Fractional uncertainty bands are shown 
beneath each PDF. The experimental and model uncertainties are shown 
separately as the red and yellow bands respectively}
\label{fig:pdfs10010000}
\end{figure}

Fig.~\ref{fig:pdfs410} shows the HERAPDF0.1 PDFs,  $xu_v,xd_v,xS,xg$, 
as a function of $x$ at 
the starting scale $Q^2=4$~GeV$^2$ and at $Q^2=10$~GeV$^2$. 
Fig.~\ref{fig:pdfs10010000} shows the same PDFs
at the scales $Q^2=100,10000$~GeV$^2$. 
Fractional uncertainty bands are shown beneath each PDF. The experimental and 
model uncertainties are shown separately. As the PDFs 
evolve with $Q^2$ the total uncertainty becomes impressively small.  

\begin{figure}[tbp]
\centerline{
\epsfig{figure=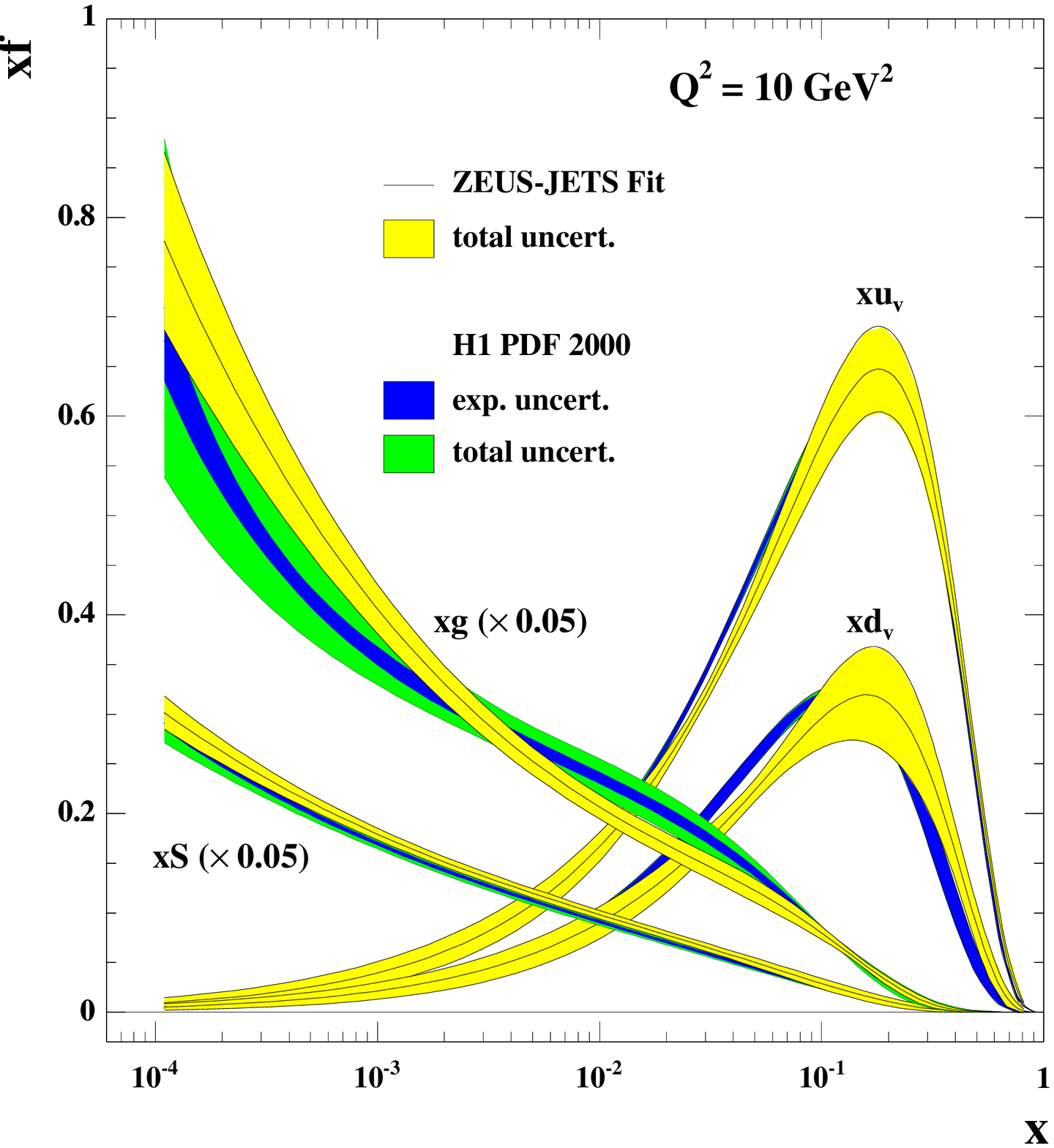,width=0.45\textwidth,height=6.1cm}
\epsfig{figure=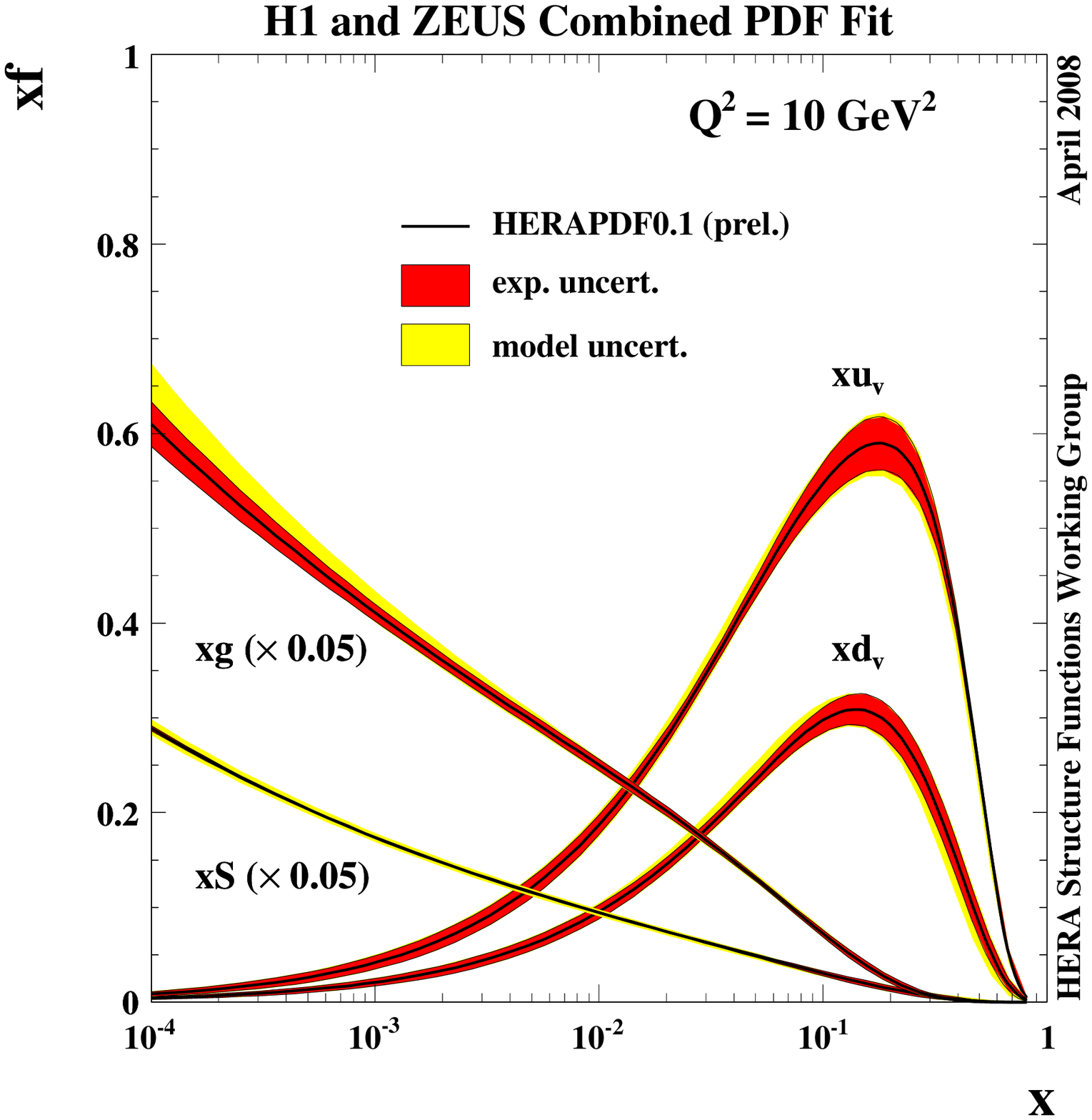 ,width=0.5\textwidth,height=6.5cm}}
\caption {Left: PDFs from the ZEUS-JETS and H1PDF2000 PDF separate 
analyses of ZEUS and H1. Right: HERAPDF0.1 PDFs
 from the analysis of the combined data set}
\label{fig:summary}
\end{figure}

\begin{figure}[tbp]
\centerline{
\epsfig{figure=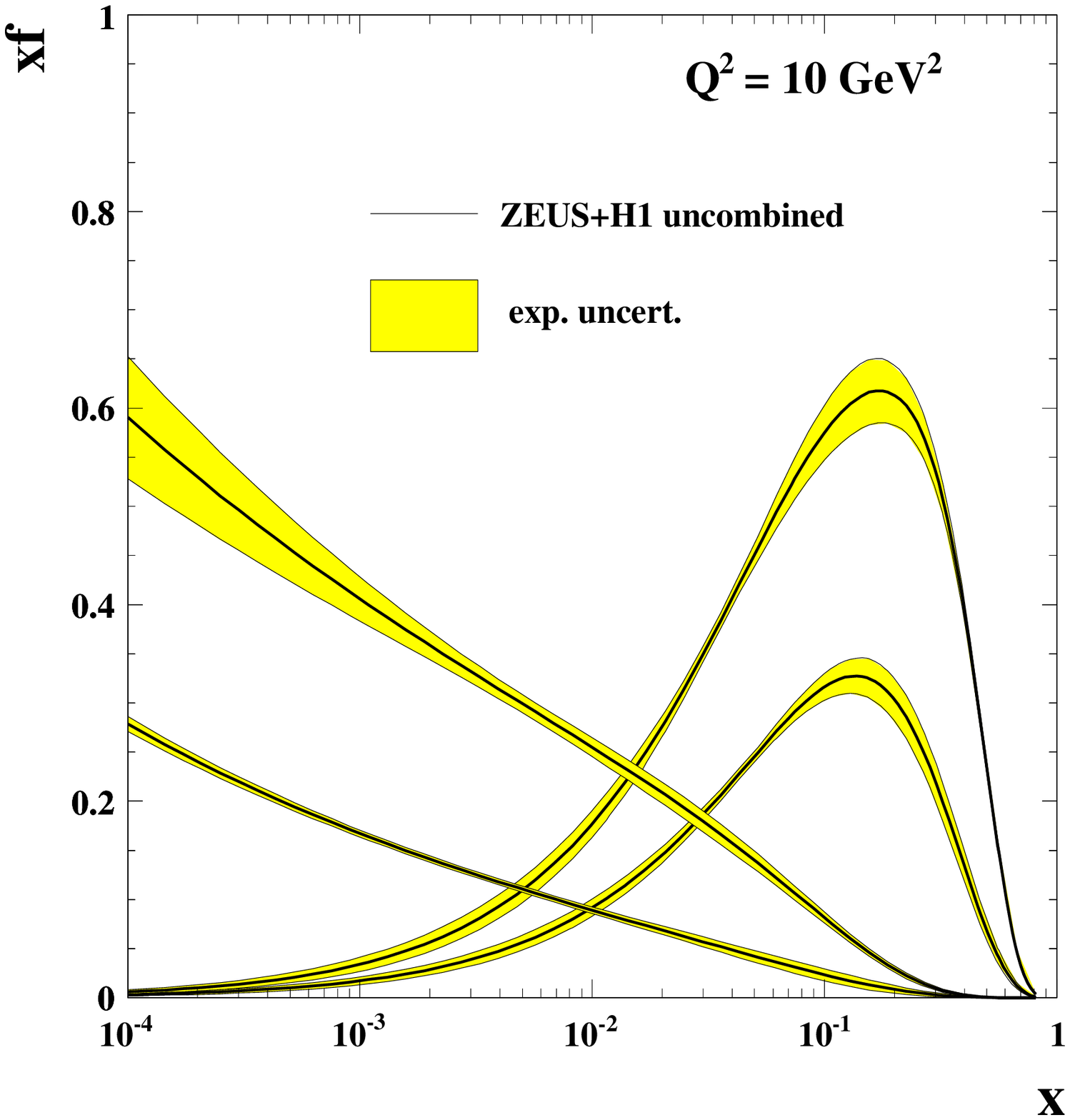,width=0.55\textwidth,height=5.5cm}
\epsfig{figure=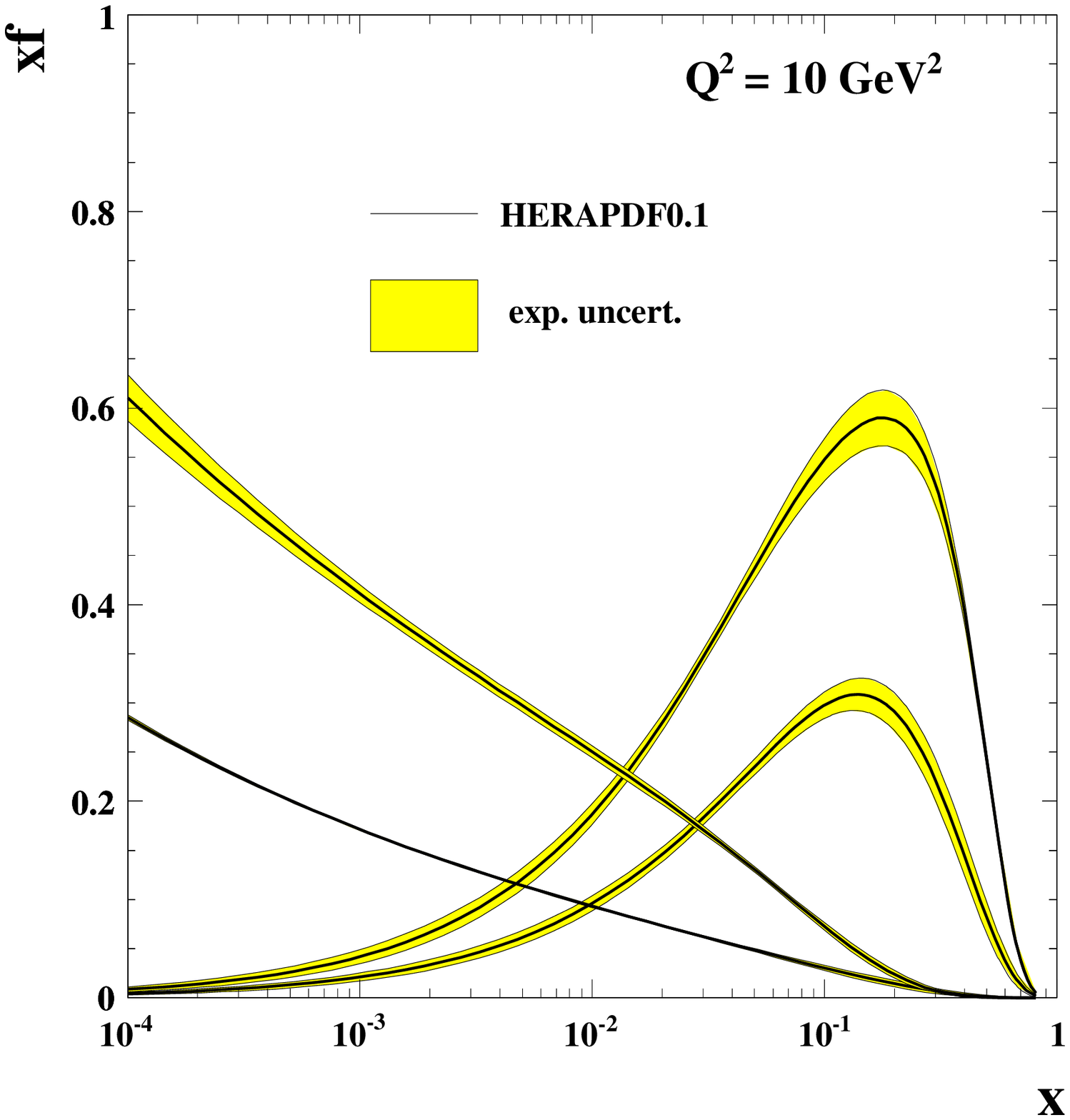 ,width=0.5\textwidth,height=5.5cm}}
\caption {Left: PDFs resulting from an analysis of the H1 and ZEUS separate 
data sets using the same parameterization and assumptions as HERAPDF0.1. 
Right: HERAPDF0.1 PDFs
 from the analysis of the combined data set (experimental uncertainties only)}
\label{fig:h1zeusuncomb}
\end{figure}

The total uncertainty of the PDFs obtained from the HERA combined data set is 
much reduced compared to the PDFs extracted from the analyses of the separate 
H1 and ZEUS data sets, as can be seen from the 
summary plot Fig.~\ref{fig:summary}, where these new HERAPDF0.1 
PDFs are compared to the ZEUS-JETS and H1PDF2000 PDFs.
It is also interesting to compare the present HERAPDF0.1 analysis of the 
combined HERA-I data set with an analysis of the separate data sets which uses
the same parameterization and assumptions. Fig~\ref{fig:h1zeusuncomb} makes 
this comparison. It is clear that it is the data combination, and not the 
choice of parameterization and assumptions, which has resulted in reduced 
uncertainties for the low-$x$ gluon and sea PDFs.

\begin{figure}[tbp]
\centerline{
\epsfig{figure=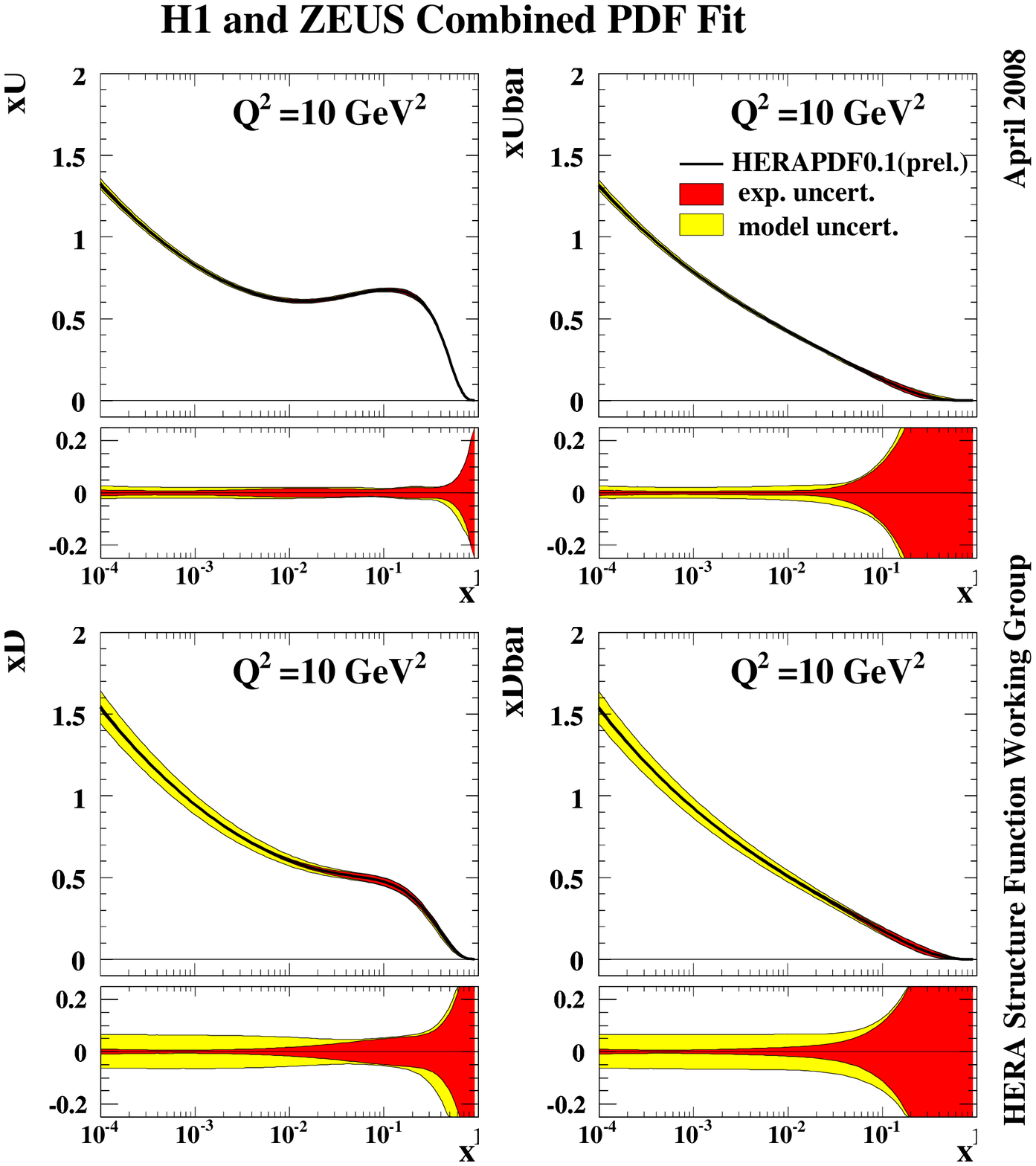 ,width=0.5\textwidth,height=6.5cm}
\epsfig{figure=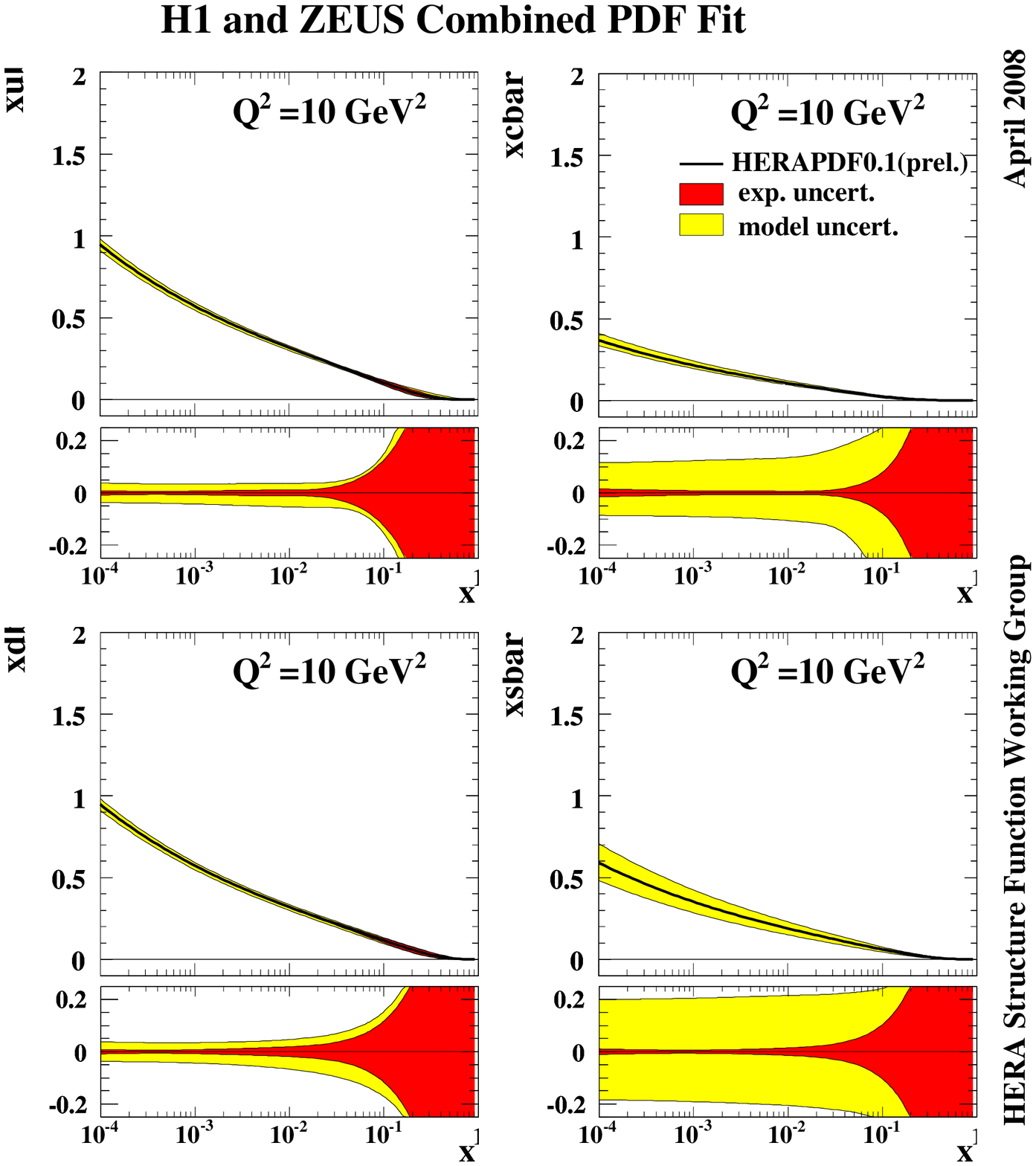 ,width=0.5\textwidth,height=6.5cm}}
\caption {HERAPDFs at $Q^2=10$GeV$^2$: (left) $xU, xD, x\bar{U}, x\bar{D}$;
 (right) $x\bar{u}, x\bar{d}, x\bar{c}, x\bar{s}$. 
Fractional uncertainty bands are shown 
beneath each PDF. The experimental and model uncertainties are shown separately as the red and yellow bands respectively}
\label{fig:pdfsflav}
\end{figure}

The break-up of the HERAPDFs into different flavours is illustrated in 
Fig.~\ref{fig:pdfsflav}, where the PDFs $xU$, $xD$, $x\bar{U}$, $x\bar{D}$ and 
$x\bar{u}, x\bar{d}, x\bar{c}, x\bar{s}$ are shown at $Q^2=10$~GeV$^2$. 
The model uncertainty on these PDFs from variation of $Q^2_{min}$, $Q^2_0$, $m_c$
and $m_b$ is modest. The model uncertainty from variation of $f_s$ and $f_c$ 
is also modest except for its obvious effect on the charm and strange quark 
distributions. 
 
\begin{figure}[tbp]
\centerline{
\epsfig{figure=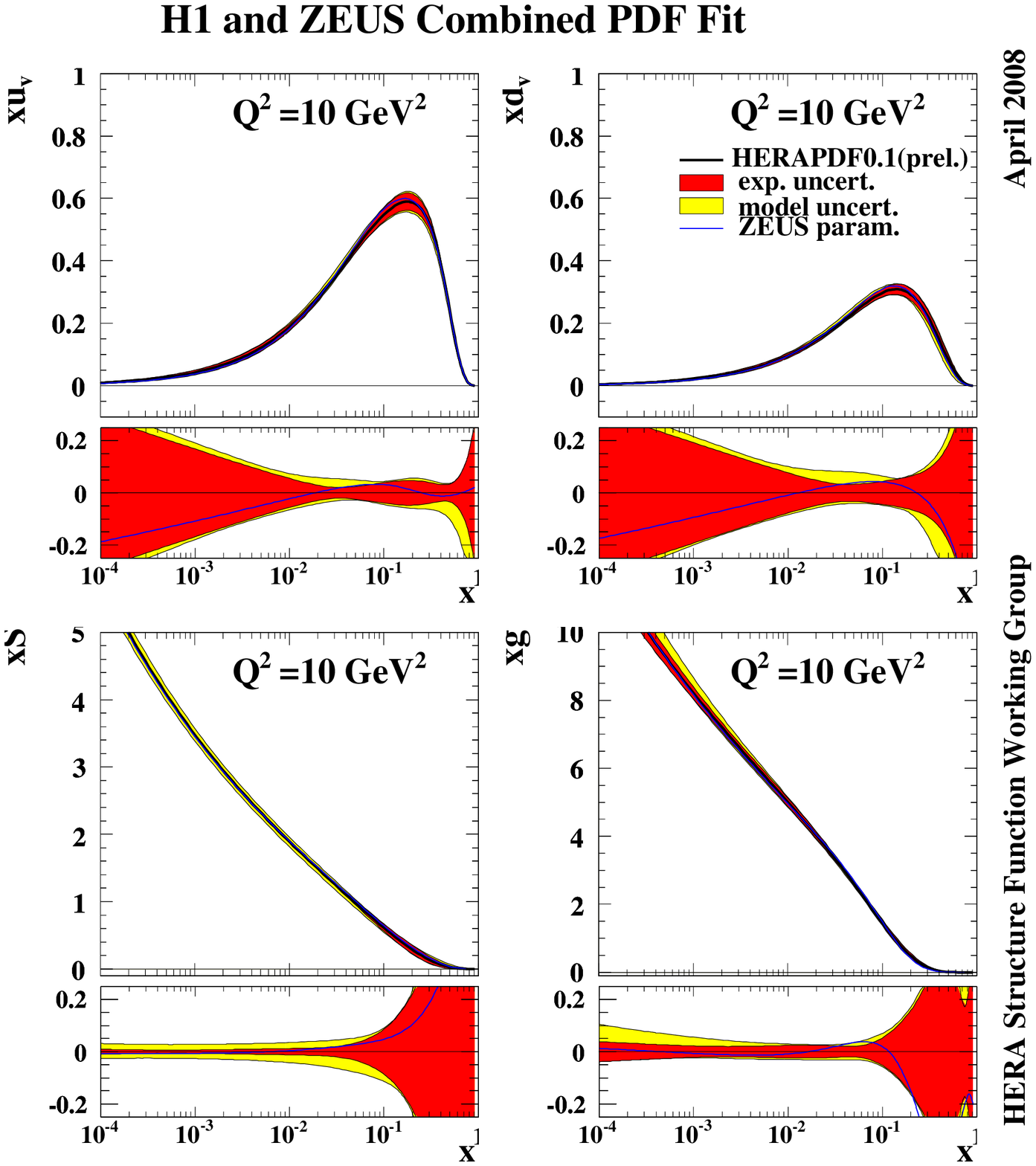 ,width=0.5\textwidth,height=6.5cm}
\epsfig{figure=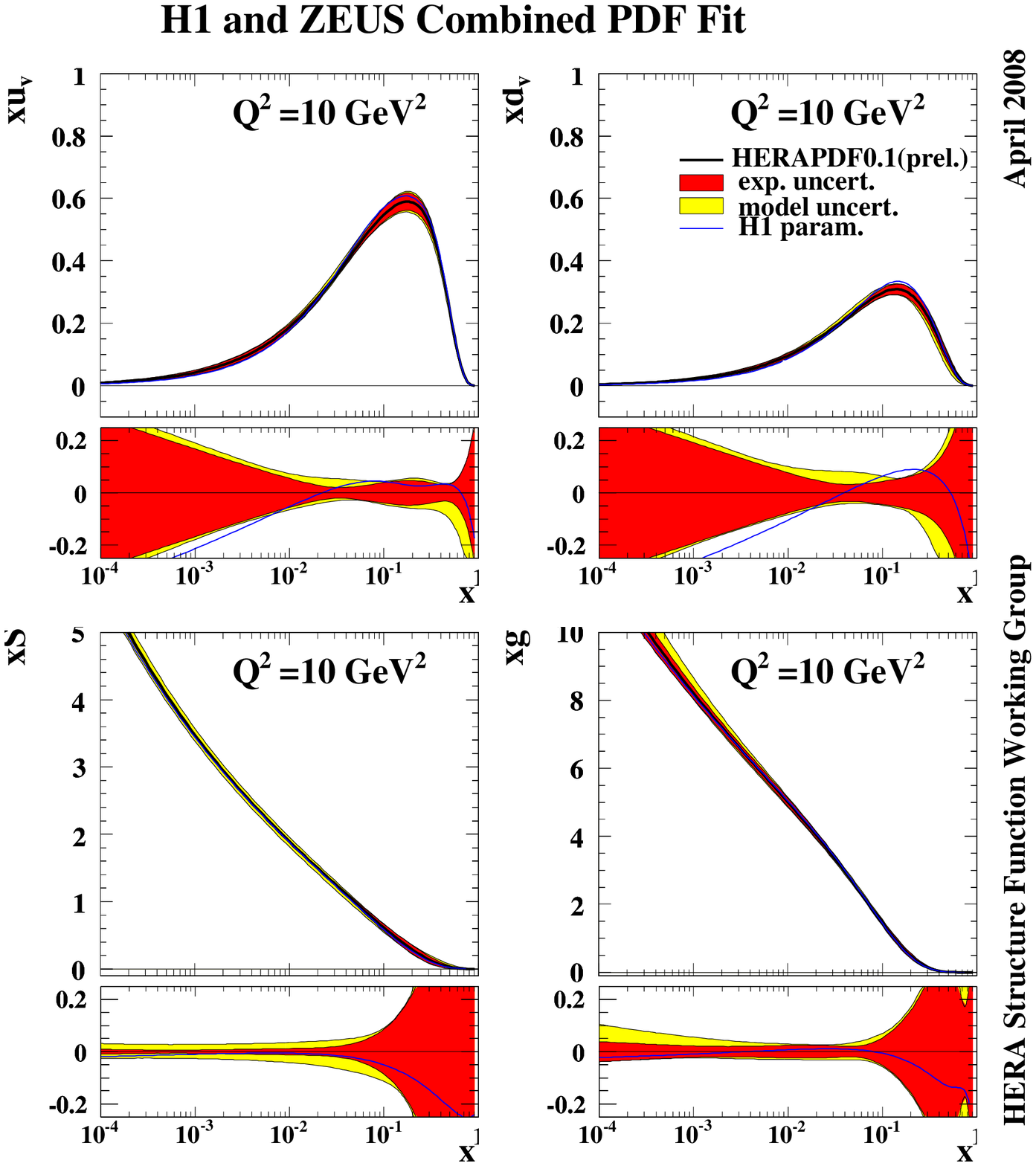 ,width=0.5\textwidth,height=6.5cm}}
\caption {HERAPDFs at $Q^2=10$GeV$^2$: with the results for the ZEUS-style 
parameterization 
(left) and for the H1-style parameterization (right) superimposed as a 
blue line.}
\label{fig:zh1chk}
\end{figure}

\begin{figure}[tbp]
\centerline{
\epsfig{figure=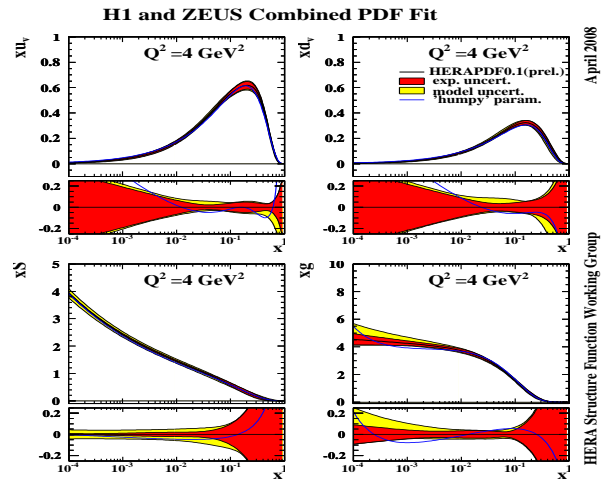 ,width=0.5\textwidth,height=6.5cm}}
\caption {HERAPDFs at $Q^2=4$GeV$^2$: with the results for the humpy version 
 superimposed as a 
blue line.}
\label{fig:humpy}
\end{figure}

It is also interesting to look at the results obtained from using the 
ZEUS-style and H1 style parameterizations described in Section~\ref{sec:anal}. 
In Fig.~\ref{fig:zh1chk} these alternative parameterizations are shown as a 
blue line superimposed on the HERAPDF0.1 PDFs.
These variations in parameterization produce changes in the resulting 
PDFs which are comparable to the experimental uncertainties in the measured 
kinematic range. A further variation of parameterization originates 
from the fact that, if the $D$ parameter for the gluon is allowed to be 
non-zero, then each type of parameterization yields a double minimum in $\chi^2$
such that the gluon may take a smooth or a 'humpy' shape. Although the 
lower $\chi^2$ is obtained for the for the smooth shape, the $\chi^2$ for the 
'humpy' shape is
still acceptable. The PDFs for the 'humpy' version of our chosen
form of parameterization are compared to the standard version
 in Fig.~\ref{fig:humpy}, where they are shown as a blue line superimposed 
on the HERAPDF0.1 PDFs. This comparison is shown at $Q^2=4$GeV$^2$, where the 
difference is the greatest. Nevertheless  the resulting PDFs 
are comparable to those of the standard choice. This explains a long-standing 
disagreement in the shape of the gluon 
obtained by the separate ZEUS-JETS and H1PDF200 analyses. The ZEUS 
data favoured the smooth shape and the H1 data favoured the 'humpy' shape. 
However the precision of the combined data set results in PDFs for these shapes
which are not significantly different in the measured kinematic 
region.

\begin{figure}[tbp]
\centerline{
\epsfig{figure=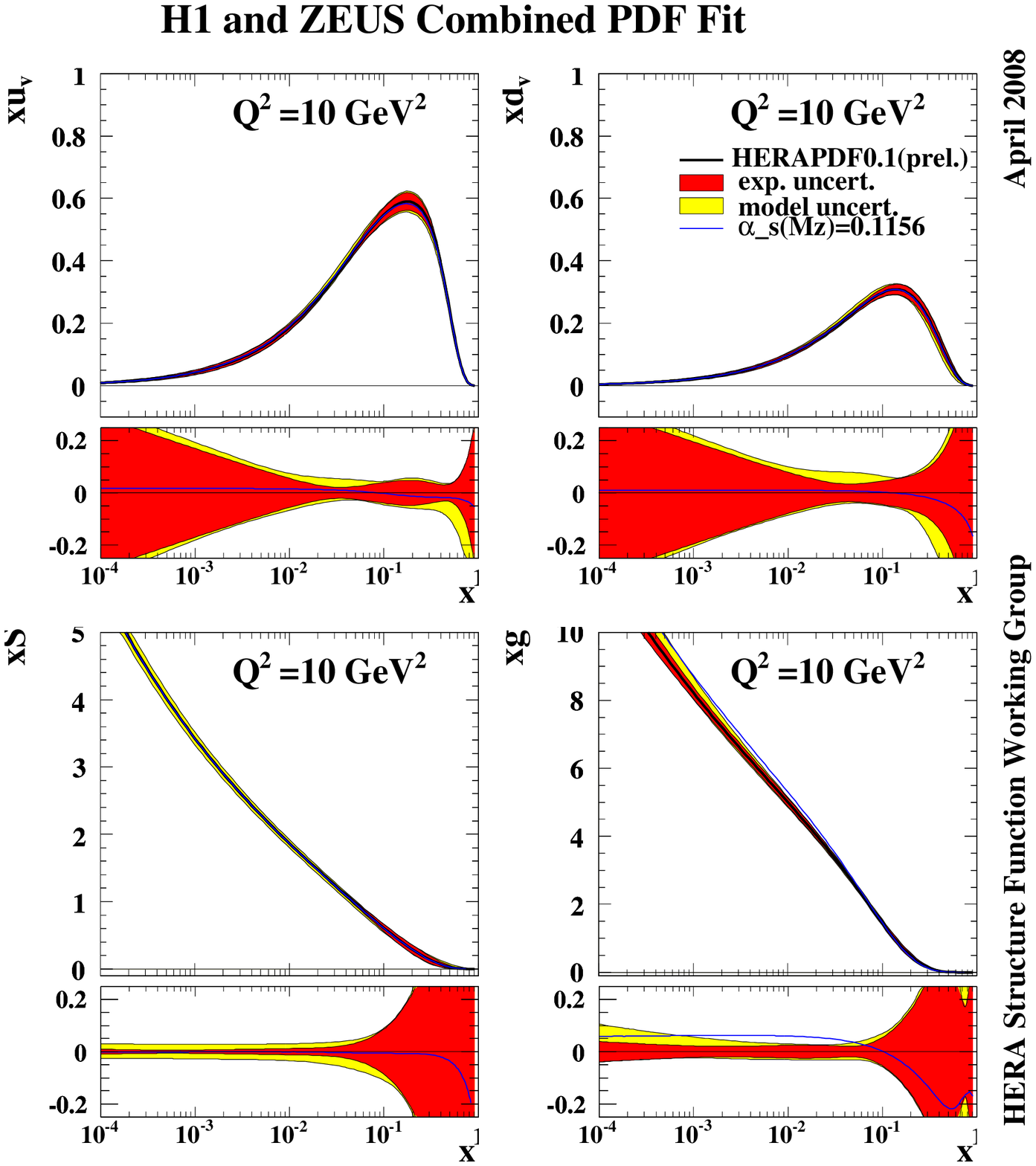 ,width=0.5\textwidth,height=6.5cm}
\epsfig{figure=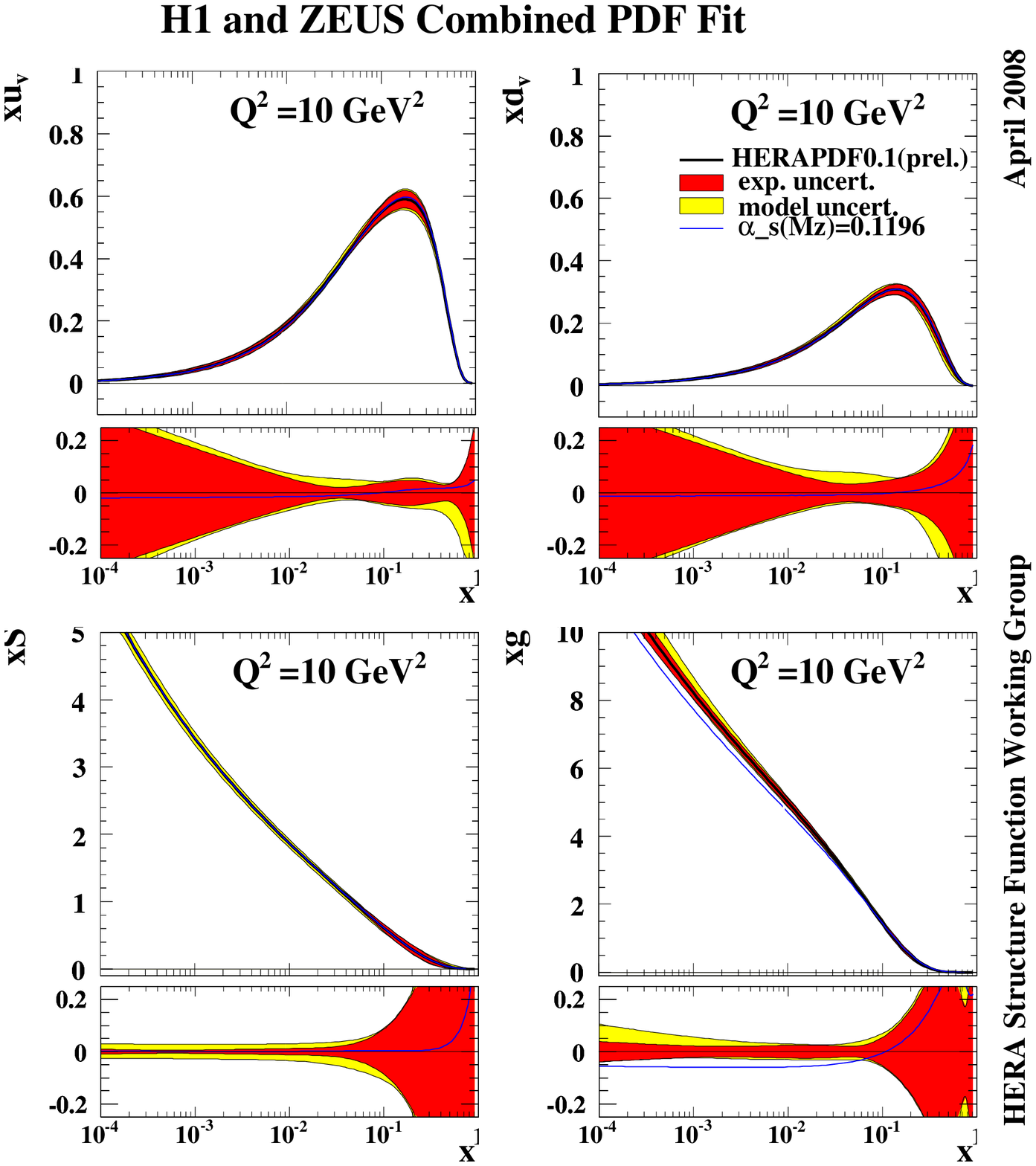 ,width=0.5\textwidth,height=6.5cm}}
\caption {HERAPDFs at $Q^2=10$GeV$^2$: with the results for $\asmz=0.1156$ 
(left) and for $\asmz=0.1196$ (right) superimposed as a blue line. 
}
\label{fig:alf}
\end{figure}

\begin{figure}[tbp]
\centerline{
\epsfig{figure=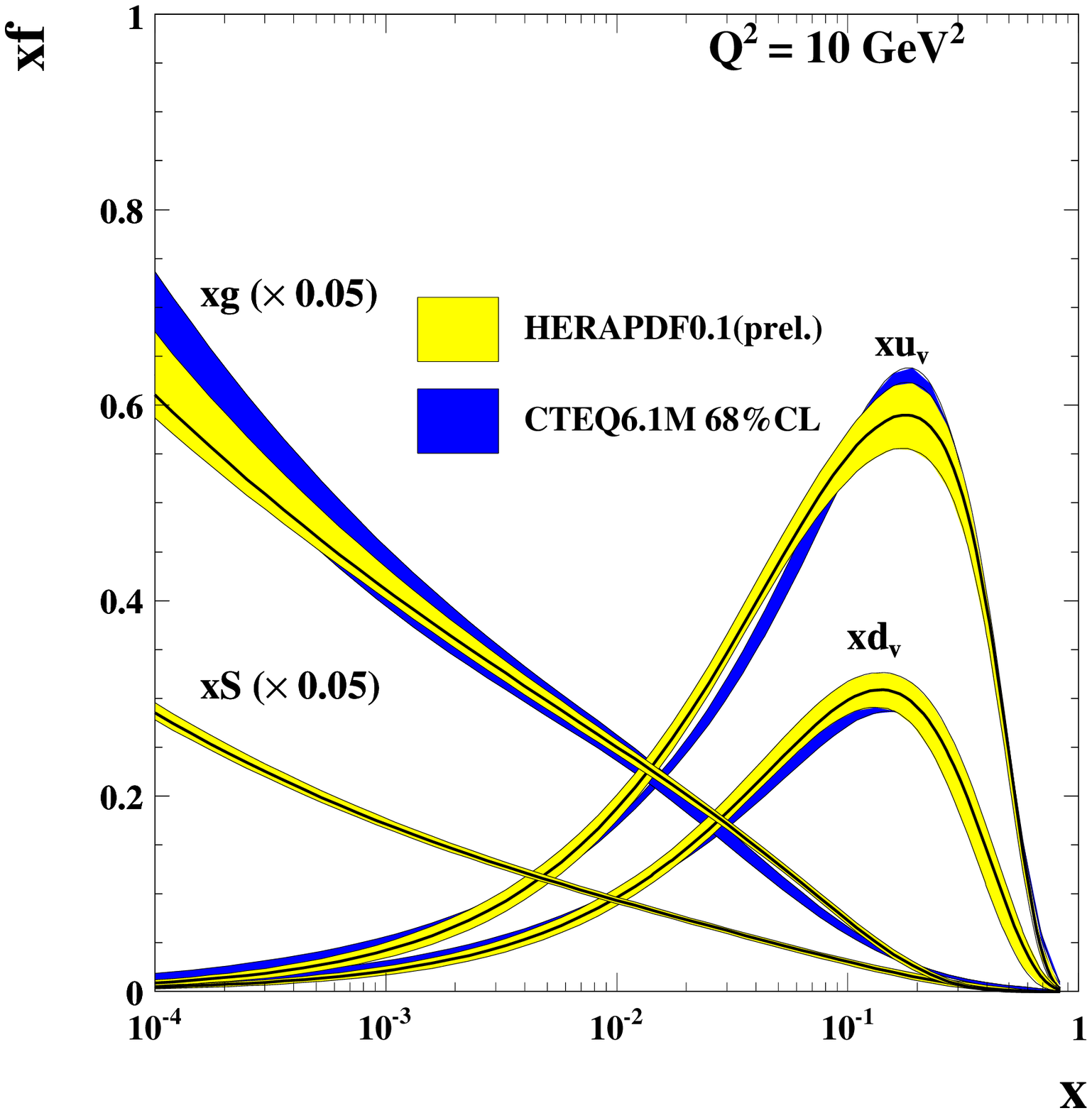 ,width=0.5\textwidth,height=6.5cm}
\epsfig{figure=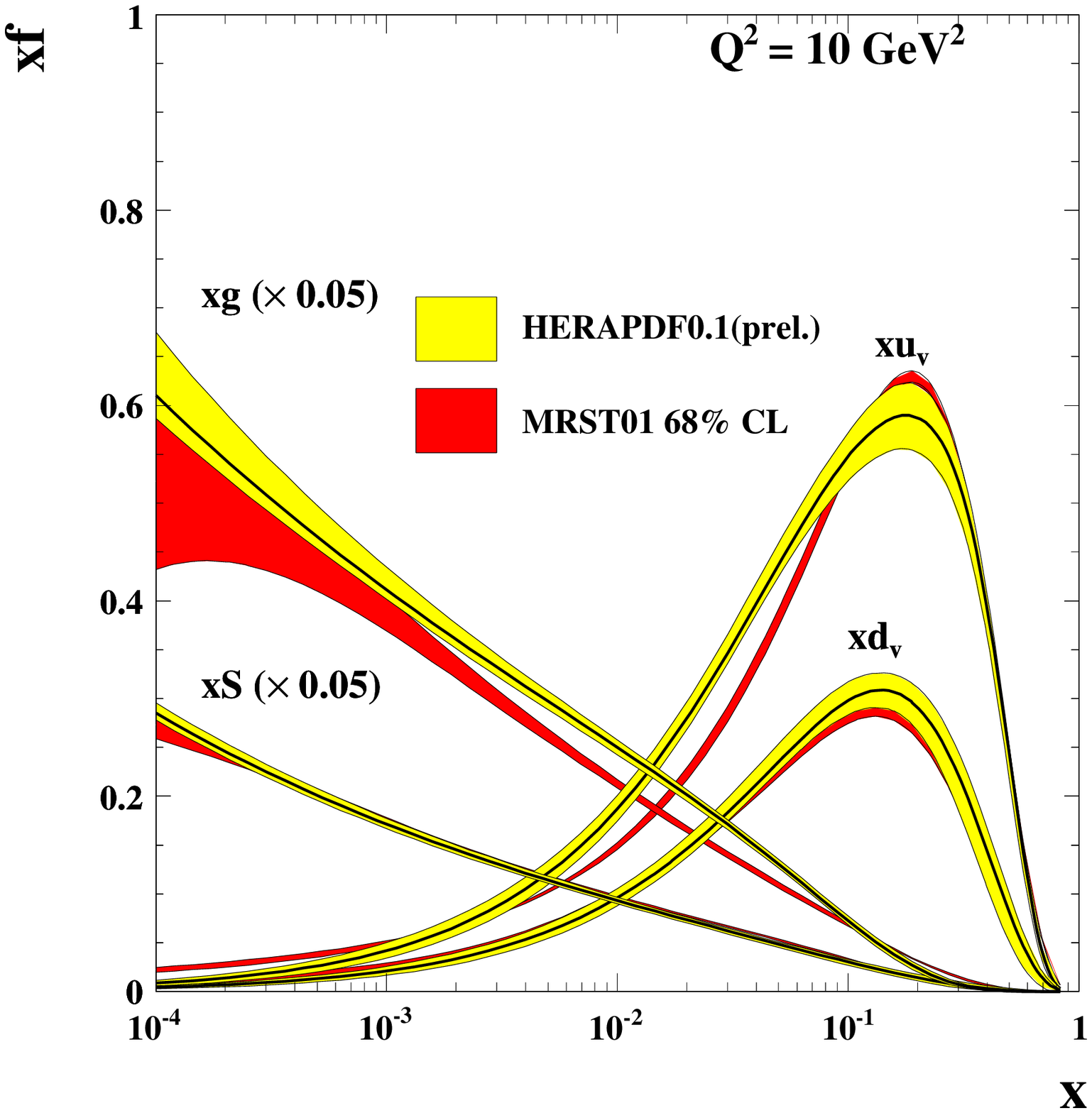 ,width=0.5\textwidth,height=6.5cm}
}
\caption {HERAPDFs at $Q^2=10$GeV$^2$ compared to the PDFs 
from CTEQ6.1 and MRST01}
\label{fig:CTEQMRSTold}
\end{figure}

\begin{figure}[tbp]
\centerline{
\epsfig{figure=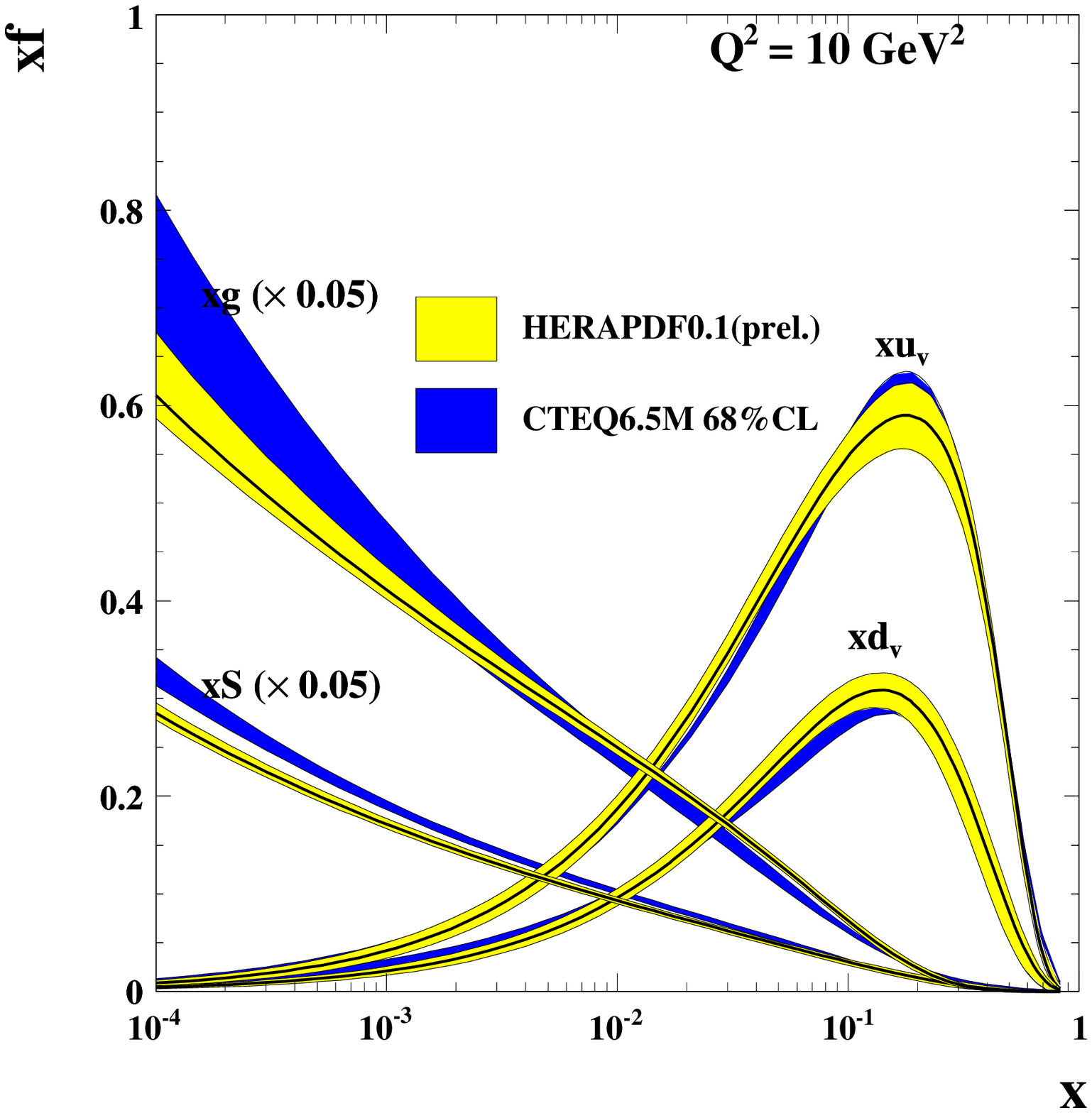 ,width=0.5\textwidth,height=6.5cm}
\epsfig{figure=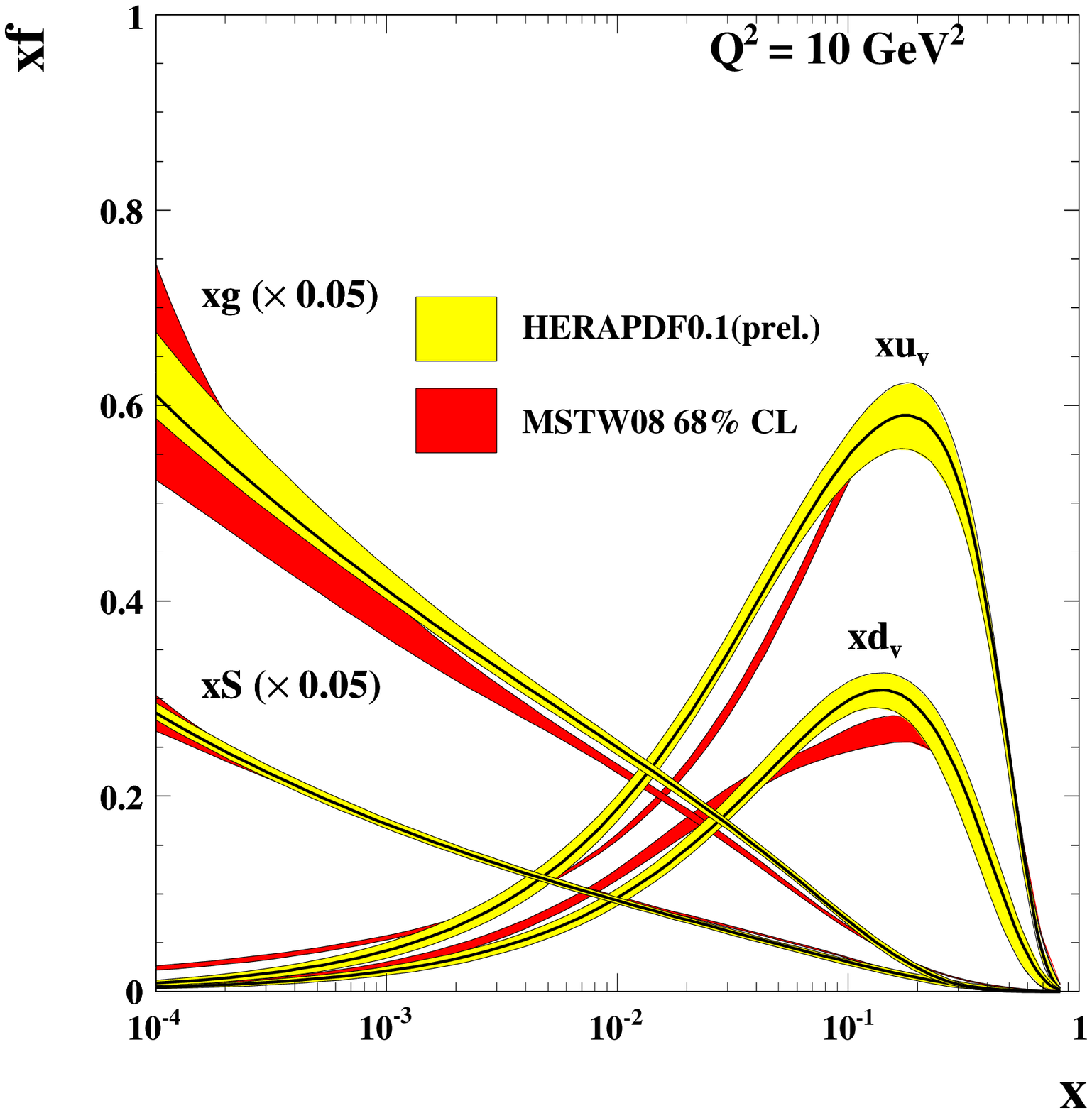 ,width=0.5\textwidth,height=6.5cm}}
\caption {HERAPDFs at $Q^2=10$GeV$^2$ compared to the PDFs 
from CTEQ6.5 and MSTW08(prel.)}
\label{fig:CTEQMRSTnew}
\end{figure}

It is also interesting to compare the PDFs for the standard choice to those 
obtained with a different input value of $\asmz$. The uncertainty on the 
current PDG value of $\asmz$ is $\pm 0.002$ and thus we vary our central 
choice by this amount. The results are shown in Fig.~\ref{fig:alf}, where we 
can see that this variation only affects the gluon PDF, such that the 
larger(smaller) value of $\asmz$ results in a harder(softer) 
gluon as predicted by the DGLAP equations. The change is 
outside total uncertainty bands of the standard fit. 
Finally, Figs.~\ref{fig:CTEQMRSTold} and ~\ref{fig:CTEQMRSTnew} 
compare the HERAPDF0.1 PDFs 
to those of the CTEQ and the MRST/MSTW groups respectively. The uncertainty 
bands of the CTEQ and MRST/MSTW analyses have been scaled to represent 
$68\%$ CL limits for direct comparability to the HERAPDF0.1. 
The HERAPDF0.1 analysis has much improved precision on the low-$x$ gluon.

\subsubsection{Summary of HERAPDF0.1 results}
\label{sec:conc}
Now that high-$Q^2$ HERA data on NC and CC
 $e^+p$ and $e^-p$ inclusive double 
differential cross-sections are available, PDF fits can be made to HERA 
data alone, since the HERA high $Q^2$ cross-section 
data can be used to determine the valence distributions and HERA low $Q^2$ 
cross-section data can be used to determine the Sea and gluon distributions. 
The combined HERA-I data set, of neutral and charged current 
inclusive cross-sections for $e^+p$ and $e^-p$ scattering, has been used as the
sole input for an NLO QCD PDF fit in the DGLAP formalism. 
The consistent treatment of systematic uncertainties in the joint data set 
ensures that experimental uncertainties on the PDFs can be calculated
without 
need for an increased $\chi^2$ tolerance. This results in PDFs with greatly
reduced experimental uncertainties compared to the separate analyses of 
the ZEUS and H1 experiments. Model uncertainties, including those 
arising from parameterization dependence, have also been carefully considered. 
The resulting HERAPDFs (called HERAPDF0.1) 
have improved precision at low-$x$ compared to the global fits. this will be 
important for predictions of the $W$ and $Z$ cross-sections at the LHC, 
as explored in the next Section.

These PDFs have been released on LHAPDF in version LHAPDF.5.6: they
consist of a
central value 
and 22 experimental eigenvectors plus 12 model alternatives.
The user should sum over Nmem=1,22 for experimental uncertainties and over 
Nmem=1,34 for total uncertainties.

\subsubsection{Predictions for $W$ and $Z$ cross-sections at the LHC using the HERAPDF0.1}
\label{sec:WZ}

\begin{figure}[tbp]
\centerline{
\epsfig{figure=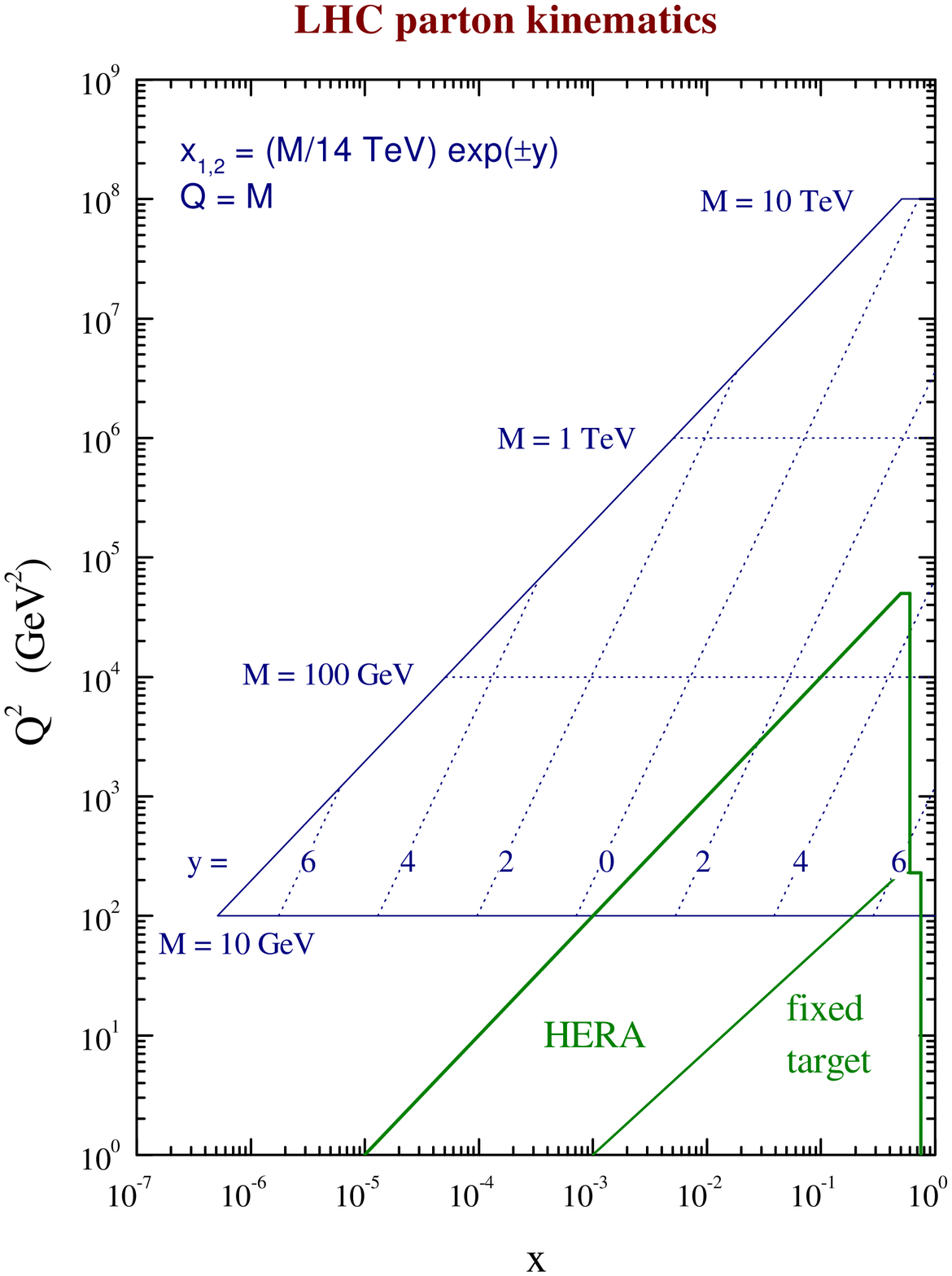,width=0.5\textwidth}
\epsfig{figure=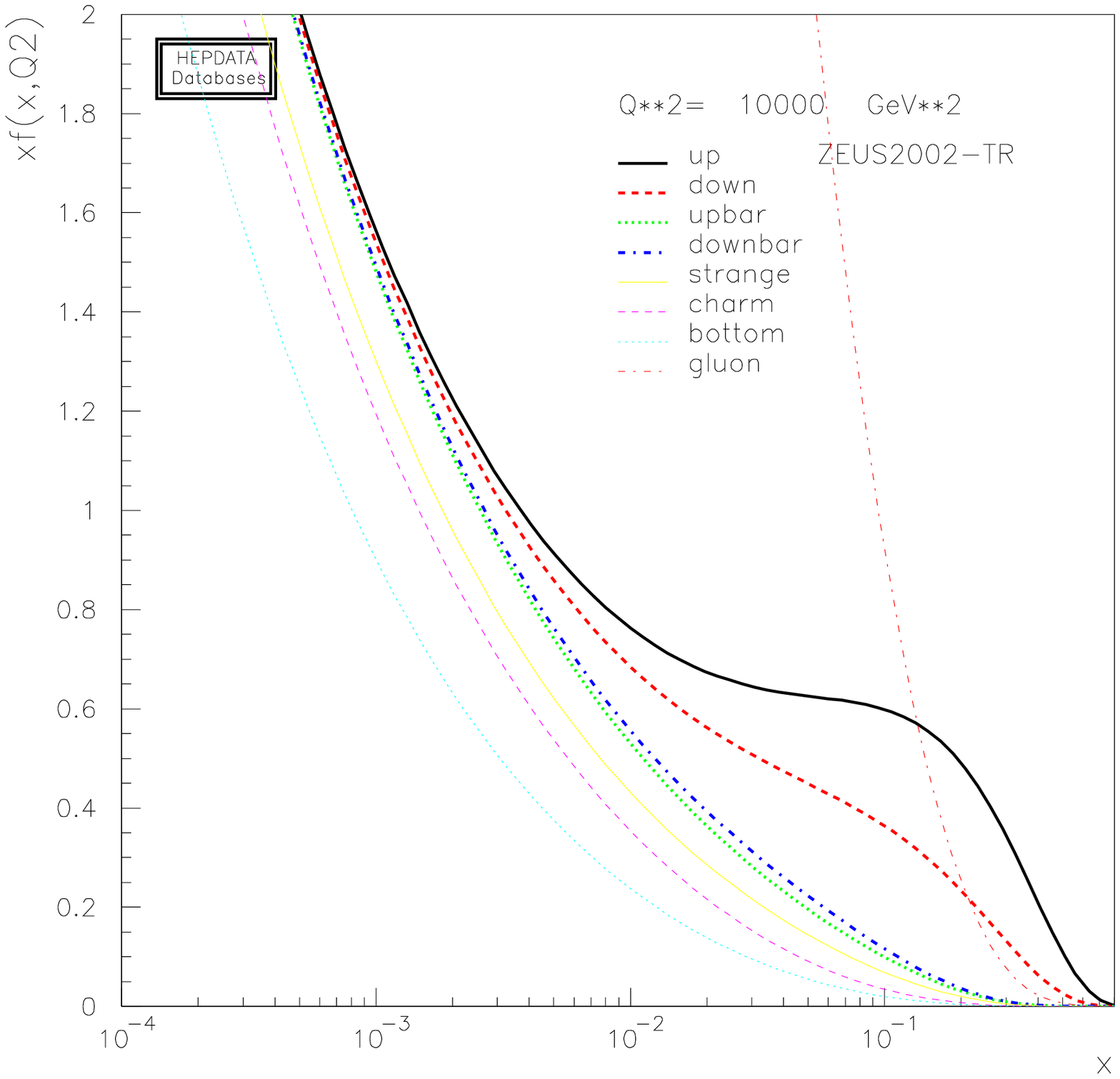,width=0.5\textwidth}}
\caption {Left plot: The LHC kinematic plane (thanks to James Stirling).
Right plot: Typical PDF distributions at $Q^2 = 10,000$~GeV$^2$.}
\label{fig:kin/pdfs}
\end{figure}

 At leading order (LO), $W$ and $Z$ production occur by the process, $q \bar{q} \rightarrow W/Z$, 
and the momentum fractions of the partons 
participating in this subprocess are given by, $x_{1,2} = \frac{M}{\surd{s}} exp (\pm y)$, where 
$M$ is the centre of mass energy of the subprocess, $M = M_W$ or $M_Z$, $\surd{s}$ is the centre of
 mass energy of the reaction  ($\surd{s}=14$ TeV at the LHC) and 
$y = \frac{1}{2} ln{\frac{(E+pl)}{(E-pl)}}$ gives the parton rapidity. 
The kinematic plane for LHC parton kinematics is shown in Fig.~\ref{fig:kin/pdfs}. 
Thus, at central rapidity, the participating partons have small momentum fractions, $x \sim 0.005$.
Moving away from central rapidity sends one parton to lower $x$ and one 
to higher $x$, but over the central rapidity range, $|y| < 2.5$, 
$x$ values remain in the range, 
$5\times 10^{-4} < x < 5\times 10^{-2}$. Thus, in contrast to the situation at 
the Tevatron, the scattering is happening mainly between sea quarks. 
Furthermore, the high 
scale of the process $Q^2 = M^2 \sim 10,000$~GeV$^2$ ensures that the gluon is the dominant 
parton, see Fig.~\ref{fig:kin/pdfs}, so that these sea quarks have mostly 
been generated by the flavour blind $g \to q \bar{q}$ splitting process. Thus the precision of 
our knowledge of $W$ and $Z$ cross-sections at the LHC is crucially dependent on the uncertainty on 
the momentum distribution of the low-$x$ gluon. 

HERA data have already dramatically improved our knowledge of the low-$x$ gluon, as 
discussed in earlier proceedings of the HERALHC 
workshop~\cite{Dittmar:2005ed}. Now that
the precision of HERA data at small-$x$ have been dramatically improved by the 
combination of H1 and ZEUS HERA-I data, we re-investigate the consequences for 
predictions of $W,Z$ production at the LHC.

Predictions for the $W/Z$ cross-sections, decaying to the lepton decay mode, 
using CTEQ, ZEUS PDFs and the HERAPDF0.1  
are summarised in 
Table~\ref{tab:datsum}. Note that the uncertainties of CTEQ PDFS have 
been rescaled to represent $68\%$ CL, in order to be comparable to the HERA 
PDF uncertainties.
\begin{table}[t]
\centerline{\small
\begin{tabular}{llllcccc}\\
 \hline
PDF Set  & $\sigma(W^+).B(W^+ \rightarrow l^+\nu_l)$ & $\sigma(W^-).B(W^- \rightarrow l^-\bar{\nu}_l)$ & 
$\sigma(Z).B(Z \rightarrow l^+ l^-)$\\
 \hline
 CTEQ6.1 & $11.61 \pm 0.34 $~nb & $8.54 \pm 0.26 $~nb & $1.89 \pm 0.05$~nb\\
 CTEQ6.5 & $12.47 \pm 0.28 $~nb & $9.14 \pm 0.22 $~nb & $2.03 \pm 0.04$~nb\\
 ZEUS-2002  & $12.07 \pm 0.41 $~nb & $8.76 \pm 0.30 $~nb & $1.89 \pm 0.06$~nb\\
 ZEUS-2005  & $11.87 \pm 0.45 $~nb & $8.74 \pm 0.31 $~nb & $1.97 \pm 0.06$~nb\\
 HERAPDF0.1& $12.14 \pm 0.13 $~nb & $9.08 \pm 0.14 $~nb & $1.99 \pm 0.025$~nb\\
 \hline\\
\end{tabular}}
\caption{LHC $W/Z$ cross-sections for decay via the lepton mode, 
for various PDFs, with $68\%$ CL uncertainties.}
\label{tab:datsum}
\end{table}
The precision on the predictions of the global fits (CTEQ6.1/5 and ZEUS-2002)
for the total $W/Z$ cross-sections 
is $\sim 3\%$ at $68\%$ CL. The precision of the ZEUS-2005 PDF fit 
prediction, 
which used only ZEUS data, is comparable, since information on the low-$x$ 
gluon is coming from HERA data alone. 
The increased precision of the HERAPDF0.1 low-$x$ gluon PDF
results in increased precision of the $W/Z$ cross-section predictions of 
$\sim 1\%$. 

\begin{figure}[tbp]
\centerline{
\epsfig{figure=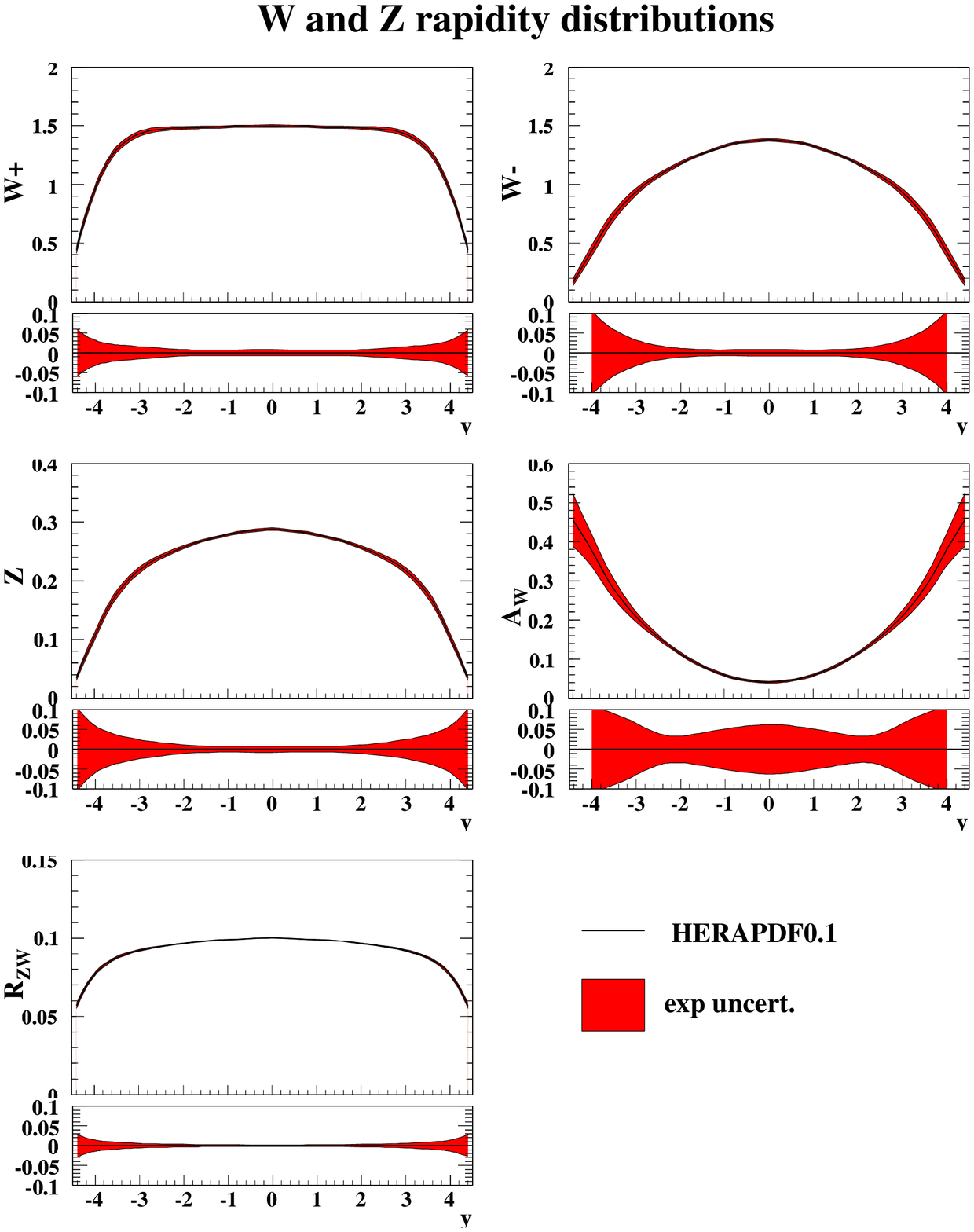,width=0.45\textwidth,height=8.0cm}
\epsfig{figure=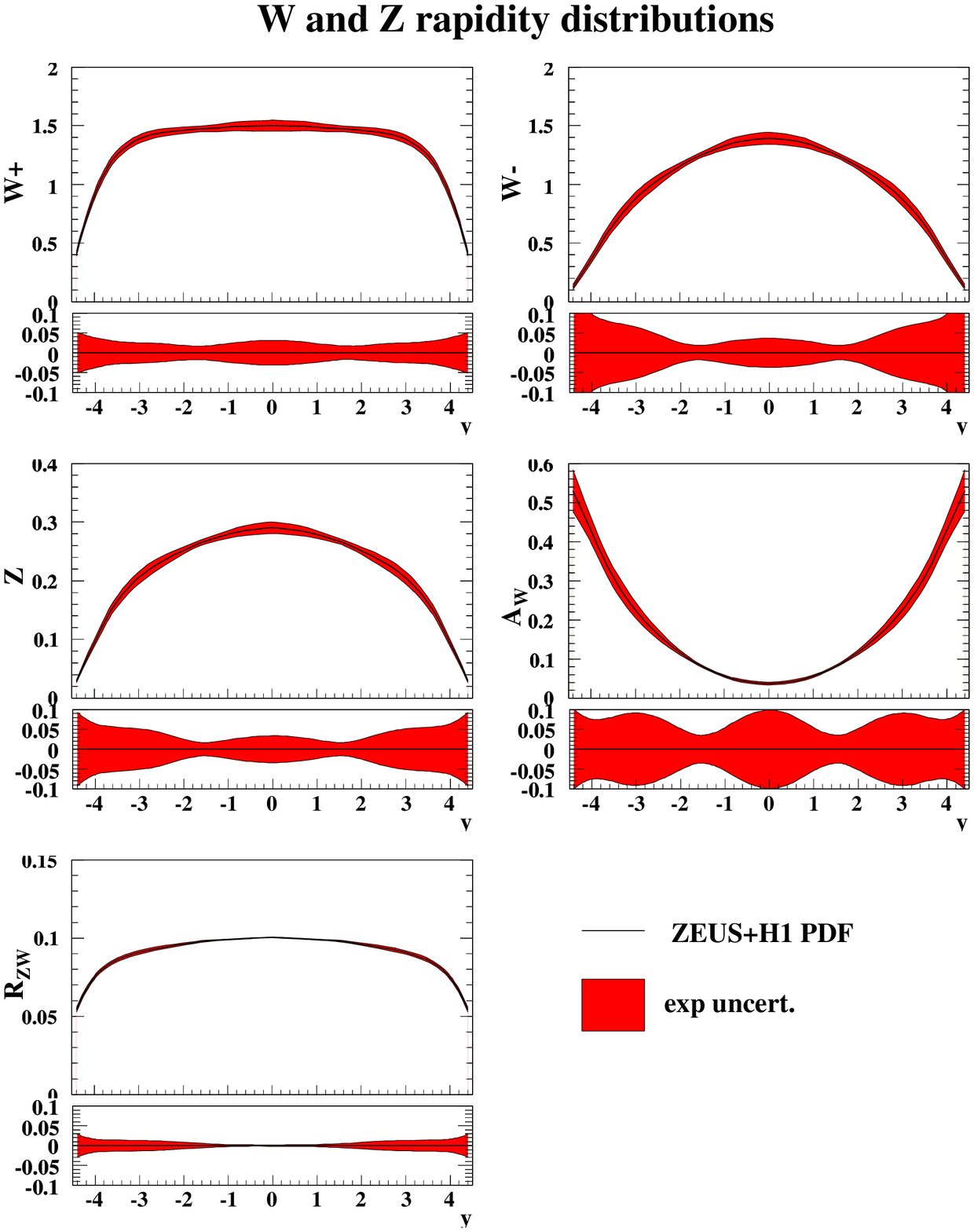,width=0.45\textwidth,height=8.0cm}}
\caption {The $W^+,W^-, Z$ rapidity distributions, $A_W$ and $R_{ZW}$ (see text) and their uncertainties as 
predicted by (left) HERAPDF0.1 (right) a similar fit to the uncombined 
ZEUS and H1 data from HERA-I. }
\label{fig:WZ}
\end{figure}

\begin{figure}[tbp]
\centerline{
\epsfig{figure=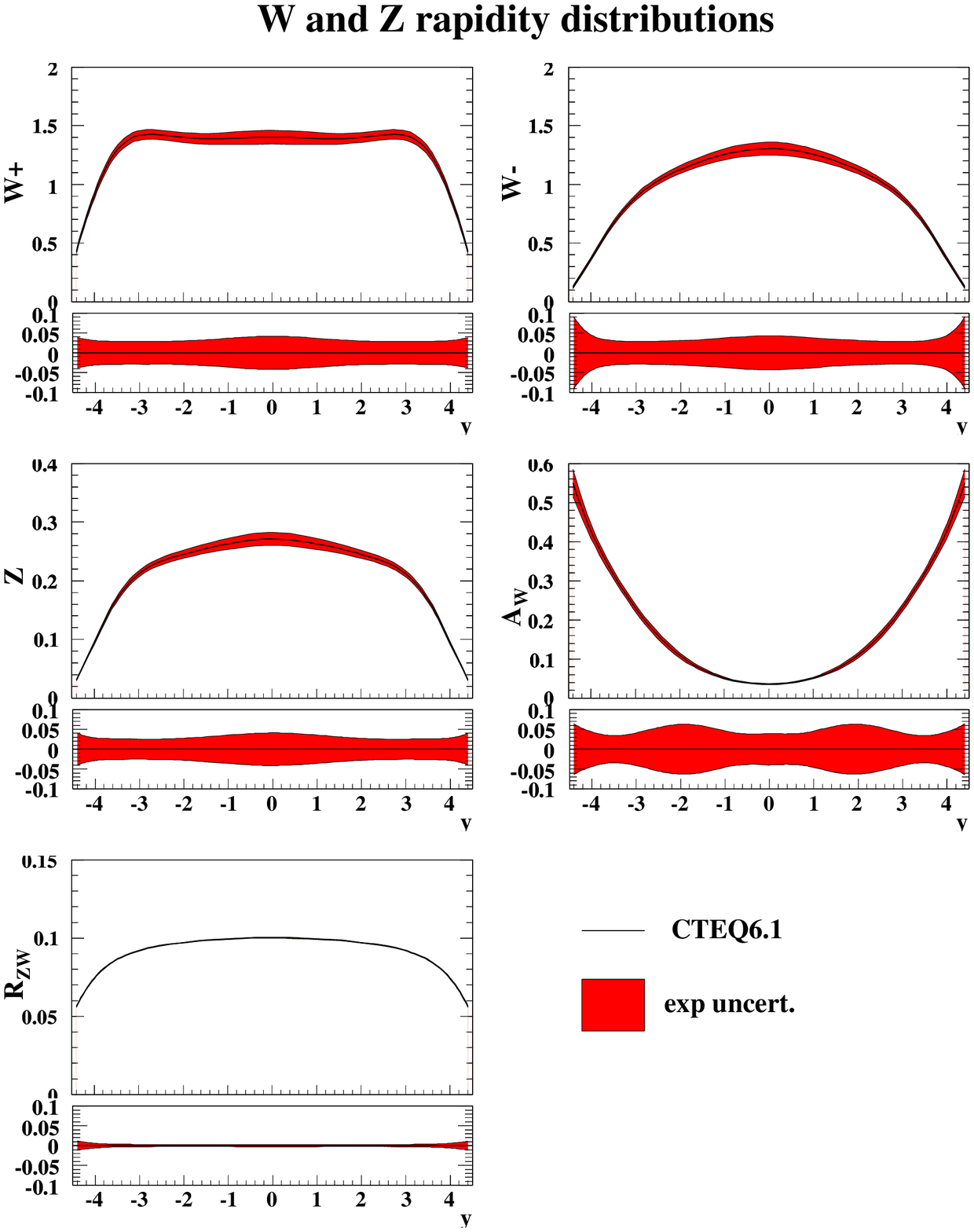,width=0.33\textwidth,height=8.0cm}
\epsfig{figure=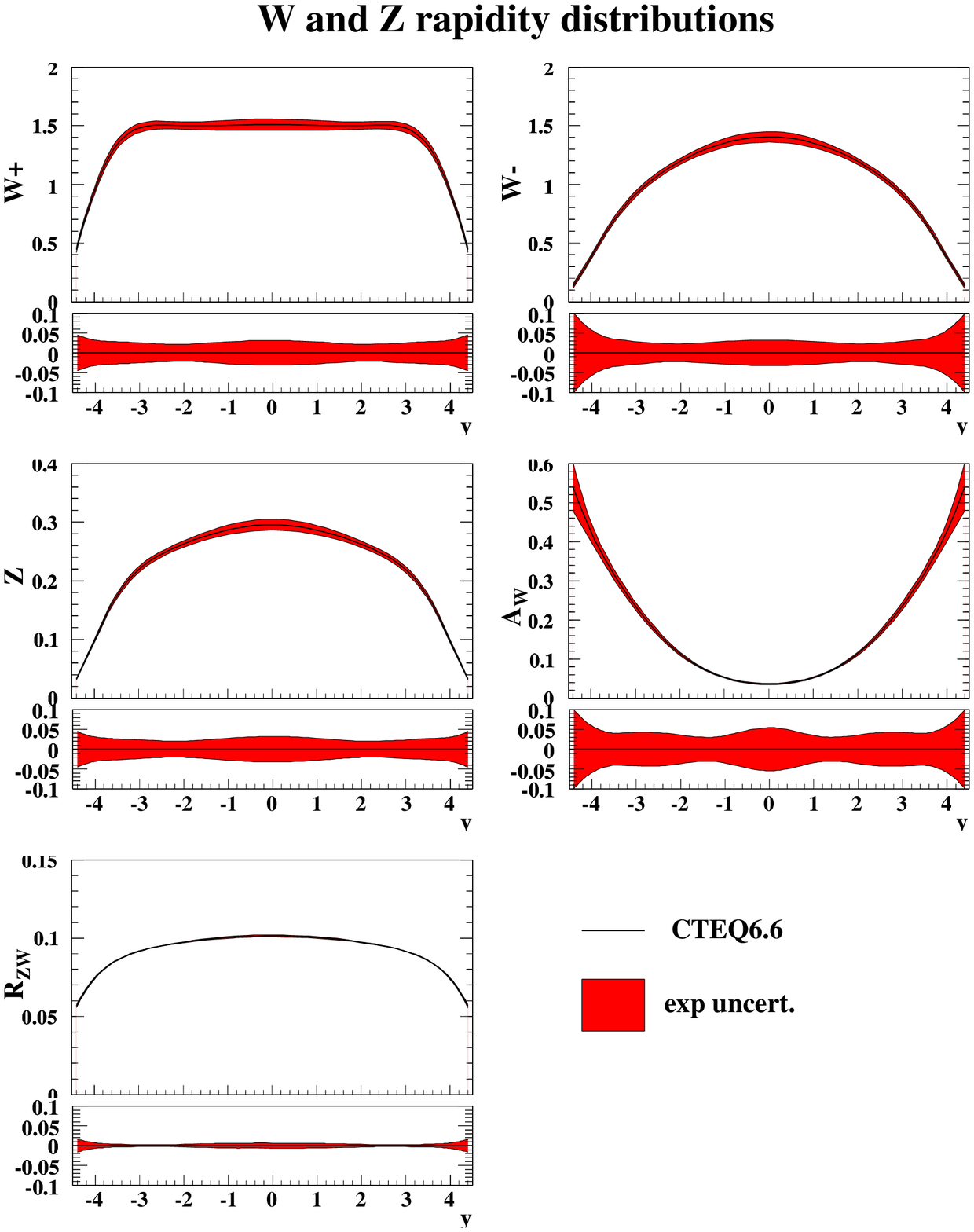,width=0.33\textwidth,height=8.0cm}
\epsfig{figure=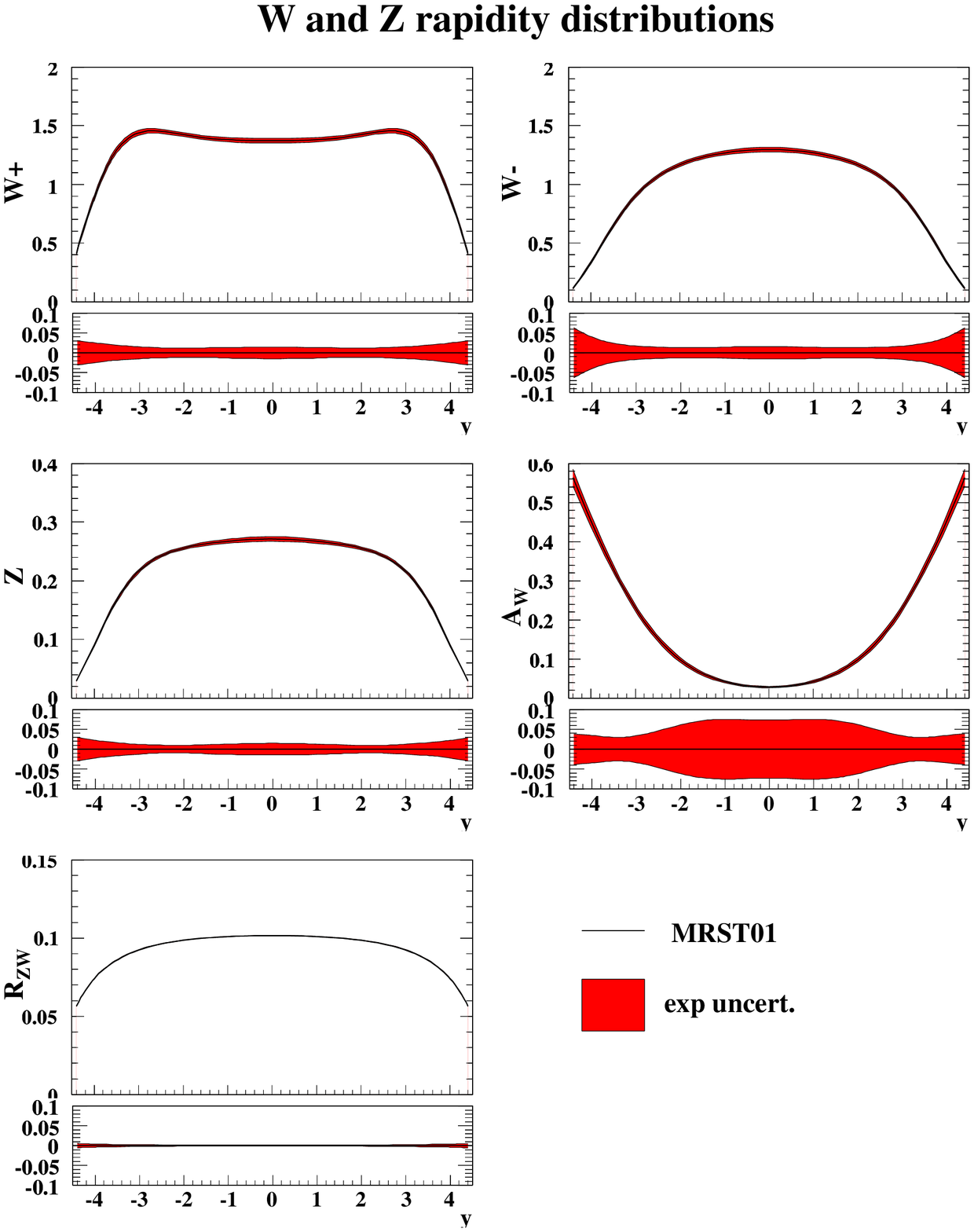,width=0.33\textwidth,height=8.0cm}}
\caption {The $W^+,W^-, Z$ rapidity distributions, $A_W$ and $R_{ZW}$ (see text) and their uncertainties (scaled to $68\%$ CL) as 
predicted by (left) CTEQ6.1, (middle) CTEQ6.6, right (MRST01 }
\label{fig:WZcteq}
\end{figure}

It is interesting to consider the predictions as a function of rapidity. 
Fig~\ref{fig:WZ} shows the predictions for $W^+,W^-, Z$ 
production as a function of rapidity from the HERAPDF0.1 PDF fit 
and compares them to the predictions from 
a PDF fit, using the same parameterization and assumptions, to the H1 and ZEUS 
data from HERA-I uncombined. The increase precision due to the combination is 
impressive. Fig.~\ref{fig:WZcteq} show the predictions for $W^+,W^-, Z$ 
production as a function of rapidity from the CTEQ6.1, 6.6 and MRST01
 PDF fits for 
comparison. The uncertainties on the CTEQ and MRST 
PDF predictions have been rescaled 
to represent $68\%$ CL limits, for direct comparability to the HERAPDF0.1 
uncertainties. At central rapidity these limits give an 
uncertainty on the boson cross-sections of 
$\sim 5\%$, ($\sim 3\%$),($\sim 2\%$)  
for CTEQ6.1, (CTEQ6.6), (MRST01) compared to $\sim 1\%$ for the HERAPDF0.1. 

\begin{figure}[tbp]
\centerline{
\epsfig{figure=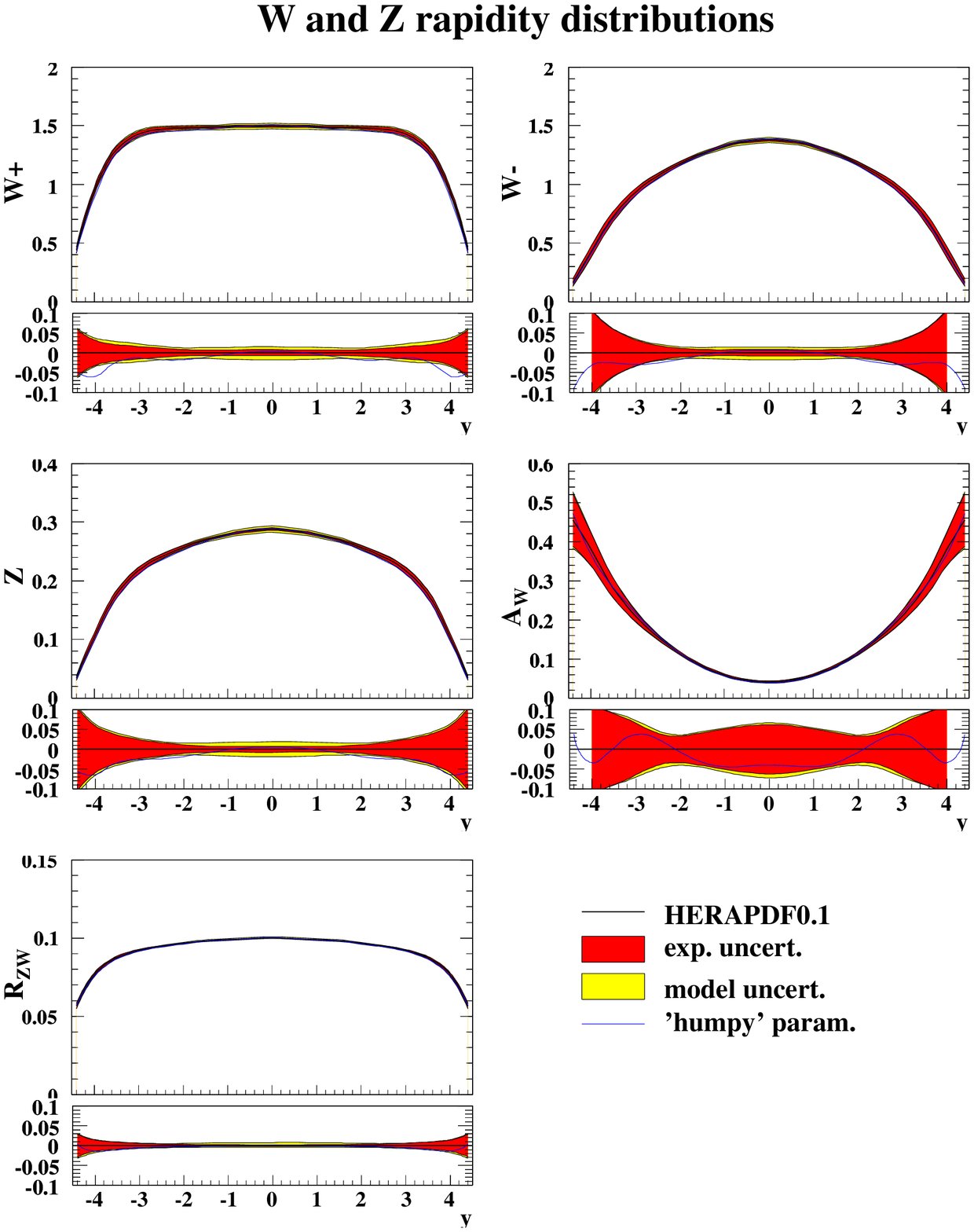,width=0.45\textwidth,height=8.0cm}
\epsfig{figure=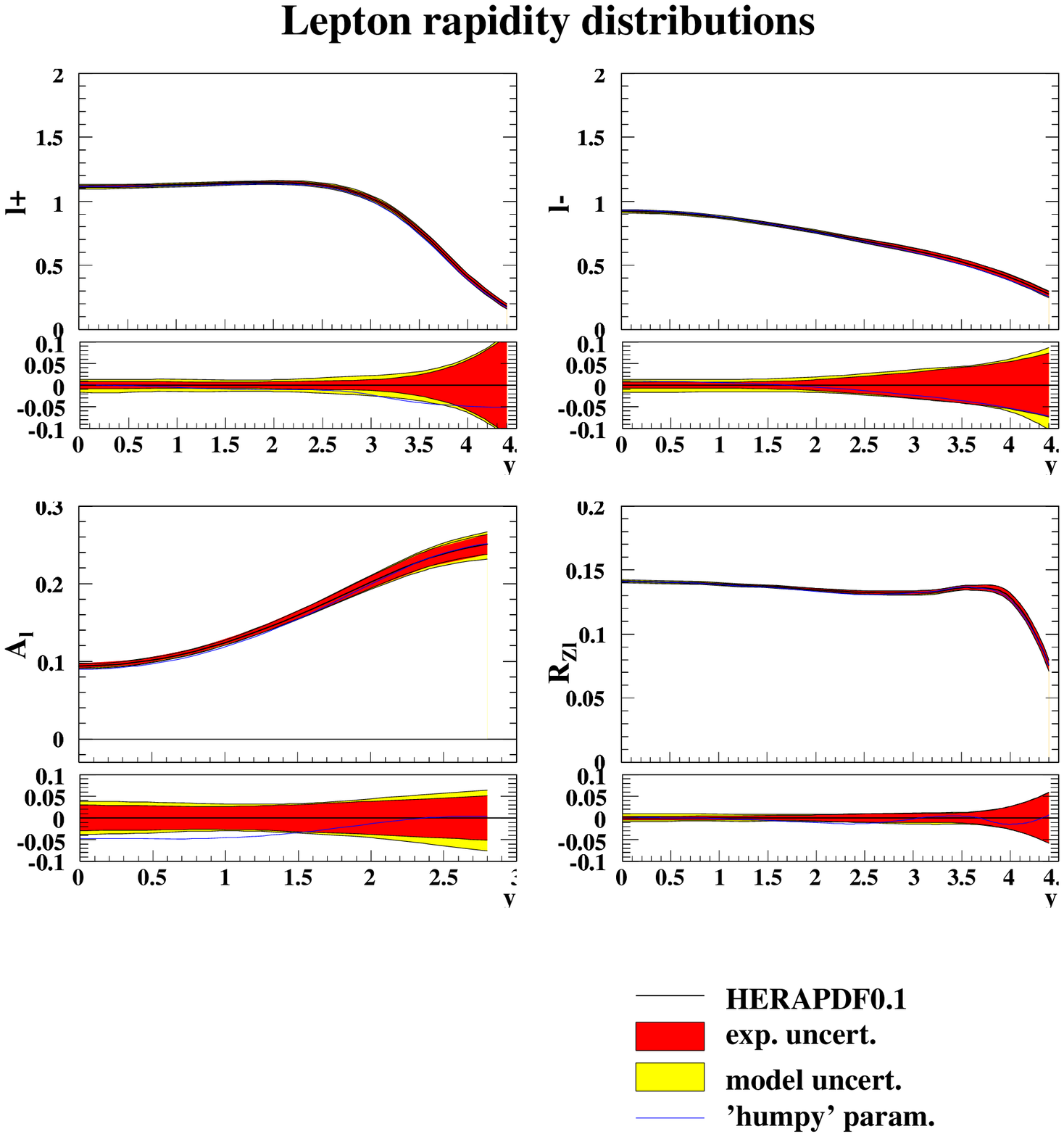,width=0.45\textwidth,height=8.0cm}}
\caption {Left: the $W^+,W^-, Z$ rapidity distributions, $A_W$, and $R_{ZW}$ (see text) and their experimental 
uncertainties (red) and model uncertainties (yellow). Right: the $l^+, l^-$ 
rapidity distributions, $A_l$ and $R_{Zl}$ (see text) and their experimental and model uncertainties. The superimposed blue line represents the results of the alternative 'humpy' gluon parameterization.}
\label{fig:WZmodel}
\end{figure}

So far, only experimental uncertainties have been included 
in these evaluations. It is also necessary to include model uncertainties.
Fig.~\ref{fig:WZmodel} shows the $W^+,W^-, Z$ rapidity distributions 
including the six sources of model uncertainty detailed in 
Section~\ref{sec:anal}. 
These model uncertainties increase the total uncertainty at central rapidity 
to $\sim 2\%$. 
Further uncertainty due to the choice of $\alpha_s(M_Z)$ is small because, 
although a lower (higher) choice results in a larger (smaller) gluon at 
low $x$, the rate of QCD evolution is lower (higher) and this largely 
compensates. Uncertainties due to the choice of 
parameterization also have little impact on the boson rapidity spectra in the 
central region as illustrated in Fig.~\ref{fig:WZmodel} by the superimposed
 blue line, which represents the alternative 'humpy' gluon parameterization
(see Sec.~\ref{sec:results}). 

Since the PDF uncertainty feeding into the $W^+, W^-$ and $Z$ production is mostly coming from the
gluon PDF, for all three processes, there is a strong correlation in their uncertainties, which can be 
removed by taking ratios. Figs.~\ref{fig:WZ},~\ref{fig:WZcteq} and \ref{fig:WZmodel} also show the $W$ asymmetry 
\[A_W = (W^+ - W^-)/(W^+ + W^-).\] 
The experimental PDF uncertainty on the asymmetry is larger 
($\sim ~ 5\%$ for both CTEQ and HERAPDFs, $\sim ~ 7\%$ for the MRST01 PDFs) 
than that on the 
individual distributions and the variation between PDF sets is also larger 
- compare the central values of the CTEQ and MRST predictions, 
which are almost $25\%$ discrepant. 
This is because the asymmetry is
sensitive to the difference in the valence PDFs, $u_v - d_v$, in the low-$x$ 
region, $5\times 10^{-4} < x < 5\times 10^{-2}$, where there is no constraint 
from current data. To see this consider that at LO,  
\[A_W \sim (u\bar{d}-d\bar{u})/(u\bar{d}+d\bar{u}+c\bar{s}+s\bar{c})\] 
and that $\bar{d} \sim \bar{u}$ at low-$x$. 
(Note that the $c\bar{s}$ and $s\bar{c}$ contributions
cancel out in the numerator). The discrepancy between the CTEQ and MRST01 
asymmetry predictions at $y=0$ can be quantitatively 
understood by considering their different 
valence PDFs (see Figs.~\ref{fig:CTEQMRSTold}, \ref{fig:CTEQMRSTnew} 
in Sec.~\ref{sec:results}). 
In fact a measurement of the asymmetry at the LHC will 
provide new information to constrain these PDFs. 

By contrast, the ratio \[R_{ZW} = Z/(W^+ +W^-),\] 
also shown in Figs.~\ref{fig:WZ},~\ref{fig:WZcteq} and \ref{fig:WZmodel}, 
has very small PDF uncertainties (both experimental and 
model) 
and there is no significant variation between PDF sets. To understand
 this consider that at LO 
\[R_{ZW} = (u\bar{u}+d\bar{d}+c\bar{c}+s\bar{s})/(u\bar{d}+d\bar{u}+c\bar{s}+s\bar{c})\] 
(modulo electroweak couplings) and that 
$\bar{d} \sim \bar{u}$ at low-$x$~\footnote{There is some small model dependence from the strange sea 
fraction accounted for in both HERAPDF0.1 and in CTEQ6.6 PDFs.}. 
This will be a crucial measurement for our understanding of 
Standard Model Physics at the LHC. 

However, 
whereas the $Z$ rapidity distribution can be fully reconstructed from its decay leptons, 
this is not possible for the $W$ rapidity distribution, because the leptonic decay channels 
which we use to identify the $W$'s have missing neutrinos. Thus we actually measure the $W$'s 
decay lepton rapidity spectra rather than the $W$ rapidity spectra. 
 Fig.~\ref{fig:WZmodel} also
shows the rapidity spectra for positive and 
negative leptons from $W^+$ and $W^-$ decay, the lepton asymmetry, \[A_l = (l^+ - l^-)/(l^+ + l^-)\] and the ratio  \[R_{Zl} = Z/(l^+ +l^-)\]
A cut of, $p_{tl} > 25$~GeV, has been applied on the decay lepton, since it will not be possible to
trigger on 
leptons with small $p_{tl}$. A particular lepton rapidity can be fed from a range 
of $W$ rapidities so that the contributions of partons at different $x$ values is smeared out 
in the lepton spectra, but the broad features of the $W$ spectra 
remain. 

In summary, these investigations indicate that PDF uncertainties, 
deriving from experimental error, on 
predictions for the $W,Z$ rapidity spectra in the central region, have
reached a precision of $\sim 1\%$, due to the input of the combined HERA-I 
data. This level of precision is maintained when using 
the leptons from the $W$ decay and gives us hope that we could
use these processes as luminosity monitors\footnote{A caveat is that the 
current study has been performed using 
PDF sets which are extracted using NLO QCD in the DGLAP formalism. 
The extension to NNLO gives small corrections $\sim 1\%$.  However, 
there may be much larger  
uncertainties in the theoretical calculations because the kinematic region 
involves  
low-$x$.  There may be a need to account for $ln(1/x)$ resummation 
or high gluon density effects.}. However, 
model dependent uncertainties must now be considered very carefully. 
The current study will be repeated using a general-mass variable-flavour 
scheme for heavy quarks.

The predicted precision on the ratios $R_{ZW}$, $R_{Zl}$  is even better  
since model uncertainties are also very small giving a total uncertainty of 
$\sim 1\%$. 
This measurement may be used as a SM benchmark. 
However the $W$ and lepton asymmetries have larger uncertainties ($5-7\%$). 
A measurement of these quantities would give new 
information on valence distributions at small-$x$.


\subsection{Measurements of the Proton Structure Function $F_L$ at
  HERA \protect\footnote{ Contributing
authors: J.~Grebenyuk, V.~Lendermann}}
\label{sec:fldet}

\subsubsection{Introduction}

The inclusive deep inelastic $ep$ scattering (DIS)
cross section can at low $Q^2$ be written in terms of
the two structure functions, $F_2$ and $F_L$, in reduced form as
\begin{equation}
 \sigma_r(x,Q^2,y) \equiv \frac{d^2\sigma}{dx dQ^2}
 \cdot \frac{Q^4 x}{2\pi \alpha^2 Y_+}
  =  F_2(x,Q^2) - \frac{y^2}{Y_+} \cdot  F_L(x,Q^2)~,
       \label{eq:sigma_red}
  \end{equation}
where $Q^2 = -q^2$ is the negative of the square of the four-momentum
transferred between the electron\footnote{The term electron is
used here to denote both electrons and positrons unless the charge state
is specified explicitly.
}
 and the proton, and $x=Q^2/2q P$ denotes
the Bjorken variable, where $P$ is the four-momentum of the proton.
The two variables are related through the inelasticity  of
the scattering process, $y=Q^2/sx$,
where $s=4 E_e E_p$ is the centre-of-mass energy
squared determined  from the electron and proton beam
energies, $E_e$ and $E_p$. In eq.\,\ref{eq:sigma_red},
$\alpha$ denotes the fine structure constant and $Y_+=1+(1-y)^2$.

The two proton structure functions $F_2$ and $F_L$ are related
to the cross sections
of the transversely and longitudinally polarised virtual photons interacting with protons,
$\sigma_L$ and $\sigma_T$, according to
$F_L \propto \sigma_L$  and $F_2 \propto (\sigma_L + \sigma_T)$.
Therefore the relation $0 \leq F_L \leq F_2$  holds.
In the Quark Parton Model (QPM),
$F_2$ is the sum of the quark and anti-quark $x$ distributions,
weighted by the square of the electric quark charges,
whereas the value of $F_L$ is zero~\cite{callan-gross}.
The latter follows from the fact that a quark with spin ${1 \over 2}$ cannot
absorb a longitudinally polarised photon.

In Quantum Chromodynamics (QCD), $F_L$ differs from zero, receiving
contributions from quarks and from gluons~\cite{ZWT,*AM}. At low $x$
and in the $Q^2$ region of deep inelastic scattering the
gluon contribution greatly exceeds the quark contribution.
Therefore $F_L$ is a direct measure of the gluon distribution
to a very good approximation.
The gluon distribution is also constrained by the scaling
violations of $F_2$ as described by the DGLAP QCD evolution
equations~\cite{Gribov:1972ri,Gribov:1972rt,Lipatov:1974qm,Dokshitzer:1977sg,Altarelli:1977zs}.
An independent measurement of $F_L$ at HERA,
and its comparison with predictions derived from the
gluon distribution extracted from the $Q^2$ evolution of
$F_2(x,Q^2)$, thus represents a crucial test
on the validity of perturbative QCD (pQCD) at low $x$.
Moreover, depending on the particular theoretical approach
adopted, whether it be a fixed order pQCD calculation,
a re-summation scheme, or a color dipole ansatz,
there appear to be significant differences in the predicted
magnitude of $F_L$ at low $Q^2$.  A measurement of $F_L$
may be able to distinguish between these approaches.

Previously the structure function $F_L$ was extracted by the H1
collaboration from inclusive data at high $y$ using indirect methods,
as discussed in Sect.\,\ref{s:indirect}.
A preliminary measurement was also presented by the ZEUS
collaboration using initial state radiation (ISR) events~\cite{zeus-isr},
although the precision of this measurement was limited.

To make a direct measurement of $F_L$, reduced cross sections must be
measured at the same $x$ and $Q^2$ but with different $y$ values.
This can be seen from eq.\,\ref{eq:sigma_red} which states that $F_L(x,Q^2)$
is equal to the partial derivative  ${\rm \partial}{\sigma_r}(x,Q^2,y)/{\rm \partial}(y^2/Y_{+})$.
Due to the relationship $y=Q^2/xs$ this requires data to be collected at
different beam-beam centre-of-mass energies, which was done in the last year
of HERA running.
To maximize the precision of this procedure, the measurable range of
$y^2/Y_{+}$ had to be maximised for each fixed $x$ and $Q^2$.
This was achieved by operating HERA at the lowest attainable centre-of-mass
energy and by measuring this data up to the highest possible value of $y$.
An intermediate HERA centre-of-mass energy was also chosen,
to improve the precision of $F_L$ extraction and to act as a consistency check.
More specifically, between March and June 2007, HERA was operated
with proton beam energies, $E_p =$\,460\,GeV and 575\,GeV, compared to the
previous nominal value of 920\,GeV.  The electron beam energy was
unaltered at $E_e=27.6$\,GeV.  Thus, three data sets, referred to the
high- (HER), middle- (MER) and low-energy running (LER) samples, were
collected with $\sqrt{s}=$\,318\,GeV, 251\,GeV and 225\,GeV, respectively.
The integrated luminosities of the data sets used by ZEUS (H1) to measure
$F_L$ are 32.8\,(21.6)\,pb$^{-1}$ for HER, 6\,(6.2)\,pb$^{-1}$ for MER and
14\,(12.4)\,pb$^{-1}$ for LER.
The specific issues of the recent H1 and ZEUS analyses are discussed in
Sect.\,\ref{s:details}, and the results are presented in
Sect.\,\ref{s:results}.

\subsubsection{Indirect $F_L$ Extraction by H1}\label{s:indirect}

H1 extracted $F_L$ from inclusive data using several indirect methods,
which exploit the turn over of the reduced cross
section at high $y$ due to the $F_L$ contribution.
The basic principle is the following.
First, the reduced neutral current cross section $\sigma_r$
is measured in a $y$ range, where the $F_L$ contribution
is negligible and thus the relation $\sigma_r = F_2$ holds very well.
Afterward, based on some theoretical assumption, the knowledge
of $F_2$ is extrapolated towards high $y$. Finally $F_L$ is extracted
from the difference between the prediction for $F_2$ and the measurement
of $\sigma_r$ at high $y$.

\begin{figure}[!tb]
\centerline{\includegraphics[width=.6\textwidth]{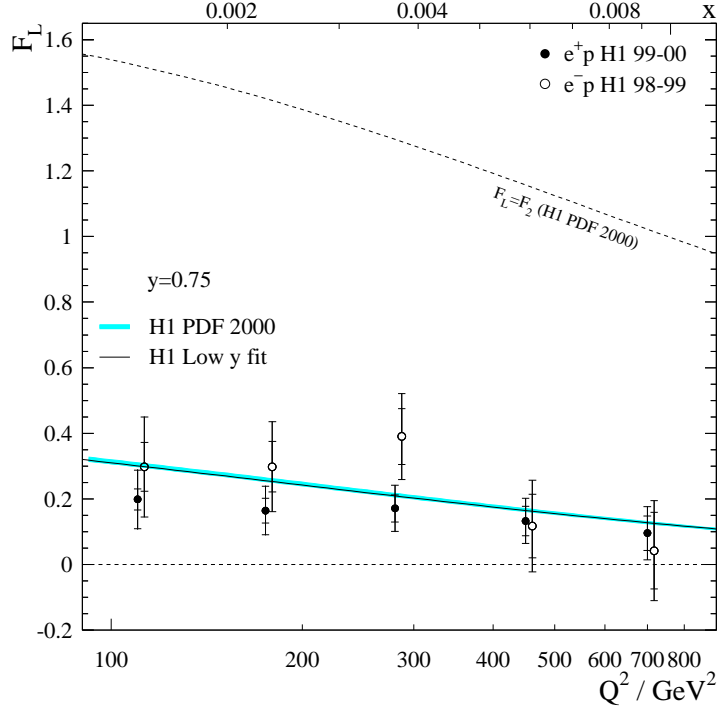}}
\caption{$F_L$ determined indirectly by H1 at a fixed $y =$\,0.75
and high $Q^2$ is shown as a function of $Q^2$ (lower scale) or
equivalently $x$ (upper scale) for $e^+ p$ (closed circles) and
$e^- p$ (open circles) data. The inner error bar represents
the statistical error, and the outer error bar also includes
the systematic error and the uncertainty arising from the
extrapolation of $F_2$.
\label{f:h1highq2}}
\end{figure}
In the analyses at
$Q^2 \gtrsim$\,10\,GeV$^2$~\cite{Adloff:1996,Adloff:2000qk,Adloff:2003uh}
the ``extrapolation'' method is used. In this method,
an NLO QCD PDF fit to H1 HERA\,I data is performed at $y <$\,0.35,
and the results are extrapolated to higher $y$ using the
DGLAP evolution equations.
$F_L$ is then extracted at a fixed $y =$\,0.75 and
at $Q^2$ up to 700\,GeV$^2$ using eq.\,\ref{eq:sigma_red}.
The extracted values are shown in Fig.\,\ref{f:h1highq2} for the
high-$Q^2$ analysis~\cite{Adloff:2003uh}.

At low $Q^2$, extrapolations of DGLAP fits become uncertain.
For $Q^2 \lesssim$\,2\,GeV$^2$, as the strong coupling constant
$\alpha_s(Q^2)$ increases, the higher order corrections to the
perturbative expansion become large and lead to the
breakdown of the pQCD calculations.
Therefore other methods
are used in the H1 low-$Q^2$ data analyses. 

The ``shape method'', as
used in the last H1 low-$Q^2$ study of HERA\,I data~\cite{h1lowq2},
exploits the shape of $\sigma_r$ in a given $Q^2$ bin.
The $Q^2$ dependence at high $y$ is driven by the kinematic factor
$y^2 / Y_+$ (eq.\,\ref{eq:sigma_red}), and to a lesser extent by $F_L(x,Q^2)$.
On the other hand, the gluon dominance at low $x$ suggests
that $F_L$ may exhibit an $x$ dependence similar to $F_2$.
Therefore it is assumed that $F_L$ is proportional to $F_2$ and
the coefficient of proportionality depends only on $Q^2$.
In the extraction procedure one uses the ratio $R$ of the cross sections
of the transversely and longitudinally polarised photons
\begin{equation}
R = \frac{\sigma_T}{\sigma_L} = \frac{F_L}{F_2 - F_L}
\end{equation}
which is thus assumed to depend only on $Q^2$.
The reduced cross section is fitted by
\begin{equation}
\sigma_{r} = F_2 \left[ 1 - \frac{y^2}{Y_+} \frac{R(Q^2)}{1 + R(Q^2)} \right]~,
\end{equation}
where some phenomenological model for $F_2$ is chosen.

\begin{figure}[!tb]
\centerline{
\includegraphics[width=.6\textwidth]{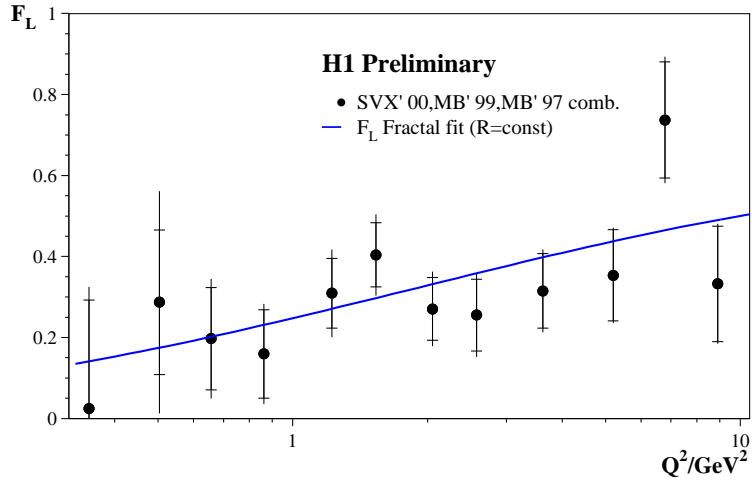}}
\caption{$Q^2$ dependence of $F_L(x,Q^2)$ at fixed $y =$\,0.75,
extracted from the preliminary H1 low-$Q^2$ data.
The solid line shows the prediction of the fractal fit with
a constant $R$.}
\label{f:h1_fl_shape_prel}
\end{figure}
An example of such an extraction using a fractal fit for $F_2$~\cite{fractal}
is shown in Fig.\,\ref{f:h1_fl_shape_prel}, where
preliminary H1 results~\cite{h1lowq2} for $F_L$ at $y =$\,0.75
in the range of 0.35\,$\leq Q^2 \leq$\,8.5\,GeV$^2$ are presented.
The data favour a positive, not small $F_L$ at low $Q^2$.
A drawback of this method is that it reveals a considerable dependence
of $R$ on the choice of the $F_2$ model.

\begin{figure}[!tb]
\centerline{\includegraphics[width=.6\textwidth]{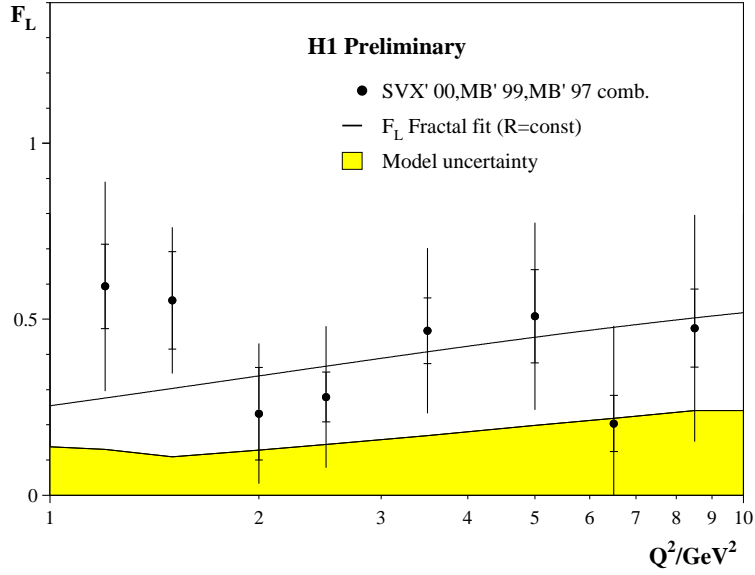}}
\caption{Structure function $F_L$ extracted by H1 using the derivative method.
The solid line shows the prediction of the fractal fit with
a constant $R$.
The inner error bars represent statistical uncertainties,
the outer error bars represent statistical and systematic
uncertainties added in quadrature.
The solid (yellow) band indicates the model uncertainty.
\label{f:h1flder}}
\end{figure}
In the derivative method~\cite{Adloff:2000qk,h1lowq2},
$F_L$ is extracted from the partial derivative of the reduced cross section
on $y$ at fixed $Q^2$
\begin{equation}
\frac{\partial \sigma_r}{\partial \ln y} \bigg|_{Q^2}
 = -x\frac{\partial F_2}{\partial x}
   - \frac{2 y^2 (2 - y)}{Y_+^2} F_L
   -x\frac{y^2}{Y_+} \frac{\partial F_L}{\partial x}
\end{equation}
which is dominated by the $F_L$-dependent term at high $y$.
The term proportional to $\partial F_L / \partial x$ is
negligible for moderately varying parametrisations of $F_L$.
For low $Q^2$ values the rise of $F_2$ is weak. The change of
the term $x \partial F_2 / \partial x$ for the two assumptions:
no rise at low $x$, i.e. $\partial F_2 / \partial x = 0$,
and $F_2 \propto x^{-\lambda}$ is numerically significantly
smaller than the experimental precision for
$\partial \sigma_r / \partial \ln y$.
Therefore the derivative methods provides a means for determining
$F_L$ at low $Q^2$ with minimal phenomenological assumption.
On the other hand,
the errors obtained with the derivative method turn out to be
significantly larger than those from the shape method.

The preliminary results of $F_L$ extraction from H1 HERA\,I
data\,\cite{h1lowq2} are presented in Fig.\,\ref{f:h1flder}.
The residual dependence of the measurement on the assumption
made for $F_2$ is estimated by a comparison with results
obtained assuming an $F_2$ which is flat in $y$.
The lower bound on $F_L$ obtained this way is depicted
as a solid band in the figure.

\begin{figure}[ht]
\centerline{
\includegraphics[width=0.5\columnwidth]{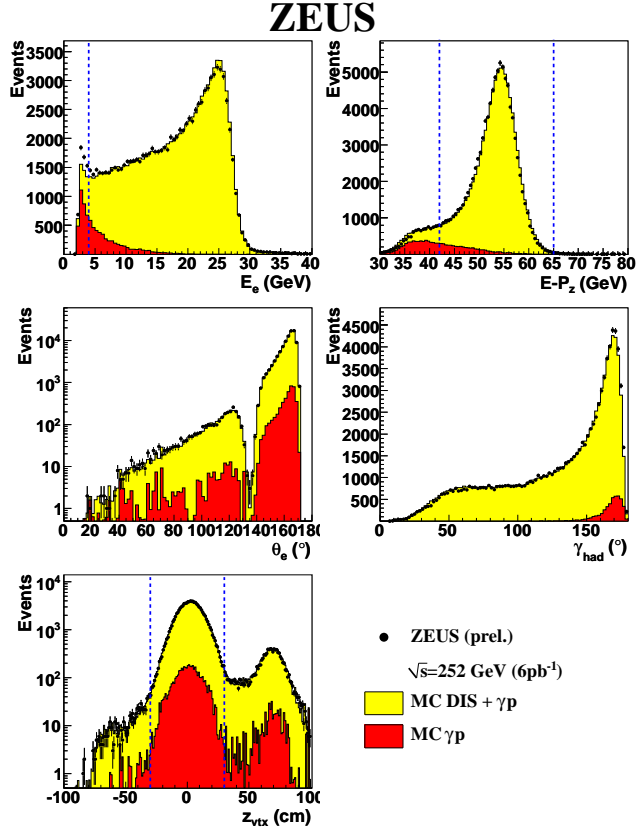}}
\caption{Comparison of 575\,GeV data with the sum of DIS and
background simulations for the energy of the scattered electron,
total $E-p_z$, theta of the scattered electron, angle of the
hadronic final state and $z$ coordinate of the vertex.
The dotted lines indicate the cuts applied.
  \label{f:zeus-control}}
\end{figure}
\subsubsection{Details of Direct $F_L$ Measurements}\label{s:details}
The H1 and ZEUS analysis procedures involve a measurement of the inclusive
cross section at $y >$\,0.1.
In this range, the kinematic variables $x$, $y$ and $Q^2$
are most accurately reconstructed using the polar angle, $\theta_e$,
and the energy, $E'_e$, of the scattered electron according to
\begin{equation}
y = 1 - \frac{E'_e}{E_e} {\sin}^2 \frac{\theta_e}{2} ~, ~~~
{Q}^2 = \frac{{E'_e}^2 {\sin}^2 \theta_e}{1-y} ~, ~~~
x = \frac{Q^2}{ys} ~.
\end{equation}
Reaching the high $y$ values necessary for the $F_L$ determination
requires a measurement of the scattered electron with energy down to a few GeV.
The electron candidate is selected as an isolated electromagnetic energy
deposition (cluster) in a calorimeter.
The crucial analysis issue at high-y region is the identification
of the scattered electron, and the estimation
of the hadronic background which occurs when a particle from the hadronic final
state mimics the electron signal.
Most of background events are photoproduction ($\gamma p$) events
with $Q^2 \approx$\,0
in which the final state electron is scattered at low angles
(high $\theta$)\footnote{The
$z$ axis of the right-handed coordinate systems used by H1 and ZEUS
is defined by the direction of the incident proton beam with the origin at
the nominal {\itshape ep} interaction vertex.
Consequently, small scattering angles of the final state particles
correspond to large polar angles in the coordinate system.}
and thus escapes through the beam pipe.

The $\gamma p$ background suppression is performed in several steps.
Firstly, calorimeter shower estimators are utilised which exploit
the different profiles of electromagnetic and hadronic showers.
Secondly, background coming from neutral particles, such as $\pi_0$,
can be rejected by requiring a track associated to the electron candidate.
Furthermore, $\gamma p$ events are suppressed by utilising
the energy-momentum conservation. For that, the variable
$E - p_z = \Sigma_i (E_i - p_{z,i})$  
is exploited, where the sum runs over energies $E_i$ and longitudinal
momentum components $p_{z,i}$ of all particles in the final state.
The requirement $E-p_z >$\,35\,(42)\,GeV in the H1 (ZEUS) analysis
removes events where the escaping electron carries a significant momentum.
It also suppresses events with hard initial state photon radiation.

However, at low $E'_e$ the remaining background contribution after
such a selection
is of a size comparable to or even exceeding the genuine DIS signal.
The further analysis steps differ for the H1 and ZEUS analyses
as discussed in the following.


\paragraph{ZEUS Analysis Procedure}

The electron candidates are selected as compact electromagnetic energy
depositions
in the Uranium Calorimeter (UCal).
The position of the candidate is reconstructed using either the Small Angle
Rear Tracking Detector (SRTD), which is a high-granularity lead-scintillator
calorimeter, or with the Hadron-Electron Separator (HES), which is a silicon
detector located in the electromagnetic section of the UCal.
The candidates are selected such that $E'_e >$\,6\,GeV\footnote{Cut of $E'_e >$\,4\,GeV is used for the event selection, although the binning for $F_L$ measurement is chosen such that $E'_e >$\,6\,GeV.}.

The candidates are validated using information from the tracking devices.
The acceptance region for ZEUS tracking is limited to polar angles
$\theta_e \lesssim$\,154$^\circ$.
The tracking detectors do provide some coverage beyond
 $\theta_e =$\,154$^\circ$,
up to $\theta_e\approx$\,168$^\circ$, however the number of tracking
layers is too sparse for full track reconstruction.
The hit information from the tracking detectors can still be used.
To do this, a ``road'' is created between the measured interaction vertex and
the position of the electron candidate in the calorimeter.  Hits in the
tracking layers along the road are then counted and compared to the
maximum possible number of hits. If too few hits are found,
the candidate is assumed to be a neutral particle and it is rejected.
To ensure the reliability of this method, the scattered electron
is required to exit the central drift chamber at a radius
$R>$\,20\,cm.  Given that $E'_e>$\,6\,GeV, this effectively limits the
maximal $y$ to $y<$\,0.8 and the minimum $Q^2$ achievable at low $y$. In
the HES analysis, events are measured down to $y=$\,0.2 roughly translating
to the $Q^2$ region, $Q^2>$\,24\,GeV$^2$.
No background treatment based on the charge of the candidate
is performed.

\begin{figure}[t]
\centerline{\includegraphics[width=0.35\columnwidth]{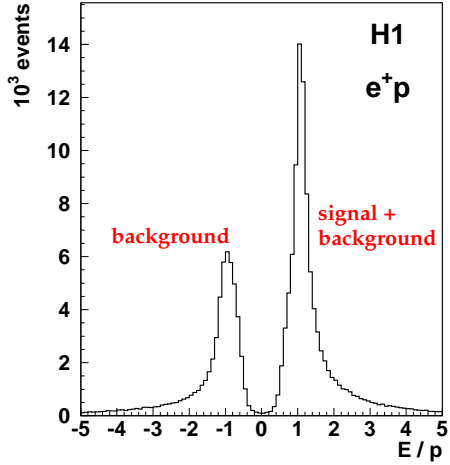}}
\caption{Distribution of energy over momentum for tracks linked to clusters
in the SpaCal with energy from $3.4$ to $10$\,GeV
that pass all the medium $Q^2$ analysis cuts.
Tracks with a negative charge are assigned a negative $E/p$.
  \label{f:eoverp}}
\end{figure}
The remaining $\gamma p$ background is estimated using Monte Carlo (MC)
simulations.
In order to minimise the model uncertainty of the $\gamma p$ simulation,
a pure photoproduction sample is selected using an electron tagger
placed close to the beam pipe about 6 meters away from the interaction point
in the rear direction.
It tags, with almost perfect efficiency and purity, the scattered
electrons in such events which are not identified in the main detector
and escape down the beam pipe. Photoproduction MC is verified against
and normalised to this sample. The normalisation factor is
found to be 1\,$\pm$\,0.1 for all data sets.


Figure~\ref{f:zeus-control} shows, as an example, comparisons
of the 575\,GeV data with simulated distributions,
for the energy of the scattered electron, total $E-p_z$, polar angle
of the scattered electron, angle of the hadronic final state and
the $z$ coordinate of the interaction vertex.
A good description of the data by the simulation is observed.
A similar level of agreement was found for both, HER and LER data sets.



A full set of systematic uncertainties is evaluated for the cross section
measurements.
The largest single contribution comes from the electron energy scale
uncertainty, which is known to within $\pm1\%$ for $E'_e >$\,10\,GeV,
increasing to $\pm$3\% at $E'_e =$\,5\,GeV. Other significant
contributions are due to the $\pm$\,10\% uncertainty in
verifying the Pythia prediction of the $\gamma p$ cross section
using the electron tagger.
The systematic
uncertainty due to the luminosity measurement was reduced by scaling the three cross sections relative to each
other. The spread of relative normalisation factor was found to be within the expected level of uncorrelated systematic uncertainty.

\begin{figure}[t]
\centerline{\includegraphics[width=.7\textwidth]{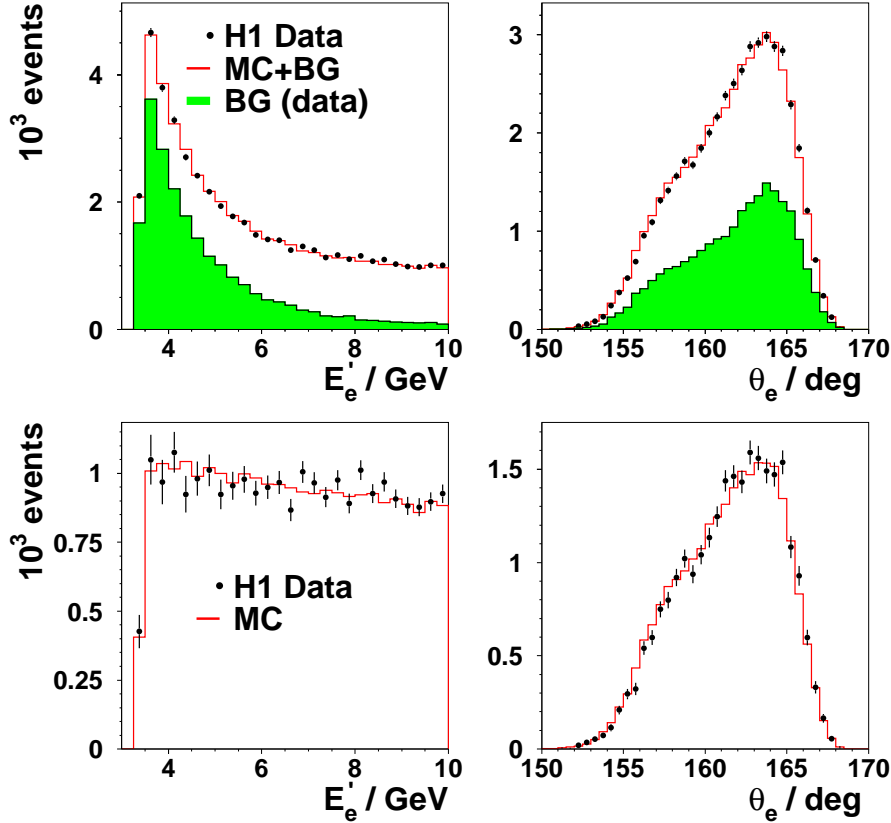}}
\caption{Top: comparison of the correct sign data
(points) with the sum (open histogram) of the DIS MC
simulation and background, determined from
the wrong sign data (shadowed histogram),
for the energy $E'_e$ (left) and the
polar angle $\theta_e$ (right) of the scattered electron,
for the 460\,GeV data with $E'_e <$\,10\,GeV. Bottom: as top but
after background subtraction.}
\label{f:h1_highy460}
\end{figure}
\paragraph{H1 Analysis Procedure}

The H1 measurements of $F_L$ are performed in separate analyses
involving different detector components and thus covering different $Q^2$
ranges.
In the high-$Q^2$ analysis the electron candidate is selected as
an isolated electromagnetic energy deposition in the
Liquid Argon (LAr) calorimeter which covers the polar angle range
4$^{\circ} < \theta <$\,153$^{\circ}$.
The selected cluster is further validated by a matching track reconstructed
in the central tracking device (CT) with an angular acceptance of
$15^\circ < \theta <$\,165$^\circ$. In the medium $Q^2$ analysis
the electron candidate is selected in the backward calorimeter SpaCal
covering the angular range 153$^{\circ} < \theta <$\,177.5$^{\circ}$ and
is also validated by a CT track.
Lower $Q^2$ values  
are expected to be accessed in the third analysis,
in which the SpaCal cluster is validated
by a track in the Backward Silicon Tracker reaching the highest $\theta$.
The first measurement of $F_L$ at medium $Q^2$ is
already published~\cite{Aaron:2008tx}, and preliminary results of the
combined medium-high-$Q^2$ analysis are available.

The remaining $\gamma p$ background is subtracted on statistical basis.
The method of background subtraction
relies on the determination of the  electric charge of the electron
candidate from the curvature of the associated track.

Figure\,\ref{f:eoverp} shows the $E/p$ distribution of the scattered electron
candidates from $e^+p$ interactions
with the energy $E$ measured in the SpaCal
and the  momentum $p$ of the linked track determined by the CT.
The good momentum resolution leads to a clear distinction between the
negative and positive charge distributions.
The smaller peak corresponds to tracks with negative charge
and thus represents almost pure background. These tracks are termed
wrong sign tracks and events with such candidates are rejected.
The higher peak, due to right sign tracks,
contains the genuine DIS signal superimposed on the
remaining positive  background.
The size of the latter to first
approximation equals the wrong sign background.  The principal
method of background subtraction, and thus of measuring the
DIS cross section up to $y \simeq$\,0.9, consists of the
subtraction of the wrong sign from the right sign event distribution
in each $x,Q^2$ interval.


The background subtraction based on the charge measurement
requires a correction for a small but non-negligible charge asymmetry
in the negative and positive background samples,
as has been observed previously by H1~\cite{Adloff:2000qk}.
The main cause for this asymmetry lies in the enhanced
energy deposited by anti-protons compared to protons at
low energies. The most precise measurement of the background charge
asymmetry  has been obtained from comparisons of samples
of negative tracks in $e^+p$ scattering with samples of positive
tracks in $e^-p$ scattering.  An asymmetry ratio of  negative
to positive tracks of $1.06$ is measured using the high statistics
$e^{\pm}p$ data collected by H1 in 2003-2006.
This result is verified using photoproduction events
 with a scattered electron tagged in a subdetector of the luminosity system.

\begin{figure}[tb]
\centerline{\includegraphics[width=.7\textwidth]{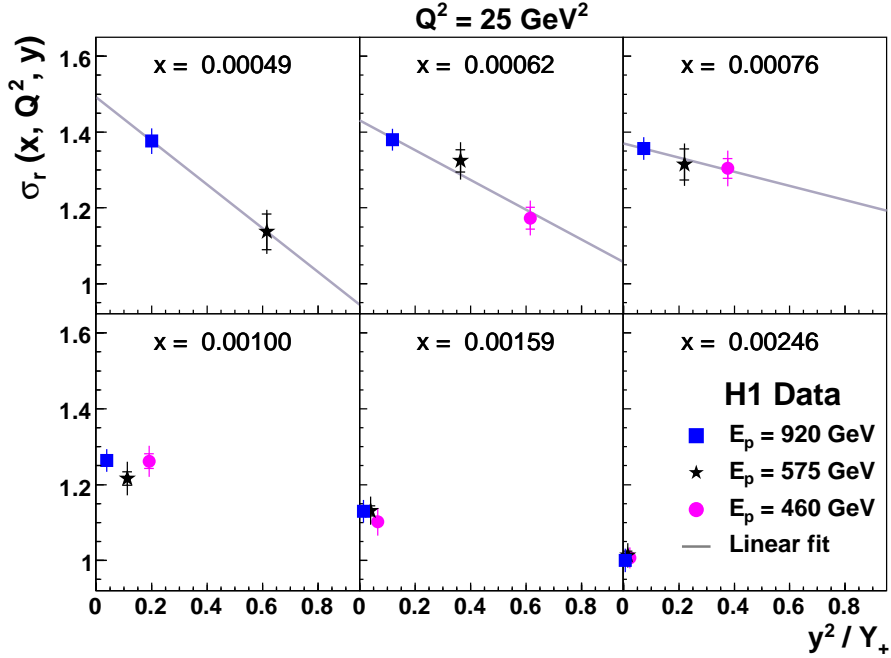}}%
\caption{The reduced inclusive DIS cross section plotted as a function of
$y^2/Y_+$ for six values of $x$  at $Q^2=$\,25\,GeV$^2$,
measured by H1 for proton beam energies of 920, 575 and 460\,GeV.
The inner error bars denote the statistical error, the full
error bars include the systematic errors. The luminosity uncertainty
is not included in the error bars.
For the first three bins in $x$, corresponding to larger~$y$,
a straight line fit is shown, the slope of which determines $F_L(x,Q^2)$.}
   \label{f:figros}
\end{figure}

Figure~\ref{f:h1_highy460} shows, as an example, comparisons
of the 460\,GeV high $y$ data with simulated distributions,
for the energy and the polar angle of the scattered electron
prior to  and after subtraction of the background,
which is determined using wrong sign data events.

The measurement of $F_L$ as described below relies on an accurate
determination of the variation of the cross section for a given $x$ and $Q^2$
at different beam energies. In order to reduce the uncertainty related
to the luminosity measurement, which presently is known to 5\% for each
proton beam energy of the 2007 data,
the three data samples are
normalised relatively to each other. The renormalisation
factors are determined at low $y$, where the cross section is determined
by $F_2$ only, apart from a small correction due to $F_L$.
The relative normalisation is known to within 1.6\%.

All correlated and uncorrelated systematic errors combined with
the statistical error lead to an uncertainty
on the measured cross sections at high $y$ of 3 to 5\%,
excluding the common luminosity error.



\subsubsection{Measurements of $F_L(x,Q^2)$ by H1 and ZEUS}\label{s:results}

\begin{figure}[tb]
\centerline{\includegraphics[width=0.7\textwidth]{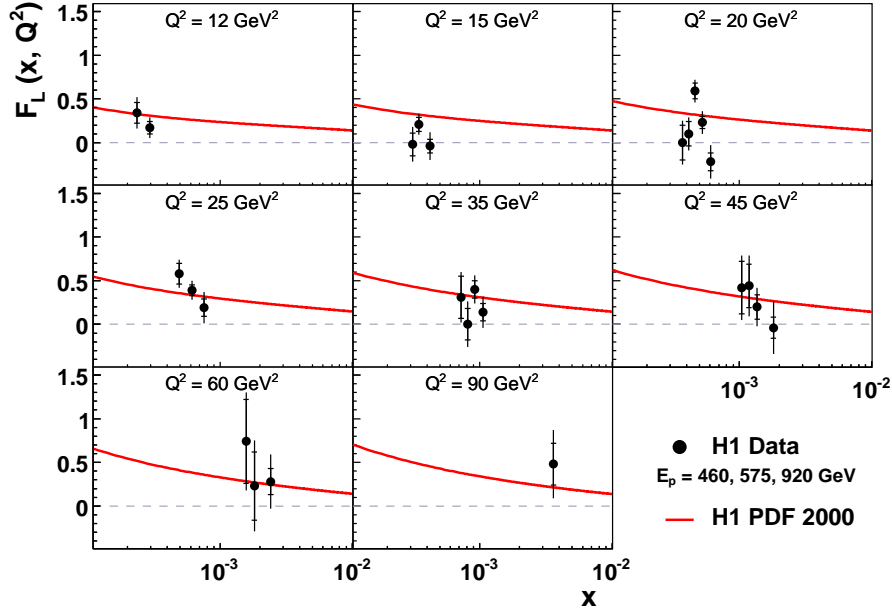}}%
\caption{The longitudinal proton structure function $F_L(x,Q^2)$
measured by the H1 collaboration.
The inner error bars denote the statistical error, the full
error bars include the systematic errors.
The curves represent the H1 PDF 2000 fit.}
   \label{f:flresults_h1}
\end{figure}

\begin{figure}[tb!]
\centerline{\includegraphics[width=0.6\textwidth]{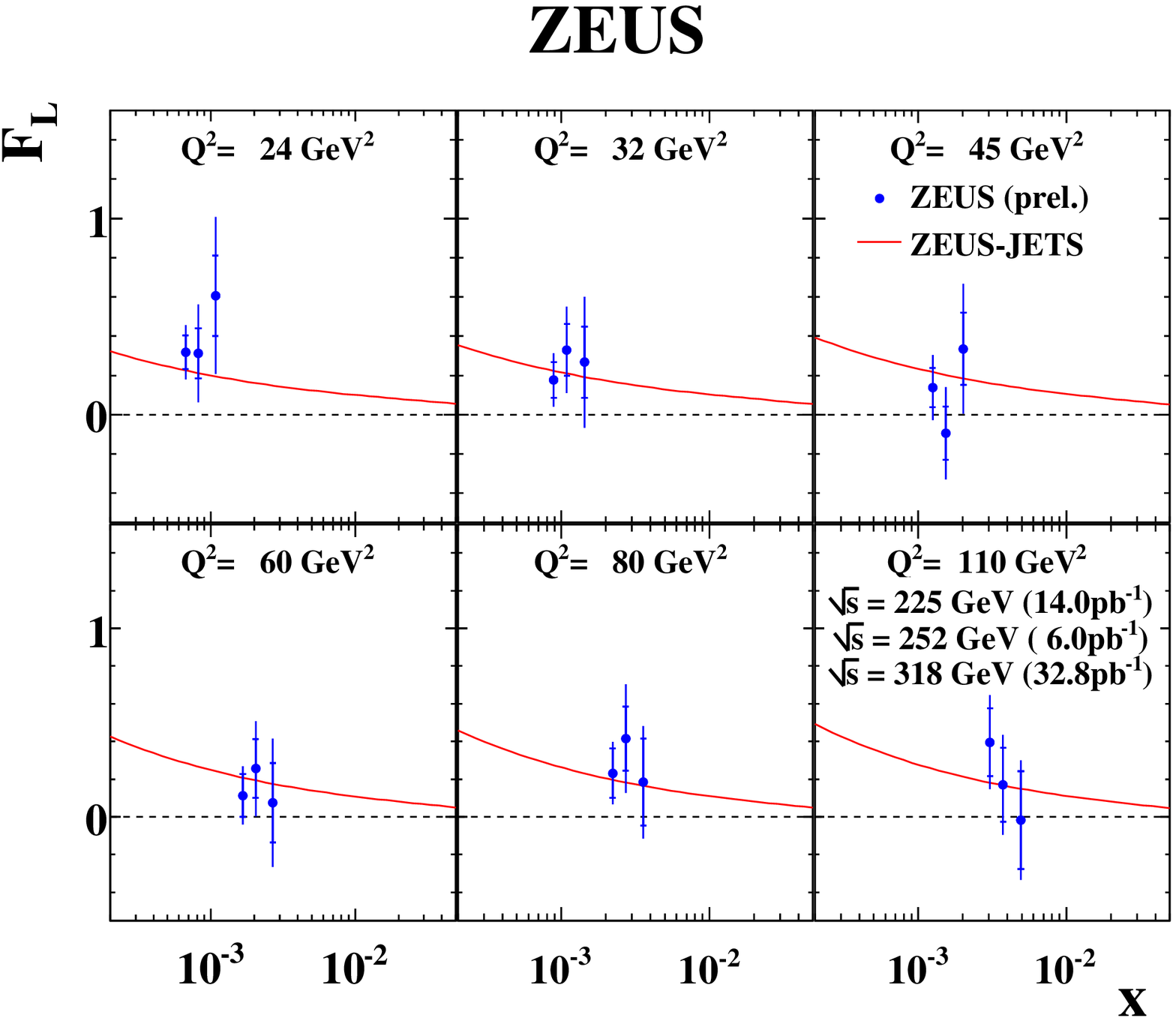}}%
\caption{The longitudinal proton structure function $F_L(x,Q^2)$
measured by the ZEUS collaboration.
The inner error bars denote the statistical error, the full
error bars include the systematic errors. The curves represent
the ZEUS-JETS PDF fit.}
   \label{f:flresults_zeus}
\end{figure}

\begin{figure}[tb!]
\mbox{%
\includegraphics[width=0.46\textwidth]{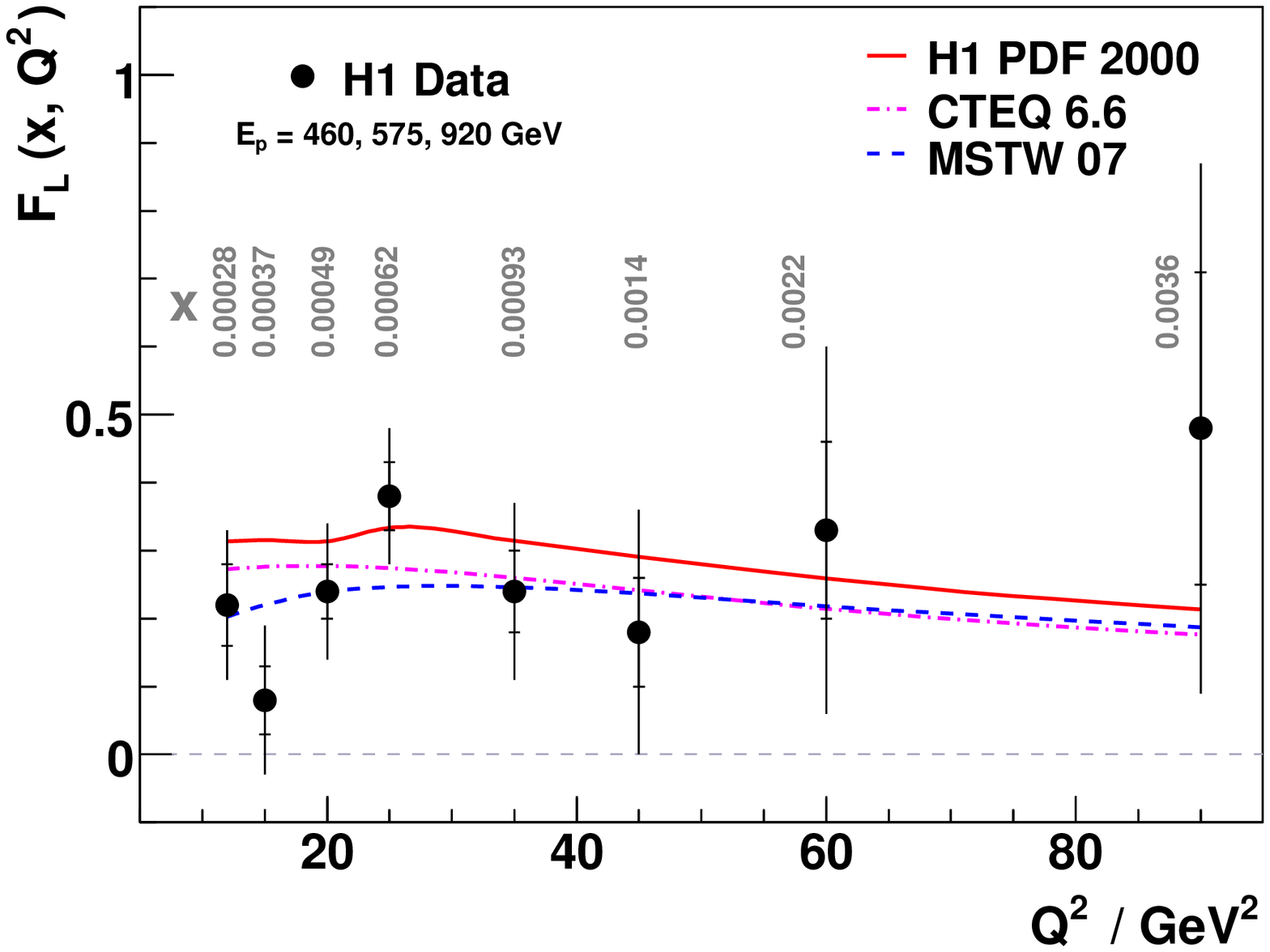}\hspace*{0.03\textwidth}%
\includegraphics[width=0.51\textwidth]{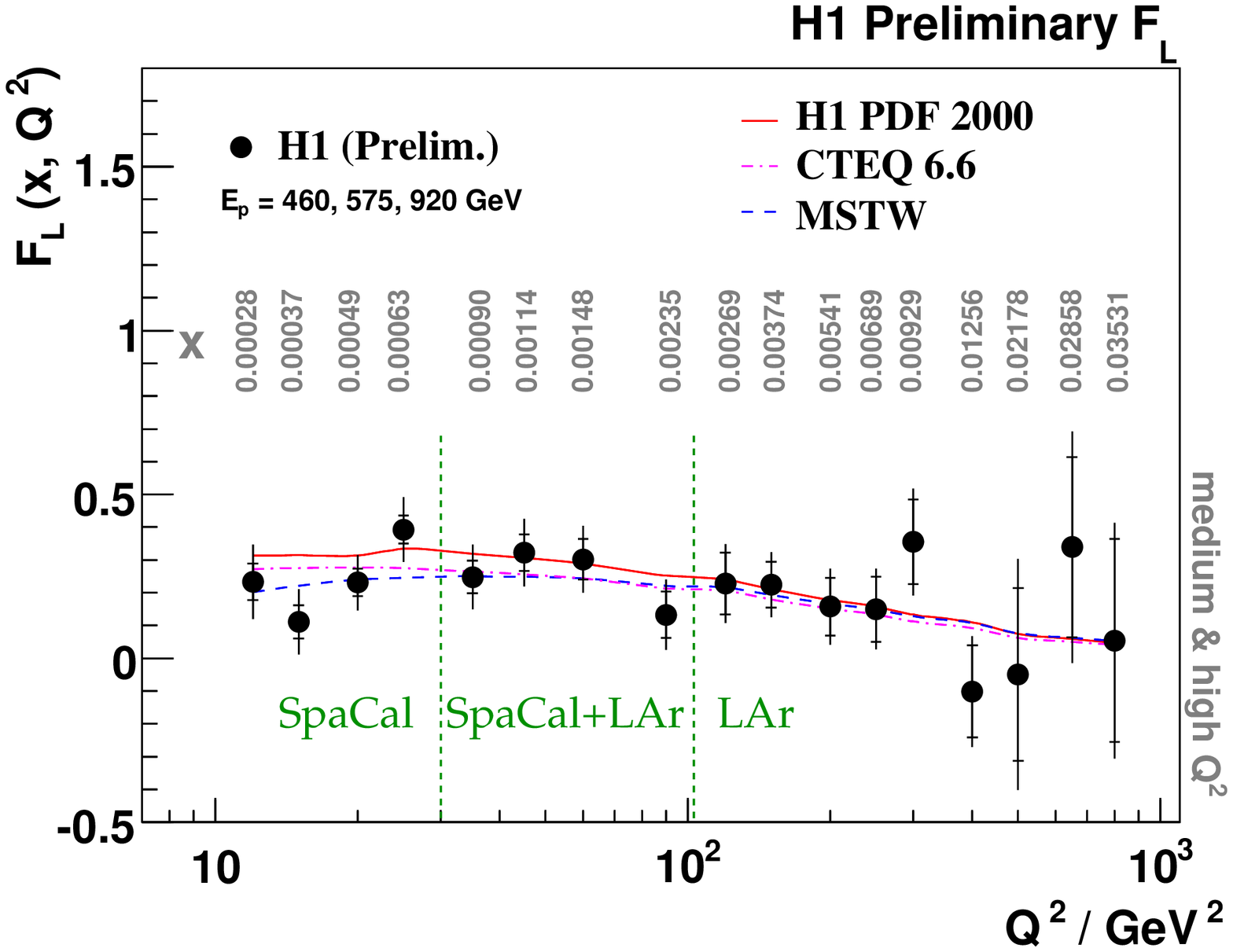}%
}
\caption{The proton structure function $F_L$
shown as a function of $Q^2$ at the given values of $x$:
a) first direct measurement at HERA by H1;
b) preliminary H1 results combining SpaCal and LAr analyses.
The inner error bars denote the statistical error, the full
error bars include the systematic errors.
The luminosity uncertainty is not included in the error bars.
The solid curve describes the expectation on $F_L$ 
from the H1 PDF 2000 fit using NLO QCD.
The dashed (dashed-dotted) curve depicts the expectation of the MSTW (CTEQ)
group using NNLO (NLO) QCD.
The theory curves connect predictions at the
given $(x,Q^2)$ values by linear extrapolation.}
\label{f:h1_flav}
\end{figure}

The longitudinal structure function is extracted from the measurements
of the reduced cross section as the slope of
$\sigma_r$ versus $y^2/Y_+$, as can be seen in eq.\,\ref{eq:sigma_red}.
This procedure is illustrated in Fig.\,\ref{f:figros}.
The central $F_L$ values are determined in straight-line fits to
$\sigma_r(x,Q^2,y)$ as a function of $y^2/Y_+$ using
the statistical and uncorrelated systematic errors.

The first published H1 measurement of $F_L(x,Q^2)$ is shown in
Fig.\,\ref{f:flresults_h1}, the preliminary ZEUS measurement
is presented in Fig.\,\ref{f:flresults_zeus}.
The H1 measured values of $F_L$
are compared with the H1 PDF 2000 fit\,\cite{Adloff:2003uh}, while
the ZEUS $F_L$ values are compared to the ZEUS-JETS PDF
fit\,\cite{Chekanov:2005nn}.
Both measurements are consistent and show a non-zero $F_L$.


The H1 results were further averaged over $x$ at fixed $Q^2$,
as shown in the left panel of Fig.\,\ref{f:h1_flav}.
The averaging is performed taking the $x$ dependent correlations
between the systematic errors into account.
The averaged values of $F_L$ are compared with H1 PDF 2000 fit and with
the expectations from global parton distribution fits
at higher order perturbation theory performed by the MSTW\,\cite{MSTW}
and the CTEQ\,\cite{cteq,Nadolsky:2008zw} groups.
Within the experimental
uncertainties the data are consistent with these predictions.
The measurement is also consistent with previous indirect
determinations of $F_L$ by H1.

In the combined medium--high $Q^2$ analysis by H1 the $Q^2$ range is extended
up to $Q^2 =$\,800\,GeV$^2$. The preliminary results are shown in the right
panel of Fig.\,\ref{f:h1_flav}.
In some $Q^2$ bins there is an overlap between
the SpaCal and LAr measurements which improves the precision of the $F_L$
extraction as compared to the pure SpaCal analysis.

\subsubsection{Summary}

Direct measurements of the proton structure function $F_L$ have been
performed in deep inelastic $ep$ scattering at low $x$ at HERA.
The $F_L$ values are extracted by the H1 and ZEUS collaborations
from the cross sections measured at fixed $x$ and $Q^2$ but different $y$
values. This is achieved by using data sets 
collected with three different proton beam energies.
The H1 and ZEUS results are consistent with each other and
exhibit a non-zero $F_L$.
The measurements are also consistent with the previous indirect
determinations of $F_L$ by H1.
The results confirm DGLAP NLO and NNLO QCD predictions
for $F_L(x,Q^2)$, derived from previous HERA data,
which are dominated by a large gluon density at low $x$.


%
%

\begin{center}
\end{center}

\normalsize
\begin{center}
\end{center}

\par
\par
\small{
}
\normalsize
  
\newpage

\section{ PROTON--PROTON  LUMINOSITY,  STANDARD CANDLES AND PDFS AT THE 
LHC\protect\footnote{Contributing authors:  J. Anderson, M. Boonekamp,
  H. Burkhardt, M. Dittmar, V. Halyo, T. Petersen}}
\label{sec:lumi}
\subsection{Introduction}
\label{sec:lumiintro}
The Large Hadron Collider (LHC) is expected to
start colliding proton beams in 2009, and is expected to reach design parameters in energy and luminosity sometime later and deliver a few
$fb^{-1}$ per year of data at the 14 TeV collision energy.
 
During the past 15 years many theoretical calculations and experimental simulations have demonstrated  
a huge potential to perform many accurate tests of the Standard Model (SM) with
LHC data, which could yield insight into new physics mechanisms.


To make these tests, the experiments identify a particular signature X and observe, using a variety of selection criteria, a certain number of events in a given 
data taking period. After correcting this event rate for backgrounds and the selection efficiency,     
the number is converted into a cross section. The cross section, $\sigma_{pp \rightarrow X}$ can be compared with 
theoretical predictions\footnote{Alternatively, one can also apply a Monte Carlo simulation to the theoretical prediction and 
compare the number of background corrected events directly.}
according to the formula:  $N_{corrected} = \sigma_{pp \rightarrow X} \times L_{pp}$ where $L_{pp}$ is the recorded proton proton luminosity. 

Besides the statistical errors of a measurement, the systematic error is related to the uncertainties from the $L_{pp}$ determination, 
the background and efficiency corrections within the detector acceptance and from extrapolations into the      
uncovered very forward rapidity regions. The interpretation of an observed cross section within the SM requires further 
the knowledge of the theoretical cross section. Thus the uncertainties of the proton 
parton distribution function (PDF) have to be considered also.
  
In this Section we describe the status and perspectives of the ATLAS, CMS and LHCb, the three LHC pp collision detectors\cite{atlascmslhcb}, 
to determine the proton proton luminosity normalization. 
The investigated methods are known and studied since many years and can be separated into the absolute (1) direct and (2) indirect proton proton 
luminosity determination. A third approach (3) tries to measure and calculate final states only relative to well understood reactions which depend 
on the parton-parton luminosity and are as such largely independent of the knowledge of the pp luminosity.

\begin{itemize}
\item Absolute, direct or indirect, proton proton luminosity normalization:
If the absolute approach is used, the interpretations of a measured reaction cross section depends still on the knowledge of   
parton distribution function (PDF), which must be obtained from other experiments. Examples are:
\begin{itemize} 
\item The proton proton luminosity normalization is based on the measurements of the beam currents and shapes. 
While the beam currents can be accurately determined using beam transformers, the beam profiles are more difficult to 
determine directly and usually constitute the dominant source of uncertainty on a luminosity measurement using this technique.
The use of the machine luminosity determination using beam parameter measurements \cite{Herr:2003em} and \cite{Burkhardt:2007zzc} will be described in Section~\ref{sec:lumiparms}.  
Alternatively one can try to measure the beam profiles also within the experiments using the precision vertex detectors. 
A short description of this idea,  currently pursued within the LHCb collaboration, is also given in Section~\ref{sec:lumiparms}.
\item The simultaneous measurements of a pair of cross sections that are connected with each
other quadratically via the optical theorem. A well known example of this is the measurement of the total inelastic cross section 
and the elastic cross section at very high pseudorapidities $ |\eta| \approx 9$ and will be described in Section~\ref{sec:lumiabsol}.

So called instantaneous or real time luminosity measurements are based on ``stable'' high rate measurements of particular final state reactions. 
Once the ratio of such reactions to the pp luminosity determination has been measured, those reactions can be subsequently used as   
independent luminosity monitors. Some possibilities are discussed in Section~\ref{sec:lumirealtime}.
\item The indirect absolute proton proton luminosity normalization is based on the theoretically 
well understood ``two photon'' reaction $pp \rightarrow pp \mu
\mu$~\cite{budnev,MartinPaper} (Section~\ref{sec:lumipplumi}).
This reaction could perhaps be considered as the equivalent of the luminosity counting in $e^{+}e^{-}$ experiments using forward Bhabha scattering.   
\end{itemize}  
\item Indirect pp luminosity measurements use final states, so called
  ``standard candles'', with well known theoretical cross sections
  (Section~\ref{sec:lumippdirindir}).\\ 
Obviously, the resulting proton proton luminosity can only be as good as the theoretical and experimental knowledge of the ``standard candle'' reaction. 
The theoretically and experimentally best understood LHC reactions are the inclusive production of W and Z bosons with subsequent leptonic decays.
Their large cross section combined with experimentally well defined final states, e.g. almost background free Z and W event samples can be selected  
over a relative large rapidity range, makes them the preferred LHC ``standard candle'' reaction.
Other interesting candidates are the high $p_{t}$ jet - boson (= $\gamma$, W or Z) final states. 
The indirect luminosity method requires also some knowledge of the PDFs, and of course, if one follows this approach, the cross section of the 
``standard candle'' reaction becomes an input and can not be measured anymore. Thus, only well understood reactions 
should be considered as candidate reactions. 
\item pp luminosity independent relative rate measurements using  ``standard candle'' reactions.\\
In addition to the above indirect pp luminosity determinations, ``standard candle'' reactions allow 
to perform luminosity independent relative event rate calculations and measurements. This approach has already been used successfully 
in the past and more details were discussed during the past HERA-LHC workshop meetings \cite{Dittmar:2005ed}.
For some reactions, this approach appears to be much easier and more accurate than standard cross section measurements and their interpretations. 
Perhaps the best known example at hadron colliders is the measurement and its interpretation of the production ratio for Z and W events, 
where Tevatron experiments have reached accuracies of about 1-2\% \cite{WZratioa,WZratiob}. Another example is related to 
relative branching ratio and lifetime measurements as used for b-flavored hadrons. 
\end{itemize}

Furthermore the rapidity distributions of leptonic W and Z decays at the LHC are very sensitive to the 
PDF parameterization and, as was pointed out 10 years ago \cite{Dittmar:1997md}, 
one can use these reactions to determine the parton luminosity directly and very accurately over a large x (= parton momentum/proton momentum) range. 
In fact, W and Z production with low transverse momentum were found in this analysis to be very sensitive to $q \bar{q}$ luminosities, 
and the jet-boson final states, e.g. the jet-$\gamma$, Z, W final states at high transverse momentum are sensitive to the gluon luminosity. 

In the following we attempt to describe the preparations and the status of the different luminosity measurements and their 
expected accuracies within ATLAS, CMS and LHCb. Obviously, all these direct and indirect methods should and will be pursued. 
In Section~\ref{sec:lumicomparing} we compare the advantages and disadvantages of the different methods. 
Even though some methods look more interesting and rewarding than others, it should be clear from the beginning 
that as many independent pp luminosity determinations as possible need to be performed by the experiments. 

We also try to quantify the systematic accuracies which might be achieved over the next few years.  
As these errors depend somewhat on the overall achieved luminosity,   
we need in addition a hypothetical working scenario for the first 4 LHC years. 
We thus assume that during the first year, hopefully 2009, data at different center of mass
energies can be collected by ATLAS and CMS. During the following three physics years 
we expect that 10 TeV will be the highest collision energy in year I and  
that at most 100 pb$^{-1}$ can be collected. We assume further that during the following two years the design energy of 14 TeV can be achieved  
and that a luminosity of about 1 fb$^{-1}$ and 10 fb$^{-1}$ 
can be collected respectively per year. During the first few years similar numbers are expected for the LHCb experiment. 
However once the LHC reaches the first and second phase design luminosity of $10^{33}$/cm$^{2}$/sec and $10^{34}$/cm$^{2}$/sec
it is expected that the LHCb experiment will run at an average luminosity of $2 \times 10^{32}$/cm$^{2}$/sec (resulting in about 2 $fb^{-1}$/per year).    


\subsection{Luminosity relevant design of ATLAS/CMS and LHCb}
\label{sec:lumiatlascmslhcb}

In the following we give a short description of the expected performance 
with respect to lepton and jet identification capabilities. 
Especially the electron and muon measurement capabilities are important for the identification 
of events with leptonic decays of W and Z bosons.  

Both ATLAS and CMS are large so called omni purpose experiments with a large acceptance 
and precision measurement capabilities for high $p_{t}$ electrons, muons and photons. 
Currently, the simulations of both experiments show very similar performance for a large variety 
of LHC physics reactions with and without jets. 
For the purpose of this Section  we focus on the possibility to identify the production of inclusive W and Z decays 
with subsequent decays to electrons and muons.   
Both experiments expect excellent trigger accuracies for isolated leptons and it is expected that 
electrons and muons with momenta above 20-25 GeV can be triggered with high efficiency and up to $|\eta|$ of about 2.5. 
The special design of the ATLAS forward muon spectrometer should allow to detect muons with good accuracy even up to $|\eta|$ of 2.7.

The operation of ALFA, a very far forward detector placed about 240 m down the beam line, is envisaged by the ATLAS collaboration 
to provide an absolute luminosity measurement, either using special optics LHC running and the use of the optical theorem 
or using the total cross section measurement from the dedicated TOTEM experiment installed near CMS; results from this device 
can be expected from 2010 and on-wards. 
In addition to absolute luminosity measurements from ALFA the two detectors LUCID and the Zero-Degree-Calorimeter (CDC) \cite{Pinfold:2005sq} are
sensitive to the relative luminosity at time scales of single bunch crossings. 

A similar approach for absolute and relative luminosity measurements is foreseen by the CMS experiment. Here it is planned that  
dedicated forward detectors, the Hadron Forward Calorimeter (HF) and the ZDC device provide similar results as the ones in ATLAS. 

Another technique that is expected to be available early on is a luminosity-independent measurement of the $pp$ total cross section.   
This will be done using a forward detector built by the TOTEM experiment \cite{Anelli:2008zz}. 
 
The LHCb experiment \cite{LHCb} has been designed to search for New Physics at the LHC through precision measurements 
of CP violating observables and the study of rare decays in the
b-quark sector. Since the\ $b\bar b$ pairs
resulting from the proton-proton collisions at the LHC will both be produced at small polar angles and in the 
same forward or backward cone, LHCb has been designed as a single-arm forward spectrometer covering the 
pseudo rapidity range $1.9<\eta<4.9$. The LHCb tracking system, 
which is composed of a silicon vertex detector, a warm dipole magnet and four planar tracking stations, 
will provide a momentum resolution of $\delta P/P=(0.3+0.0014 P/GeV)\%$ \cite{LHCbReOpt}. 
Muon identification is primarily achieved using a set of five planar multi-wire proportional chambers, one 
placed in front of the calorimeter system and four behind, and it is expected that for the momenta range 3-150GeV/c
an identification efficiency of $\sim$98\% and an associated pion dis-identification rate of $\sim$1\% 
will be achieved. The reconstruction of primary and secondary vertices, a task of 
crucial importance at b physics experiments, will be virtually impossible in the high particle multiplicity 
environment present with the nominal LHC running luminosity of $10^{34}cm^{-2}s^{-1}$ - LHCb has therefore 
been designed to run at the lower luminosity  of $2\times10^{32}cm^{-2}s^{-1}$. 

Recent LHCb simulations have shown that leptonic W and Z decays to muons can be identified 
with a small background in the forward and very forward rapidity region starting from $\eta$ of 1.9 and up to values larger than 4. 
As will be discussed later in more detail, the common muon acceptance region for the three LHC experiments between 1.9 and about 2.5 
will allow to cross check and normalize the W and Z measurements in this region. Consequently the unique large rapidity from 2.5 to 4.9 
can be used by LHCb to investigate the very low x range of the PDFs for the first time. 

The absolute luminosity at LHCb will be obtained either directly, by making measurements of the beam parameters, 
or indirectly via a measurement of the event rate of an accurately predicted physics process.

As will be explained in the following Sections, all experiments will try to perform as many as possible direct and 
indirect absolute and relative luminosity measurements and will, if available, at least during the first years, 
also use luminosity numbers from the machine group.

\subsubsection{Lepton triggering and W/Z identification.}
\label{sec:lumitriggering}

Generally, the lepton trigger selections depend on the instantaneous
luminosity and some pre-scaling might eventually needed. However, current simulations by all experiments show that 
the envisaged $|\eta|$ and $p_{t}$ thresholds will not limit the measurement accuracies of leptons originating from W and Z decays.

The lepton trigger selections that generally perceived to be used for
most W and Z related analysis are very similar in ATLAS and CMS as indicated in Table~\ref{tab:selections}. 

\begin{table}[htbp!]
\begin{center}
\begin{tabular}{lcccc}
  \hline
  \hline
                 &\multicolumn{2}{c}{Trigger selection $e$}
                                            &\multicolumn{2}{c}{Trigger selection $\mu$}\\
    Experiment   &$p_T$     &$|\eta|$       &$p_T$       &$|\eta|$\\
  \hline
    ATLAS        &25 \ggev   &2.5            &20  \ggev     &2.7       \\
  \hline
    CMS          &20 \ggev   &2.5            &20  \ggev     &2.1       \\
  \hline
    LHCb$^{*}$        & --       &   --          &2.5 \ggev     & 1.9-4.9 \\

  \hline
\end{tabular}
\end{center}
  \caption{For ATLAS and CMS the lepton trigger/selection $p_{t}$ thresholds are given for single isolated leptons.
 $^{*}$For the LHCb threshold is given for the muon pair mass instead of single muons and only positive values of $\eta$ are covered.}
  \label{tab:selections}
\end{table}
   
Trigger and reconstruction efficiencies for leptonic W and Z decays within the acceptance of the detectors 
have been estimated for ATLAS to be 97.7\% and 80.0\% for electrons and 84.3\% and 95.1\%
for muons, respectively. The reconstruction efficiency includes the trigger efficiencies and the off-line 
electron and muon selections used later to identify clean inclusive W and Z event samples \cite{Besson:2008zs}.

The current equivalent trigger and off-line efficiencies for CMS are about 85\% and 77\% for electrons 
and combined about 85\% for single muons\cite{cmspasemu}. 
Similar efficiency numbers for muons from W and Z decays are expected within the LHCb acceptance region \cite{lhcbmu}.
Current simulations show that these numbers can be determined with high accuracies, reaching perhaps 1\% or better,
at least for isolated leptons\footnote{As isolated high $p_{t}$ photons are triggered essentially like electrons  
similar accuracies for both particle types can be assumed.}  
which have a transverse momentum some GeV above the trigger thresholds.   
For lower momenta near the thresholds or for additional special trigger conditions 
somewhat larger systematic uncertainties can be expected. 


\subsection{Direct and indirect absolute pp luminosity measurements}
\label{sec:lumidirindir}

Three different absolute proton proton luminosity measurements are discussed in this Section.
(1) The machine luminosity determination using beam parameter measurements \cite{SchmicklerTalkAtlas2005},    
(2) the luminosity independent total pp cross section measurement combined with the measurement of the elastic pp scattering 
rate \cite{Anelli:2008zz} and (3) the measurement of the ``two photon'' reaction $pp \rightarrow pp \mu \mu$~\cite{budnev,MartinPaper}.
As will be discussed in more detail in Section~\ref{sec:lumicomparing}, only method (3) can be performed during the 
normal collision data taking. For method (1) some special methods, which take the actual detector performance during each run into account,  
need to be developed. Method 2 uses a two phase approach (a) a special machine optics run with low luminosity to determine the 
total cross section and (b) a normalization to some high rate final state reactions which can be counted during normal physics runs.

\subsubsection{Proton-proton luminosity from machine
  parameters\protect\footnote{Contributing author: H.~Burkhardt}}
\label{sec:lumiparms}

The luminosity for colliding beams can be directly obtained from geometry and numbers of particles flowing per time unit\,\cite{Herr:2003em}.
This can be used to determine the absolute LHC luminosity from machine parameters without prior knowledge of pp scattering cross sections.
The principle is briefly outlined here. More details can be found in\,\cite{Burkhardt:2007zzc}.

\begin{figure}[htpb]
\center{\includegraphics[width=8cm]{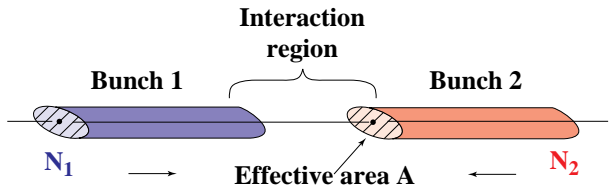}}
  \vspace{-2mm}\caption{Luminosity from particles flux and geometry.}
  \label{plot:lumi1}
\end{figure}

For two bunches of $N_1$ and $N_2$ particles colliding head-on in an interaction region as sketched in Fig.\ref{plot:lumi1} 
with the frequency $f$ the luminosity is given as
\begin{equation}
{\cal L} =\frac{N_1\,N_2\,f}{A_{\rm eff}}\;.
\label{eq:Lumi1}
\end{equation}
$A_{\rm eff}$ is the {\em effective transverse area} in which the collisions take place. For a uniform transverse particle 
distribution, $A_{\rm eff}$ would be directly equal to the transverse beam cross section. More generally, 
the effective area can be calculated from the overlap integral of the two transverse beam distributions $g_1(x,y)$, $g_2(x,y)$ according to
\begin{equation}
\frac{1}{A_{\rm eff}} = \int\;g_1(x,y)\,g_2(x,y)\;dx\,dy~.
\end{equation}
For equal Gaussian beams
\begin{equation}
g_1 = g_2 = \frac{1}{2\pi\sigma_x\sigma_y}~\exp~\Biggl[-\frac{x^2}{2\sigma_x^2}-\frac{y^2}{2\sigma_y^2}\Biggl]
\end{equation}
we obtain for head-on collisions
$ A_{\rm eff} = 4 \pi \, \sigma_x\sigma_y $
so that
\begin{equation}
{\mathcal{L}} = \frac{N_1\,N_2\,f}{4\pi\sigma_x\sigma_y}\;.
\end{equation}
The collision frequency $f$ is accurately known. The number of particles circulating in a storage ring is measured using beam current 
transformers to roughly 1\% precision\,\cite{SchmicklerTalkAtlas2005}. 

\begin{figure}[htpb] 
\includegraphics[width=5cm,trim=0 -40 0 0]{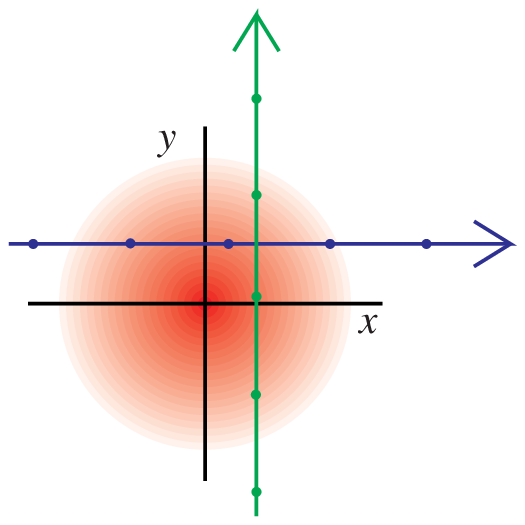}\hfill\includegraphics[width=8cm]{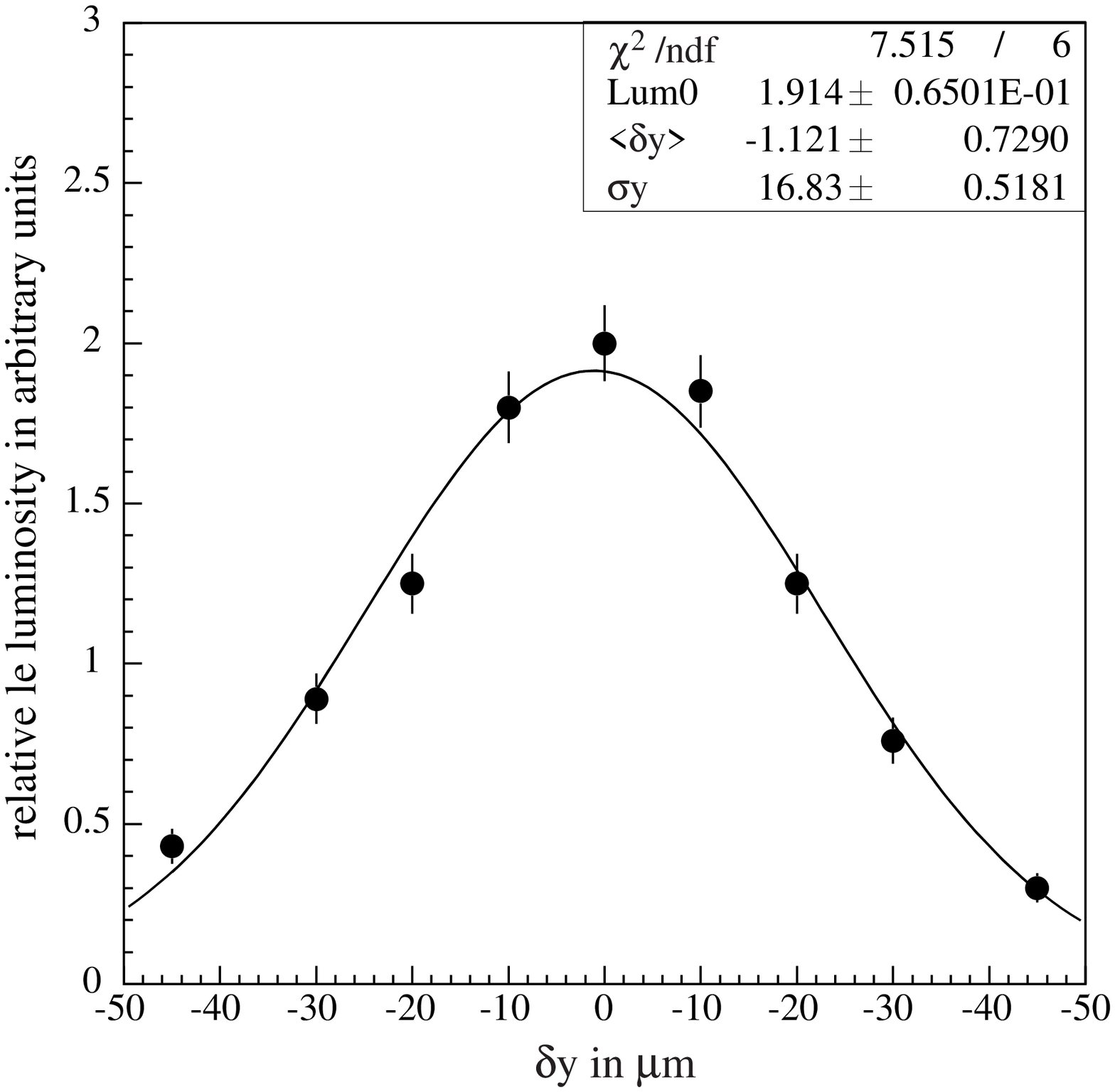}
  \vspace{-2mm}\caption{Schematic view of the steps involved in an orthogonal separation scan proposed for the LHC (left) and a possible 
result in one direction (based on early LEP data) shown on the right.}
  \label{plot:ScanStrategy}
\end{figure}

The main uncertainty in the absolute luminosity determination from machine parameters is expected to originate in the knowledge 
of the transverse beam dimensions.
Safe operation of the LHC requires a rather good knowledge of the optics and beam sizes and we expect that this should already allow a 
determination of the luminosity from machine parameters to about $20 - 30$ percent.
A much better accuracy can be obtained when the size of the overlap region at the interaction points is determined by measuring the relative 
luminosity as a function of lateral beam separation, as illustrated in Fig.\,\ref{plot:ScanStrategy}.
This technique was pioneered at the ISR\,\cite{VanDerMeerISR1968} and allowed to reduce the uncertainty to below 1\%, \cite{Potter:1992jy,Carboni:1984sg}.

For the more complicated LHC and early operation, a 10\% overall uncertainty in the absolute LHC machine luminosity calibration should be a realistic goal. 
The actual precision will depend on the running time and effort which is invested. A relatively small number of scans under favorable beam conditions will 
in principle be sufficient to obtain and verify the reproducibility in the absolute luminosity calibration.
While fast scans may always be useful to optimize collisions, we assume that any dedicated, detailed luminosity scans will become obsolete when the other, 
cross section based luminosity determinations described in these proceedings allow for smaller uncertainties.

Optimal running conditions are moderate bunch intensities, large bunch spacings, no crossing angle and $\beta^* = 2\,{\rm m}$ or larger.
These conditions are in fact what is proposed anyway for the initial LHC operation with 43 -- 156 bunches per beam.
Statistics are not expected to be a problem. For early operation at top energy (10 - 14 TeV) with 43 bunches and $4\times 10^{10}$ particles per bunch, 
before beams are squeezed. at a $\beta^* = 11\,{\rm m}$, we already expect luminosities of the order of $10^{30}\,{\rm cm}^{-2}{\rm s}^{-1}$ 
resulting in event rates of $10^4\,{\rm Hz}$, for a cross section of 0.01\,barn as typical for the low angle luminosity monitors.

From the LHC injectors, we expect bunch by bunch variations of about 10\% in intensity and 20\% in emittance. For the large spacing between 
bunches in the operation with up to 156 bunches, there is no need for crossing angles at the interaction points. Parasitic beam-beam effects will be 
negligible. All bunches in each beam will follow the same equilibrium orbit and collide at the same central position.

Calibration runs require good running conditions and in particular good beam lifetimes. Bunch by bunch differences are not expected to 
change significantly during a scan. Storing bunch intensities at the beginning and end of a scan and using one set of timed averaged bunch 
intensities for a scan should be sufficient. To avoid any bias, it will be important to use the correct pairing of bunch intensities and 
relative luminosities in the calculation of absolute bunch luminosities according to Eq.\,\ref{eq:Lumi1}, before any summing or averaging over 
different bunches.

We are currently preparing an on-line application for automatic luminosity scans\footnote{Done by Simon White, as part of his PhD thesis 
work on the LHC machine luminosity determination}. Scan parameters like range, step size and duration can be set before the start of the scan.
Once the parameters are defined, it is possible to launch automatic horizontal and vertical separation scans in the LHC interactions regions. 
For a detailed scan, we may choose a range from -4 to +4 $\sigma$ in nominal beam size in steps of 0.5 $\sigma$, resulting in 17 equidistant points. 
If we wait 1\,s between points to allow for the magnets to change and for 2\,s integration time, the scan time would still be below a minute per plane. 
Details are currently being worked out in close collaboration with the experiments. Exchanging all data bunch-by-bunch at a 1\,Hz rate between the 
machine control room (CCC) and the experiments would be rather demanding and risks to saturate current capacities.

For the initial running, it will be sufficient to exchange average values at about 1\,Hz rate.
It allows quality monitoring and the determination of the peak position. For the detailed off line analysis,
we only have to rely on local logging and timing information synchronized to at least 1\,s precision at the beginning of the scan. 
With fixed time interval defined and saved before the scan, this allows for off-line synchronization of the detailed data and 
a complete bunch by bunch analysis.

\subsubsection{Direct measurements of the absolute luminosity at LHCb} 
\label{sec:lumilhcb}

LHCb plans to measure the absolute luminosity 
using both the Van Der Meer scan, \cite{VanDerMeerISR1968}, and beam-gas techniques following a  
more recently proposed method \cite{MassiPaper}. Here one tries to 
determine the transverse beam profiles at colliding beam experiments utilizing the precision vertex
detectors found at modern HEP experiments to reconstruct beam gas interactions near
the beams crossing point. The vertex resolution in the transverse direction at LHCb can be parameterized by the relation 
\begin{equation}
\sigma_{x,y}=\frac{100\mu m}{\sqrt{N_{tracks}}}
\end{equation}
where $N_{tracks}$ is the number of tracks originating from the vertex. Since the nominal transverse bunch size
at LHCb will be $100\mu m$, the reconstruction of beam-gas vertices's, which will have a track multiplicity of $\sim10$,
will enable the measurement of the colliding bunch profiles and the beam overlap integral. This method is currently
under investigation by the LHCb collaboration and is expected to result in a luminosity measurement with an
associated uncertainty of 3-5\%. 

\subsubsection{Absolute pp luminosity from specialized detectors and from the total cross section measurement}
\label{sec:lumiabsol}

ATLAS and CMS are planning to perform absolute and relative pp luminosity measurements 
using dedicated luminosity instruments. 

Three particular luminosity instruments will operate around the ATLAS interaction point.\\
The absolute luminosity measurement will be provided by ALFA \cite{Pinfold:2005sq}
placed 240m down the beam line and due to operate in 2010. This measurement requires some 
special optics low luminosity running of the LHC and should be able to
measure the very low angle Coulomb scattering reaction. The expected precision is
of the order 3\%, depending on yet unknown LHC parameters during running.
The ALFA detector can also measure the absolute luminosity using the optical theorem
if the Coulomb region can not be reached. Extrapolating the elastic
cross section to very low momentum transfer $t=0$ and using the total cross section as measured by
TOTEM \cite{Anelli:2008zz} (located at the CMS interaction point) current simulations indicate that a 
precision of about 3\% might also be reached with this method.
%
%
In addition to absolute luminosity measurements from ALFA, LUCID and a Zero-Degree-Calorimeter (ZDC) 
\cite{Pinfold:2005sq} are sensitive to the relative single bunch crossings luminosity. 
LUCID and ZDC will however not give absolute measurements.

A similar approach is currently foreseen by the CMS collaboration \cite{VHalyo}. 

\subsubsection{Real time relative luminosity measurements}
\label{sec:lumirealtime}

A large number of instantaneous relative luminosity measurements have been discussed during the past years by ATLAS, CMS and LHCb
and more details can be found in the three presentations given during the ``standard candle'' session
of this workshop \cite{standardcandlesession}.
As an example we outline in the following some ideas discussed within CMS.
 
Multiple  techniques capable of providing suitable luminosity
information in real time have been identified in CMS.   One technique
employs signals from the  forward hadron calorimeter (HF) while
another, called the Pixel Luminosity Telescope (PLT), uses a set
of purpose-built particle tracking telescopes based on single-crystal
diamond pixel detectors.  At this writing, the PLT has not been
formally approved, but is under study.  The methods based on signals 
from the HF described are the ones being most vigorously pursued.  

Two methods for extracting a real-time relative instantaneous luminosity
with the HF have been studied. The first method is based on
``zero counting,'' in which the average fraction of empty towers is used to
infer the mean number of interactions per bunch crossing. The second method 
called ``EtSum method'' exploits the linear relationship between the average transverse energy per tower and the luminosity. 

Outputs of the QIE chips used to digitize the signals from the HF PMTs 
on a bunch-by-bunch basis are routed to a set of 36 HCAL Trigger and 
Readout (HTR) boards, each of which services 24 HF physical channels.
In order to derive a luminosity signal from the HTR, an additional 
mezzanine board called the HF luminosity transmitter (HLX) 
is mounted on each of the HTR boards.
The HLX collects channel occupancy and $E_T$ sum data to 
create eight  histograms: two sets of three occupancy histograms, one $E_T$-sum 
histogram, and one additional occupancy histogram.
These histograms comprise about 70~KB of data, which is transmitted at 
a rate of approximately 1.6~Mbps to a dedicated luminosity server
 via an Ethernet switch that  aggregates the data from multiple HLX boards
 for further processing. 

Although all HF channels can be read by the HLX, MC studies indicate 
that the best linearity is obtained using only the inner four $\eta$ rings. 
The algorithm has been optimized to minimize sensitivity to 
pedestal drifts, gain changes and other related effects.  
Both ``Zero Counting'' and the ``EtSum'' method have demonstrated linearity 
up to LHC design luminosity.  A statistical error of about $1 \%$ will be
achieved at $\mathrm{few times} \times 10^{31} \mathrm{cm}^{-2} \mathrm{s}^{-1}$ 
Hence the dominant error on the absolute luminosity will result 
from the normalization of the online relative luminosity.

\subsubsection{Proton-proton luminosity from the reaction $\mathbf{pp \rightarrow pp \mu\mu}$}
\label{sec:lumipplumi}

The QED process $pp\rightarrow pp\mu^{+}\mu^{-}$, where a $\mu^{+}\mu^{-}$ pair is produced via photon-photon
scattering, was first proposed for luminosity measurements at hadron colliders in~\cite{budnev}. At the LHC such pairs 
will be predominantly produced with small transverse momenta, at small polar angles and in the same forward or backward cone.

All three experiments are considering to use the well calculated $pp \to pp \mu
\mu$ process for measuring absolute luminosity. The theoretical understanding of this QED photon-photon scattering 
reactions is considered to be accurate to better than 1\%. Consequently this final state is thus often considered to be 
the perfect theoretical luminosity process. 
However, the experimental identification of this process 
requires to select muon pairs with low mass and within a well understood acceptance.  
The measurement of this reaction at a hadron collider appears to be much more 
difficult than the corresponding measurements of the reaction $ee \rightarrow ee \mu\mu$ at LEP.
The systematic measurement error for example in L3 and   
after several years of data taking was about $\pm 3$\%\cite{Achard:2004jj} 

Current simulations by the three LHC experiments indicate that the final state 
can be identified using straight forward criteria. For ATLAS and CMS one finds that 
about 1000 accepted events could at best be expected for 
an integrated luminosity of 1 fb$^{-1}$, resulting in a statistical error of about $\pm$ 3\%. 

%
For example the ATLAS study selects oppositely charged back-to-back
muon tracks with $p_T > 6\,\ggev$ and $|\eta| < 2.2$ with an invariant
mass less than 60 \ggev and a common vertex with no other tracks
originating from it (isolation), yields a cross section of 1.33 pb.
Thus, about 1300 events can be expected for running periods  
with a luminosity of 1 fb$^{-1}$ and yielding a potential statistical error of 3\%.
However, backgrounds not only from pile up events will be a critical issue.
Some proton tagging with high luminosity roman pots is currently investigated 
but this will certainly reduce the accepted cross section and introduce additional acceptance errors.
Similar conclusions have been reached by simulations performed within the CMS collaboration.
Consequently, both experiments expect that, during the coming years, this reaction will be mainly used as a cross check of the other methods.

The cross section for this process where both muons lie inside the LHCb acceptance and have a combined invariant mass greater 
than 2.5GeV is $\approx 88$ pb. The expected uncertainty is perhaps 1\% or smaller and comes mainly from rescattering corrections \cite{MartinPaper}, 
i.e. strong interactions between the interacting protons. 

The feasibility of using the elastic two photon process $pp\rightarrow p+\mu^{+}\mu^{-}+p$ to make
luminosity measurements at LHCb was first explored in \cite{Shamov:2002yi} and has recently been 
investigated in more detail by members of the LHCb collaboration \cite{Luminote}. A variety of 
background processes have been studied: dimuons produced via inelastic two-photon fusion and double pomeron 
exchange; low mass Drell-Yan pairs; QCD processes such as $b\bar{b}\rightarrow\mu^{+}\mu^{-}+X$; and the 
combinatoric backgrounds caused by K/$\pi$ mis-identification. A simple
offline selection has been developed that requires: the dimuon pair transverse momentum to be less than 
50MeV/c; the dimuon invariant mass to be in the range $2.5GeV/c^{2}<M_{\mu\mu}<20GeV/c^{2}$; and a charged
particle multiplicity of less than 3 (i.e. the event should contain a $\mu^{+}\mu^{-}$ pair and no other
charged particles). These criteria select $\sim27\%$ of the signal events that pass the trigger and are 
reconstructed and result in a background contamination that is $(4.1\pm0.5(stat.)\pm1.0(syst.))\%$ of the 
signal level with the dominant contribution due K/$\pi$ mis-identification. Overall it is expected that $\sim10^{4}$
$pp\rightarrow p+\mu^{+}\mu^{-}+p$ events will be triggered, reconstructed and selected at LHCb during one nominal
year of data taking ($2fb^{-1}$). Systematic uncertainties on a luminosity measurement at LHCb using this
channel are estimated to be $\sim1.31\%$ and are dominated by the uncertainty on the predicted
cross section for events containing dimuons produced via double pomeron exchange, an uncertainty
that is expected to be reduced in the near future. A measurement of the absolute luminosity at LHCb using this
channel and a dataset of $2fb^{-1}$ will therefore be possible with an associated uncertainty of $\sim1.5\%$.


In summary, the accurate measurement of this theoretically well understood reaction looks like an interesting challenge for 
the LHC experiments. Interesting results can be expected once integrated luminosities of 5 fb$^{-1}$ and more can be 
accumulated for ATLAS and CMS and about 1 fb$^{-1}$ for LHCb. 
Of course, it remains to be proven, if the systematic uncertainties under real data taking conditions 
can indeed be reduced to the interesting 1\% level.

\subsection{Indirect and relative pp luminosity measurements}
\label{sec:lumippdirindir}

The methods to measure the absolute proton proton luminosity 
and their limitations have been described in the previous chapter.

In this Section we will describe the possibilities to measure the luminosity indirectly using well 
defined processes, so called ``Standard Candles'' and their use to further constrain the PDFs 
and discuss the possibility to ``measure'' directly the parton-parton luminosities. 

Before describing the details of these indirect approaches, a qualitative comparison of luminosity measurements at  
$e^{+}e^{-}$ colliders and hadron colliders might be useful. The most important difference appears to be 
that in the $e^{+}e^{-}$ case one studies point like parton parton interactions. In contrast, 
at hadron hadron interactions one studies the collision of protons and other hadrons made of 
quarks and gluons. As a result, in one case the Bhabha elastic scattering reaction $e^{+}e^{-} \rightarrow e^{+}e^{-}$
at low $Q^{2}$ reaction can be calculated to high accuracy and the observed rate can be used as a luminosity normalization tool.
In contrast, the elastic proton proton scattering cross section can not be calculated at the LHC nor at any other hadron colliders.
As a consequence, absolute normalization procedures depend always on the 
measurement accuracy of the pp total cross section. Even though it is in principle possible to 
determine the pp total cross section in a luminosity independent way using special forward detectors  
like planned by the TOTEM or the ALFA experiments, the accuracy will be limited ultimately and after a few years 
of LHC operation to perhaps a few \%.  

Furthermore, as essentially all interesting high $Q^{2}$ LHC reactions are parton parton collisions,  
the majority of experimental results and their interpretation require the knowledge of parton distribution functions 
and thus the parton luminosities.
 
Following this reasoning, more than 10 years ago, the inclusive production of W and Z bosons with subsequent leptonic decays 
has been proposed as the ultimate precision parton parton luminosity monitor at the LHC \cite{Dittmar:1997md}. The following points summarize 
the arguments why W and Z production are indeed the ideal ``Standard Candles'' at the LHC. 
\begin{itemize}
\item The electroweak couplings of W and Z bosons to quarks and leptons are known from the LEP measurements to accuracies smaller than 1\%  
and the large cross section of leptonic decays W and Z bosons allows that these final states can be 
identified over a large rapidity range with large essentially background free samples. 
\item Systematic, efficiency corrected counting accuracies within the detector acceptance of 1\% or better might be envisioned 
during the early LHC running. In fact it is believed that the relative production rate of W and Z can be measured within  
the detector acceptance with accuracies well below 1\%. 
\item Theoretical calculations for the W and Z resonance production are the most advanced and accurately 
known LHC processes. Other potentially more interesting LHC reactions, like various diboson pair production final states 
are expected to have always larger, either statistical or systematic, experimental and theoretical uncertainties than the 
W and Z production.
\item The current PDF accuracies, using the latest results from HERA and other experiments demonstrate that the 
knowledge of the quark and anti quark accuracies are already allowing to predict the W and Z cross 
at 14 TeV center of mass energies to perhaps 5\% or better. 
The measurable rapidity and $p_{t}$ distributions of the Z boson and the corresponding ones for the charged leptons from 
W decays can be used to improve the corresponding parton luminosity functions.
\end{itemize}

Obviously, the use of W and Z bosons as a luminosity tool requires that the absolute cross section 
becomes an input, thus it can not be measured anymore. As a result this method has been 
criticized as being ``a quick hack at best". In contrast,       
advocates of this method point out that this would not be a noticeable loss for the LHC physics program.
  
\subsubsection{Using the reaction $pp \rightarrow Z \rightarrow \ell^{+} \ell^{-}$ to measure $L_{pp}$}  
\label{sec:lumiusing}

Very similar and straight forward selection criteria for the identification of leptonic $Z$ decays, depending 
somewhat on the detector details and the acceptance region, are applied by ATLAS, CMS and LHCb.
In the following the current selection strategy in ATLAS and LHCb are described.

\subsubsection{Measuring Z and W production, experimental approaches in ATLAS}
\label{sec:lumimeasatlast}

The ATLAS W and Z cross section measurements are based on the
following selections in the electron and muon channels:\\
\begin{itemize}
\item A typical selection of $W \to e \nu$ requires that events with ``good" 
electrons have to fulfill the additional kinematic acceptance criteria: \\  
$p_T > 25\,\ggev$, $|\eta| < 1.37$ or $1.52 < |\eta| < 2.4$.\\
The criteria for $W \to \mu \nu$ muons are similar where $p_T > 25\,\ggev$ and $|\eta| < 2.5$. is required.
Furthermore, in order to classify the event as a $W$ event, the reconstructed missing transverse momentum and the transverse mass
should fulfill $E_{T}(miss) > 25 GeV$ and $m_T(W) > 40\,\ggev$.
\item The selection of $Z \to ee$ and $Z \to \mu \mu$ requires that 
a pair of oppositely charged electrons or muons is found. Due to lower background 
the electrons should have $p_T > 15\,\ggev$ and $|\eta| < 2.4$ and their invariant mass 
should be between 80-100 GeV. \\
Similar criteria are applied for the muons with $p_T > 15\,\ggev$ and $|\eta| < 2.5$.
The reconstructed mass should be between 71-111 GeV.
\end{itemize}

Following this selection and some standard Monte Carlo simulations, 
the expected number of reconstructed events per $\mbox{10 pb}^{-1}$ at $\sqrt{s} = 14$ TeV are about 
45000, 5500 for W and Z decays to electrons and 60000, and 5000 for the decays to muons, respectively. 
Thus, even with a small data sample of only 10 $\mbox{pb}^{-1}$, the statistical
uncertainty for the $Z$ counting comes close to 1\% in each channel. 

Systematic uncertainties from the experimental selection are dominated by 
the Z efficiency determination and from backgrounds in the W selection.
Other sources of uncertainties originate from the knowledge of energy scale and the resolution.
The lepton efficiencies are evaluated by considering $Z \to \ell \ell$
events and using the so called ``tag and probe'' method, like for example described by the D0 experiment \cite{WZratioa,WZratiob}. The
efficiency uncertainty associated with the precision of this method has been estimated for 
a data sample of 50 $\mbox{pb}^{-1}$ (1 $\mbox{fb}^{-1}$) of data
to be 2\% (0.4\%) for W and 3\% (0.7\%) for Z events.
The backgrounds for W events are of the order 4\% in the electron
channel and 7\% in the muon channel. The main contributions are from
other W or Z decays, and are thus well understood, leading to
background uncertainties of the order 4\% for both channels if a sample 50
$\mbox{pb}^{-1}$ is analyzed. For much larger samples it is expected that uncertainties at or below 1\% can be achieved.
The backgrounds for the Z decays are very small, and can be determined accurately from
mass spectrum, and hence does not carry any sizable uncertainty.
It has been demonstrated, that the detector scales and resolutions can
be determined very accurately \cite{Besson:2008zs}, and the associated
uncertainties are therefore also close to negligible.\\
Some detailed studies demonstrate that eventually the systematic error between 1-2\% or even smaller might be achieved for the 
W and Z counting and within the detector acceptance up to rapidities 
of about 2.5.



In order to use this number for the pp luminosity determination 
the total inclusive W and Z cross-section at NNLO can be used. 
These have been calculated to be 20510 pb and 2015pb, respectively \cite{Anastasiou:2003ds}. 
Variations in models, floating parameters, and other theoretical uncertainties lead to
significant variations in the estimates. The uncertainties on these
calculation are estimated to be 5\% or smaller. This uncertainty appears to be currently dominated 
by the PDF uncertainties needed to extrapolate to the experimentally uncovered large rapidity region.  
More discussions about these uncertainties can be found for example at \cite{phy11} and \cite{phy12}.

It can be assumed that the detailed studies of the rapidity distributions within the acceptance region 
with W and Z decays might eventually lead to further error reductions.

%

\subsubsection{Measuring Z production, experimental approach in LHCb}
\label{sec:lumimeaslhcb}

The uncertainty on the predicted Z production cross section at the LHC comes from two sources: the uncertainty
on the NNLO partonic cross section prediction \cite{Anastasiou:2003ds}, which contributes an uncertainty of $<1\%$, and 
uncertainties in our understanding of the proton Parton Distribution Functions (PDFs) which, for the latest 
MSTW fit~\cite{Thorne:2007bt}, contribute an uncertainty of $\sim3\%$ for Z bosons produced with rapidities in the range $-5<y<5$. 

A measurement of the Z production rate at LHCb via the channel $Z\rightarrow\mu^{+}\mu^{-}$, which
provides a final state that is both clean and fully reconstructible, can be achieved with high 
efficiency and little background contamination. In addition, since the dimuon trigger stream at LHCb~\cite{TriggerTDR} 
requires two muons with an invariant mass larger than 2.5GeV and a summed transverse momentum 
($P^{1}_{T}+P^{2}_{T}$) greater than 1.4GeV, a high trigger efficiency of $\sim95\%$ is expected for 
these events. A variety of background sources for this channel have been investigated: other electroweak processes 
such as $Z\rightarrow\tau^{+}\tau^{-}$ where both taus decay to muons and neutrinos; QCD processes 
such as $b\bar{b}\rightarrow\mu^{+}\mu^{-}+X$; and events where two hadrons with an invariant 
mass near the Z mass are both mis-identified as muons. To deal with these backgrounds an off-line
selection has been developed \cite{LuminoteZ} that requires: the dimuon invariant mass to be within 20 GeV of the
Z mass; the higher and lower transverse momentum muons to be greater than 20 GeV and 15 GeV respectively;
the impact parameter of both muons is consistent with the primary vertex; and both muons have associated 
hadronic energy that is less than 50 GeV. For $Z\rightarrow\mu^{+}\mu^{-}$ events that are triggered and 
reconstructed at LHCb, these off-line selection criteria will select $91\pm1\%$ of the signal events while 
reducing the background to $(3.0\pm2.9)\%$ of the signal level with the dominant contribution due to the
combinatoric backgrounds from pion and kaon mis-identification. It is expected that these backgrounds can be
well understood from real data or removed using muon isolation criteria. Overall it is expected that 
$Z\rightarrow\mu^{+}\mu^{-}$ events will be triggered, reconstructed and selected at LHCb at a rate of 
$\sim190evts/pb^{-1}$. Systematic uncertainties have also been investigated and it is expected that with 
as little as $5pb^{-1}$ of data the experimental efficiency (trigger, tracking, muon identification etc.) can be 
measured with an uncertainty of $\sim1.5\%$ enabling a luminosity measurement with an uncertainty of $\sim3.5\%$.

\subsubsection{PDF and relative parton-parton luminosity measurements}
\label{sec:lumipdfrel}

Theoretically well understood reactions at the LHC offer the possibility
to use their rapidity distributions to improve todays knowledge of PDFs.
Especially the resonance production of W and Z bosons with leptonic decays
with low and high transverse momentum and the production of isolated high $p{_t}$ $\gamma$-Jet events 
have been demonstrated to be very sensitive to the relative parton distribution functions. 
Simulations from ATLAS and CMS have shown that experimental errors on these rapidity regions 
up to $|y|$ of about 2.5 can probably performed with accuracies eventually reaching perhaps 1\% or better. 
The possibility to cross-check the measurements with W and Z decays to (a) electron(s) and (b) muon(s)  
and between both experiments will of course help to reach the accuracy.

During the past years simulation studies from the LHCb collaboration have shown that 
the experiment has a unique potential to extend the acceptance region from ATLAS and CMS 
for muons up to rapidity values at least up to 4.5. Furthermore, the existing overlap region 
for y between 1.9 and 2.5 should allow to reduce normalisation uncertainties.
Obviously, these rapidity values are understood as being reasonably accurate but qualitative values and more precise  
values will be defined once real data will allow to define a well understood fiducial volume of the detectors. 

In addition, the LHCb collaboration has investigated the possibility to identify 
clean samples of very low mass Drell-Yan mu-pair events.    
The results indicate that such pairs can be measured within their acceptance region 
down to masses of 5 GeV. Such a measurement would in principle allow to 
measure PDFs for $x$ values approaching extremely low values of $10^{-6}$ for the first time \cite{McNulty}.  

It should be clear that such measurements, which are known to be very sensitive to quark, antiquark and gluon 
relative parton luminosities will not allow an absolute PDF normalisation.
Such an improvement of absolute PDF normalisation would require the accurate knowledge of the proton-proton luminosity 
to better than todays perhaps $\pm$ 3\% PDF accuracy obtained from the HERA measurements over a large x range and obviously lower $Q^{2}$.   
The alternative approach to combine the relative parton luminosities over the larger $x, Q^{2}$ range using the sum rules 
has, to our knowledge, so far not been studied in sufficient detail.     

A more detailed analysis of the different experimental approaches to improve the PDFs are interesting but are beyond the scope of this note
about the luminosity. Nevertheless we hope that the experimentalists of the three collaboration will start to combine their efforts 
and will pursue the PDF measurements, in direct collaboration with theorists, during the coming years.   

\subsection{Comparing the different pp luminosity measurements} 
\label{sec:lumicomparing}

A relatively large number of pp luminosity measurements has been proposed and the most relevant have been discussed in this 
note. Here we try to give a critical overview of the different methods and their potential problems.
Despite these advantages and disadvantage it should be clear that it is important to perform as many as possible independent 
luminosity methods during the coming years.

\begin{itemize}
\item {\bf The machine luminosity determination using beam parameters:} \\
This method will be pursued independently of the experiments and its main purpose will be to 
optimize the performance of the LHC and thus providing a maximum number of physics collisions for the experiments.
The potential to use this number as an almost instantaneous absolute luminosity number
with uncertainties of perhaps $\pm$ 10\% (and eventually $\pm$ 5\%), assuming that non gaussian 
tails of the beam can be controlled to this accuracy will certainly be useful to the experiments.
Of course the experiments would lose somewhat their ``independence" and 
still need to combine this number with their actual active running time.

However, one should remember that the Tevatron experiments did not use this method for their measurements.

The method to determine the beam size using the LHCb precision vertex detector look  
very promising and it is hoped that their approach might result in a pp luminosity measurement with an
associated uncertainty of 3-5\%.
 
\item {\bf Total cross section and absolute luminosity normalisation with specialized far forward Detectors:} \\
The luminosity independent total pp cross section measurement is planned by the TOTEM collaboration and by the ALFA detector. Using these 
numbers both ATLAS and CMS plan to obtain the pp luminosity from the counting of the pp elastic scattering counting numbers from the forward detectors
which thus depend on the knowledge of the total cross section measurement. In order to obtain this number some few weeks of special 
optics and low luminosity LHC running are required. As all LHC experiments are very keen to obtain as quickly as possible some
reasonable luminosity at 14 TeV center of mass energy it is not likely that those special LHC data taking will happen 
during the first year(s) of data taking. Furthermore, despite the hope that the total cross section can be determined 
in principle with an interesting accuracy of $\pm$ 1\%, it remains to be demonstrated with real LHC running. 
In this respect it is worth remembering that the two independent measurements of the total cross section at the Tevatron 
differed by 12\% while much smaller errors were obtained by the individual experiments. As a result the 
average value with an error of $\pm 6$\%  was used for the luminosity normalisation.  
\item {\bf Luminosity determination using $Z \rightarrow \ell \ell$}: \\
This method provides an accurate large statistic relative luminosity number. It will be as accurate as the theoretical cross section calculation, 
which is based on the absolute knowledge of the PDFs from other experiments, from unknown higher order corrections 
and their incomplete Monte Carlo implementation. Todays uncertainties are estimated to be about 5\%. 
It has been estimated, assuming the experiments perform as expected, that   
the potential Z counting accuracy within the acceptance region
including efficiency corrections might quickly reach  
$\pm 1$\%. The extrapolation to the uncovered rapidity space, mainly due to the worse knowledge of the PDFs in this region, 
increases the error to perhaps 3\%. Taking other theoretical uncertainties into account an error of $\pm$5\% is currently estimated.
Of course, advocates of the Z normalisation method like to point out that the real power of this method starts once relative measurements, 
covering similar partons and similar ranges of the parton distribution functions 
will be performed with statistical errors below 5\%. 
Examples where such a normalization procedure looks especially interesting 
are the relative cross section measurements of $N(Z)/N(W)$, $N(W^{+})/N(W^{-})$, high mass Drell-Yan events with respect to Z events and 
diboson final states decaying to leptons. 
Of course, correlations and anticorrelations between quark and gluon dominated production rates exist and need to be carefully investigated 
before similar advantages for the gluon PDFs can eventually be exploited. The loss of an independent Z cross section measurement would of course be 
a fact of life.  
\item {\bf pp luminosity from the reaction $\mathbf{pp \rightarrow pp \mu\mu}$:} \\
A measurement of this reaction offers in principle a direct and theoretically accurate proton proton luminosity value. 
Unfortunately current simulations from the experiments indicate that 
the accepted cross section is relatively small and only a few 1000 events can be expected per fb$^{-1}$.
The different simulation results indicate that the backgrounds can be suppressed sufficiently 
without increasing the experimental systematics too much.  The current simulation results indicate that
small systematic errors of perhaps 1-2\% might eventually be achievable\footnote{It might be interesting to study the experience 
from similar measurements at the experimentally ideal conditions of LEP, where uncertainties above $\pm$ 3\% have been reported\cite{Achard:2004jj}.}
once a yearly luminosity of 5-10 fb$^{-1}$   
in ATLAS and CMS (2 fb$^{-1}$ for LHCb) might be recorded. 
It remains to be seen if muons with transverse momenta 
well below 20 GeV can indeed be measured as accurately as muons with transverse momenta above 25 GeV.
\end{itemize}

\subsubsection{Which luminosity accuracy might be achievable and when}
\label{sec:lumiwhich}

Of course the potential time dependent accuracy of the different luminosity methods can only be guessed 
today as such numbers depend obviously on the LHC machine performance during the coming years.
For the purpose of this Section we are mainly interested in measurements at the 14 TeV center of mass 
energy and assume that the following ``data samples" would define such ``years". 
Of course, it could be hoped that the luminosity and energy increase would go much faster 
resulting in ``some" shorter LHC years. Thus we assume that the first 14 TeV year, currently expected to be 2010, will correspond to
0.1 fb$^{-1}$, followed by a 1 fb$^{-1}$ year. During the third and fourth year 
ATLAS and CMS expect to collect about 5 fb$^{-1}$ and 10 fb$^{-1}$ while LHCb expects to collect roughly 2 fb$^{-1}$ per year. 
We assume further that the special optics low luminosity 
data taking periods requiring perhaps a few weeks for TOTEM and similar for ALFA will take 
only place during the year when more than 1 fb$^{-1}$ per year or more can be expected. 

As a result, for the first two 14 TeV running years, realistic luminosity numbers 
could come from (1) the machine group and (2) from the indirect method using the inclusive 
production of Z events with leptonic decays.   

As has been pointed out in Section~\ref{sec:lumiparms} the method (1) would,  
without any additional efforts by the machine group, allow a first estimate 
with a $\pm$ 20-30\% luminosity accuracy. We assume however that, due to the delay of the real 14 TeV start
to 2010,  enough resources could be found that people within the machine group could carefully prepare for the necessary beam parameter measurements 
and that the experiments will do the corresponding efforts to correct such a machine luminosity number
for real detector data taking one could hope for a 10\% measurement for 2010 and a 5\% accuracy for 2011.

In contrast, method (2) would by definition be an integrated part of any imaginable experimental 
LHC data taking period. In fact, if enough attention is put into the $Z$ counting method, the data expected during 2010 running 
might already reach statistical errors of $\pm$ 2\% per 5 pb$^{-1}$ periods. 
Thus perhaps about 10-20 such periods could be defined during the entire year and 
systematic errors for the lepton efficiency correction within the detector acceptance could reach 
similar $\pm$ 2-3\% accuracies. During the following years these errors might decrease further to 1\% or better.
Once the rate of any ``stable'' simple high rate final states and even trigger rates relative to 
the $Z$ counting rate has been determined, such relative event rates can be used subsequently to track the ``run'' luminosity and even the real time
luminosity with similar accuracy.   

Theoretical limitations of the cross section knowledge, not expected to improve without LHC data taking, 
would limit the accuracy to about $\pm$ 5\%. The expected detailed analysis of the 
2010 rapidity distributions of W, Z and $\gamma$-jet events will allow some improvements
for the years 2011 and beyond. We can thus expect that appropriate ratio measurements 
like the cross section ratio measurements of $Z/W^{\pm}$ and $W^{-}/W^{+}$ will already  
reach systematic accuracies of $\pm$ 1-2\% during 2010 and 1\% or better in the following years.
Measurement of b physics, either in LHCb or in ATLAS and CMS  might in any case 
prefer to perform luminosity independent measurements and
relate any of the ``new" measurements to some relatively well known and measurable 
B-hadron decays. 

It is also worth pointing out that currently no other high $Q^{2}$ reaction 
has been envisioned, which might be measurable to a systematic precision of better than 5-10\% 
and a luminosity of up to 1fb$^{-1}$. In addition, most of the interesting high $Q^{2}$ electroweak final
states will unfortunately even be limited for the first few LHC years to statistical accuracies to 5\% or more. 

The prospect for the other luminosity measurements start to become 
at earliest interesting only once a few 100 pb$^{-1}$ can be recorded.
Consequently one can expect to obtain a statistical interesting accuracy from the reaction
$pp \rightarrow pp \mu\mu$ after 2010. Similar, it looks unlikely that 
low luminosity special optics run will be performed before 2011.
Consequently one might hope that few \% accurate total cross section numbers become available 
before the 2012 data taking period will start. 

\subsection{Summary and Outlook}
\label{sec:lumisumm}

A large variety of potentially interesting pp luminosity measurements, 
proposed during the past 10-15 years, are presented in this Section. 

Realistically only the machine luminosity measurement and the counting of the Z production 
might reach interesting accuracies of 5\% before 2011. For all practical purposes 
it looks that both methods should be prepared in great detail before the data taking 
at 14 TeV collision energies will start in 2010. 

We believe that a working group, consisting of interested members of the three pp collider 
experiments and interested theorists, should be formed to prepare the necessary 
Monte Carlo tools to make the best possible use of the soon expected W and Z data, 
not only for the pp luminosity normalization but even more for the detailed investigations  
of the parton parton luminosity determination and their use to predict other 
event rates for diboson production processes and high mass Drell-Yan events.

\pagebreak
\section{OUTLOOK: THE PDF4LHC INITIATIVE\protect\footnote{Contributing author: A.~de~Roeck}}
\label{sec:pdflhc}

This document demonstrates the vast amount of 
progress that has taken place
in the last years on pinning down the PDFs of the proton, as well as
the dramatic increase in 
awareness of the impact of PDFs on  the physics program
of  LHC experiments. The HERALHC workshop 
has acted as a regular forum for working meetings between the experiments,
PDF phenomenologists and theorists. In the course of this workshop, 
it was realized 
that the momentum on the PDF studies should be kept and perhaps even 
focused more on the LHC,  in order to continue the discussions, 
investigations and
further work towards improving our knowledge on the PDFs.

Clearly, LHC will need the best PDFs,
 especially for precision measurements, setting of limits in searches, 
and even for discoveries.
Ideally the ATLAS and CMS (and LHCb and ALICE) analyses should follow a common
procedure for using PDFs and their uncertainties in their key analyses. 
Such a common procedure, across the experiments, is being used in 
other contexts, such as significance estimates in searches. 
Also, changing frequently the PDFs in the software of the experiments, 
e.g. for cross--checks or the determination  of
error bands, is often non-trivial (e.g. due to the inter-connection 
with parameter choices for underlying event modeling, showering
parameters and so on) and sometimes impractical if CPU intensive 
detector simulations are involved. 
LHC studies therefore will need both good central values for the PDFs 
to start with, and a
good estimate of the associated uncertainties. 

This has triggered the so called PDF4LHC initiative. PDF4LHC offers a 
discussion forum for PDF studies and information exchange between all 
stake-holders in the field: the PDF global fitter groups, such as CTEQ and 
MSTW; the current experiments, such as the HERA and Tevatron ones; QCD
theorists and the LHC experimental community. The PDF4LHC initiative  
started in 2008. More details and links to the meetings so far
can be found on the PDF4LHC web site~\cite{pdf4lhcweb}.

The mission statement of PDF4LHC is: 
\begin{itemize}
\item Getting the best PDFs, including the PDF uncertainties, based on the 
present data.
\item Devise strategies to use future LHC data to improve the PDFs.
\end{itemize}
All this needs a close collaboration between
theorists and those that are preparing to make the measurements.
In order to reach the first goal, the PDF4LHC forum aims to
stimulate discussions and trigger further comparison exercises across the PDF 
community, in order to select one or a limited number of possible 
strategies that can be adapted to determine and use PDFs.
For the second goal, PDF4LHC should also be a forum for discussions on how to  
include measurements from the LHC to constrain PDFs: what should be
measured at LHC, and correspondingly calculated in theory. 
Such measurements include $W$ and $Z$ production and asymmetries, di-jet 
production, hard prompt photons, Drell-Yan production, 
bottom  and top quark production, Z-shape fits and Z+jets measurements.
One expects that some of these channels can 
already be studied  with first data, hence we need to prepare for that well in 
advance. 

The following issues are part of the program for in depth discussions via 
topical workshops, some of which took place already in 
2008~\cite{pdf4lhcweb}.
\begin{itemize}
\item
Data to be included in the PDFs. Would we get better results with 
a selection of data to be used? New data will become available such as 
$F_L(x,Q^2)$, and combined data from H1/ZEUS. Can we 
 extract more from the data?
\item
Determination of PDF uncertainties, 
including the statistical treatment of the data.
\item
Theoretical uncertainties and regions/processes where they matter: 
higher--order corrections; heavy flavour  treatment; low-$x$ (and high-$x$) 
resummation; 
other PDFs like  unintegrated PDFs (and GPDs). 
\item
PDFs for usage Monte Carlo generators.
\end{itemize}

One can expect that the LHC experiments most likely  
will be using for most of their studies the PDF
sets and errors that are delivered by either one of the CTEQ or MSTW family.
Hence it is important that the lessons learned from exercises on studies of
the systematics on PDFs will be adapted by these main global PDF providers.
PDF4LHC aims to advice the experiments in the use for PDFs for the LHC, 
based on the discussions, results and future consensus at the forum.
The experience and results from 
HERAPDFs, and PDFs from other groups, like the Neural Net or Alekhin ones are 
extremely valuable in this discussion 
and will serve as crucial input in studies to demonstrate
how well we actually know the  
parton distributions. Several important benchmark exercises have been already
performed and are reported in section 3 of this report.

A special case are the PDFs for Monte Carlo generators. For experiments it 
is important that generated events be kinematically distributed
close to the distribution of the real data, such that the simulated and 
reconstructed Monte Carlo events can be used in a straightforward way 
to calculate efficiencies 
for e.g. experimental cuts in an analysis. 
In case the initially generated distribution 
does not resemble the data close enough, the Monte Carlo samples need to be
reweighted, with all its possible drawbacks. 
Since calculations based on LO Matrix Elements 
and LO PDFs are known not to describe the data well, 
and NLO Matrix Element based
generators to date have so far only a restricted number of processes 
implemented, 
studies are ongoing on so called ``improved LO'' PDFs, which try to cure some
of the LO PDF drawbacks. Examples are given in~\cite{Sherstnev:2008dm}. 
This is yet another part
of the discussions in the PDF4LHC forum

In short, it is crucial that the work started here continues, with
discussions 
and studies on PDFs and their uncertainties, the impact of the upcoming
data on 
future PDF determinations and more, all with special focus on the needs for 
the LHC. The PDF4LHC initiative will offer a framework to do all this.
\bigskip
\bigskip
\bigskip
\bigskip

\begin{center}
{\bf ACKNOWLEDGEMENTS}
\end{center} 
This work was supported in part by the following grants and agencies:
the European network HEPTOOLS under contract
MRTN-CT-2006-035505; ANR-05-JCJC-0046-01 (France)
PRIN-2006 (Italy); 
MEC FIS2004-05639-C02-01;
(Spain) and the Scottish Universities Physics Alliance (UK).


\bibliographystyle{heralhc}

{\raggedright

\bibliography{heralhc}

\begin{mcbibliography}{100}
\bibitem{LuminoteZ}
\protect{More details about the LHCb approach using $\sigma_{Z}\cdot
  Br(Z\rightarrow\mu^{+}\mu^{-})$, are given in the talk by J.~Anderson at the
  HERA-LHC Workshop ``standard candles''.}\relax
\relax
\end{mcbibliography}

}

 \end{document}